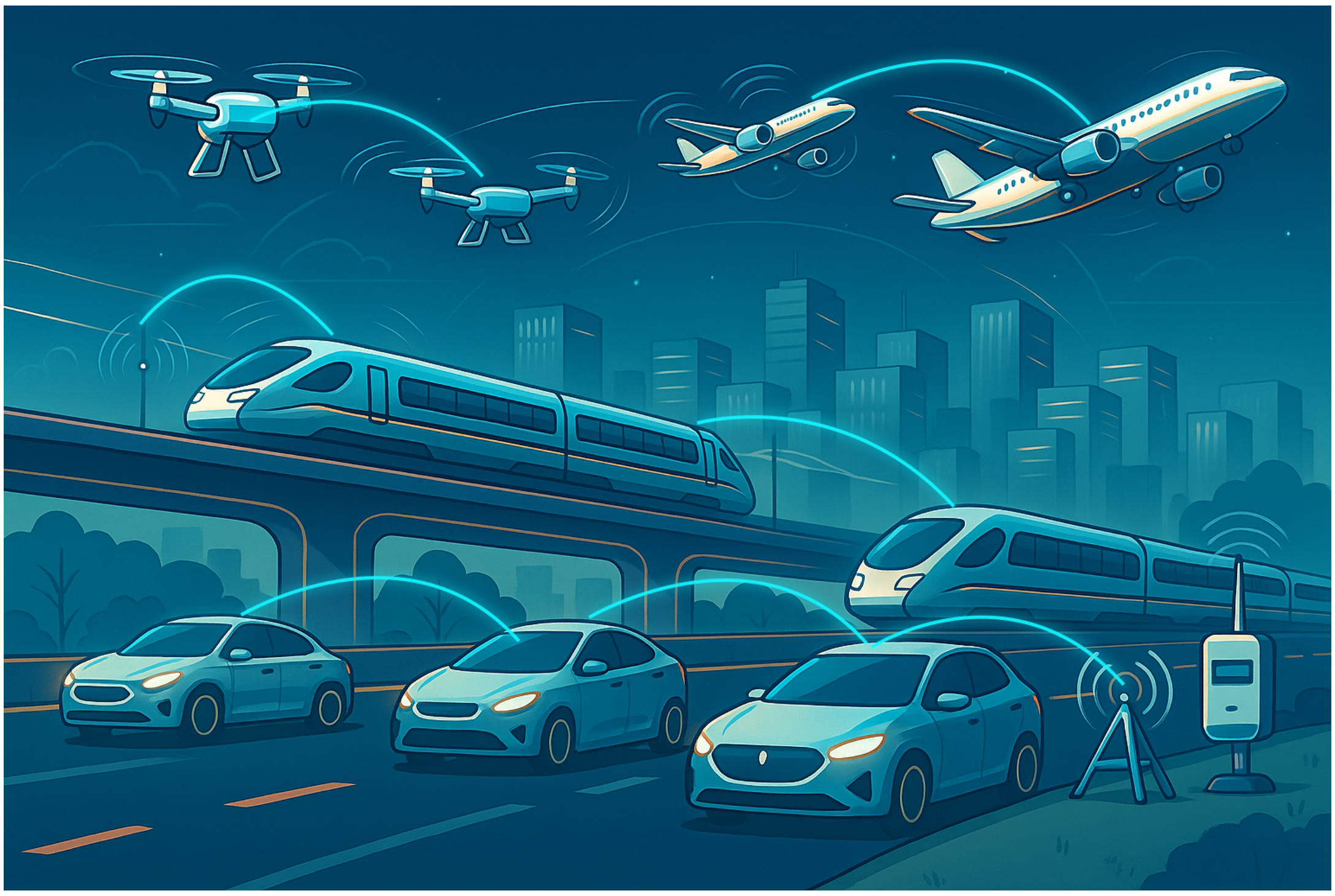

# Communication Technologies for Intelligent Transportation Systems:

## From Railways to UAVs and Beyond

*VT2 White Paper*

Edited by: Shrief Rizkalla and Adrian Kliks

Funded by the European Union

# Contributing Authors

(in alphabetical order)

- **Nila Bagheri** – sec. 7.2, 7.3

  Instituto de Telecomunicações – DEM
  Universidade da Beira Interior
  Covilhã, Portugal
  https://orcid.org/0000-0002-0893-6482

- **Miguel A. Bellido-Manganell** – sec. 4

  German Aerospace Center (DLR)
  Wessling, Germany
  https://orcid.org/0000-0002-0668-3477

- **Aniruddha Chandra** – sec. 6

  National Institute of Technology Durgapur
  Durgapur, WB, India
  https://orcid.org/0000-0002-9247-9009

- **Anja Dakić** – sec. 5.4

  AIT Austrian Institute of Technology
  Vienna, Austria
  https://orcid.org/0000-0003-2088-8935

- **Laura Finarelli** – sec. 3.4, 7.7

  Technische Universität Berlin
  Berlin, Germany
  HES-SO Valais-Wallis
  Sion, Switzerland
  https://orcid.org/0009-0000-8349-420X

- **Davy Gaillot** – sec. 7.1

  Université de Lille
  Villeneuve d'Ascq, France
  https://orcid.org/0000-0003-3455-5824

- **Matti Hämäläinen** – sec. 3.3

  University of Oulu
  Oulu, Finland
  https://orcid.org/0000-0002-6115-5255

- **Ruisi He** – sec. 2, 5.1

  Beijing Jiaotong University
  Beijing, China
  https://orcid.org/0000-0003-4135-3227

- **Markus Hofer** – sec. 2

  AIT Austrian Institute of Technology
  Vienna, Austria
  https://orcid.org/0000-0002-1915-9869

- **Sandaruwan Jayaweera** – sec. 3.3

  University of Oulu
  Oulu, Finland
  https://orcid.org/0009-0002-0066-0263

- **Adrian Kliks** – sec. 1.2, 1.3, 8

  Poznan University of Technology
  Poznań, Poland
  https://orcid.org/0000-0001-6766-7836

- **Francesco Linsalata** – sec. 3.5, 7.4, 7.6

  Politecnico di Milano
  Milan, Italy
  https://orcid.org/0000-0002-6725-3606

- **Konstantin Mikhaylov** – sec. 3.3

  University of Oulu
  Oulu, Finland
  https://orcid.org/0000-0002-5518-6629

- **Jon M. Peha** – sec. 7.2, 7.3

  Carnegie Mellon University
  Pittsburgh, PA, USA
  https://orcid.org/0000-0003-4915-306X

- **Ibrahim Rashdan** – sec. 5.2

  German Aerospace Center (DLR)
  Oberpfaffenhofen, Germany
  https://orcid.org/0000-0001-9208-8078

- **Shrief Rizkalla** – sec. 1.1, 8

  Silicon Austria Labs (SAL)
  Linz, Austria
  https://orcid.org/0000-0003-1037-1844

- **Gianluca Rizzo** – sec. 3.4, 5.3

  HES-SO Valais, Switzerland
  Università degli Studi di Torino
  Turin, Italy
  https://orcid.org/0000-0001-7129-4972

- **Abdul Saboor** – sec. 3.1, 3.2, 3.6

  KU Leuven
  Leuven, Belgium
  https://orcid.org/0000-0002-6512-1562

- **Martin Schmidhammer** – sec. 7.5

  German Aerospace Center (DLR)
  Wessling, Germany
  https://orcid.org/0000-0002-9345-142X

- **Michał Sybis** – sec. 7.2

  Poznan University of Technology
  Poznań, Poland
  https://orcid.org/0000-0002-7149-2386

- **Fredrik Tufvesson** – sec. 7.1

  Lund University
  Lund, Sweden
  https://orcid.org/0000-0003-1072-0784

- **Paul Unterhuber** – sec. 2

  German Aerospace Center (DLR)
  Wessling, Germany
  https://orcid.org/0000-0002-0564-9066

- **Fernando J. Velez** – sec. 7.2, 7.3

  Instituto de Telecomunicações – DEM
  Universidade da Beira Interior
  Covilhã, Portugal
  https://orcid.org/0000-0001-9680-123X

- **Evgenii Vinogradov** – sec. 3.1, 3.2, 3.6

  Universitat Politècnica de Catalunya
  Barcelona, Spain
  https://orcid.org/0000-0002-4156-0317

- **Michael Walter** – sec. 4

  German Aerospace Center (DLR)
  Wessling, Germany
  https://orcid.org/0000-0001-5659-8716

- **Thomas Zemen** – sec. 5.4

  AIT Austrian Institute of Technology
  Vienna, Austria
  https://orcid.org/0000-0002-9392-9155

- **Haibin Zhang** – sec. 7.5

  Netherlands Organisation for Applied
  Scientific Research
  The Hague, The Netherlands
  https://orcid.org/0000-0002-5854-1841

- **Zhengyu Zhang** – sec. 2

  Beijing Jiaotong University
  Beijing, China
  https://orcid.org/0000-0002-4888-4023

# Contributors' Affiliations

SAL SILICON AUSTRIA LABS

POZNAN UNIVERSITY OF TECHNOLOGY

DLR Deutsches Zentrum für Luft- und Raumfahrt German Aerospace Center

UNIVERSITY OF OULU

Hes·so VALAIS WALLIS School of Management

KU LEUVEN

Carnegie Mellon University

北京交通大學 BEIJING JIAOTONG UNIVERSITY

it instituto de telecomunicações

UNIVERSIDADE BEIRA INTERIOR

AIT AUSTRIAN INSTITUTE OF TECHNOLOGY TOMORROW TODAY

POLITECNICO MILANO 1863

TNO

UPC UNIVERSITAT POLITÈCNICA DE CATALUNYA BARCELONATECH

Université de Lille

UNIVERSITÀ DI TORINO

NATIONAL INSTITUTE OF TECHNOLOGY DURGAPUR

LUNDS UNIVERSITET

BEIJING INSTITUTE OF TECHNOLOGY

Technische Universität Berlin

# Contents

# Chapter 1: Introduction

## 1.1 Background on Intelligent Transportation System

By **Shrief Rizkalla**

THE International Organization for Standardization defines Intelligent Transportation System (ITS) as systems in which information and communication technologies are applied in the field of road transport, including infrastructure, vehicles, and users, and in traffic management and mobility management [1]. More broadly, the World Road Association characterizes ITS as control and information systems that use integrated communications and data processing technologies to meet transport policy goals and objectives. While these definitions historically focused on road transport, the concept has expanded considerably. Today's ITS encompasses rail systems, aviation, and emerging aerial mobility platforms, forming an integrated multi-modal ecosystem that brings together information and communication technologies with transportation infrastructure to enable real-time data exchange, automated decision-making, and cooperative operation across different transport modes.

At their core, ITS rely on reliable, low-latency communication links between various platforms, infrastructure, and control systems. This includes Vehicle-to-Vehicle (V2V) and Vehicle-to-Infrastructure (V2I) communication for road transport, Aircraft-to-Aircraft (A2A) links for aviation, Train-to-Train (T2T) and Train-to-Ground (T2G) communication in railways, and Unmanned Aerial Vehicle (UAV) connectivity for drone operations. When combined with sensors, artificial intelligence, and edge computing, these communication systems transform isolated transport networks into cooperative, intelligent systems capable of autonomous operation.

Railways have actually been pioneers in communication-based operations. Communication based Train Control (CBTC) systems have controlled urban metros since the 1980s. The broader European Rail Traffic Management System (ERTMS) framework, now operates on over 60.000 km of track in Europe and is being adopted worldwide. T2G communication via Global System for Mobile Communications – Railway (GSM-R) is used for moving block operations, while T2T communication is utilized for train integrity monitoring. The railway sector is now transitioning to Future Railway Mobile Communication System (FRMCS), which leverages Fifth Generation (5G) technology. Deployment begins in 2025, with full migration from GSM-R expected by 2040.

Several large-scale deployments demonstrate how ITS concepts translate into practice. In Europe, the C-Roads Platform has coordinated ITS deployment across member states since 2016 [2]. The initiative has equipped 20.000 km of roads with short-range ITS – 5 GHz (ITS-G5) communication, where ITS-G5 is the European standard for vehicular communications. These systems provide hazardous location warnings, road works notifications, and emergency vehicle alerts across borders [2].

The drone sector has seen particularly rapid ITS development. Zipline operates what is now the world's largest autonomous delivery system, having completed over one million commercial deliveries in the past two years [3]. The company delivers medical supplies in Rwanda and Ghana, and has expanded to retail deliveries for Walmart in the Dallas-Fort Worth area. In emergency medical services, Swedish trials demonstrated that drones delivering automated external defibrillators arrived before ambulances in 57% of cardiac arrest cases [4]. Abu Dhabi has implemented a drone-based medical supply network with 40 stations operating around the clock, creating the Middle East's first such system [5].

Looking ahead, several emerging technologies promise to address current limitations. Cell-free Multiple-Input Multiple-Output (MIMO) eliminates traditional cell boundaries by distributing many access points that jointly serve all users [6]. This provides consistent quality regardless of location, which is particularly valuable for high-speed rail and highways [7]. Reconfigurable Intelligent Surfaces (RIS) panels offer programmable control over wireless propagation by dynamically adjusting reflection and refraction properties [8]. Deploying these in tunnels, stations, and along roads can eliminate coverage dead zones [8]. Integrated Sensing and Communications (ISAC) systems use the same hardware and waveforms for both communication and sensing, to simultaneously provide high-capacity links and high-resolution radar [9]. This is relevant for autonomous vehicle perception and railway obstacle detection [10]. Finally, spectrum constraints require advanced sharing techniques. Dynamic spectrum sharing allows multiple technologies to coexist in the same band through dynamic allocation.

## 1.2 Objectives and Scope

By **Adrian Kliks**

This white paper aims to comprehensively analyze and consolidate the state of the art in communication technologies supporting modern and future Information and Communication Technology (ICT). Its primary objective is to establish a common understanding of how communication solutions enable automation, safety, and efficiency across multiple transport domains, including railways, road vehicles, aircraft, and unmanned aerial vehicles. The document seeks to identify key communication requirements and technological enablers necessary for interoperable and reliable ITS operation. It also assesses the limitations of current systems and proposes pathways for integrating emerging technologies such as 5G, Sixth Generation (6G), and Artificial Intelligence (AI)-driven network control.

The white paper also intends to support harmonization between different transport modes through a unified framework for communication modeling, testing, and standardization. It highlights the importance of accurate channel modeling and empirical validation to design efficient, robust, and scalable systems.

Another objective is to explore the use of reconfigurable intelligent surfaces, integrated sensing and communication, and digital twin concepts within ITS. The document emphasizes the role of spectrum management and standardization efforts in ensuring interoperability among diverse communication systems.

Finally, the paper seeks to stimulate collaboration among academia, industry, and standardization bodies to advance the design of resilient and adaptive communication infrastructures for future transportation systems.

## 1.3 Structure of the Document

By **Adrian Kliks**

The whole paper consists of eight chapters. Following the Introduction, the remainder of the paper is structured as follows:

- Chapter 2 focuses on Railway Communication, covering train-to-ground and train-to-train links, channel modeling, and experimental validation.
- Chapter 3 addresses UAV Communication, emphasizing air-to-ground channel modeling, cooperative awareness, drone-mounted base stations, and localization methods.
- Next, chapter 4 discusses Aircraft-to-Aircraft Communication, including propagation models, system challenges, and validation with measurement data.
- Chapter 5 examines V2V and V2X Communication, presenting models for vehicle-to-vulnerable-user links, learning-assisted content delivery, and testing methodologies.
- Chapter 6 deals with Interoperability and Standardization, describing key ITS standards (IEEE, 3GPP, ISO, SAE) and spectrum allocation policies.
- Chapter 7, outlines Future Trends and Research Directions such as cell-free massive MIMO, intelligent surfaces, digital twins, integrated sensing and communication, and long-term ITS visions in the 6G era.
- The document concludes with Chapter 8, which summarizes the main findings and perspectives for future research and standardization.

# Chapter 2: Railway Communication

By **Markus Hofer, Paul Unterhuber, Ruisi He, Zhengyu Zhang**

## 2.1 Overview

Modern railways are hard to imagine without communication systems and technologies. Train control and traffic management rely heavily on Transmission-Based Signaling (TBS) systems such as the ERTMS and CBTC. Both are enabled by robust T2G communication links: GSM-R currently supports ERTMS (to be replaced by the FRMCS in the 2030s), while CBTC metro lines typically employ Institute of Electrical and Electronics Engineers (IEEE) 802.11 and, to a lesser extent, Fourth Generation (4G)/5G technologies.

It is impossible to have fully-automated (this is, Grade of Automation 4 (GoA4)) lines without real-time communications between ground and onboard systems, Closed-Circuit Television (CCTV) cameras for passenger surveillance, intercoms to address emergencies or sensor networks onboard the train (for example, to detect derailments or any other issue). All these systems require a high-capacity and reliable T2G communication system.

Usually, railway communications services are explained by dividing them into three categories: Safety-Oriented Services (SOS) which usually comprise only TBS systems, Operator-Oriented Services (OOS), which are not safety-related but provide added value to railway operators and infrastructure managers (among other stakeholders), like CCTV, sensor networks, remote maintenance, etc., and Passenger Oriented Services (POS) which are focused only on the end user, where good examples are internet access and infotainment applications. However, this taxonomy is far from perfect, as there are some blurry lines. For example, CCTV is assumed in general terms not to be safety-related (which means that it is an OOS). But if an infrastructure manager uses this system to detect a person or an object on the tracks, maybe we have a safety-related service.

This chapter explores the foundations and advancements of railway communication systems, with a focus on T2G and T2T communications. It highlights the importance of robust, low-latency, and reliable wireless links for applications such as autonomous trains, Virtually Coupled Train Sets (VCTS), remote operations, and next-generation Train Control and Management System (TCMS). VCTS represent an emerging railway concept in which multiple train sets operate as if they were physically coupled, but without mechanical connections. This is achieved through ultra-reliable, low-latency T2T communication and advanced control systems that synchronize braking and acceleration in real time. By enabling trains to run with minimal headways and to dynamically join or split during operation, VCTS can significantly increase line capacity and operational flexibility.

The chapter begins with an overview of communication architecture, including T2G, intra-consist, and T2T links as shown in Fig. 2.1 focusing on safety critical communication scenarios. It describes specific technologies and standards, such as the almost obsolete GSM-R, FRMCS, IEEE 802.11p/bd, and 3rd Generation Mobile Group (3GPP) LTE/5G Vehicle-to-Everything (V2X). Subsequently, channel sounding results, channel modeling techniques, and simulation/test data collected in real-world railway environments are presented. The chapter ends with the description of key challenges such as modeling of highly dynamic and non-stationary wireless environments, optimizing for various terrains, and managing seamless transitions between different communication domains (T2T and T2G). Emerging standards and ongoing measurement campaigns will help shape future strategies for safer, faster, and smarter railway systems.

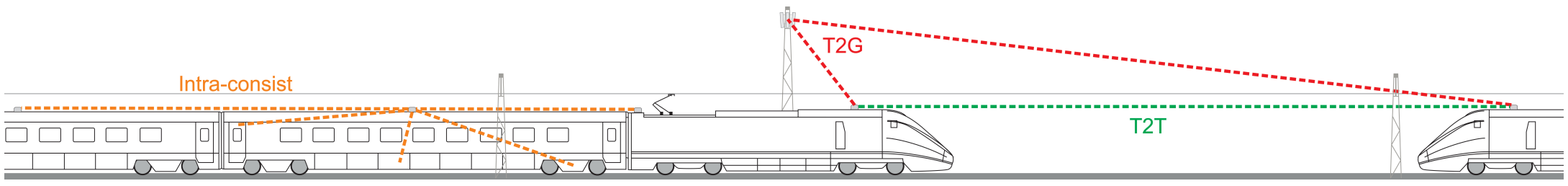

Figure 2.1: Railway communications: T2G, T2T, and intra-consist communication links.

## 2.2 Train-to-Ground Communication

T2G communication systems serve as the foundation for railway operations and passenger services, supporting applications such as train dispatching, signaling control, emergency communications, onboard Wi-Fi, and multimedia services. Advanced communication technologies enhance railway safety and operational efficiency by enabling comprehensive monitoring and control of rolling stock, track infrastructure, and station facilities. Key operational scenarios include Automatic Train Operation (ATO), real-time interlocking signaling, driver supervision, and remote dispatching.

The most widely used system is GSM-R, which is an improved version of the traditional GSM network [11]. GSM-R operates in dedicated frequency bands for uplink (876 − 880 MHz) and downlink (921 − 925 MHz), with a bandwidth of only 0.2 MHz, supporting voice communication, data transmission, and emergency messaging. The low frequency bands and narrow bandwidth affect railway channel modeling, where statistical large- and small-scale channel models in some specific railway environments (such as viaducts, cuttings, and tunnels) are developed. GSM-R has served global users for over 20 years, known for its reliability, efficiency, and safety. However, limitations of GSM-R have gradually become apparent, such as its limited data rate, inability to support high-bandwidth multimedia data transmission, and insufficient flexibility and scalability, preventing it from offering broadband internet, low-latency communications, and other services required by modern railway systems.

With the maturation and commercialization of 5G technology, leveraging the advantages of 5G and applying it to railway communication systems, i.e., 5G-Railway (5G-R), has become a potential evolution path for GSM-R [12]. In Europe, the project 5GRAIL tested FRMCS based on 5G as the successor to GSM-R [13]. Railway companies in Japan and South Korea have also deployed 5G infrastructure on some railway lines to verify its performance. China is shifting its research focus from GSM-R directly to 5G-R [12], [14].

The dedicated frequency bands for 5G-R have yet to reach international consensus. The Union Internationale des Chemins (UIC) has been actively working to secure frequency allocations for 5G in European railways. The Electronic Communications Committee (ECC) has approved a draft recommendation, designating 5.6 MHz in the 900 MHz band and 10 MHz in the 1900 MHz band for European railways. Additionally, the 5.9 GHz band, allocated for ITS, is being considered to support reliable communication links for T2T, intra-train, and T2G connections. Meanwhile, in China, the potential frequency band for 5G-R for train control applications is identified as 1.9/2.1 GHz with a 10 MHz bandwidth. Drawing on measurements in a dedicated 5G-R railway test track, [15] characterized the 2.16 GHz radio-propagation behavior under high-mobility conditions. Specifically, it quantified path loss, power delay profile, delay spread, Ricean $K$-factor, angle of arrival estimation and angular spread, etc. Furthermore, a Tapped-Delay Line (TDL) channel model based on a first-order two-state Markov chain is developed in [16], thereby enabling realistic link-level simulation of 5G-R propagation.

Besides safety critical railway communications, communication networks also support a wide range of non-safety-critical services for passengers, such as providing high-speed internet access on trains and platforms, real-time passenger information systems, and onboard multimedia entertainment. These services significantly enhance passenger comfort and travel efficiency. Additionally, wireless communication enables data collection and transmission for trackside and onboard sensors, monitoring train health, track conditions, and environmental factors, thus supporting predictive maintenance and safety alerts. Overall, both operational control and passenger service applications demand higher bandwidth, lower latency, and improved reliability, necessitating an integrated communication platform that seamlessly connects trains, trackside base stations, and various smart devices.

In summary, T2G communications are rapidly transitioning from GSM-R-based single networks to multi-bearer, intelligent, all-IP networks. Innovations such as FRMCS, 5G/6G integration, satellite communication, and AI-based intelligent networking will drive the future development of safer, faster, and smarter rail systems.

## 2.3 Train-to-Train Communication

Innovative operational solutions and future applications for railways will rely on wireless communications. Low latency and highly reliable wireless communication enable autonomous driving trains, VCTS, Self Driving Freight Wagons (SDFW), remote train operation, Europe's future TCMS and many more.

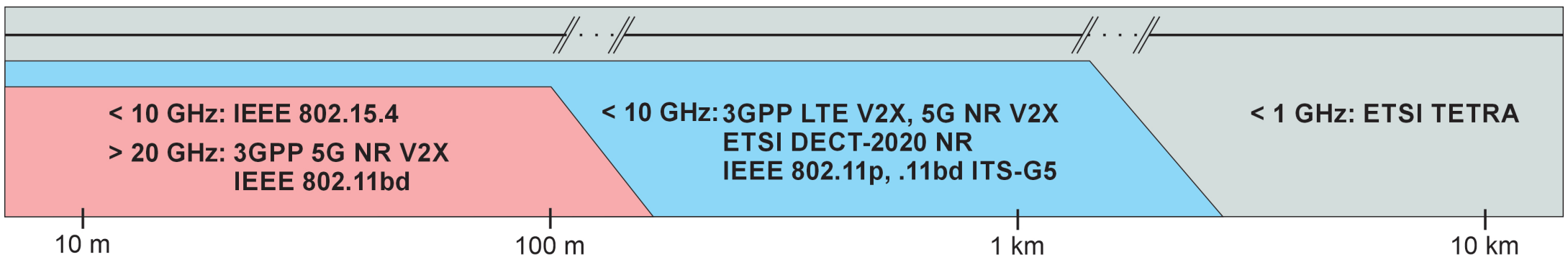

Figure 2.2: Communication standards for different T2T communication ranges.

Wireless links within a consist[1], between consists and between trains[2] are required and have to support data exchange for safety critical and control applications, sensor and actuator networks, and surveillance and information services.

For all kind of wireless links in consists or trains, the changing environment along the track plays an important role. As a consist or train moves, it passes through varying environments and encounters different objects along the railway line. In addition to the static parts of the environment, parts of the environment can move as well, e.g. other trains, or road vehicles in the near vicinity.

A differentiation of future wireless links can be made by links which are facing a change and links which are facing a constant geometry over time. On the one hand, the geometry for wireless links within the consist most likely remains stable during operations. Effects on the Line-of-Sight (LoS) and the cause of the most prominent multipaths will be defined by the vehicle body and therefore remain static. Investigations on the communication inside one consist and the position of communication nodes were presented for high speed trains in [17] and for metros in [18], [19].

On the other hand, for wireless links between consists or between trains, the geometry is likely to change significantly over time. The communication range changes dramatically from several meters to several kilometers, e.g., the composition of the VCTS changes, or communicating trains driving in opposite directions. Furthermore, the relative velocity between the communication nodes can vary from zero to several hundreds of kilometers per hour. Hence, considering the time variant behavior plays an important role for those wireless communication links. Due to the mentioned distinctions the wireless links within one consist can be grouped as intra-consist communications, while the links between consists and between trains can be grouped as T2T communications.

The use of existing communication standards for railway communications has been investigated in the last decades. In detail, intra-consist and T2T communication have been not foreseen as use or business case during standardization. Hence, the applicability, arising limitation and potential adaptations have to be investigated. The most promising communication standards are 3GPP Long Term Evolution (LTE) V2X and IEEE 802.11p and their successors 3GPP 5G New Radio (5G NR) Rel 18 and IEEE 802.11bd, respectively. For short range communication, IEEE 802.15.4, for mid range European Telecommunications Standards Institute (ETSI) DECT-2020 NR and for long range ETSI TETRA have to be mentioned as suitable candidates. Fig. 2.2 relates the proposed communication standards, the frequency band used and the T2T communication range. In general, all communication standards mentioned in Fig. 2.2 are applicable for railway communication. Depending on the actual use case, the fulfillment of detailed requirements has to be tested individually. Tests of different communication standards for railway applications in typical railway environments are presented in Section 2.6.

ITS standards are presented in details in Chapter 6. In the following, the different railway-related standards are briefly outlined.

### 2.3.1 IEEE 802.11 V2X Standards

In the US, Dedicated Short Range Communication (DSRC) was introduced for the use in vehicular communication. The PHY and MAC layers were standardized by IEEE 802.11p, which is not part of the baseline IEEE 802.11-2020 standard. In Europe, the IEEE 802.11p standard was adopted by ETSI, extended and standardized as ITS-G5. Higher layers for V2X communication are defined by ETSI, IEEE and the Society of Automotive Engineers and are also used for LTE V2X. 802.11p is a derivative from the 802.11a standard with the main difference that bandwidth and carrier spacing are halved. Two pilot symbols are used for channel estimation at the beginning, which makes it not suitable for highly mobile scenarios.

[1] A consist can be a single vehicle or a set of vehicles, which is not changed during operations.
[2] Several consists can be coupled to one train.

Its successor, IEEE 802.11bd, introduced improvements not just for road applications. In fact, railway use cases were introduced for T2T and vehicle to train communication, and vehicular positioning and location [20]. Compared to IEEE 802.11p, .11bd introduces additional pilot symbols during transmission and uses low-density parity check codes for channel coding. In addition, radio-based ranging and operation at millimeter wave (mmWave) frequencies were adopted for the use of highly dynamic use-cases on road and rail. In detail, for T2T communication use-case .11bd defined the following requirements: relative speed of 500 km/h (with directional antennas 800 km/h), distance of 2000 m, data rates of at least 1 Mbit/s, a ranging accuracy of 1% of distance and latency of 10 ms. IEEE 802.11bd is backward compatible with 802.11p. That means, a communication based on the mode of .11p is possible with .11bd units.

Following the publication of IEEE Standard 802.11bd - 2022 Amendment 5, this standard has also been adapted by ETSI for the European market and published as the ITS-G5 Access layer Release 2 in ETSI EN 303 797 V2.1.1 (2024-02).

### 2.3.2 3GPP V2X Standards

LTE V2X was developed under Rel 14 and was further enhanced in Rel 15. LTE V2X operates in the ITS band at 5.9 GHz. LTE V2X offers two modes (modes 3 and 4) for Side Link (SL) broadcast communications.

5G NR was started with Rel 15, where the frequency range 2 was introduced and multiple bands from 24 GHz to 52 GHz were allocated. With the introduction of SL communication for 5G NR in Rel 16, 5G NR-based V2X was possible. Rel 16 defines two new modes (modes 1 and 2) for the selection of subchannels in NR V2X SL communications using the NR V2X interface. In comparison to LTE V2X, NR V2X supports broadcast, groupcast, and unicast SL communications.

In Rel 17 the V2X communication is enhanced and the frequency range expands up to 71 GHz. In 5G NR mode 2 and LTE mode 4 User Equipment (UE) can operate without network coverage. However, 5G NR mode 2 radios and LTE mode 4 radios are incompatible. With the Rel 18, the beginning of 5G advanced is marked. Among others, major improvements for SL communication were presented. Release 18 introduces features that enhance reliability, low latency, and scalability for vehicular communication and direct links between vehicles. With these enhancements, remote driving and information sharing for full automated driving are supported. Furthermore, ranging-based services and SL positioning were introduced.

ETSI has adopted LTE V2X as an access layer for ITS in the 5 GHz (ITS-G5) in ETSI TS 103 794 V1.1.1 (2021-04) and is currently working on Draft ETSI EN 303 798 V1.2.4 (2023-09) to specify both LTE V2X and 5G NR V2X as access layers for ITS-G5 in Release 2.

## 2.4 Channel Sounding in Railway Environments

As the propagation conditions for intra-consist, T2T, T2G and V2X communications clearly differ, several channel sounding measurement campaigns were conducted in the last years. An intra-consist channel sounding measurement campaign was performed in metro trains in [18]. The difference between continuous trains (e.g. metro line, where you can walk between the waggons) and non-continuous (e.g. waggons where you cannot change or you have to open a door between the waggons) trains, the influence of passengers and the propagation from inside to outside were investigated at 2.6 GHz.

The propagation along the roof of a high speed train from one to the next coach and the one after were investigated in [21]. The German Aerospace Center (DLR) RUSK channel sounder with a bandwidth of 120 MHz was used at 5.2 GHz. The train was operated on a high speed track between Naples and Rome in Italy for one night. The resulting TDL channel model was used for LTE SL communication simulations.

The worldwide first propagation measurements with two high speed trains for T2T communications was conducted in Italy as well. The two high speed trains performed different maneuvers in the same direction representing VCTS or traveling in opposite direction achieving a maximum relative speed of 560 km/h. For these measurements, the DLR RUSK channel sounder was set to a bandwidth of 120 MHz at 5.2 GHz. Both trains operated always on two parallel tracks on the high speed line between Naples and Rome. The detailed setup is described in [21]. These measurements were used as basis for the Geometry-based Stochastic Channel Model (GSCM) for T2T communication representing open field railway station and hilly terrain with cutting [22], [23].

The authors of [24] present a wideband channel sounding campaign for both T2G (in the paper called train-to-infrastructure) and T2T communications using a 150 MHz bandwidth at 5.8 GHz. The measurements were performed at moderate speeds (max. 20 km/h) in a shunting yard with varying distances between trains. Key findings highlight how environment geometry, antenna polarization, and surrounding traffic significantly affect delay spread, Doppler spread, and path loss, which are critical for future 5G-based railway communication systems. The Channel sounding results of massive MIMO measurements at two frequencies (1890 MHz and 748 MHz) for a railway station scenario were presented in [25]. The results show differences in Root Mean Square (RMS) delay spread, hardening factor and path loss coefficients between the two frequency bands.

A channel sounding measurement with the DLR RUSK channel sounder at 60 GHz with 120 MHz bandwidth was conducted with two experimental trains [26]. The two trains were driving on a shunting track back and forth or one train parked and the other one driving with communication ranges up to 130 m. The mmWave front-ends were installed on the coupler. Obviously, this mounting causes a pronounced ground reflection. But the major finding of the measurements was strong double reflections between the train fronts. The signal was reflected back from the receiving train to the transmitting train and back to the receiver at the receiving train.

In April 2025, DLR performed tests with IEEE 802.15.4 communication equipment installed on two trains driving on the same track behind each other with distances up to 300 m. The Ultra WideBand (UWB) system with a bandwidth of 500 MHz at channel 2 (3.99 GHz) and in parallel on channel 5 (6.49 GHz) recorded the channel state information, as well as statistics for data exchange and time-based distance estimation between the two driving trains.

## 2.5 Channel Models

Wireless channel models play a crucial role in the design, evaluation and optimization of railway communication systems. Nowadays, traditional railway communication faces increasing challenges, and it requires communication systems to adapt to different environments and to provide high data rate, low latency, and high reliability. Channel models provide the necessary theoretical support and practical guidance for these demands.

As railway communication scenarios are complex and train speed increases significantly, the demand for 5G-R channel models has evolved, where statistical methods based on measurements, geometric-based methods, and deterministic methods have all been extensively explored [27].

At this stage, more attention is being focused on modeling of rapid time-varying characteristics of the channel, such as non-stationarity, multipath birth-death, and Doppler shifts and spreads. Some recent results of 5G-R channel measurements and modeling can be found in [16], [28], which fills in the research blank. In addition to the traditional sub-6 GHz frequency bands, the mmWave band has received some attention in 5G-R theoretical research [29]. Some studies focus on dynamic Doppler effects, spatial correlation, T2T interference, and the impact of motion states on channel performance [30].

Beside T2T communication, T2G communication is the core of railway communications. Characterization of signal propagation and multipath effects in railway-specific T2G scenarios is important for channel modeling. The authors of [31] present the findings of the IEEE P1944 site-specific channel representation standardization group, that argues for a geometry based double-directional multi-path component based channel representation for challenging railway scenarios.

For railway station communication, the signal propagation is influenced by several unique factors including the architecture of the station, moving trains, and various obstacles like lifts and underpasses. The complexity of these environments necessitates advanced modeling techniques to ensure reliable communication. One study conducted a ray-tracing simulation to evaluate 5G coverage in railway stations, particularly in medium-sized stations. The study used the 3.7 GHz frequency, as this band provides a good balance between coverage and capacity. The simulation revealed significant coverage gaps due to the interference from metal structures such as trains and lifts, which create shadow areas with poor signal reception [25].

Another study focused on massive MIMO measurements at two frequencies (1890 MHz and 748 MHz) for a railway station scenario. It was observed that the train's movement from LoS to Non Line-of-Sight (NLoS) significantly impact the channel characteristics. The results indicate that path loss increases more significantly in the higher frequency band (1890 MHz), and the delay spread was smaller at lower frequencies, thanks to stronger LoS components [32].

Meanwhile, T2T communication as an emerging communication need is receiving increasing attention. Early T2T channel models mainly consider LoS communication link between trains and relatively simple environments. However, due to the large inter-train distances in T2T communications, NLoS links and complex propagation dynamics may occur, and the model must remain applicable across diverse terrain conditions and extended communication ranges. Hence, the latest channel models for T2T communication represent the time-variant propagation channel between two communication nodes mounted on rail vehicles.

The following characteristics distinguish these channel models from those developed for T2G or V2X communications:

- **Environment:** The environment is changing continuously with regular and sparse objects appearing along the railway track.
- **Movement:** The trains are driving on the same track or on parallel tracks in the same direction at different distances with low relative velocity (e.g. for the VCTS concept), or on parallel tracks in opposite direction with fast changing distances and very large possible relative velocities (this would reflect the concept of the railway collision avoidance system).
- **Antenna position:** Each node is mounted on a train on a similar height, either in the coupling area or on the roof of the trains.

Based on high speed train measurements in Italy 2016, several channel models were derived. First, stochastic model parameters for railway station, open field and hilly terrain environments were derived in [33]. Parameters like path loss exponent, log-normal shadow fading distribution, $K$-factor, root mean square delay and RMS Doppler Spread were extracted. Second, a TDL model was derived in [34]. The model was used to analyze IEEE 802.11bd capabilities for T2T communication in typical railway environments.

Third, a GSCM for T2T communications was designed. For open field environment, the GSCM was presented in [22], [23]. The GSCM was also designed for railway stations and in hilly terrain with cutting environments. The model considers objects like rails and the overhead line system including regular appearing masts, sparsely occurring close buildings, cross bridges, cellular radio masts and trees, as well as large walls and cuttings close to the tracks. The environment can be adapted as in Fig. 2.3 and the movement model of the trains can be varied, which allows the analysis of various scenarios and the time-variant behavior of the T2T propagation channel. Antenna position and characteristics can be added to the model.

A comparison of the local scattering function of T2T channel between the conducted measurements and derived GSCM is shown for three different time stamps in Fig. 2.4. The LoS signal and the MultiPath Components (MPCs) caused by the surrounding objects are clearly visible. Four MPCs are highlighted with A, B, C and D in Fig. 2.4 and their temporal evolution is presented.

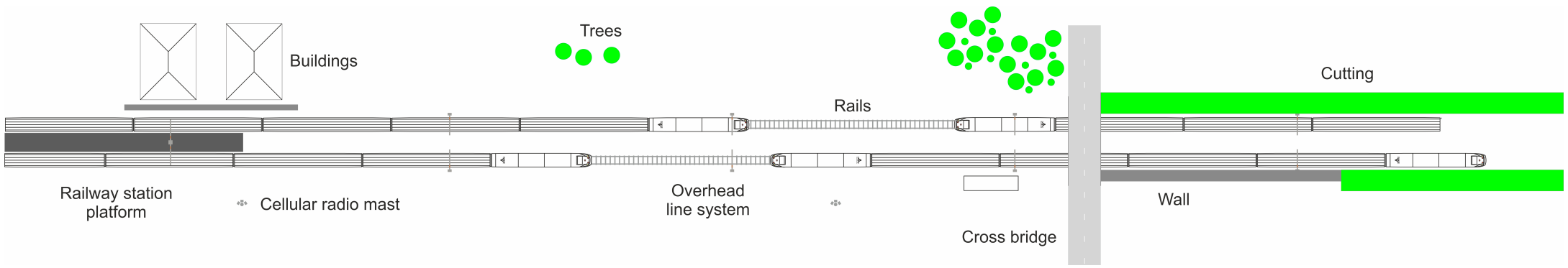

Figure 2.3: Architecture of the GSCM for T2T communications.

In railway communications, channel modeling mainly faces the following challenges:

- **Dynamic and Non-Stationary Channels**: The complex dynamic characteristics of railway environments are a major challenge. Most existing channel models are based on static or simplified assumptions such wide-sense-stationarity, whereas for future railway communications, rapid variations in channels lead to complex non-stationarity and multipath dispersions in space-time-frequency domains. Accurate modeling and simulation technologies in railway environments remain a pressing issue.
- **Multi-Band Channel Modeling**: With the development of 5G and 6G, railway communication will involve more frequency bands, especially the coordinated use of low, medium, and mmWave bands. Modeling of multi-band channels, considering interference, correlation, and delay characteristics among different bands, will be an important research direction.

- **Seamless Handover and Connectivity**: Seamless connection and handover between T2T and T2G communications will be necessary in future railway communication. Designing methods that ensure both efficient communication and stable connections for T2T and T2G, and avoiding signal interruptions and interference during high-speed motion, are important issues that need to be addressed in the future.

## 2.6 Simulation and Test Results

In the last decade, tests in railway environments were conducted with different communication standards. The simulation and test results reflect the evolution of communication standards, the trend to use frequencies above 20 GHz and the need for MIMO antenna systems.

For long distance T2T communication coverage, tests were performed with TETRA radios in direct mode operation. Results on the communication performance and the implementation in a railway collision avoidance system were presented in [35].

In recent years, ITS-G5 communication modules based on IEEE 802.11p were widely used for tests in railway environments. The general focus of those tests was to investigate the possibility of transferring the ITS-G5 standard, designed for road applications, to rail applications:

- [17] presents a study on intra-consist communication. ITS-G5 units were installed on multiple positions inside a high speed train, either in one coach (LoS conditions) or in following coaches (NLoS conditions).
- In [36], a T2T communication between a commuter train and a car was investigated on secondary lines up to 600 m. For the major part of the tests, the car, emulating a second train, was driving next to the railway track in the same direction or in opposite direction. Some tests were performed for level crossings and collision avoidance.
- In parallel to channel sounding measurements, tests with Cohda ITS-G5 units were performed between two high speed trains on the high speed track between Naples and Rome in Italy. [37] presents the ITS-G5 T2T link analysis up to 2 km communication range and velocities up to 300 km/h absolute and 560 km/h relative.

The interference between CBTC based on IEEE 802.11a or LTE V2X (3GPP R14) with ITS-G5 V2X communication based on IEEE 802.11p was investigated in [38]. The authors perform a four-day measurement campaign in Berlin to evaluate radio interference between urban road and rail C-ITS operating in the 5.9 GHz band with one train, two cars and one base station. The study involved various V2X communication technologies (ITS-G5, LTE C-V2X, and IEEE 802.11a) and analyzed their performance under different interference scenarios using the cars and the train equipped with high-precision measurement equipment. The results show that even adjacent-channel interference can significantly degrade performance.

Simulations for T2T communication based on different T2T TDL models are presented in [34]. The authors derived six TDL models for typical railway scenarios and compared the performance of IEEE 802.11p and .11bd for different modulation and coding schemes.

In comparison to standard ITS communication systems, IEEE 802.15.4 offers a large bandwidth and enables very accurate time-based distance estimation and object detection. Those UWB systems are designed for indoor applications. Nevertheless, there is a high potential for communication systems with 500 MHz bandwidth on short distances for railway application. A test campaign to verify the applicability of IEEE 802.15.4 for T2T communication was performed in April 2025. Two commuter trains were equipped with communication nodes based on the Qorvo DW1000 chipset as shown in Fig. 2.5. The UWB communication units were installed in the silver/gray boxes on the couplers of the trains. Two UWB links in parallel, one using a omni-directive antenna, one using a directive antenna were installed. The communication performance and the time-based distance estimation were tested in railway environments with both trains moving on the same track back and forth.

Open source 3GPP 5G NR stacks can be implemented in software defined radios for research and testing. Especially, the use of the 5G frequency range 2 up from 26.5 GHz in combination with antenna arrays offers great potential for ISAC. 5G preliminary tests at 62 GHz were performed by [26]. Next to channel sounding, tests were conducted to estimate the distance between two trains.

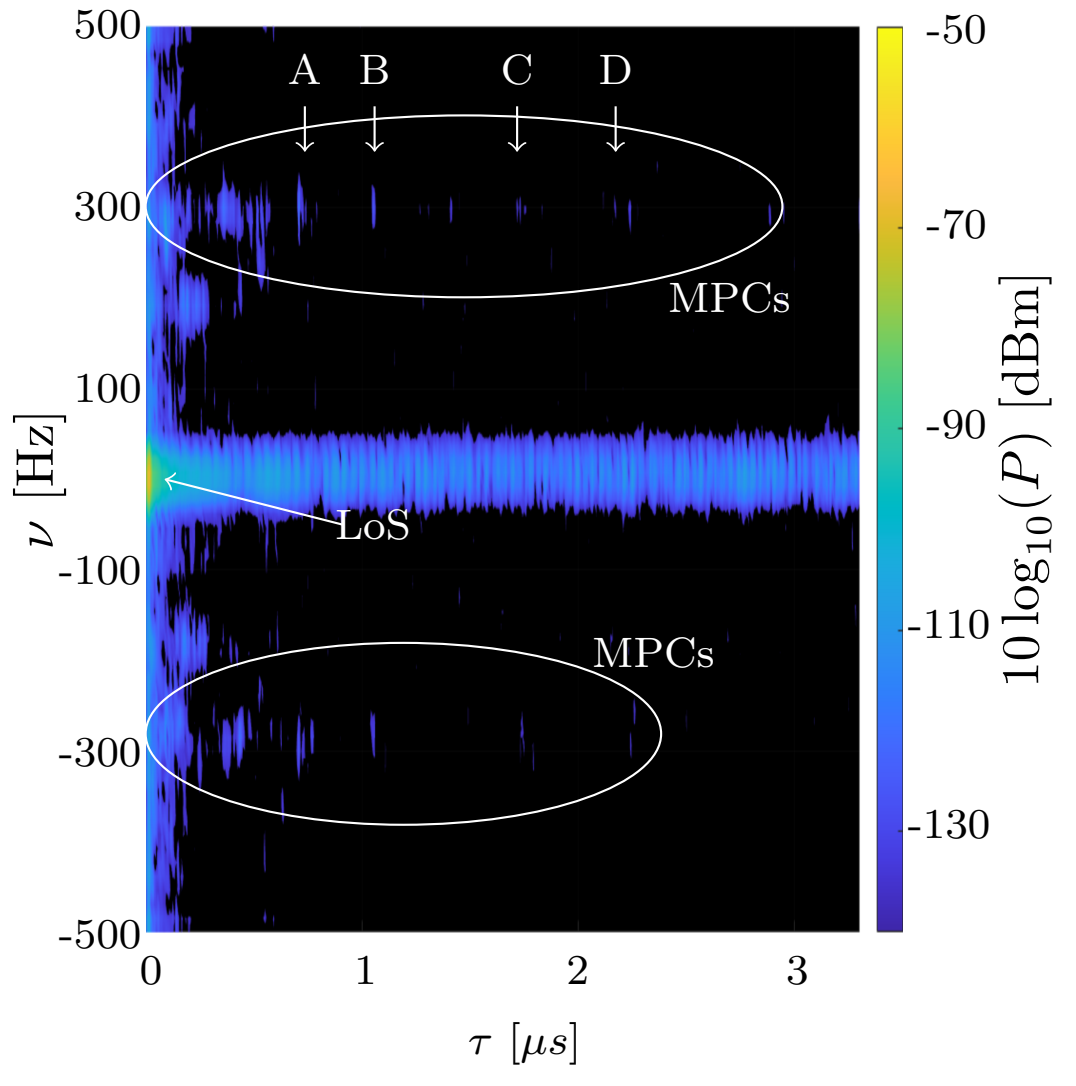

(a) $t = t(d = d_{min})$.

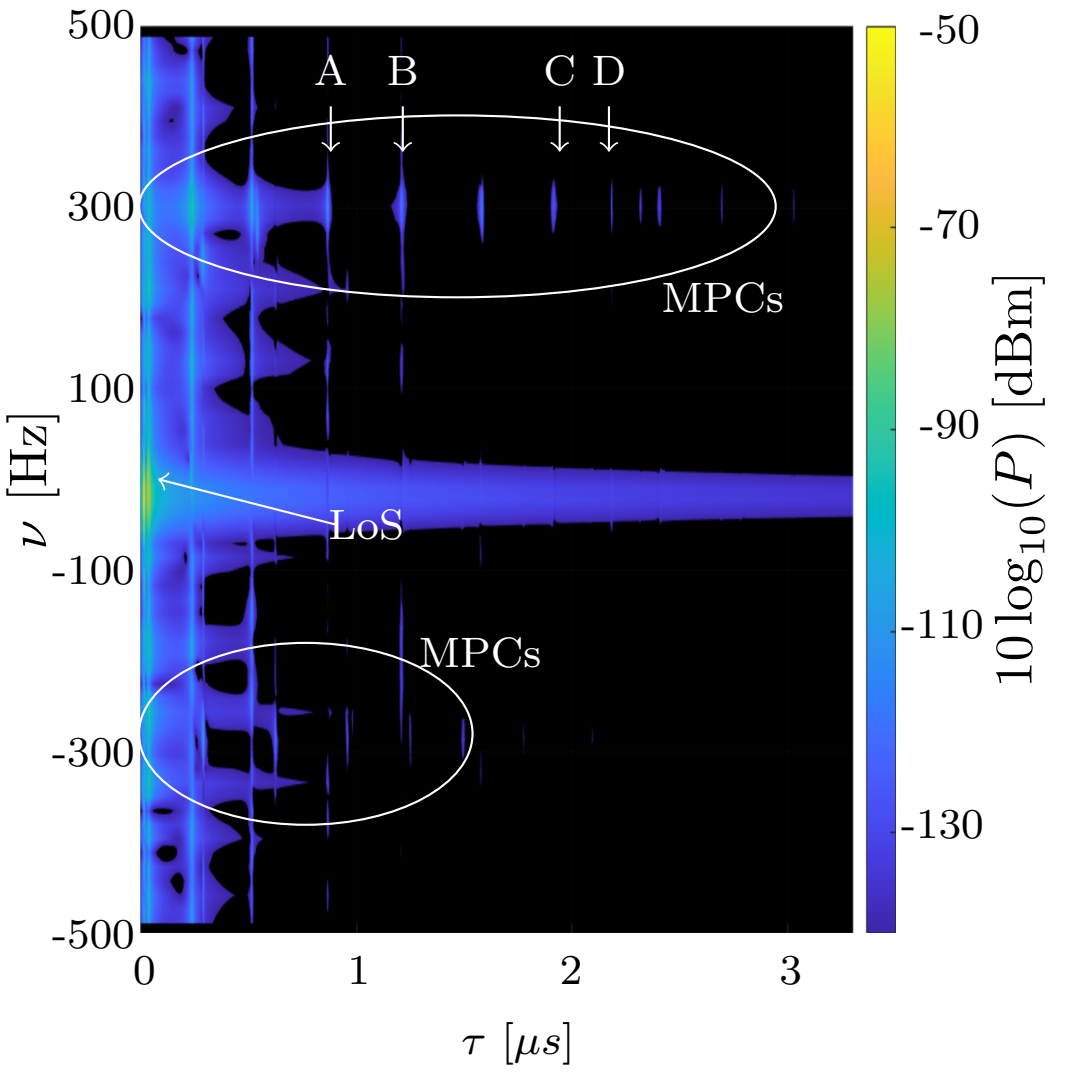

(b) $t = t(d = d_{min})$.

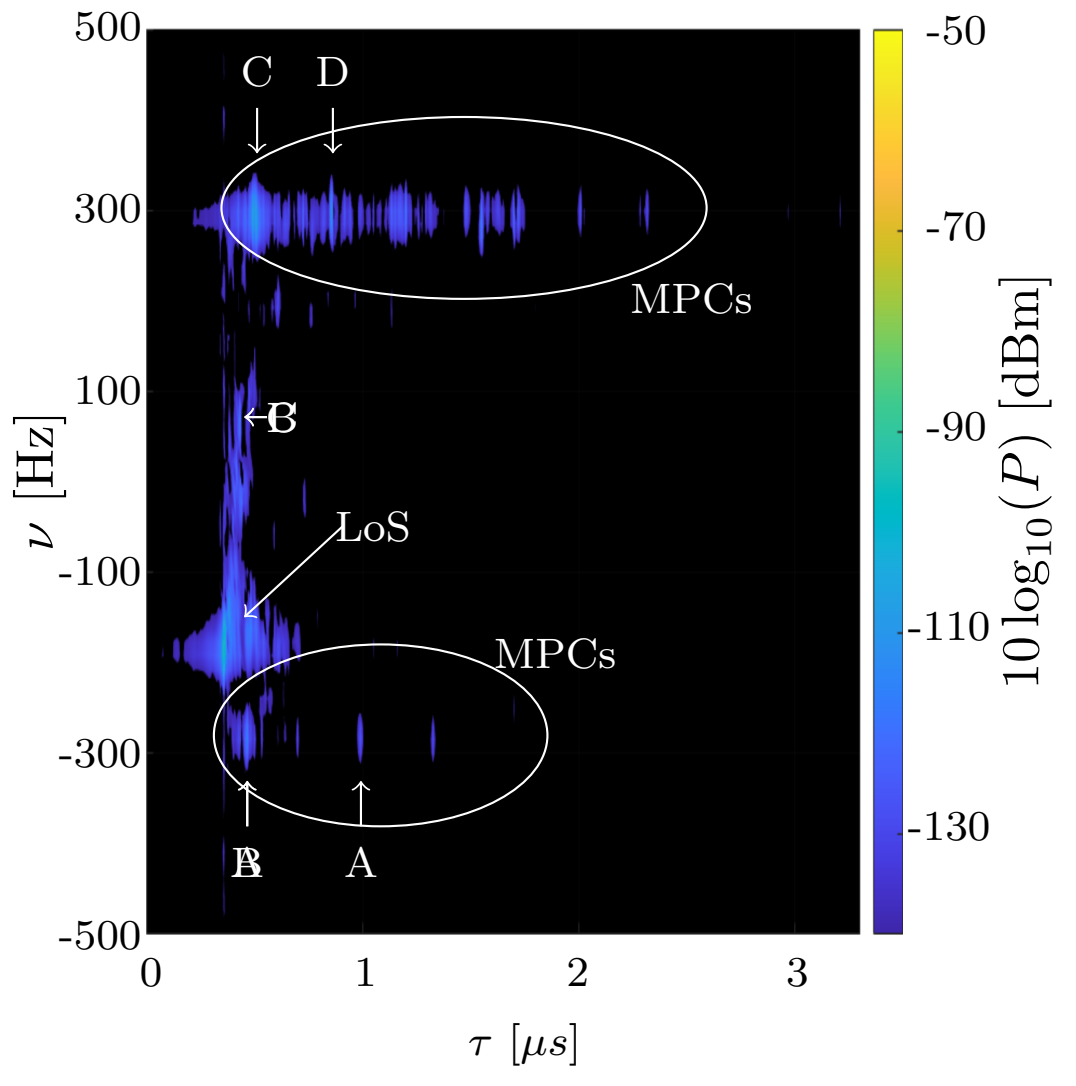

(c) $t = t(d = d_{min}) + 10s$.

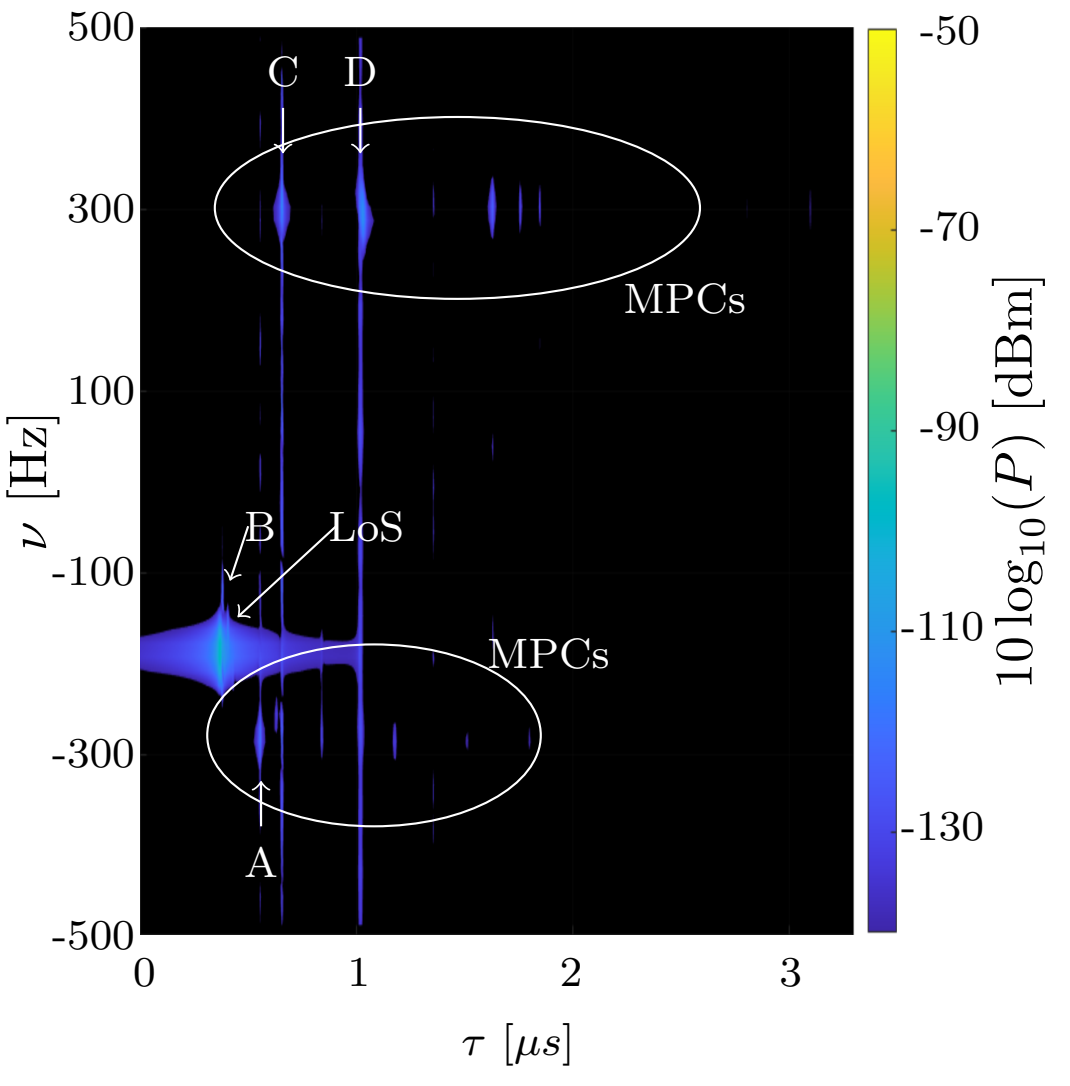

(d) $t = t(d = d_{min}) + 10s$.

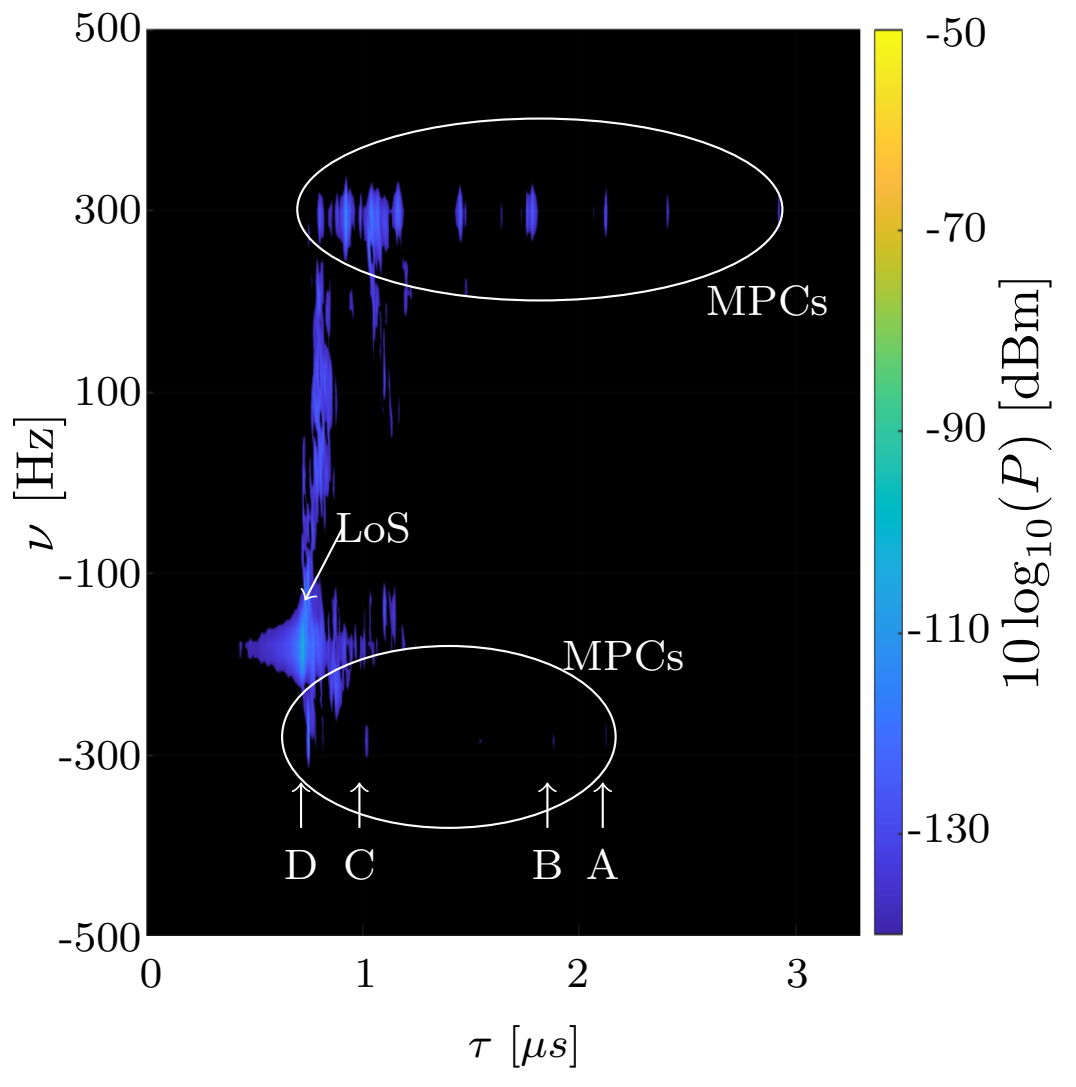

(e) $t = t(d = d_{min}) + 20s$.

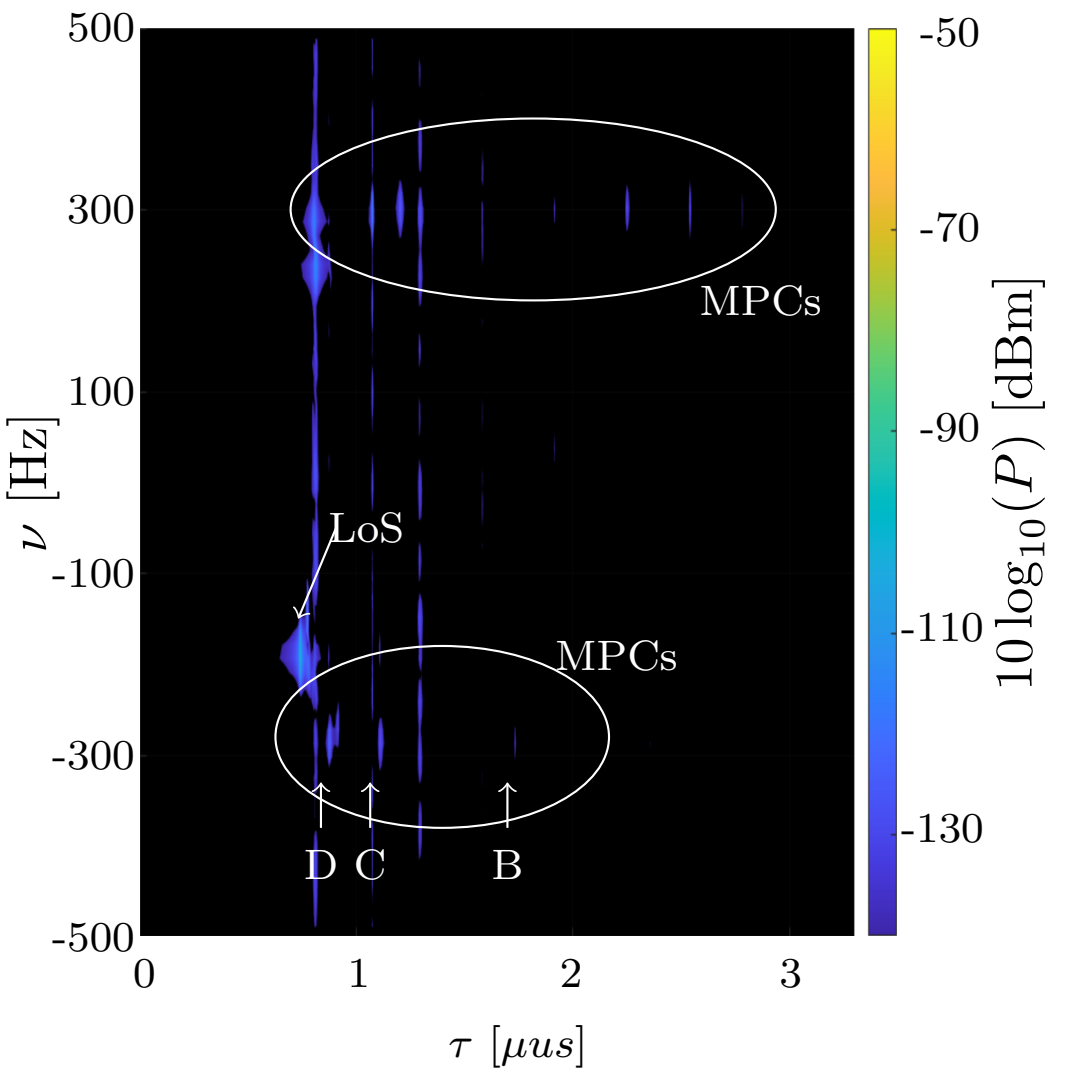

(f) $t = t(d = d_{min}) + 20s$.

Figure 2.4: Local Scattering Function for T2T communication in hilly terrain with cutting environment for three time stamps. Left (a, c, e) the results from the measurements; right (b, d, f) the results of the GSCM.

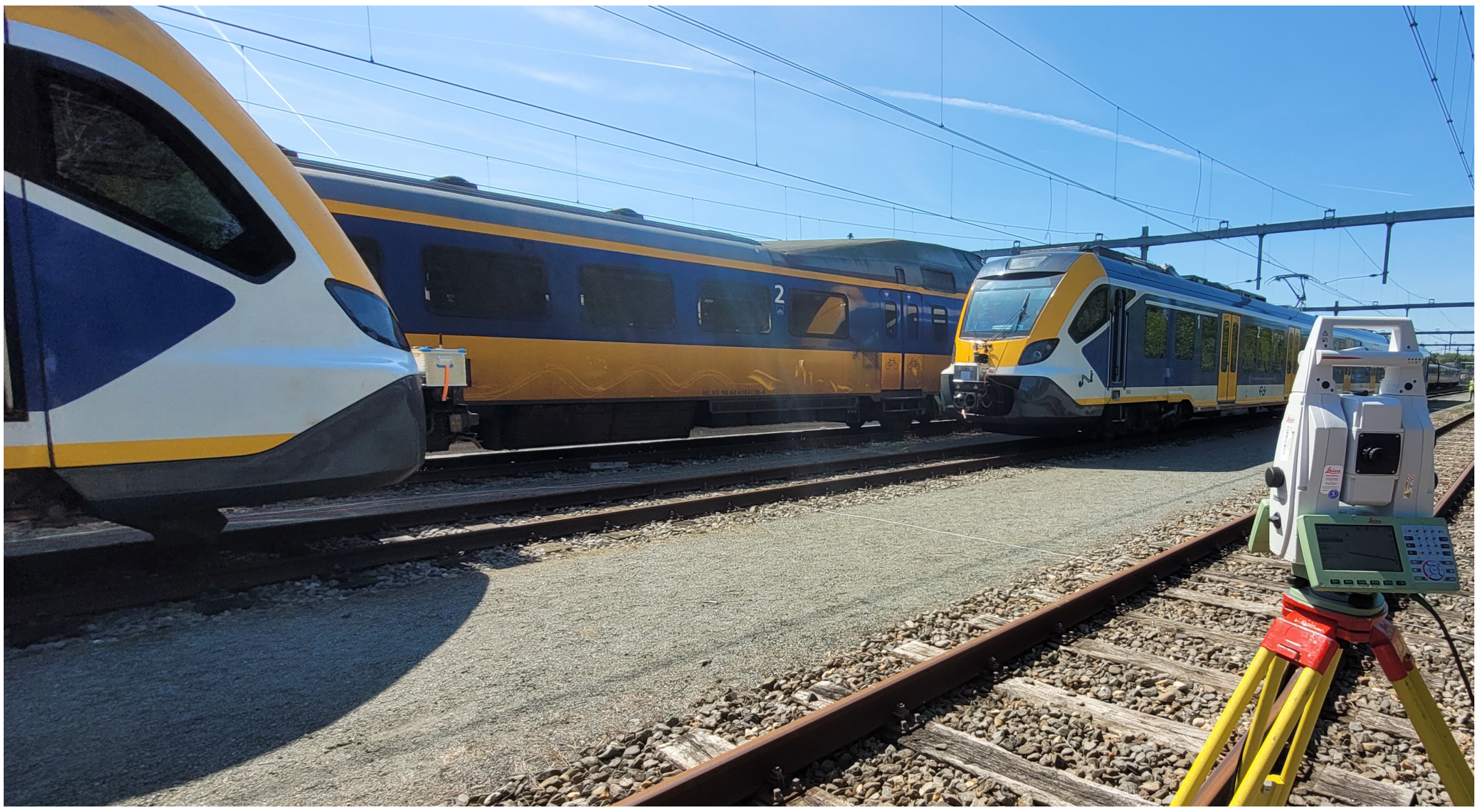

Figure 2.5: T2T Communication and ranging tests using IEEE 802.15.4.

## 2.7 Challenges and Open Questions

Modern T2T and T2G communication systems are evolving rapidly with the advent of 5G, mmWave, and even future THz technologies to meet stringent requirements on data rate, latency, reliability, and capacity. However, these advancements introduce significant challenges in handling very high mobility and Doppler effects, ensuring seamless handover across heterogeneous networks, and providing reliable and low latency communications for SOS. The major challenges can be summarized as follows:

- **Propagation conditions:** Shielding effects of train carriages, obstructions from objects along the railway track (especially the masts of the overhead line system) and electromagnetic interference from overhead catenary systems exacerbate the path loss. Furthermore, high-speed trains (e.g., over 350 km/h) induce significant Doppler shifts, leading to frequency offsets and rapid time-varying channels. These challenges increase significantly with the increase of the operation frequency above 20 GHz. Leveraging mmWave and THz communication for T2G links has not yet been investigated in more detail.

  For SOS in different railway scenarios, the data rates are typically low. Hence, mmWave or THz bands are not necessarily required. Furthermore, mmWave and THz bands are highly susceptible to LoS blockage and are strongly affected by increased Doppler shifts resulting from train vibrations. The higher path loss requires narrow beams with precise beam steering and consequentially, fast beam finding to allow for a proper link budget between transmitter and receiver. For a T2T or an intra-consist scenario, mmWave communication links might be useful but still have to be investigated in more detail.

- **Channel modeling:** Existing channel models predominantly rely on static or low-speed assumptions, lacking empirical data for dynamic changing channels. The impact of the mentioned effects differs for T2G, T2T and intra-consist links. Hence, the measurements and models have to differ. High-precision modeling methods based on channel sounding data and calibrated ray tracing tools are urgently needed for potential railway frequency allocations.

- **Low latency and reliability:** SOS such as ERTMS, remote train operation or autonomous driving trains require low latency and highly reliable T2G communication. Designing protocols for fading and blocked links, possibly leveraging network coding or edge computing, is an open question for ensuring deterministic performance under diverse channel conditions. Direct communication between train units based on T2T communications can reduce end-to-end latencies below 10 ms and may provide reliability exceeding 99.999%. Hence, this kind of T2T communication enables collision avoidance systems and VCTS even for high speed trains.

- **Seamless handover and coverage:** Maintaining uninterrupted connectivity at speeds over 500 km/h demands ultra fast handover strategies and multi connectivity across track side 5G NR cells, FRMCS base stations, and satellite links. Optimizing handover decision criteria to balance packet loss, latency, and signaling overhead remains a critical challenge, especially in borderless cross operator scenarios.

The future generations of wireless communication systems will cover novel capabilities in addition to standard data communication. With the introduction of ISAC in combination with wideband communication systems for railway communication, ranging between train units, absolute positioning of trains and sensing of the railway environment are possible. Those potential technologies will enable new railway applications in the future.

To gain the confidence of the rail industry, operators, and passengers, specific safety levels must also be ensured for the communication system. New communication technologies that are intended to be used in a safety critical railway context have to undergo safety evaluation and approval by the corresponding authorization bodies. This process is often challenging and time-consuming.

However, resilient communication systems with redundant and parallel links need to be designed for T2G, T2T and intra-consist communications. Furthermore, cyber-security aspects need to be taken into account at all levels of future railway communication systems.

# Chapter 3: UAV Communication

## 3.1 Overview

By **Abdul Saboor, Evgenii Vinogradov**

UAV have the ability to revolutionize the existing transport system by introducing an additional (third) mobility dimension due to their high speed and maneuverability [39]. By integrating UAVs with ITS, we can envision a new transportation paradigm called Three-dimensional (3D) ITS, where aerial and ground transportation work together to improve the whole transport experience and make it efficient [40]. While traditional ITS mainly relies on terrestrial infrastructure, introducing UAVs adds a vertical dimension to overcome the constraints of existing terrestrial infrastructure like congestion, thus leading to better urban and interurban mobility.

### 3.1.1 Advanced Air Mobility

With recent advancements in UAV technology, Advanced Air Mobility (AAM) is becoming a critical component of the future of transportation. AAM generally includes drone-based urban mobility systems, such as air taxis and cargo drones, designed to reduce traffic congestion and provide fast and efficient transportation in urban areas [41].

In this context, various companies/startups like Joby Aviation and Archer are working on Electric Vertical Take-Off and Landing (eVTOL)-based air taxis, which can carry several people around in the sky by avoiding traffic congestion and burning nasty fuel [42]. Similarly, Joby and British airline Virgin Atlantic are bringing air taxis with a top speed of 300 km/h with a distance of 240 km on a single battery charge. This project will reduce the journey time from the London airport to Canary Wharf to eight minutes, compared to 80 min by car [43]. Additionally, UAVs can help with last-mile delivery, where they can transport passengers and goods over short to medium distances, mainly in highly congested, dense urban environments.

In addition to air transportation, UAVs can provide several other use cases in ITS due to their mobility, camera, and on-board sensing technology, such as:

- The imaging technologies of UAVs can provide a real-time aerial view of road conditions to assess congestion or the severity of emergency incidents [40].
- UAVs can help deliver medical supplies faster compared to ground vehicles, such as ambulances. Similarly, they can assist in search-and-rescue missions [44].
- UAVs can be incorporated into the law enforcement elements of ITS, like police, to increase surveillance by providing an aerial view.
- Lastly, UAVs can be used to inspect roads, bridges, and tunnels better than traditional methods, resulting in reduced maintenance costs and enhanced safety.

### 3.1.2 UAV-enabled Connectivity in ITS

Integration of UAVs into ITS brings many advantages in terms of better mobility, real-time traffic monitoring, and improved emergency response [40]. However, one of the most important use cases of UAVs in ITS is providing wireless connectivity.

In ITS, UAVs can act as Aerial Base Stations (ABSs) to extend network coverage to areas with limited terrestrial infrastructure. The key advantage of ABS is the ability to provide a LoS connection due to its maneuverability compared to fixed terrestrial infrastructure [39]. The LoS availability is critical for smart transportation in ITS, where connected vehicles rely on stable and fast communication for safe and efficient operation. Furthermore, ABSs will facilitate temporary coverage of locations such as large gatherings or high-traffic urban areas experiencing network congestion. As ITS continues to evolve, UAV-based connectivity will be the key to intelligent and effective transportation systems.

## 3.2 Air-to-Ground Channel Modeling

By **Abdul Saboor, Evgenii Vinogradov**

One of the most important requirements for integrating UAVs into ITS is efficient and reliable communication between UAVs and ground-based users or UE. Compared to terrestrial wireless networks, where signal propagation is largely impacted by buildings, terrain, or other obstructions like vehicles or vegetation, UAV-based Air-to-Ground (A2G) communication introduces a dynamic altitude parameter that offers higher elevation angles and Probability of LoS ($P_{LoS}$). However, it also faces different challenges like mobility, interference, and Path Loss (PL) sensitivity, depending on the location or elevation angle of the UAV. Therefore, accurate modeling of the A2G channel is important to optimize the placement of the UAV for the best coverage.

Furthermore, channel modeling is important to mitigate signal degradation caused by PL, fading, and interference to achieve high throughput and Ultra-Reliable Low Latency Communication (URLLC) to support mobility in ITS. This subsection will provide an overview of some existing A2G channel models and different components.

### 3.2.1 Components of A2G Channel Modeling

The A2G channel modeling consists of several components, each affecting the wireless signal distinctively. This section will discuss four major components along with common models used in the literature and standards.

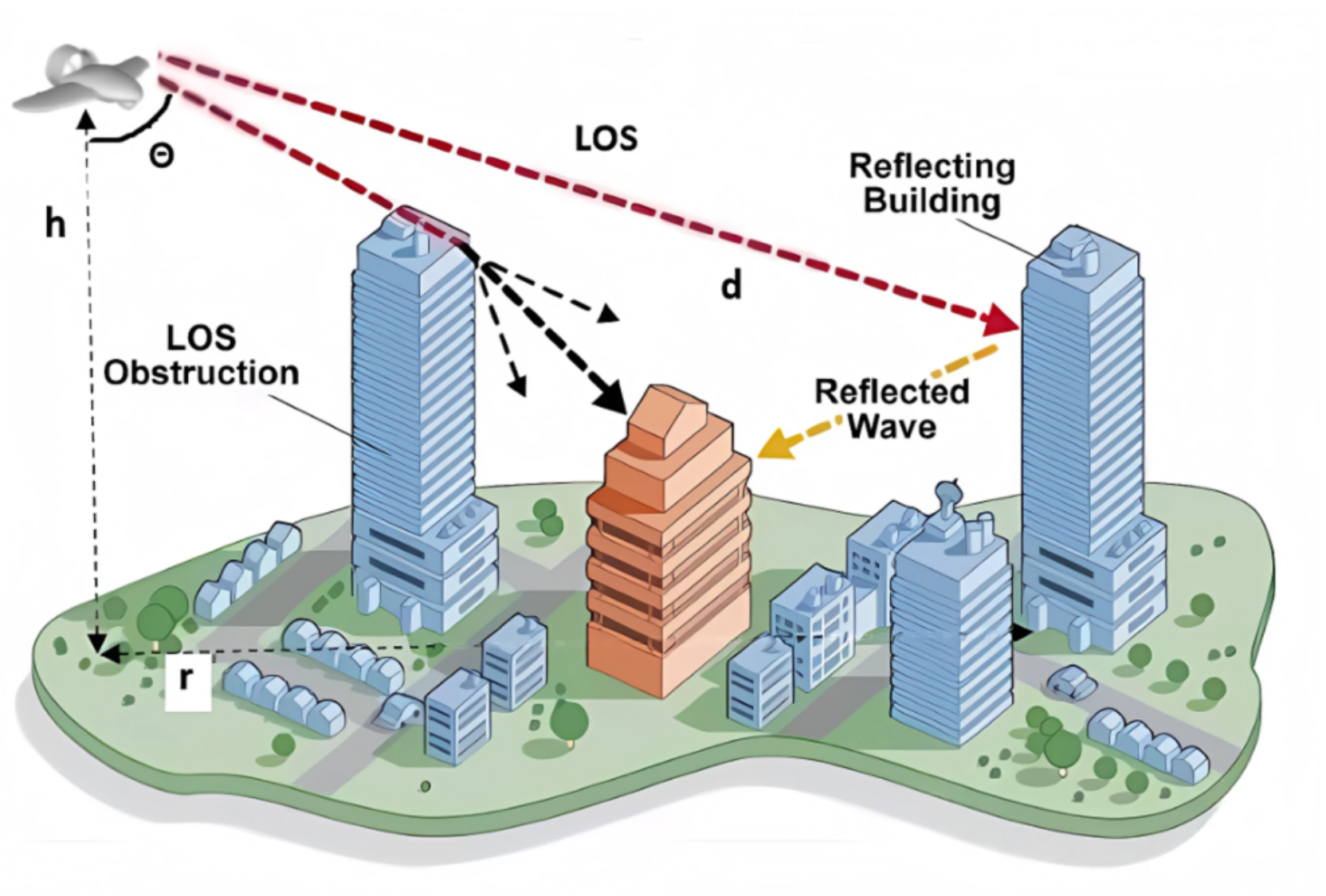

Figure 3.1: Example of LoS and NLoS links [45].

#### 3.2.1.1 Line-of-Sight modeling

The LoS refers to a direct and unobstructed path between the transmitter (UAV) and the receiver, as illustrated by the red line in Fig. 3.1. Generally, a LoS link offers a stronger signal strength and lower PL. Therefore, $P_{LoS}$ modeling is the most attractive feature of UAV-enabled communication, as it directly impacts the quality and reliability of the wireless link. A higher $P_{LoS}$ ensures more stable connectivity and data rates. It is important for efficient coverage planning, especially in dense urban environments where NLoS conditions can significantly degrade performance. In the literature, we can broadly divide $P_{LoS}$ modeling into Manhattan-based $P_{LoS}$ models and non-Manhattan-based $P_{LoS}$ models.

**Manhattan-based $P_{LoS}$ Models:** To facilitate early performance evaluation of wireless communication systems in urban environments, the International Telecommunications Unit (ITU) introduced the Manhattan grid-based urban layout [46]. This layout can be built using three parameters or built-up parameters $(\alpha, \beta, \gamma)$, as described below:

- $\alpha$**:** Ratio of the area of the buildings to the total area of the land (dimensionless);
- $\beta$**:** Average number of buildings/km$^2$;
- $\gamma$**:** Rayleigh-based parameter for controlling the building height distribution.

Table 3.1 lists the built-up parameters for the four standard urban environments defined by the ITU. It is a simplified urban layout consisting of square buildings of the same width $W$, with uniform space between them, also called streets $S$, as illustrated in Fig. 3.2. This structured environment makes it easy to analytically model propagation characteristics, such as $P_{LoS}$ as a function of the built-up parameters. Due to simplicity and regularity, numerous UAV-based A2G $P_{LoS}$ models use this structure to estimate $P_{LoS}$ in different urban environments.

| **Environment** | $\alpha$ | $\beta$ **(buildings/km$^2$)** | $\gamma$ **(m)** |
|---|---|---|---|
| Suburban | 0.1 | 750 | 8 |
| Urban | 0.3 | 500 | 15 |
| Dense Urban | 0.5 | 300 | 20 |
| Urban High-rise | 0.5 | 300 | 50 |

Table 3.1: Built-up parameters for standard environments.

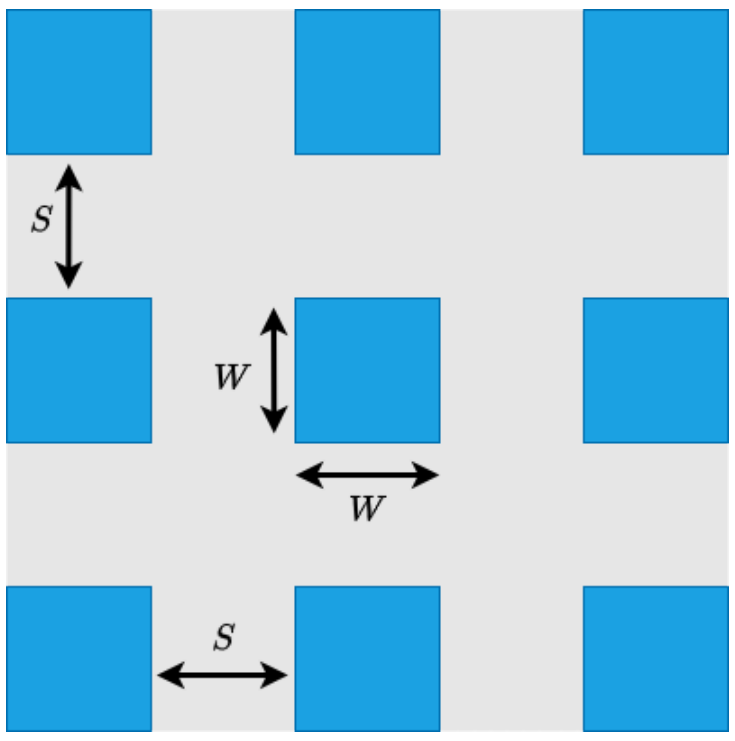

Figure 3.2: Top view of ITU-defined Manhattan layout.

Based on the above-mentioned grid structure, the ITU proposed one of the earliest analytical $P_{LoS}$ for urban environments [46]. The model estimates potential building obstructions between a UAV and a ground user by assuming evenly spaced buildings. Hence, $P_{LoS}$ is computed using:

$$P_{LoS} = \prod_{b=1}^{b_r} P(\text{building_height} < h_{LoS}) \tag{3.1}$$

Where $b_r$ is the total number of obstructing buildings between the transmitter and receiver, and $h_{LoS}$ is the height of the LoS ray at the obstruction point. The simplified expression of the ITU model is given in (3.2).

$$P_{LoS} = \prod_{n=0}^{m} \left[ 1 - \exp\left( - \left( \frac{h_{TX} - \frac{(n+\frac{1}{2})(h_{TX}-h_{RX})}{m+1}}{\sqrt{2}\gamma} \right)^2 \right) \right] \tag{3.2}$$

In the above equation, $h_{TX}$ and $h_{RX}$ are the heights of the transmitter and receiver, respectively. Whereas $m$ is the number of obstructing buildings that can be computed using $m = \text{floor}(r\sqrt{\alpha\beta} - 1)$ and $r$ is the Two-dimensional (2D) distance between the transmitter and the receiver.

Al-Hourani et al. [47] approximated the $P_{LoS}$ expression in Equation (3.2) using a simple Sigmoid function (S-curve) with parameters $a$ and $b$, as given in Equation (3.3).

$$P_{LoS}(\theta) = \frac{1}{1 + a \cdot \exp(-b[\theta - a])} \tag{3.3}$$

The $P_{LoS}$ models, presented in Equations (3.2) and (3.3), consider the direct optical path between the transmitter and receiver. Pang et al. [48] presented a modified version of the ITU $P_{LoS}$ model, where the authors consider the impact of the Fresnel zone based on the signal frequency. The final expression is given in Equation (3.4):

$$P_{LoS}(r, h_{TX}, h_{RX}, \psi, \lambda) = \prod_{i=0}^{m} \left[ 1 - \exp\left( - \frac{\left[ h_{TX} - \frac{d_i(h_{TX}-h_{RX})}{d_{TR}} - \frac{\sqrt{\lambda r}\cdot \min(d_i, r-d_i)}{\sqrt{r^2+(h_{TX}-h_{RX})^2}} \right]^2}{2\gamma^2} \right) \right] \tag{3.4}$$

where $\psi$ represents built-up parameters $(\alpha, \beta, \gamma)$, $\lambda$ is the wavelength and $d_i$ is the distance to $ith$ obstructing building.

Imran et al. [49] emphasize the importance of UAV locations or azimuth $\varphi$ on $P_{LoS}$ and derive an empirical $P_{LoS}$ expression for the four standard urban environments in the form of modified S-curve, as given in Equation (3.5).

$$P_{LoS}(\theta) = \frac{1}{1 + e^{a\theta^3 + b\theta^2 + c\theta + d}} \tag{3.5}$$

In the above equation, $a, b, c$ and $d$ are the S-curve fitting parameters for the standard urban environments given in Table 3.2.

| **Environment** | **a** | **b** | **c** | **d** |
|---|---|---|---|---|
| Suburban | $-2.791 \times 10^{-5}$ | 0.004 | $-0.2193$ | 2.2839 |
| Urban | $-2.397 \times 10^{-5}$ | 0.0034 | $-0.1985$ | 3.7876 |
| Dense Urban | $-1.984 \times 10^{-5}$ | 0.0026 | $-0.1583$ | 4.0667 |
| High-rise Urban | $-1.130 \times 10^{-5}$ | 0.000947 | $-0.0596$ | 3.5939 |

Table 3.2: $P_{LoS}$ S-curve fitting parameters for Equation (3.5).

The same authors presented another model where $P_{LoS}$ is determined only by the first obstructing building between the ground user and the UAV [50]. The closed-form expression of their first-building $P_{LoS}$ exponential model is given by Equation (3.6).

$$P_{LoS}(r, h_{TX}) = 1 - \hat{\lambda}_1 \sqrt{\pi} e^{\frac{\hat{\lambda}_1^2}{4\rho}} \cdot \frac{1}{2\sqrt{\rho}} \left[ \operatorname{erf}\left( \frac{2\rho r + \hat{\lambda}_1}{2\sqrt{\rho}} \right) - \operatorname{erf}\left( \frac{\hat{\lambda}_1}{2\sqrt{\rho}} \right) \right] \tag{3.6}$$

Here, $\rho = \frac{h^2}{2\gamma^2 r^2}$ is a scaling factor that captures the height and elevation distribution of the building. In contrast, $\hat{\lambda}_1$ is a rate parameter of the exponential distribution to model the distance to the first building. It mainly depends on the built-up parameters, which can be computed using:

$$\hat{\lambda}_1 = \frac{1}{S} \left( \frac{613}{753} - \frac{901}{2116} \cdot \frac{S}{W} + \frac{1258}{8477} \cdot \left( \frac{S}{W} \right)^2 - \frac{239}{10712} \cdot \left( \frac{S}{W} \right)^3 \right) \tag{3.7}$$

Saboor et al. [51] highlighted the key limitations of the existing models and proposed a new model incorporating both the elevation angle $\theta$ and the azimuth angle $\phi$, along with the locations of users on streets and intersections, to obtain a more accurate and generalizable 3D model $P_{LoS}$. In addition to high accuracy, the key benefit of this model is that it works for any arbitrary urban environments defined by the ITU built-up parameters, making it more realistic and applicable to real-world layouts.

Their final 3D closed-form $P_{LoS}$ expression is presented in Equation (3.8):

$$P_{\text{LoS}}(\theta, S, W) = \frac{SW}{A}(P_{\text{LoS}}^{\text{R1}} + P_{\text{LoS}}^{\text{R2}}) + \frac{S^2}{A}(P_{\text{LoS}}^{\text{R3}}). \tag{3.8}$$

In the above equation, $A$ is the total area occupied by three regions ($A = (S+W)^2 - W^2$), and $P_{\text{LoS}}^{\text{Ri}}$ is the $P_{LoS}$ at $ith$ region that can be estimated using:

$$P_{\mathrm{LoS}}^{\mathrm{Ri}}(\theta, S, W, \varphi) = \prod_{i=1}^{n} P_{\mathrm{LoS}}^{\mathrm{s}}(\theta, S'_{Ri}, W')_i \tag{3.9}$$

$P_{\mathrm{LoS}}^{\mathrm{s}}$ represents the LoS probability in the presence of single *ith* obstructing building that can be approximated using Equation (3.10).

$$P_{\mathrm{LoS}}^{\mathrm{s}}(\theta, S, W)_i = 1 - \sqrt{\frac{\pi}{2}}\frac{\gamma}{h_{\mathrm{Bmax_1}}} \times \left(\mathrm{erf}\left(\frac{h_{\mathrm{Bmax_i}}}{\sqrt{2}\gamma}\right) - \mathrm{erf}\left(\frac{h_{\mathrm{Bmin_i}}}{\sqrt{2}\gamma}\right)\right) \tag{3.10}$$

where the $\kappa 1_i$, $\kappa 2_i$, $h_{\mathrm{Bmin_i}}$ and $h_{\mathrm{Bmax_i}}$ are calculated for the $i$-th building using:

$$\begin{aligned} \kappa 1_i &= (i-1) \times (S+W), \\ \kappa 2_i &= \kappa 1_i + S, \end{aligned} \tag{3.11}$$

and

$$\begin{aligned} h_{\mathrm{Bmin_i}} &= \kappa 1_i \tan(\theta), \\ h_{\mathrm{Bmax_i}} &= \kappa 2_i \tan(\theta). \end{aligned} \tag{3.12}$$

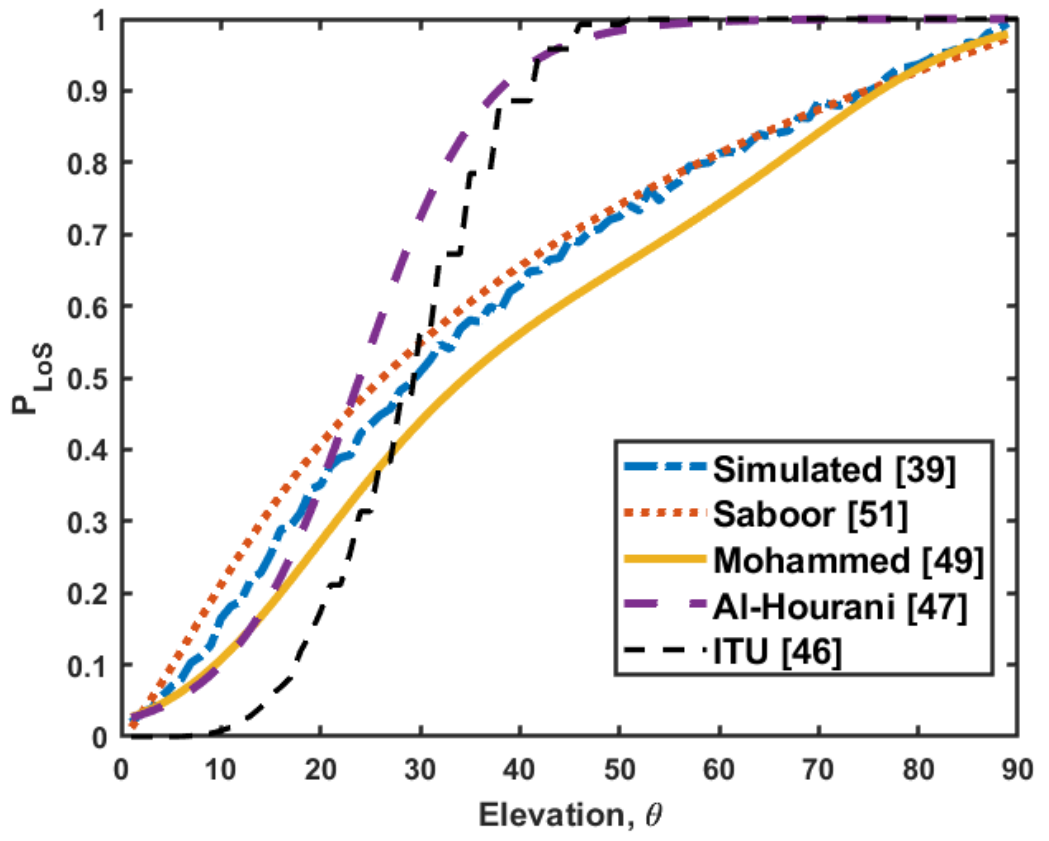

(a) Urban Environment

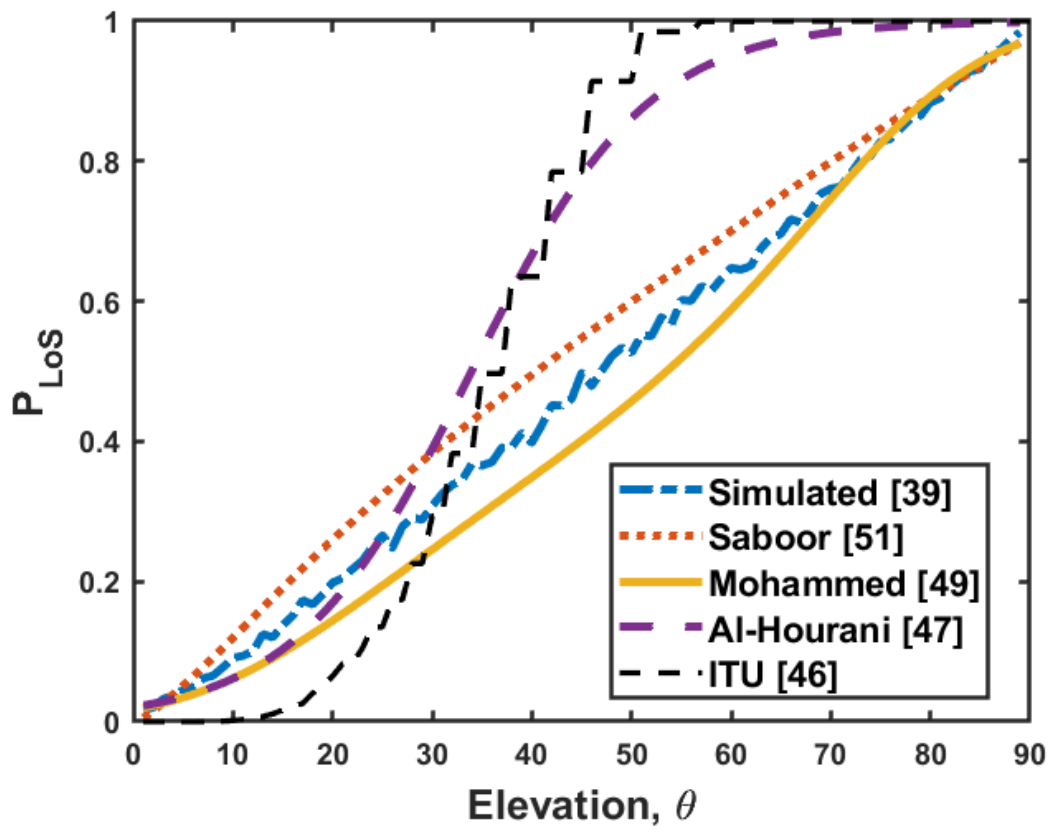

(b) Dense Urban Environment

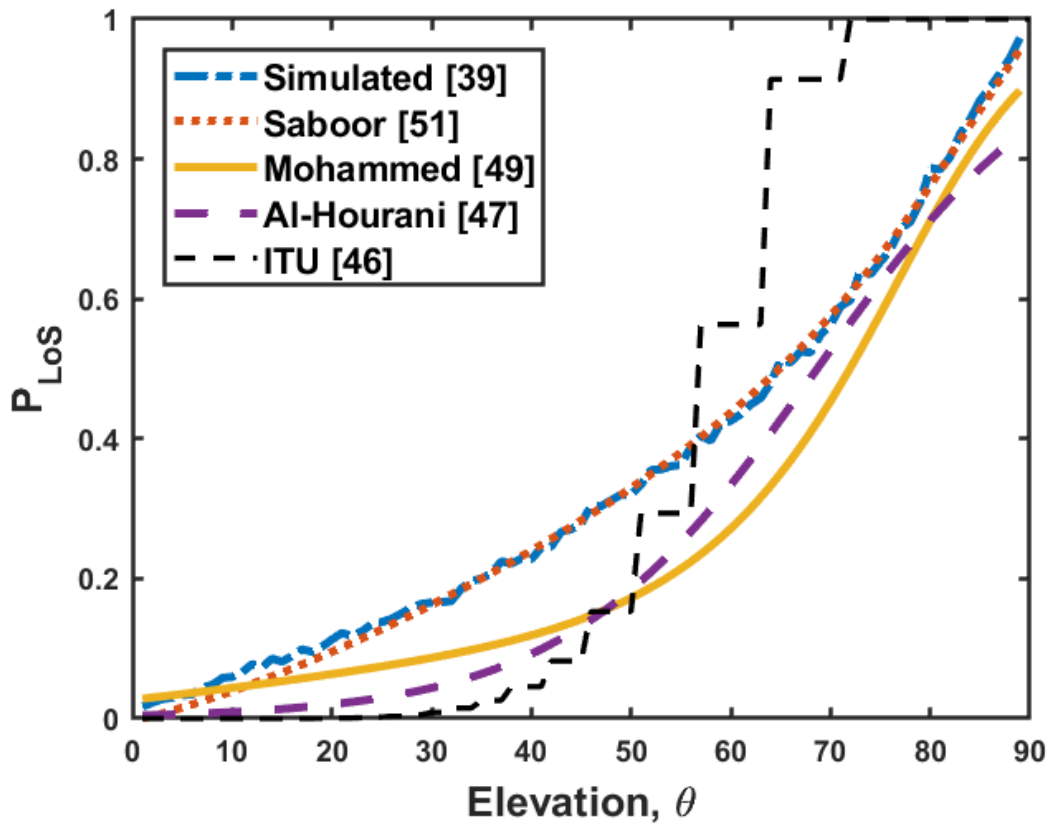

(c) High-rise Environment

Figure 3.3: $P_{LoS}$ comparison of various analytical models with the 3D simulator for standard urban environments.

Fig. 3.3 compares the $P_{LoS}$ results extracted via a 3D simulator presented in [39] with some analytical models [46], [47], [49], [51]. The model presented in [51] shows better alignments with the simulated results, demonstrating its higher precision due to the incorporation of the impact of elevation and azimuth angles and user locations.

The authors in [52] further extended the $P_{LoS}$ model in [51] that can distinguish between roads and sidewalks. Their proposed model incorporates $\varphi$ and models individual regions, sidewalks and roads to offer a more realistic model for pedestrian and vehicle users. The final $P_{LoS}$ expression is presented in Equation (3.13).

$$P_{LoS}^{\kappa}(\theta, S, W) = \frac{1}{M}\sum_{m=1}^{M}(\chi_1^{\kappa} P_{LoS}^{\chi_1,\kappa} + \chi_2^{\kappa} P_{LoS}^{\chi_2,\kappa} + \chi_3^{\kappa} P_{LoS}^{\chi_3,\kappa}) \tag{3.13}$$

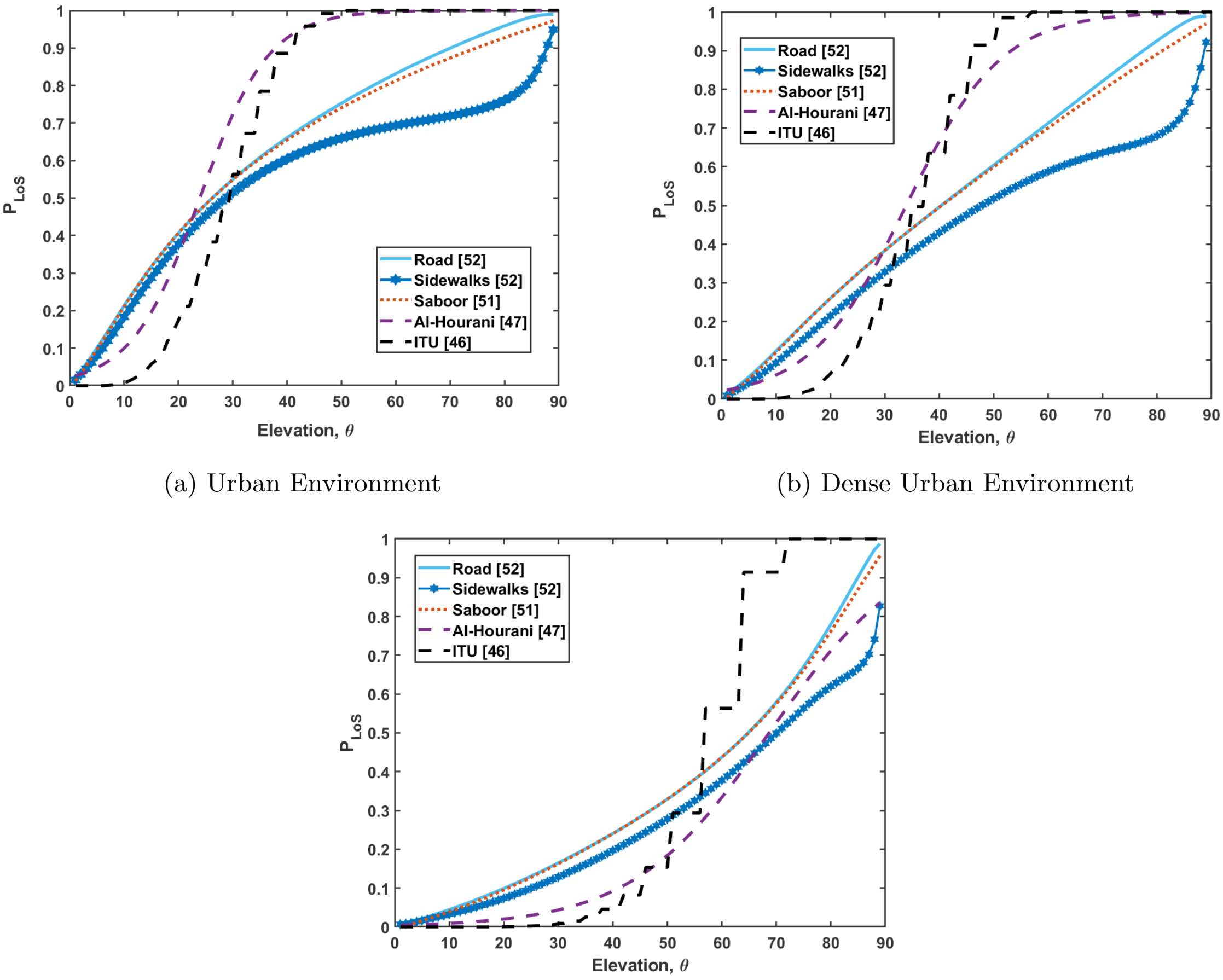

(a) Urban Environment

(b) Dense Urban Environment

(c) High-rise Environment

Figure 3.4: $P_{LoS}$ comparison of road and sidewalks compared to the existing analytical models.

Lastly, Fig. 3.4 compares the $P_{LoS}$ results proposed in [52] for sidewalk and road users against existing analytical models, highlighting a significant gap in $P_{LoS}$ for sidewalk users. In ITS, pedestrians and vehicles experience mainly different link conditions. Therefore, using application-specific models is important to ensure accurate performance evaluation and reliable communication in ITS.

**Non-Manhattan-based $P_{LoS}$ Models:** This section discusses popular $P_{LoS}$ models that do not follow the Manhattan grid environment/layout. One of the most popular models is proposed by the 3GPP for Urban Macro (UMa), Urban Micro (UMi), and Rural Macro (RMa) environments in their technical report 36.777 [53]. These models define $P_{LoS}$ as a function of horizontal 2D distance $r$ and UAV height $h_{TX}$, as given in Equation (3.14).

$$P_{LoS}^{env} = \begin{cases} \frac{d_1}{r} + \exp\left(-\frac{r}{p_1}\right)\left(1 - \frac{d_1}{r}\right), & r > d_1 \\ 1, & r \leq d_1 \end{cases} \tag{3.14}$$

where $env$ defines a particular environment and ($p_1$, $d_1$) are empirical parameters for a particular environment, which are presented in Table 3.3. These models are applicable within specific UAV height ranges, with $h_{TX}$ typically varying from 1.5 m up to 300 m depending on the scenario [53].

| **Environment** (*env*) | $p_1$ | $d_1$ |
|---|---|---|
| RMa | $\max(15201\log_{10}(h_{TX}) - 16053,\ 1000)$ | $\max(1350.8\log_{10}(h_{TX}) - 1602,\ 18)$ |
| UMa | $4300\log_{10}(h_{TX}) - 3800$ | $\max(460\log_{10}(h_{TX}) - 700,\ 18)$ |
| UMi | $233.98\log_{10}(h_{TX}) - 0.95$ | $\max(294.05\log_{10}(h_{TX}) - 432.94,\ 18)$ |

Table 3.3: Environment-specific empirical parameters for 3GPP $P_{LoS}$ model.

Al-Hourani [54] used a stochastic geometry approach, where they modeled buildings as cylinders with log-normally distributed heights. Their proposed model accounts for both equal-height and arbitrary-heights for transmitter and receiver, making it applicable for both UAV-UAV and UAV-users $P_{LoS}$ estimation. If the arbitrary heights of the transmitter and receiver ($h_1$ and $h_2$) are separated by a horizontal distance $r$, the final $P_{LoS}$ can be computed using the Equation 3.15.

$$P_{LoS}(r, h_1, h_2) = \exp\left(-2r_o\lambda_o \int_0^{d-\frac{\pi}{2}r_o} G(h)\,dx\right) \tag{3.15}$$

where the height of the LoS ray at distance $x$ along the link is:

$$h = \frac{x}{r}(h_2 - h_1) + h_1 \tag{3.16}$$

and $G(h)$ is the Complementary Cumulative Distribution Functions (CCDF) (tail) of the log-normal height distribution:

$$G(h) = 1 - F(h) = \frac{1}{2} - \frac{1}{2}\,\mathrm{erf}\left(\frac{\ln h - \mu_o}{\sqrt{2}\sigma_o}\right) \tag{3.17}$$

In the above model, $r_o$ is the average radius of the building, $\lambda_o$ is the buildings/m$^2$, $\mu_o$ and $\sigma_o$ are logarithmic normal parameters for the height of the building.

Lastly, the authors in [55] proposed a simulator that generates three different urban layouts based on ITU-defined built-up parameters. By simulating UAV–UE links in these layouts, the authors provide $P_{LoS}$ expression in the form of two Sigmoid functions, similar to Equations (3.3) and (3.5). However, the key conclusion is that Manhattan-based $P_{LoS}$ models can still provide reasonably accurate results for different non-Manhattan urban layouts as long as they use the same ITU-defined built-up parameters. In addition to the above-mentioned models, several other $P_{LoS}$ models, such as [56], [57], have not been discussed in this white paper.

#### 3.2.1.2 Path Loss modeling

Precise PL modeling is vital for evaluating wireless link performance, particularly for UAV-based communications in urban environments where links may alternate between LoS and NLoS. Therefore, PL modeling is critical in ITS, where reliability, high throughput, and low latency are critical to various applications. For example, applications such as connected vehicles or traffic management need a precise estimate of signal attenuation to provide reliable A2G communication.

The above sections covered various $P_{LoS}$ for urban environments. These $P_{LoS}$ models serve as a critical input to PL models, which typically incorporate both LoS and NLoS conditions with the expression given in Equation (3.18).

$$PL(d) = P_{LoS}(d) \cdot PL_{LoS}(d) + (1 - P_{LoS}(d)) \cdot PL_{NLoS}(d) \tag{3.18}$$

where $PL_{LoS}(d)$ and $PL_{NLoS}(d)$ are the path loss values under LoS and NLoS conditions, respectively, as functions of the link distance $d$.

Several approaches have been proposed in the literature to model these two components depending on the application scenario and frequency band. Some models rely on empirical measurements and curve fitting, while others adopt geometry-based formulations derived from analytical approximations. LoS conditions generally follow a Free Space Path Loss (FSPL) model given in equation (3.19).

$$FSPL(d) = 20\log_{10}(d) + 20\log_{10}(f_c) + 20\log_{10}\left(\frac{4\pi}{c}\right) \tag{3.19}$$

In contrast, NLoS path loss incorporates additional losses due to diffraction, reflection, or penetration. This subsection discusses a few popular PL models proposed in the literature. 3GPP proposed standardized path loss models for UAVs in RMa, UMa, and UMi environments in their technical report 36.777 [53]. These models provide separate PL expressions for LoS and NLoS conditions based on the 3D link distance $d_{3D}$, UAV height $h_{TX}$, and frequency $f_c$ in GHz.

**3GPP Path Loss Model:**

$$PL_{LoS}^{RMa} = \max\left(23.9 - 1.8\log_{10}(h_{TX}), 20\right)\log_{10}(d_{3D}) + 20\log_{10}\left(\frac{40\pi f_c}{3}\right) \tag{3.20}$$

$$PL_{NLoS}^{RMa} = \max\left(PL_{LoS}^{RMa},\ -12 + (35 - 5.3\log_{10}(h_{TX}))\log_{10}(d_{3D}) \quad +20\log_{10}\left(\frac{40\pi f_c}{3}\right)\right) \tag{3.21}$$

$$PL_{LoS}^{UMa} = 28.0 + 22\log_{10}(d_{3D}) + 20\log_{10}(f_c) \tag{3.22}$$

$$PL_{NLoS}^{UMa} = \ -17.5 + (46 - 7\log_{10}(h_{TX}))\log_{10}(d_{3D}) + 20\log_{10}\left(\frac{40\pi f_c}{3}\right) \tag{3.23}$$

$$PL_{LoS}^{UMi} = \max\left(FSPL,\ 30.9 + (22.25 - 0.5\log_{10}(h_{TX}))\log_{10}(d_{3D}) + 20\log_{10}(f_c)\right) \tag{3.24}$$

$$PL_{NLoS}^{UMi} = \max\left(PL_{LoS}^{UMi},\ 32.4 + (43.2 - 7.6\log_{10}(h_{TX}))\log_{10}(d_{3D}) + 20\log_{10}(f_c)\right) \tag{3.25}$$

**Log-distance Path Loss Model:** One of the most commonly used path loss models is the log-distance path loss model [58], [59], [60], [61], which expresses signal attenuation as a logarithmic function of distance. The general form of this model is given by:

$$PL(d) = PL(d_0) + 10n\log_{10}\left(\frac{d}{d_0}\right) + X_\sigma \tag{3.26}$$

where $PL(d_0)$ is the reference FSPL at distance $d_0$ and frequency $f_c$. The term $10\,n\log_{10}\left(\frac{d}{d_0}\right)$ is the excess PL dependent on the presence of LoS/NLoS link. $n$ is the path loss exponent that quantifies how quickly a signal's strength decreases with distance in a particular environment. $X_\sigma$ is a Gaussian random variable with standard deviation $\sigma$ that represents shadow fading, which will be discussed in the following subsection.

Yanmaz et al. [58] analyzed A2G communication by measuring Receiver Signal Strength (RSS) over varying horizontal and vertical distances in open space. The authors used the log-distance PL model and computed the path loss exponent $n \approx 2.01$ for A2G links. In contrast, the authors in [59] analyzed A2G PL using 3D ray-tracing simulations in an urban environment. Their proposed model mainly separates LoS and NLoS conditions and computes the path loss exponent across UAV heights ranging from 6 m to 150 m for each condition. Lastly, they used the log-distance path loss model to fit the data for both conditions, where $n$ ranges from 2 to 2.6 for LoS and from 5.6 to 8.8 for NLoS.

Ruoyu et al. [60] presented A2G channel measurements for mountainous and hilly terrain, using over 18 million Power Delay Profile (PDP) from the National Aeronautics and Space Administration (NASA). The authors develop a modified log-distance PL model, incorporating flight direction effects and terrain roughness to better capture real-world signal behavior. Their key findings indicate that $n$ ranges from 2.2 to 3.5 depending on elevation and terrain. Finally, Coene et al. [61] performed a measurement-based comparison of the log-distance model with the sin-log-elevation model [62] and showed that the log-distance model provides equal fitting, while remaining simpler.

| Log-Distance Path Loss Model | | | | |
|---|---|---|---|---|
| **Study** | **Environment** | **Freq. (GHz)** | $PL(d_0)$ | **Exponent** $n$ |
| [58] | Open Field | 5 | 46.4 | 2.01 (U2G), 2.03 (U2U) |
| [59] | Urban | 2–6 | – | 2.2 (LoS), 6 (NLoS) |
| [60] | Hilly / Mountainous | 0.968 (L-band)<br>5.060 (L-band) | – | 1.6–2.2 (L), 1.5–2.2 (C) |
| **AB Path Loss Model** | | | | |
| **Study** | **Environment** | **Freq. (GHz)** | $A$ (LoS/NLoS) | $B$ |
| [63] | Dense urban | 28<br>73 | 61.4/72.0<br>69.8/86.6 | 2.0 / 2.92<br>2.0 / 2.45 |
| [64] | Sub-Urban (Victoria) | 28<br>73 | 84.64 / 113.63<br>93.63 / 115.40 | 1.55 / 1.16<br>1.52 / 1.43 |
| | Urban (Paris) | 28<br>73 | 82.54 / 97.81<br>90.86 / 100.83 | 1.68 / 1.87<br>1.69 / 2.09 |
| | Dense Urban (Mumbai) | 28<br>73 | 78.58 / 98.05<br>85.71 / 105.37 | 1.85 / 1.86<br>1.90 / 1.91 |
| | High-Rise Urban (NYC) | 28<br>73 | 88.76 / 66.25<br>102.10 / 88.76 | 1.68 / 3.30<br>1.92 / 2.22 |
| [65] | Urban | 28 | 43.90 / 46.55 (No Trees / Trees) | 3.38 |
| | Dense Urban | 28 | 40.83 / 43.52 (No Trees / Trees) | 3.75 |
| | High-Rise | 28 | 38.64 / 40.26 (No Trees / Trees) | 4.26 |

Table 3.4: Overview of parameters in Log-Distance and AB Path Loss Models.

**Alpha-Beta (AB) Path Loss Model:** Alpha-Beta (AB) model is a variation of the log-distance $PL$ model commonly used in A2G communication. Unlike the log-distance model that fixes the path loss at a reference distance (e.g., free-space at 1 m), the AB model introduces two flexible parameters: $B$ (the slope) and $A$ (intercept). The AB PL model is given in Equation (3.27).

$$PL(d) = A + B \cdot 10 \log_{10}(d) + X_\sigma \tag{3.27}$$

Akdeniz et al. [63] used the above-mentioned AB PL model for mmWave communication at 28 GHz and 73 GHz in dense urban environments. The model estimates the omnidirectional PL using measurement data collected in New York City. Later, the authors modeled LoS and NLoS conditions separately. At 28 GHz, the LoS scenario yields $A = 61.4$, $B = 2.0$, while the NLoS gives $A = 72.0$, $B = 2.92$. Similarly, for 73 GHz, the LoS parameters are $A = 69.8$, $B = 2.0$, whereas the NLoS are $A = 82.7$, $B = 2.69$.

From the findings, we can observe that stronger attenuation is observed in NLoS mmWave links. In contrast, the authors in [64] propose a Ground-to-Air (G2A) PL model for mmWave (28 and 73 GHz) for four standard urban environments: suburban, urban, dense urban, and high-rise. Based on simulations conducted in cities like Victoria, Paris, Mumbai, and New York, the model fits the PL data using AB model. At 28 GHz in the dense urban environment, the LoS parameters are $A = 78.58$, $B = 1.85$, while the NLoS parameters are $A = 98.05$, $B = 1.86$. More detail is presented in Table 3.4.

Saboor et al. [65] use simulated data to model A2G PL in urban environments with irregular building shapes, non-uniform spacing, and random obstacles such as trees and streetlights. Using this, the authors derive an empirical log-distance AB model for standard urban, dense urban, and high-rise environments at 28 GHz, highlighting how foliage impacts AB (see Fig. 3.5). Their findings indicate that obstacles like trees increase the intercept $A$ due to added attenuation. At the same time, the slope $B$ remains stable, as given in Table 3.4.

**Two-Ray Path Loss Model:** The Two-Ray PL model considers surface-reflected paths between a transmitter and a receiver in addition to the direct LoS link. It is particularly relevant in open and flat G2A scenarios, where ground reflections significantly shape the received signal. The model assumes that the total propagation path satisfies $r_1 + r_2 \approx d$ and is expressed as:

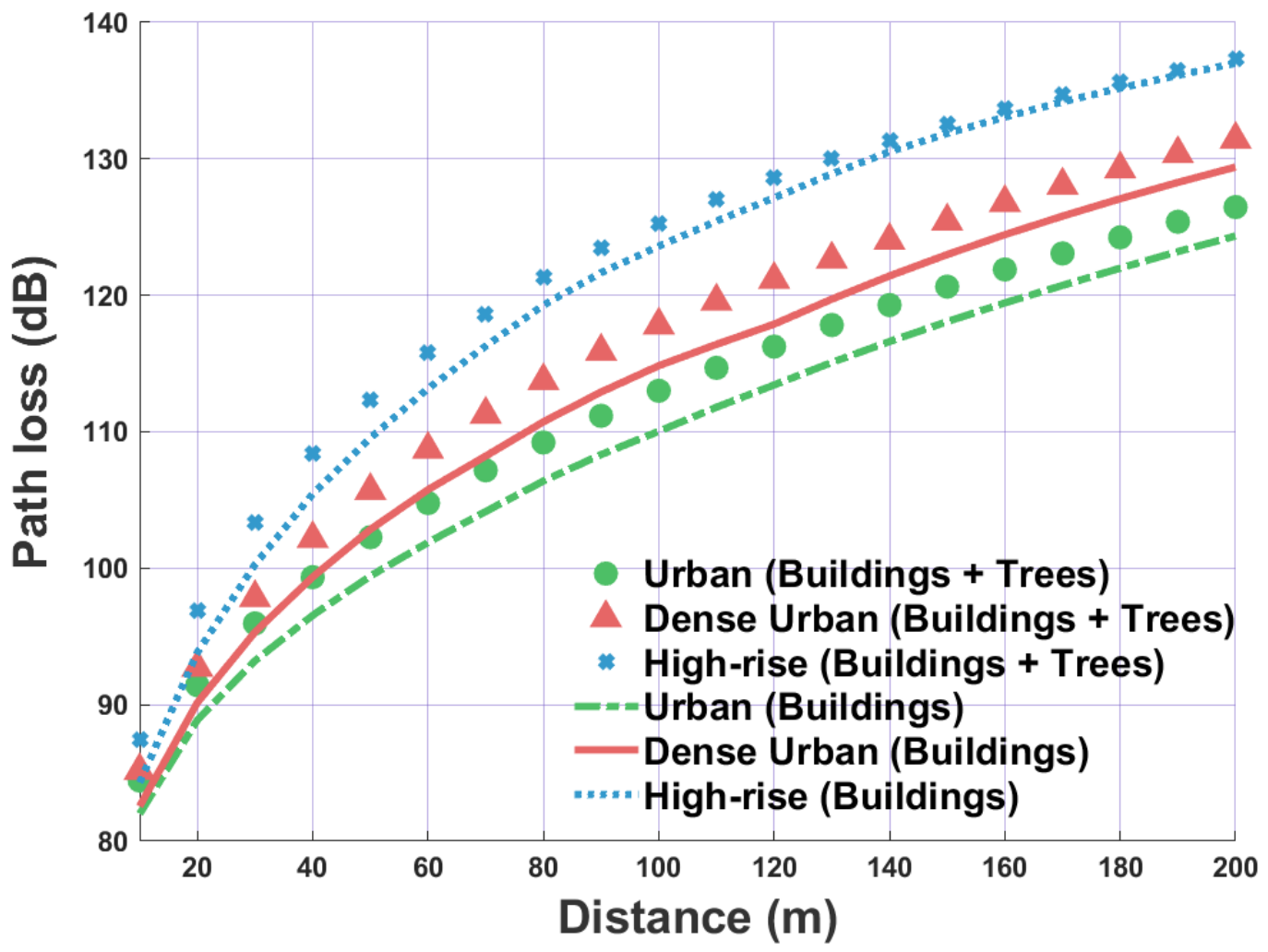

Figure 3.5: PL in standard urban environments with and without 200 trees.

$$PL(d) = -10\log_{10}\left\{\left(\frac{\lambda}{4\pi r}\right)^2\left[2\sin\left(\frac{2\pi h_{RX}h_{TX}}{\lambda r}\right)\right]^2\right\} \tag{3.28}$$

This model captures constructive and destructive interference patterns due to path differences. In addition to these models, various other models like the Two-Ray, Knife-Edge Diffraction, air-to-air, and modified 3GPP PL models exist in the literature [66], [67], [68], [69], which are not discussed in this white paper.

#### 3.2.1.3 Shadow Fading Modeling

Shadow fading represents the large-scale signal variation due to obstructions such as buildings, trees, and other structures. This fading effect can be modeled (in dB) as a zero-mean Normal variable, denoted by $\xi$ described by its standard deviation (it is often LoS-dependent). Moreover, when an UAV communicates with a mobile user, we need to take into account the spatial domain. Hence, the shadow fading is spatially correlated with an autocorrelation function:

$$R(\Delta d) = e^{-\frac{\Delta d}{d_{\mathrm{decorr}}}}, \tag{3.29}$$

where $\Delta d$ is the distance between two points and $d_{\mathrm{decorr}}$ is the decorrelation distance. The decorrelation distance indicates the spatial extent over which the shadowing effect is correlated, and it may vary depending on the environment, UAV altitude, and whether the link is LoS or NLoS.

In suburban or rural areas with fewer obstructions, we could expect a long LoS decorrelation distance, however, the measurements report rather short values below 5 m [70] in contrast to 9.5 – 12.9 m in urban environment [71]. NLoS shadow fading decorrelation distances are typically shorter than in the LoS case. In urban environments, the decorrelation distance $d_{\mathrm{decorr}}$ may become long in NLoS scenarios, since the shadow effect is caused by large buildings. An overview of the parameters extracted from relevant measurement campaigns can be found in Table 3.5.

#### 3.2.1.4 Small-Scale Fading Modeling

Small-scale fading is caused by multipath propagation resulting in multiple signal copies arriving at the receiver with different delays, phases, and amplitudes. In A2G communication, the propagation environment and UAV altitude strongly influence the fading characteristics.

The Rician distribution is the most commonly adopted model for small-scale fading in A2G propagation [72], [73], since A2G channels often exhibit a more dominant LoS component compared to terrestrial links. The Rician $K$-factor, which denotes the ratio of power in the LoS path to the scattered paths, is typically higher in A2G than in terrestrial scenarios. In environments where the LoS component is obstructed, such as urban canyons or deep NLoS conditions, the Rayleigh fading model provides a better fit [73]. Other statistical distributions have also been explored in the literature, including Nakagami-m, Chi-squared ($\chi^2$), and Weibull [72].

| Ref. | Environment | Freq. GHz | Std. Dev. [dB] | Decorr. Dist. [m] |
|---|---|---|---|---|
| [53] | RMa | | LoS: $4.2\exp(-0.0046h_{TX})$ | - |
| | | | NLoS: 6 | - |
| | UMa | 0.5 - 100 | LoS: $4.64\exp(-0.0066h_{TX})$ | - |
| | | | NLoS: 6 | - |
| | UMi | | LoS: $max(2, 5\cdot\exp(-0.01h_{TX}))$ | - |
| | | | NLoS: 8 | - |
| [60] | Hilly / | 0.968 | 3.2 – 3.9 | - |
| | Mountainous | 5.060 | 2.2 – 2.8 | - |
| [70] | Suburban | 2.4 | LoS: 2.13 – 2.71 | LOS: 2.32 – 5.15 |
| | | | NLoS: 2.54 – 3.29 | NLoS: 0.79 – 1.38 |
| | | 5.9 | LoS: 2.42 – 3.5 | LoS: 1.31 – 1.92 |
| | | | NLoS: 2.19 – 2.58 | NLoS: 0.54 – 0.62 |
| | Urban | 2.4 | NLoS: 3.17 – 3.4 | NLoS: 59.1 – 65.2 |
| [71] | Urban | 1.8 | 2 – 4 | 9.5 – 12.9 |

Table 3.5: Reported Shadow Fading Parameters in A2G Channels.

3GPP has proposed several methods for incorporating small-scale fading into UAV-specific propagation models [53]. One simplified approach assumes a fixed Rician $K$-factor of 10 or 20 dB for NLoS and LoS, respectively, and reuses all other parameters (e.g., delay and angular spreads, number of clusters, correlation matrices) from terrestrial models [74]. Analogously, such a hybrid model can rely on the channel model parameters from other vehicular small-scale fading models, for instance, such as those presented in Chapter 5 and detailed in Table 5.1.

Although the importance of modeling nonstationary channel behavior is widely recognized [75], only a few geometry-based channel models deal with this aspect [76], [77]. Moreover, there is a lack of work dedicated to measurement-based stationarity definition. Colpaert et al. [78] evaluate temporal (can be converted to distance), frequency, and spatial stationarity regions based on Massive MIMO measurements. They report that (i) the stationarity distance ranges from 2.2 – 3.6 m; (ii) larger RMS delay spread, corresponding to a narrower coherence bandwidth (8 – 9 MHz), is observed when the altitude of the UAV increase; (iii) spatial stationarity is comparable with terrestrial channels.

## 3.3 Cooperative Awareness

By **Sandaruwan Jayaweera, Konstantin Mikhaylov, Matti Hämäläinen**

The ETSI technical committee on ITS has made significant contributions to vehicular communications, particularly through the specification of the Cooperative Awareness (CA) basic service as part of the Basic Set of Applications (BSA) [79]. This service enables the continuous exchange of dynamic status information between vehicles and infrastructure, enhancing situational awareness and supporting safety-critical applications in connected and autonomous transportation systems.

CA service is mandatory for any actor within the ITS. It allows road users and roadside infrastructure to share information about each other's position, speed, acceleration, and many different dynamics and attributes. In a highly dynamic environment with short-life communications and severe fading effects, systems require periodic broadcasts to keep the information updated. ETSI proposes a Cooperative Awareness Message (CAM) structure for ground vehicles and infrastructure containing the aforementioned information fields to broadcast at a rate between 1 to 10 Hz [79].

The high-level structure of this ETSI CAM is depicted in Fig. 3.6a. These information fields are specifically designed for ground vehicle applications, making the ETSI CAM less effective for UAV-based CA. As a result, most ITS-UAV research has focused on utilizing UAVs as communication relays rather than fully integrated ITS stations. This limitation prevents UAVs from directly decoding CAM or other ITS messages, restricting their ability to participate in cooperative vehicular communication beyond simple data forwarding.

One of the key aspects often overlooked in UAV-focused V2X research is the potential benefits UAVs can gain from full integration into ITS. ITS provides a standardized framework of communication protocols designed to enhance safety and efficiency for all users. Leveraging ITS is essential for advancing UAV operations within complex airspaces and diverse operational environments.

CA, a fundamental component of ITS, plays a crucial role in enabling UAVs to operate safely, efficiently, and collaboratively in shared airspace. CA involves sharing real-time data about UAV positions, velocities, intentions, and environmental conditions among UAVs and other entities, such as ground control stations and manned aircraft. CA enhances the situational awareness of UAVs, facilitating collision avoidance and efficient navigation. Furthermore, this capability is beneficial for coordinated UAV operations such as swarm deployments, urban air mobility, and air traffic management, where seamless integration with existing aviation infrastructure is required.

In addition, CA optimizes energy consumption, minimizes operational inefficiencies, and supports regulatory compliance by enabling adherence to airspace restrictions and safety protocols. As UAVs become increasingly autonomous, CA will play a pivotal role in ensuring resilience to disruptions, such as Global Positioning System (GPS) jamming or communication failures, while fostering public trust through improved transparency and accountability. Ultimately, this approach contributes to the broader integration of UAVs into controlled airspace, enabling scalable and sustainable aerial operations across various applications, including disaster response, precision agriculture, and urban logistics.

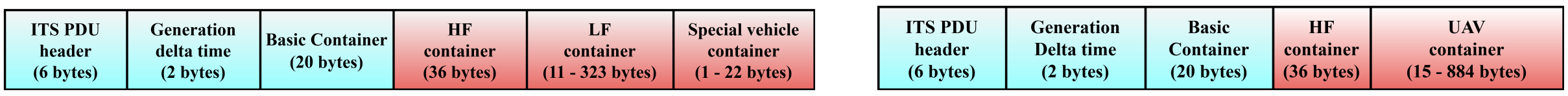

(a) (b)

Figure 3.6: The high-level structure of (a) ETSI CAM, (b) modified UAV-CAM.

Therefore, modifications to the existing ETSI CAM specification are required to provide ITS benefits to UAVs. One major shortcoming of the current CAM structure is its inability to account for the 3D mobility of UAVs. Addressing this shortcoming, a modified UAV-CAM specification that supports 3D mobility, path planning, and other UAV-specific dimensions has been introduced in [80]. The high-level structure of this suggested UAV-CAM is shown and compared against conventional CAM in Fig. 3.6.

Established vehicular communication standards such as IEEE 802.11p, IEEE 802.11bd, LTE-V2X, and NR-V2X offer high data rates, low latency, and reliable connectivity, making them promising candidates for UAV-CAM [81]. However, their size, power consumption, and infrastructure dependencies may pose challenges for smaller UAVs or those operating in constrained environments. Consequently, alternative communication technologies such as Bluetooth Low Energy (BLE) present a viable solution for lightweight UAVs, ensuring efficient cooperative awareness while maintaining energy efficiency.

The feasibility of utilizing BLE for UAV-CAM in miniature UAVs has been demonstrated through BLE-based proof-of-concept implementations [82]. The reported experiment demonstrates the feasibility of sharing Cooperative Information between UAVs operating up to 90 m apart while utilizing BLE 2M Physical layer (PHY). In a separate evaluation of BLE performance for UAV-CAM, a communication range of up to 75 m was achieved using the BLE 1M PHY [83]. As an extension of this work, the experimental (Exp.) Packet Error Rate (PER) performance of BLE Coded PHY against distance is depicted in Fig. 3.7. The results are compared against analytical models from literature [83].

The analytical models for PER have been derived from path loss models in the literature. As no universally accepted A2A path loss model exists in the literature, four path loss models are applied to provide a more reliable estimation of communication performance. They are,

- a probabilistic two-ray path loss (Prob. TR) model which has been derived in [84] for urban building-heavy environments,
- the two-Ray (TR) Ground Reflection Model from [85] derived for large separation distances,
- the FSPL model from Equation (3.19),
- the A2A Two-Ray (A2AT-R) Model from [86], which has been proposed to model the effect of ground reflection component accurately.

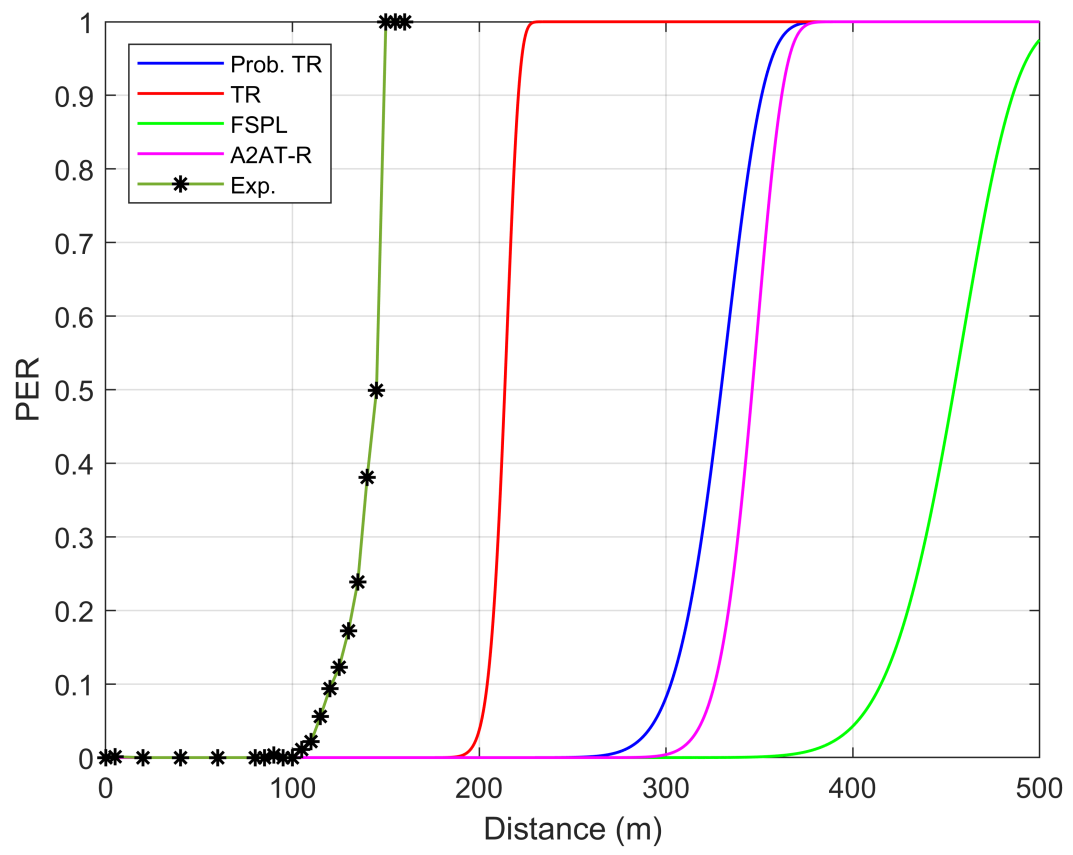

Figure 3.7: PER vs distance for UAV-CAM operating over BLE Coded PHY.

The analytical models predicted communication ranges between 200 − 400 m, depending on the propagation assumptions. Specifically, the TR model indicated a range of approximately 200 m, while the Prob. TR and A2AT-R models predicted ranges closer to 300 m. The FSPL model provided the most optimistic estimate, with expected ranges of around 400 m. In contrast, the experimental results revealed a considerable discrepancy from these predictions. While the models suggested a minimum achievable range of 200 m, the measurements showed that reliable communication could only be maintained up to about 100 m.

Despite this reduction, the experimentally observed range still holds practical significance for miniature UAV operations. At approximately 100 m, two drones approaching each other at the maximum speed permitted under EU regulations (19 m/s) [87] would still have a time window of about 2.5 s to exchange 50 UAV-CAM at a 10 Hz rate.

Future research should focus on achieving interoperability between diverse communication technologies, enabling seamless data exchange across different UAV platforms. Such advancements will be essential for realizing a comprehensive CA framework that maximizes the benefits of ITS for UAV operations across a wide range of applications. Successful deployment of UAV-CA will facilitate many other UAV-based applications, such as drone-mounted base station deployment, which will be discussed in the next Subsection.

## 3.4 Drone-mounted Base Stations

By **Laura Finarelli, Gianluca Rizzo**

Network densification, already a cornerstone of 5G technology [88], addresses the challenge of growing traffic demand by deploying a denser network infrastructure of small cells. However, this comes at a significant cost. Densification leads to a double whammy on network operators, driving up both Capital Expenditure (CAPEX) and Operational Expenditure (OPEX). Additionally, densification often relies on over-provisioning, deploying excess network resources to cope with unpredictable traffic fluctuations. This approach, while ensuring network stability and high service availability, is inherently inefficient.

To overcome these limitations, a promising avenue lies in the concept of Moving Networks (MNs) [89], [90]. This paradigm is based on the integration of Mobile Base Stations (MBSs), i.e., small cells mounted on vehicles or other mobile platforms, which can be strategically deployed to provide additional capacity exactly where and when needed, catering to localized traffic surges and thus potentially reducing over-provisioning needs.

Several works have focused on optimizing MBS positioning and addressing mobility-related challenges. [91] proposes an optimization algorithm to ensure fairness among UEs, while maximizing the throughput in two-tier networks with macro cells and cognitive micro cells. However, they do not account for the impact of backhauling.

A similar analysis with relay nodes, instead of MBS, was conducted in [92]. All these works, however, consider static scenarios and thus do not account for MBSs' ability to dynamically and naturally adapt the base station density to both spatial and temporal variations in user density and traffic patterns, particularly in urban settings [93], thanks to the correlation between cell user densification and vehicular densification patterns. [94] characterizes the correlation between patterns of mobile user densities and those of vehicles in a set of realistic urban scenarios. This is not even necessary if we consider drones that are directly operated by mobile network operators.

However, all these works do not allow for quantifying the reduction in the amount of network resources required to serve users with a target Quality of Service (QoS), which MNs enable with respect to traditional, static base station deployments (thus potentially reducing also the overall energy footprint of the network).

UAVs are emerging as promising platforms for antenna deployment thanks to their unique operational capabilities. For example, UAVs can serve as MBSs, providing flexible and rapidly deployable communication infrastructure.

A substantial body of literature investigates the integration of UAVs within cellular networks [95], focusing on their potential to enhance network coverage, improve signal quality, and serve as relay nodes in multi-hop communication scenarios. These studies illustrate the advantages of leveraging UAVs for dynamic and adaptable network configurations.

The main limitation of UAVs is their finite battery capacity, which must provide sufficient energy for both communication and flight operations. To address this challenge, a viable solution involves the establishment of dedicated landing zones for UAVs. By utilizing these landing sites, UAVs can minimize energy expenditure associated with hovering and simultaneously facilitate the recharging of their batteries at integrated charging stations within these designated areas. This approach is supposed to enhance operational efficiency and extend the operational range of UAVs. Additionally, the sustainability of recharging batteries, instead of fixed-capacity batteries, makes the research on smart charging scheduling approaches worthwhile [96].

Anyway, introducing moving 6G antennas to avoid densification and enhance capacity while reducing carbon footprint introduces some challenges in the game. For instance, mobility requires wireless BackHauling (BH). Integrated Access and Backhaul (IAB) has been proposed as a way to adopt the same technologies and resources that are already allocated for cellular networks to provide BH connectivity.

To address these critical issues, we adopt a novel framework for deriving a first-order quantitative evaluation of the infrastructure savings achievable in MNs in IAB scenarios. Such an analytical approach allows determining the optimal configuration in terms of Static Base Station (SBS) and MBS density that minimizes overall infrastructure requirements, while guaranteeing the desired QoS for all users. Crucially, this approach accounts for the target QoS and resource utilization of wireless backhauling links.

## Analytical Framework

We consider a hybrid network of SBSs and MBSs mounted on vehicles (e.g., drones or cars), with UE, representing BroadBand (BB) terminals, distributed via a Poisson Point Process. The system aims to optimize deployment and resource allocation under QoS constraints over time ($J$ time slots) and space (the overall area is divided into $Z$ regions). MBSs use in-band wireless backhauling, with SBSs acting as access points.

We adopt a distance-based path loss model and a random frequency reuse scheme with reuse factor $k$. All Base Stations (BSs) transmit at power $P$. Thus, users connect to the closest BS, which provides the highest Signal to Interference plus Noise Ratio (SINR). The downlink capacity for a user at distance $r$ is $C(r, P, I)$ computed according to Shannon's capacity law. BSs implement a Weighted Processor Sharing (WPS) scheme to allocate airtime fairly among users. SBSs serve both BB users and MBSs, each with its own WPS weight. MBSs have unit weight, while BB users have weight $\phi_j^z \geq 0$, constant across all SBSs in region $z$ and interval $j$. The utilization of an SBS serving $N_M$ MBSs and $N_U$ UEs is $U_s = \frac{N_M+\phi_j^z N_U}{N_M+\phi_j^z N_U+\beta_s}$, and for an MBS serving $N_U$ users $U_m = \frac{N_U}{N_U+\beta_m}$ where $\beta_s$ and $\beta_m$ are the idle time fractions.

To evaluate QoS, we adopt the *per-bit delay*, defined as the inverse of the user throughput. The system tunes $\phi_j^z$ and BS utilizations to satisfy two conditions: (i) the Palm expectation of per-bit delay for a typical BB user $\bar{\tau}_m^{j,z}$ and $\bar{\tau}_m^{j,z}$ served by a moving or a static BS respectively, matches the target $\tau_0$, and (ii) the *violation probability* $V_j^z$ for MBS backhaul traffic remains below a threshold $\delta$.

We formulate the optimization problem to determine, for a given mean user density, the optimal densities of static and mobile BSs, along with round-robin coefficients, that minimize deployment costs while ensuring QoS for both fronthaul and backhaul traffic. The detailed derivation of all the expressions can be found in [97].

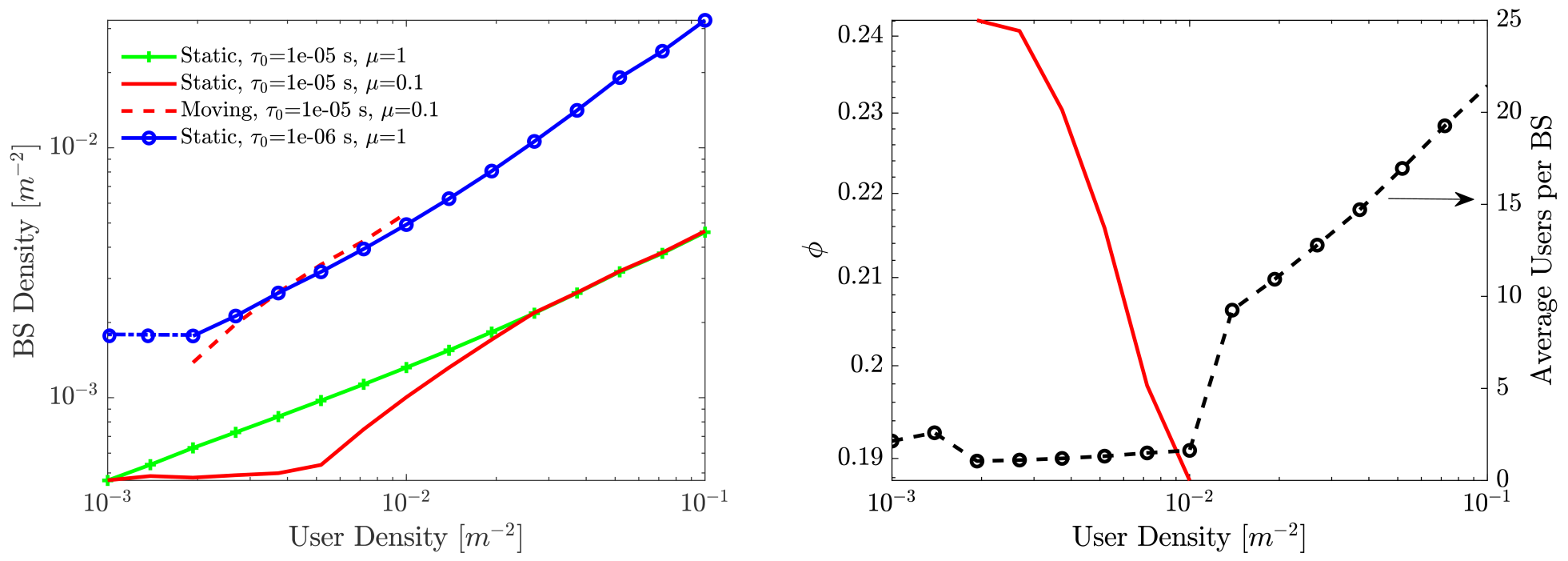

(a) (b)

Figure 3.8: Optimal network configurations as a function of user density, for different target QoS values, and different relative BS unitary cost $\mu$ (ratio between the unitary cost of MBSs and SBSs). (a) Optimal BS densities; (b) Round robin weight $\phi$ at optimum for backhauling links.

For the sake of realism, we assume that the total number of MBSs remains constant over time, i.e., $\sum_{z=1}^{Z} \lambda_{m,j}^{z} E_z = M, \quad \forall j$, where $E_z$ denotes the area of region $z$. The study aims to determine the optimal configuration of base stations for both static and moving infrastructures that minimizes the overall installation cost across the entire area throughout the observation window:

**Minimization of BS deployment costs:**

$$\underset{\lambda_{m,j}^{z}, \lambda_{s}^{z}, \phi_{j}^{z}}{\text{minimize}} \; \mu M + \sum_{z=1}^{Z} \lambda_{s}^{z} E_z$$

Subject to the constraints on QoSs and number of vehicles. Here, $\mu$ reflects the unitary cost ratio between MBSs and SBSs, and it is in general $< 1$ since static infrastructures typically require urban modifications, expensive authorizations, and rent for the location.

The optimization results are presented in Fig. 3.8 and Fig. 3.9. For the first set of experiments, we focused on a single region and time slot to investigate the system behavior when varying the density of users connected to the network and the relative cost $\mu$.

As shown in Fig. 3.8(a), the optimal density of active BSs increases with user density across all scenarios, following a power-law trend. This growth is steeper under stricter QoS targets, due to the increased demand on both fronthaul and backhaul resources, which negatively affects the backhaul traffic link. Due to the cost of in-band backhauling, in most cases, the optimal solution involves only SBSs. However, when the relative cost of MBSs drops to 0.1 or below, a hybrid deployment appears optimal over a specific user density range, with MBSs serving users and SBSs supporting backhaul links.

For very low and high user densities ($\geq 10^{-2}$ users/m$^2$), the backhaul cost becomes prohibitive, and the system reverts to SBS-only deployments. As seen in Figure 3.8(b), the MBS density and round-robin weights adapt to maintain a constant low user load per BS, minimizing backhaul overhead. The transition to SBS-only configurations occurs, when round-robin weights become too small, indicating excessive backhaul usage, making MBS-based setups less efficient.

These trends highlight the significant impact of QoS-aware in-band wireless backhauling on network resource efficiency. This is mitigated by considering the reusability of MBS and further reducing the relative cost between static and moving infrastructures. Notably, for $\tau_0 = 10^{-6}$ s (corresponding to a higher required throughout) and low user densities, BS density does not drop further due to strict QoS constraints. Even if many BSs are temporarily idle, their presence is needed to meet latency targets. While dynamic BS activation based on user mobility could help, such mechanisms are beyond this study and would not alter the required deployment density.

In a second set of experiments, we considered a simplified urban model composed of two regions, representing a residential and an office area. The objective is to illustrate how our theoretical framework can be leveraged to quantify the potential savings, particularly in terms of the number of deployed BSs and associated CAPEX, enabled by reusing MBSs across different city areas at different times of the day. We assumed that the user density in each region alternates between a high and a low level (denoted in Fig. 3.9 as $L_h$ and $L_o$). To model user mobility within the urban scenario, we consider both regions to experience the same high and low user density levels, but in opposite time slots: when one region is at its peak, the other is at its minimum. During nighttime hours (from 8 p.m. to 4 a.m.), both areas are assumed to be at their low user density level (Fig. 3.9(a)).

While this setup is idealized, it still captures key performance dynamics that occur in practical urban deployments. To account for differences in user spatial distribution between the two regions, and particularly the typically lower user density in residential areas compared to office/business districts, we assume that the office area spans 1 km$^2$, while the residential region spans $\gamma$ km$^2$, where $\gamma \geq 1$ is the *area ratio*. Both MBSs and SBSs were assumed to have the same unitary deployment cost.

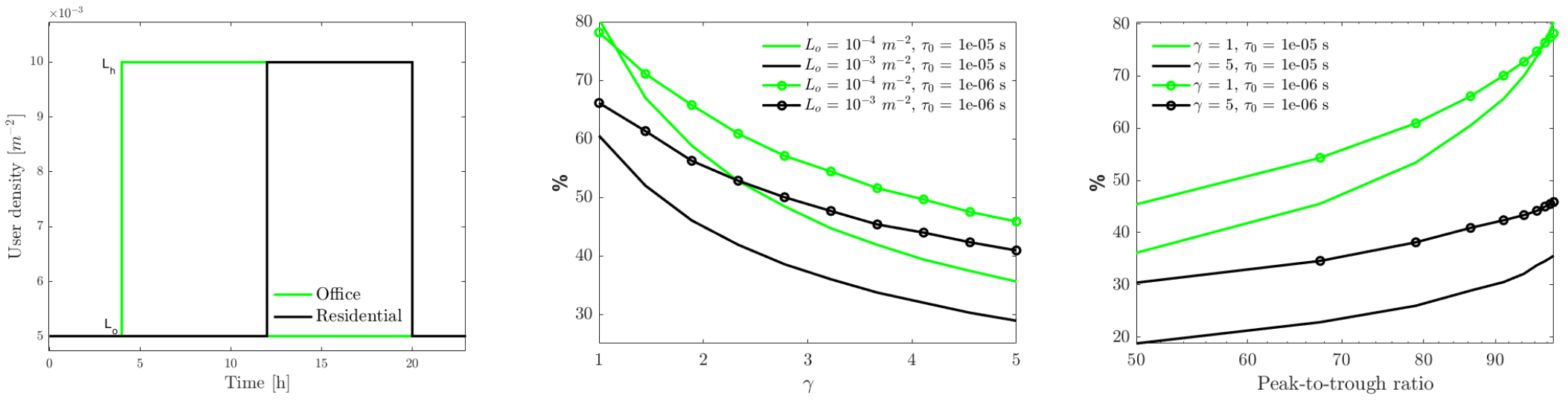

(a) (b) (c)

Figure 3.9: (a) User density profile over 24 h for the residential and office district; (b) percentage of BS reused as a function of the area ratio $\gamma$, for a 90% and 95% peak-to-trough ratio, (c) and of peak-to-trough ratio for different values of area ratio.

Fig. 3.9(b) illustrates the proportion of MBSs that are reused, that is, serving users in both regions during different time intervals, expressed as a percentage of the total BSs deployed. This proportion directly translates into CAPEX savings compared to a static deployment scenario where MBS users are instead served by SBSs. Such reuse is particularly beneficial in urban settings, where the use of wired backhaul is often constrained by cost and feasibility.

The figure highlights that the highest gains are obtained when the area ratio $\gamma$ equals 1, i.e., both regions have the same area. As $\gamma$ increases, users in the residential region become more spatially dispersed, which raises the required BS density to meet QoS targets and reduces reuse opportunities. Additionally, the figure shows that reuse gains increase with higher target user performance levels. This is because stringent QoS targets necessitate deploying more BSs, when transitioning from low to high user densities (as evidenced by the steeper slopes in Fig. 3.8 for higher QoS values). This results in a greater pool of BSs, that are idle during off-peak hours and thus available for reuse.

Similarly, maintaining a constant peak-to-trough ratio[1], while increasing the low-density level also improves reuse, as it reduces the magnitude of the BS deployment swing between time intervals. Ultimately, these patterns stem from the fact that the marginal cost of supporting higher user loads or improving QoS grows with user density. This is reinforced in Fig. 3.9(c), which shows that an increase in the peak-to-trough ratio significantly boosts the fraction of MBSs, that are redundant during low-traffic periods and can thus be reused, resulting in a reduced total number of BSs.

Notably, these gains increase faster than linearly with the peak-to-trough ratio. These insights suggest that, in light of the anticipated growth in traffic burstiness and QoS demands from emerging 6G use cases, such as Augmented Reality (AR)/Extended Reality (XR) and Six Degrees of Freedom (6DoF), the moving network paradigm could play a crucial role in enhancing both the resource efficiency and financial viability of future wireless infrastructures.

The presented novel analytical framework for quantifying the potential advantages of the MN paradigm within urban environments shows that a significant amount of moving base stations can be reused in most of the considered setups, suggesting the overall validity of the MN approach to mitigate the need for dense BS deployments. The proposed model's accuracy has been confirmed through Monte Carlo simulations. Future research should also emphasize the energy savings achieved through the deployment of the MN paradigm.

[1] The peak-to-trough ratio is defined as the ratio between the maximum traffic demand during peak hours and the minimum demand during off-peak hours.

## 3.5 Localization

By **Francesco Linsalata**

The convergence of radar sensing and wireless communication, referred to as ISAC, is emerging as a key enabler of next-generation wireless systems. In particular, 6G networks envision a native integration of sensing functionalities into communication infrastructures, allowing radio devices not only to exchange information but also to perceive and interpret the physical environment. Among the most promising applications of this paradigm is the use of UAVs in search and rescue scenarios, where real-time imaging and victim localization can significantly enhance the effectiveness of emergency response missions [98].

Traditional methods rely heavily on optical or thermal imagery and often require dedicated hardware such as avalanche beacons or passive reflectors. These solutions are either limited by environmental visibility or by the assumption that victims carry specific equipment. ISAC-equipped UAVs offer a flexible and scalable alternative by simultaneously providing connectivity and sensing capabilities using standard communication waveforms.

This section presents a comprehensive ISAC framework deployed on a UAV platform, capable of performing both passive environmental imaging and active victim localization. The proposed system is based on Orthogonal Frequency Division Multiplexing (OFDM), a waveform widely adopted in modern wireless standards such as 5G NR. The overall localization process is divided into two phases:

- **Passive phase**: The UAV generates Synthetic Aperture Radar (SAR) images of the environment using reflected communication signals [99].
- **Active phase**: The UAV estimates the position of a victim by analyzing uplink synchronization signals or Receiver Signal Strength Indicator (RSSI) values [100].

### 3.5.1 Passive Localization

In the passive phase, the UAV performs SAR imaging by transmitting a standard OFDM downlink signal and collecting its echoes from the ground. The goal is to evaluate the feasibility of high-resolution radar imaging under strict constraints dictated by communication standards, such as limited bandwidth, power restrictions, and defined signal structure.

The OFDM signal is defined in the discrete frequency-time domain as:

$$\Lambda_{FT} = \{(n\Delta f, kT) \mid n = 0, \ldots, N-1;\; k = 0, \ldots, K-1\}, \tag{3.30}$$

where $\Delta f$ is the subcarrier spacing, $T$ is the symbol duration, $N$ is the number of subcarriers and $K$ is the number of OFDM symbols per frame.

The baseband time-domain representation of an OFDM symbol is:

$$x(t) = \sum_{m=0}^{M-1} s(m) e^{j2\pi m \Delta f t} g(t), \tag{3.31}$$

where $s(m)$ denotes the complex modulation symbols (e.g., QPSK, 64-QAM) mapped onto subcarriers and $g(t)$ is the pulse-shaping filter, defined as:

$$g(t) = \begin{cases} 1, & t \in [-T_{\text{CP}}, T], \\ 0, & \text{otherwise}, \end{cases} \tag{3.32}$$

with $T_{\text{CP}}$ is the cyclic prefix duration.

**Pulse Compression and SAR Imaging.** The received signal is processed through pulse compression to estimate the time delay of the echoes [99]. Two methods are compared:

- **Matched Filtering (MF)**:
$$X_{\text{RC}}(f) = X^*(f) \cdot X_{\text{RX}}(f), \tag{3.33}$$
- **Zero Forcing (ZF)** with regularization:
$$X_{\text{RC}}(f) = \frac{1}{X(f) + \kappa} \cdot X_{\text{RX}}(f), \tag{3.34}$$

where $X(f)$ and $X_{\mathrm{RX}}(f)$ are the Fourier transforms of the transmitted and received signals, and $\kappa$ is a regularization term to prevent numerical instability.

SAR focusing is achieved using the Time-Domain Back Projection (TDBP) algorithm. For a pixel located at position $(x, y)$, the focused image is [99]:

$$F(x,y) = \int x_{\mathrm{RC}}\left(t = \frac{R(\tau; x, y)}{c}, \tau\right) \cdot e^{j\frac{4\pi}{\lambda}R(\tau;x,y)} d\tau, \tag{3.35}$$

where $R(\tau; x, y)$ is the distance from the UAV to the pixel at time $\tau$, $c$ is the speed of light, and $\lambda$ is the carrier wavelength.

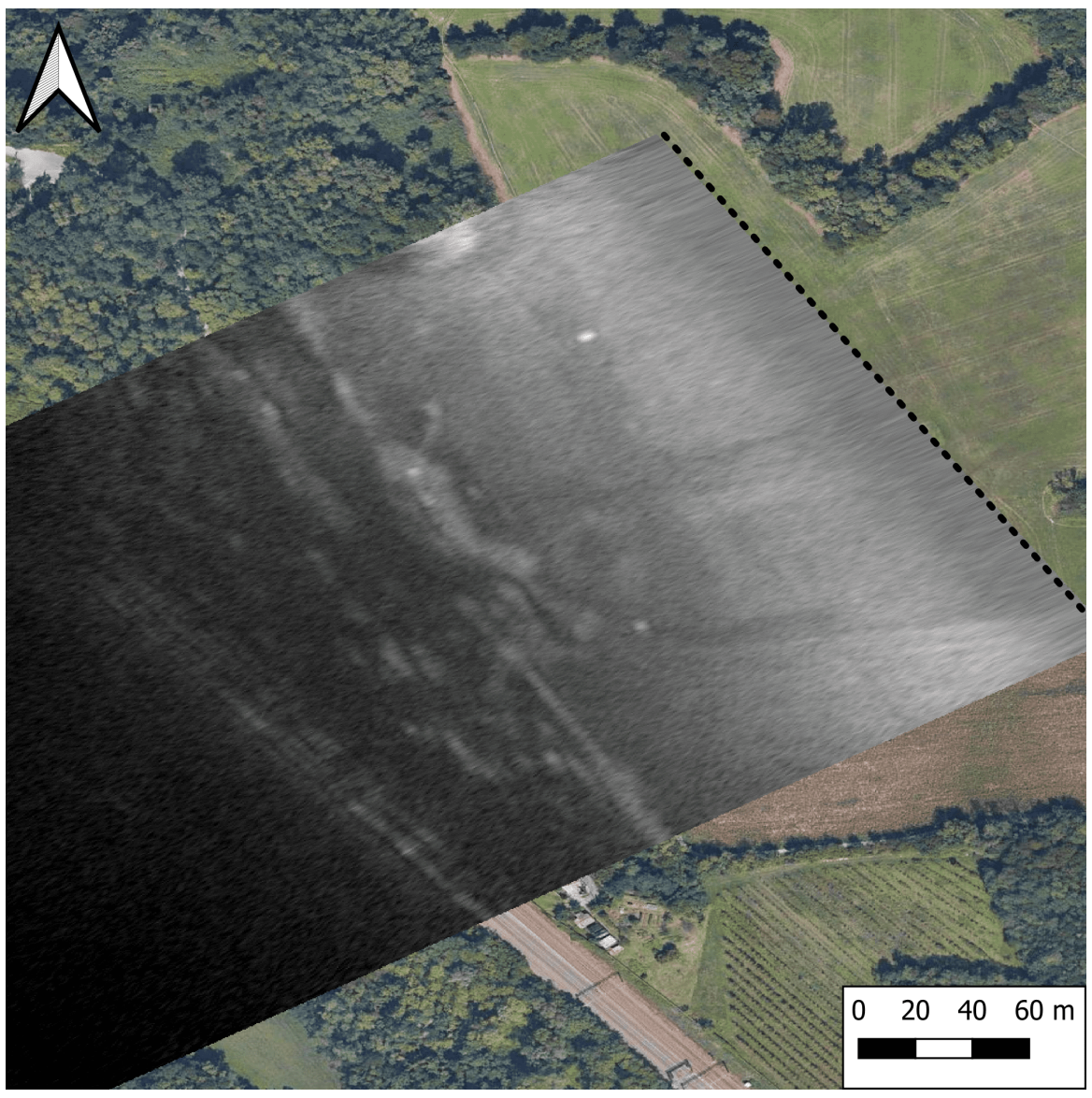

Figure 3.10: SAR image generated by back-projecting the range-compressed OFDM signal. The black dotted line represents the trajectory of the drone [98].

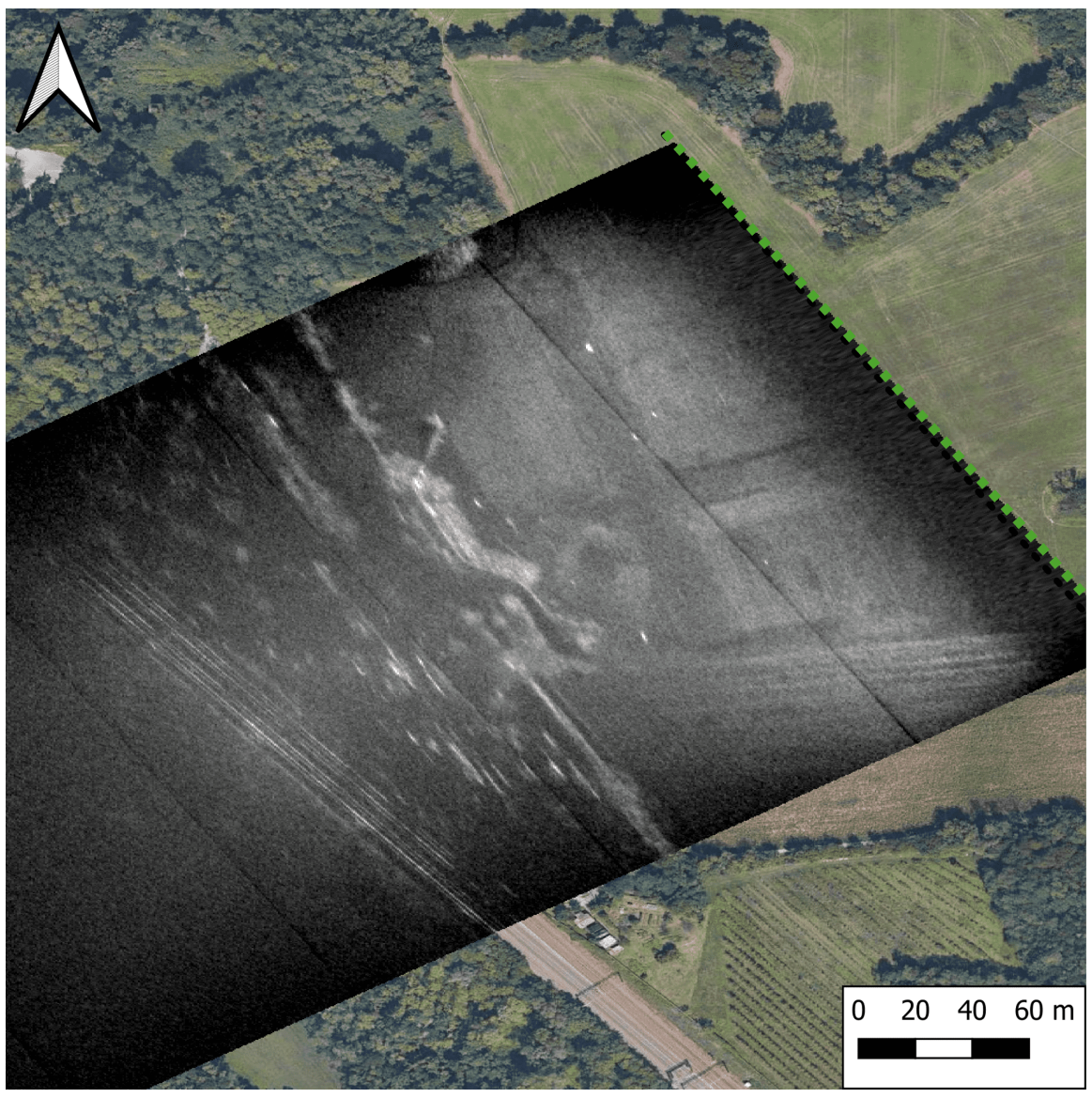

Figure 3.11: SAR image generated with the original FMCW radar setup with full bandwidth. The green dotted line represents the trajectory of the drone [98].

The scene in Fig. 3.10 and Fig. 3.11 corresponds to an airfield near Milan, easily identifiable by the characteristic cross-shaped runway. Several corner reflectors were deployed in the area, serving as calibration references due to their well-defined impulse response in both azimuth and range.

The imagery also reveals natural and man-made features such as a tree line, residential structures, and a railway track towards the western portion of the scene. A direct comparison between the OFDM-based image (Fig. 3.10) and the higher-resolution Frequency Modulated Continuous Wave (FMCW) radar image (Fig. 3.11) highlights the resolution trade-off: the original FMCW system uses a 400 MHz bandwidth, yielding a theoretical resolution of approximately 37 cm, whereas the OFDM image, acquired with only 40 MHz bandwidth, exhibits a resolution around 3.7 m.

It is worth noting that Fig. 3.11 displays a low-intensity artifact aligned with the UAV flight path, introduced by a notch filter used to suppress interference in the FMCW radar. This feature does not correspond to a physical object. Both images were processed using a multisquint algorithm and normalized with respect to their maximum intensity for visual clarity.

### 3.5.2 Active Localization

The active phase complements the imaging process by exploiting uplink transmissions from UE for localization. A common approach is based on the RSSI, which provides a coarse estimate of the range by measuring the signal attenuation.

The RSSI-based model is given by:

$$P_{\mathrm{RX}} = P_{\mathrm{ref}} - 10\, n_p \log_{10}\left(\frac{d}{d_{\mathrm{ref}}}\right) + n, \tag{3.36}$$

where $P_{\text{ref}}$ is the power at a known reference distance $d_{\text{ref}}$, $n_p$ is the path loss exponent (typically between 2 and 4), $n$ is the log-normal shadowing (Gaussian noise in dB) and $d$ is the estimated distance from UAV to target.

To estimate the position $\mathbf{u} = (x, y)$ of the emitter, maximum likelihood estimator can be used:

$$\hat{\mathbf{u}} = \arg\max_{\mathbf{u}} \prod_{i=1}^{N} \frac{1}{\sqrt{2\pi\sigma_i^2}} \exp\left(-\frac{[\rho_i - h(\mathbf{u}, \mathbf{s}_i)]^2}{2\sigma_i^2}\right), \tag{3.37}$$

where $\rho_i$ is the RSSI measurement, $\sigma_i$ is the standard deviation of the noise, $\mathbf{s}_i$ is the UAV position, and

$$h(\mathbf{u}, \mathbf{s}_i) = P_{\text{ref}} - 10\, n_p \log_{10}\left(\frac{\|\mathbf{u} - \mathbf{s}_i\|}{d_{\text{ref}}}\right). \tag{3.38}$$

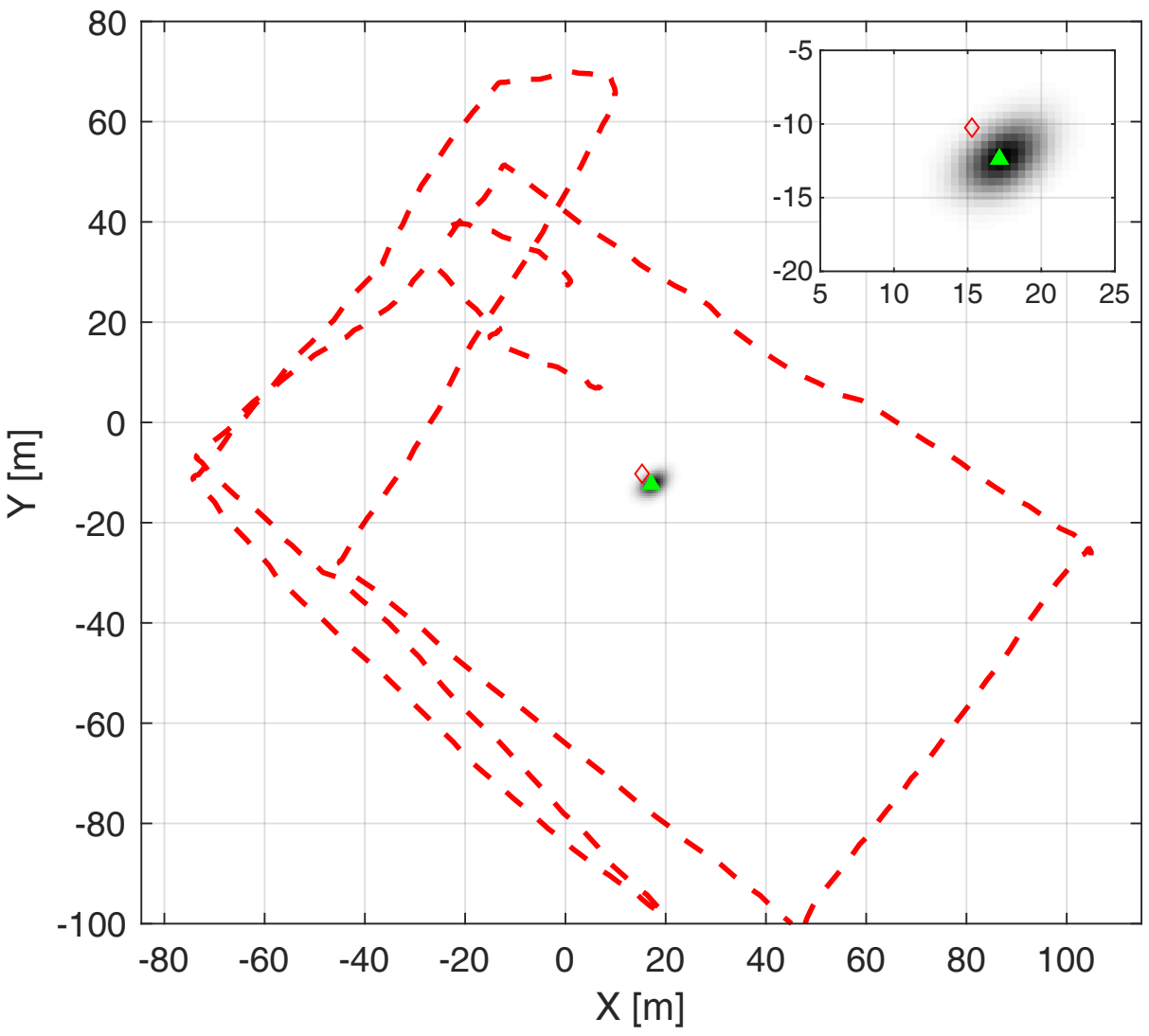

Figure 3.12: Localization likelihood map generated from RSSI observations. The estimated position is marked in green, ground truth in red [98].

Fig. 3.12 presents the likelihood map obtained by applying the maximum likelihood estimation method described in Equation 3.37. To generate synthetic RSSI measurements, the QuaDriGa channel simulator was employed in combination with a UAV trajectory recorded during an actual flight, depicted as a red dotted line in the figure. The estimated position of the ground transmitter corresponds to the global maximum of the likelihood function, resulting in a localization error of approximately 2.8 m in this scenario. The green marker indicates the estimated position, while the red diamond denotes the true location of the transmitter.

Although the accuracy of RSSI-based localization is lower than that of radar imaging, the information it provides is highly valuable during the target classification stage. SAR images often contain multiple strong scatterers that may not correspond to actual victims, introducing ambiguity. The integration of RSSI-based positioning allows for effective filtering of false positives, thereby improving the reliability and operational efficiency of search and rescue missions.

## 3.6 Operational Challenges

By **Abdul Saboor, Evgenii Vinogradov**

Integrating UAVs into ITS has numerous benefits and applications, as emphasized throughout this chapter. For example, ABSs based on UAVs can improve vehicular connectivity for dense urban environments, provide coverage in hard-to-reach locations, assist AAM operations, and improve real-time sensing capabilities for traffic detection and hazard detection. This section outlines the main enablers for integrating UAVs to complement ground-based ITS components, such as A2G models, cooperative awareness mechanisms, and localization frameworks. Transitioning, however, from terrestrial infrastructure to a more dynamic 3D heterogeneous ITS system brings several operational challenges that must be carefully addressed.

- **Licensing and Airspace Regulations:** Licensing and registration criteria for UAV operations vary significantly from nation to nation, causing ambiguity in regulation for worldwide ITS deployments. In many countries, such as Singapore or the United States, operators require certification and registration of UAVs based on weight and function, which can delay deployment [101]. Similarly, different countries have strict airspace policies to protect the ground population and commercial aviation. These policies' structure airspace using no-fly zones, predefined flight altitudes, and keep-in geofences, which limit UAV operational flexibility, especially in urban or sensitive environments [102]. These challenges can be resolved by deploying UAV traffic management systems [103] simplifying flight authorization systems and aerial corridor operations, which enable structured and predictable flight paths, while minimizing conflicts with commercial aircraft and restricted zones.

- **Privacy and Data Protection:** Privacy and data protection are major challenges in UAV-based 3D ITS, particularly with the increasing use of high-resolution cameras and other sensors. The concern becomes more relevant in urban areas, where surveillance capability can unintentionally capture personal data. Therefore, regulatory frameworks must be established that force people to consent and encrypt data securely to maintain their privacy. Besides, UAV introduction brings about new attacks, such as GPS spoofing, jamming, interception, and physical capture, which can be resolved by advanced methods, such as blockchain-based security systems for offering data integrity, decentralization, and transparency in UAV-enabled 3D ITS [104].

- **Size, Weight, and Power (SWaP) Constraints:** SWaP is a major limitation in ABS operations, because it must carry communication payloads (antennas, transceivers, power amplifiers), while keeping low weight and power budgets. The more capable the onboard equipment, the greater the energy demand, which results in limited flight time and ABS service. Therefore, UAVs must balance the trade-off between SWaP and coverage area/time. These constraints increase the importance of lightweight, energy-efficient hardware and smart deployment strategies to ensure effective ABS's operations within ITS. Limited energy availability is yet another operational challenge of ABSs in ITS. In UAVs, propulsion, hovering, and demanding tasks, such as wireless communication or real-time data processing, are the primary sources of energy depletion. Most UAVs are powered by batteries, which typically offer less than one hour of flight duration, making them insufficient for long-term or large-area coverage tasks. However, in missions such as traffic monitoring or disaster response, UAVs may need to remain airborne for longer periods, constrained by limited onboard battery. Therefore, it is important to address the energy/battery constraints of ABSs. Some research directions include strategically located recharging stations, deploying wireless power transfer systems, and controlled UAV swarms to distribute load and maximize energy efficiency.

- **Spectrum Allocation:** ABSs in ITS depend on reliable access to wireless spectrum for data transmission and control. However, allocating spectrum for UAV operations is challenging, due to their dynamic 3D mobility. High mobility of drones, particularly in high-rise urban environments, introduces rapid variations in link quality, leading to frequent handovers. Secondly, the aerial deployment of base stations can lead to increased interference with existing terrestrial networks and other UAVs due to dominant LoS links [105]. Therefore, in urban ITS scenarios, it is important to use adaptive and intelligent spectrum management techniques for reliable and secure communication for UAVs [106].

In addition to the key challenges discussed above, several other operational issues must be addressed to integrate UAVs into ITS safely. For example, multi-UAV coordination to prevent redundant coverage, while avoiding collisions, and real-time path planning to conserve resources and avoid obstacles [107]. Extreme temperatures, wind, and rain could influence UAV stability and communication performance in ITS [108]. Other challenges are scalability, fleet management, and integration with existing ITS infrastructure. Furthermore, public acceptance remains a key challenge, as noise, safety, and visual data collection can impact the deployment of UAVs in ITS [109]. Finally, logistical aspects such as UAV maintenance, battery replacement, firmware updates, and safe landing zones add further complexity to day-to-day operations.

# Chapter 4: Aircraft-to-Aircraft Communication

By **Michael Walter, Miguel A. Bellido-Manganell**

## 4.1 Overview

A2A communication plays an increasingly important role within the broader ITS framework, particularly as the scope of ITS expands beyond terrestrial modes to include aeronautical and satellite communications. By enabling real-time exchange of position, velocity, and trajectory information, A2A communication supports cooperative air traffic management, enhances situational awareness, and facilitates autonomous or semi-autonomous flight operations. These capabilities mirror V2V communication in terrestrial ITS, contributing to improved safety, efficiency, and scalability of airspace usage. Furthermore, A2A communication is integral to emerging applications such as Unmanned Aerial Systems (UAS) traffic management and urban air mobility, making it a vital component of future multimodal, connected, and intelligent transportation systems.

The design of modern A2A communication systems relies on the availability of information on the A2A propagation channel. Mobile-to-Mobile (M2M) propagation channels, including A2A channels, have garnered increasing attention in recent years, driven by the development of advanced communication systems for a wide range of mobile platforms, including aircraft, drones, cars, and maritime vessels.

However, most existing channel models are designed for average or worst-case conditions and typically overlook the specific characteristics of the environment in which the Transmitter (TX) and Receiver (RX) operate. Thus, they lack the ability to accurately capture the channel behavior in non-stationary scenarios. This limitation is especially evident in the treatment of scattering components, which are often neglected or modeled using coarse approximations derived from unrelated stochastic environments.

For M2M channels one of the most severe problems when modeling the channel is its non-stationarity. This is especially true for A2A channels, since the velocities of both TX and RX are higher compared to other M2M channels. Another important difference in the more complex geometry, where a full 3D model is needed, whereas for V2V channels often 2D models are adequate. A typical application of A2A communication would be a coordinated approach to reduce scheduling times as shown in Fig. 4.1.

Channel models are used to take the propagation effects into account that affect the transmitted signal on its way to the receiver and we look at those models in more detail in the next section.

## 4.2 Channel Model

A2A channel models are incorporated into the ITS framework to primarily support real-time aeronautical applications such as flight planning, collision avoidance, and direct air-to-air communication. The importance of such models stems from the ability to precisely model high Doppler frequencies, multipath fading, and non-stationary behavior in complex environments like mountainous terrain or low altitude flying. The accuracy however is necessary to ensure reliable communication and optimizing 4D trajectories.

In general, channel models are used to mathematically describe the propagation effects that affect the transmitted signal on its way to the receiver. The most important effects are reflection, scattering, and diffraction. In comparison to other M2M channels, the A2A propagation channel presents some peculiarities that separate it from both conventional terrestrial and air-to-ground channels: generally comparatively long LoS distances, high relative velocities, which lead to large Doppler spreads, occasional rapid transitions between LoS and obstructed LoS during maneuvers due to the fuselage, and altitude-dependent multipath components from the ground.

Measurement campaigns state that, when a clear LoS exists, the large-scale path loss often follows free-space or near free-space behavior, but that small-scale statistics like delay spreads, Doppler spreads, and Ricean $K$-factors, vary strongly with geometry, altitude, and the presence of reflecting surfaces, e.g., buildings or sea surface, making empirical characterization indispensable [110], [111] and [112].

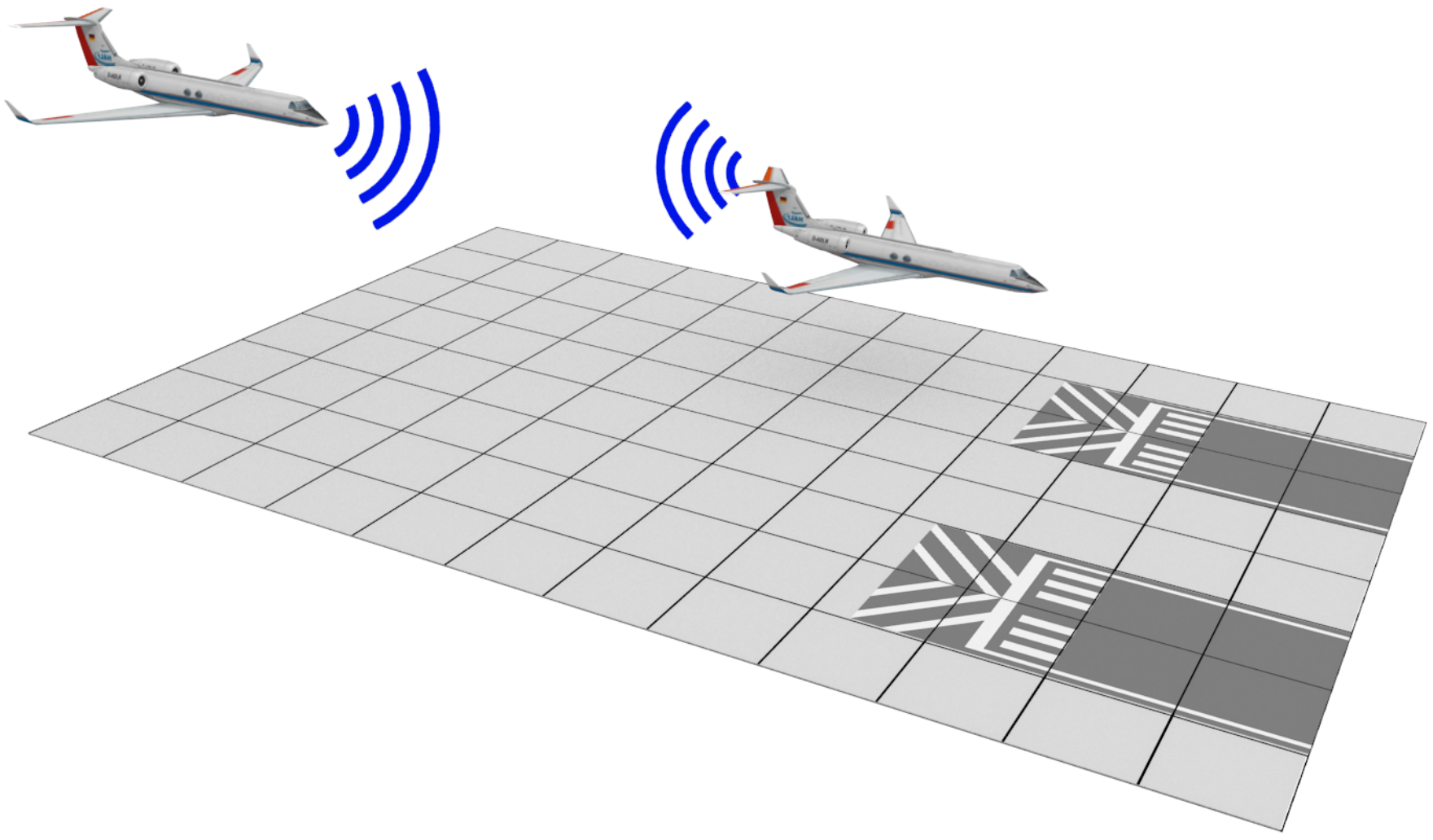

Figure 4.1: Schematic aircraft-to-aircraft communication link while landing on parallel runways to enable tighter scheduling.

Measurement show that A2A channels can have either concentrated multipath around the direct path and specular path, which results in fading, e.g., lower or higher Ricean $K$-factor of the LoS and Specular Reflection (SR) components, or widely-spread multipath when the nearby ground causes scattering, leading to different Doppler spectra. The delay or Doppler spectra depend on the flight phase and relative velocity.

For aircraft flying at cruising altitude, there is mostly only the LoS component present, although the specular reflection can become significant if the flight takes place over water and the geometry leads to a strong reflection. For aircraft flying at low altitudes, the specular reflection and the scattering from the ground have a greater effect on the propagation. This is mainly the case for take-off and landing in civil aviation, but can be for longer periods of time in military aviation.

Modeling approaches in the literature fall into three general categories: statistical, deterministic, and GSCM. Many contemporary works adopt hybrid models to capture the behavior of the A2A channel. Statistical models like the TDL and Wide-Sense Stationary Uncorrelated Scattering (WSSUS) models are useful for system-level simulations and for matching measured delay and Doppler spectra, while deterministic ray-tracing captures site-specific geometric relationships like reflections and scattering.

The Wide-Sense Stationary (WSS) assumption often breaks down in high mobility, low altitude or complex environment scenarios [113]. GSCM methods provide a mix between stochastic and pure geometric modeling: they represent scatterers with geometric primitives to yield realistic angular, delay, and Doppler statistics, while remaining efficient enough for Monte Carlo performance studies. Non-stationary 3D GSCMs that allow time-evolving clusters have been proposed and validated against aircraft and drone measurements to better model the rapid temporal evolution seen in many A2A situations [114], [115].

The main effects for propagation between two aircraft include reflection, scattering, and diffraction. The A2A channel is quite different from the other M2M channels. For aircraft flying at high altitudes, the LoS component generally is stronger than the other components, although the SR component from the ground becomes significant in some specific geometries. For aircraft flying at low altitudes, the SR component and the scattering components from the ground have a bigger effect on the propagation.

According to Bello, the classical channel model commonly used in the aeronautical community represents the channel as a sum of the LoS component $h_{\mathrm{los}}(t, \tau)$, the SR $h_{\mathrm{sr}}(t, \tau)$, and the scattered component $h_{\mathrm{scat}}(t, \tau)$ [116]. Both the LoS component and the SR can be modeled by means of ray tracing.

This can be directly extracted from the geometry or from the positions of a global navigation satellite system. The LoS component is mainly characterized by a complex gain determined by the free space loss and the antenna patterns. For the specular reflection, the complex gain describes the effects of transmitting and receiving antenna gain in the specular reflection direction and free space path loss. Fading statistics for the LoS and SR components can be modeled according to [117] and [118]. Especially for aircraft flying at low altitudes like military aircraft, the scattering components have enough power to disturb the communication [119]. Similar problems and challenges arise for air-to-ground channels as shown in [120] and [121].

For modeling non-stationary scattering, the authors present in [114] a theoretical channel model for M2M propagation that utilizes an explicit geometric representation of the environment to derive the time-varying channel between two mobile stations.

The proposed method is applied to a range of scenarios including A2A, UAV-to-UAV (U2U), V2V, and Ship-to-Ship (S2S) communications and its performance is evaluated against empirical data from multiple channel measurement campaigns, both at 250 MHz and 5.2 GHz. Across all scenarios, the model demonstrates a high accuracy in reproducing the non-stationary channel characteristics, including LoS component, SR component, and scattering components.

The best match between model and measurements is observed in environments, where scatterers can be effectively approximated by finite and infinite planar surfaces, i.e., when their geometric irregularities are minor relative to the overall structure. Furthermore, the model enables to simulate the non-stationary behavior of the channel as the TX and RX move, and provides insight into the specific reflectors or scatterers responsible for distinct features in the delay and Doppler domain.

Current and future research directions indicate the need for more empirical datasets, including long-range en-route flights and dense multi-aircraft environments, to parameterize channel models across all flight phases. Furthermore, non-stationary models that capture rapid changes during flight maneuvers, especially for military aviation, are required.

In summary, the state of A2A channel modeling is able to model the main physical propagation mechanisms like high power of the LoS and SR components, geometry-dependent multipath from the ground, which causes unique delay Doppler spectra, but still evolving in terms of broadly accepted parameter sets for different frequency bands, flight phases, and environments.

Best practice for researchers using A2A models is to select a modeling approach matched to the simulation use-case e.g., system-level or site-specific. Secondly, to consider hybrid GSCMs for scenarios where non-stationarity and scattering from complex environment strongly affect signal quality.

## 4.3 Model Verification with Measurement Data

In order to characterize the A2A channel, a measurement campaign was conducted with two research aircraft from the DLR. A RUSK channel sounder was used to record the time-variant frequency response of the channel. The measurement parameters are summarized in Table 4.1. The carrier frequency is 250 MHz and bandwidth of $B = 20\,\text{MHz}$ is used. The length of the signal period in delay direction is 25.6 µs and the signal is recorded at the receiver every $\Delta t = 2.048\,\text{ms}$. More information on the measurement campaign can be found in [110].

The TX was located in a Cessna C-208B (D-FDLR), which was flying in front and the RX was located in a Dornier 228-101 (D-CODE), which was trailing the Cessna. The flights took place in the south of Germany and for characterizing the scattering, both aircraft were flying at a very low altitude of 600 m above ground. The distance between the aircraft was roughly 1500 m, which corresponds to a delay of $\tau = 5\,\mu\text{s}$.

The flight route is shown in Fig. 4.2a. For the verification of our geometric modeling approach, the flights marked in the alpine valley scenario close to the city of Garmisch-Partenkirchen are evaluated, where the aircraft were flying so low that the mountaintops were higher than the aircraft. The location of this scenario is indicated in yellow in Fig. 4.2a. This particular setup allowed to receive rich scattering components not only from the ground, but also from the mountainsides.

To recreate the propagation geometry of the A2A channel, the valley scenario was recreated by using 38 finite planes to model the mountainsides and one infinite plane to model the ground, as shown in Fig. 4.2b. Details of the simulation setup and the analysis of the results can be found in [114], [115].

According to [114], the following can be summarized: The A2A valley scenario for different positions of the aircraft within the valley is simulated. Fig. 4.3 shows the simulation results, which are compared with the measurement data obtained in the flight campaign. Fig. 4.3c is obtained shortly after both aircraft enter the valley and Fig. 4.3d describes the flight situation approximately 30 s later, with both aircraft flying in the middle of the valley. In the two figures, we recognize the strong LoS component centered at about 5 µs and with a Doppler shift slightly above 0 Hz. After the LoS component, we can see the strong specular reflection component from the ground centered at approximately 6.5 µs with a Doppler shift slightly below the LoS Doppler shift.

| Parameter Name | Variable | Value |
|---|---|---|
| Carrier frequency | $f_{\mathrm{c}}$ | 250 MHz |
| Bandwidth | $B$ | 20 MHz |
| Frequency resolution | $\Delta f$ | 39.1 kHz |
| Signal period | $\tau_{\mathrm{max}}$ | 25.6 µs |
| Delay resolution | $\Delta\tau$ | 50 ns |
| Time period | $t_{\mathrm{max}}$ | 2.1 s |
| Measurement time grid | $\Delta t$ | 2.048 ms |
| Max. Doppler frequency | $\nu_{\mathrm{d,max}}$ | ±244 Hz |
| Doppler resolution | $\Delta\nu$ | 0.5 Hz |

Table 4.1: Measurement and evaluation parameters.

In addition, both the LoS and SR component are leaked in delay and Doppler directions, because of the limited sampling time and bandwidth. The LoS and specular reflection components are very well recreated by the proposed analytical model in Figs. 4.3a and 4.3b.

Right next to the SR component, one can observe the scattering components coming from the ground and from the mountains surrounding the aircraft. Overall, the channel simulated by the mathematical model matches the measurements very accurately. Specifically, one can identify many interesting effects that are recreated by the presented channel model.

The power density of the scattering components is particularly high directly after the SR component, i.e., between 6.5 µs and 8 µs. The lower parts of the mountains and the area around the specular reflection point lead to many scattering components with very similar Doppler shifts and delays, which increases the power density in a narrow Doppler shift range directly after the specular reflection component.

Furthermore, the proposed channel model is capable of accurately predicting the Doppler and delay regions, where no scattering components are expected, as well as the isolated clusters of scattering components appearing at higher delays. In Fig. 4.3a, the scattering components span most Doppler frequencies until a delay of approximately 15 µs. For higher delays, the scattering components are mainly concentrated on the limiting Doppler frequencies, i.e., around ±100 Hz. In this region, only isolated clusters can be seen, like the one observed in Fig. 4.3a between 23 µs and 25.6 µs with a Doppler shift between 20 Hz and 60 Hz. These isolated scattering components are also accurately recreated by our channel model as well as the wider distribution of the scattering components at −100 Hz, compared to its narrower distribution at 100 Hz.

In Fig. 4.3b, we look at the channel 30 s later. Not only the LoS and SR components are well recreated, but also the shape of the scattering components and their distribution in the delay Doppler domain. Furthermore, the model predicts that the scattering components below 12 µs span most Doppler frequencies, whereas those components are mainly concentrated around the limiting frequencies of ±100 Hz at higher delays. We can also stress that the model can truthfully recreate most of the isolated clusters of scattering components. For instance, the clusters of scattering components in the range from −40 to −80 Hz at 15 − 17 µs, and from 20 to 90 Hz at 17 − 22 µs, as well as the two small clusters spanning from 20 Hz down to −60 Hz at 21 − 25.6 µs.

It is important to understand that the presence of the scattering components at certain Doppler frequencies and delays, or the absence thereof, is deterministic mainly by the geometry between the

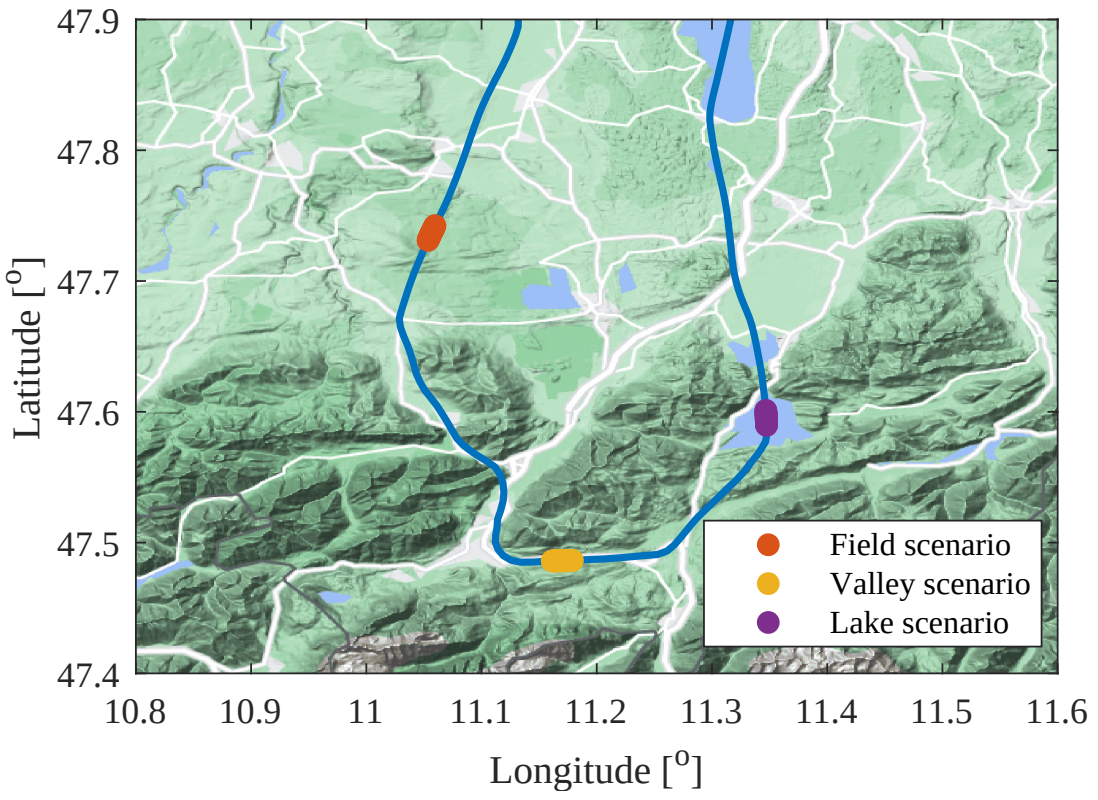

(a) Flight route of the A2A channel measurement campaign in southern Germany [114].

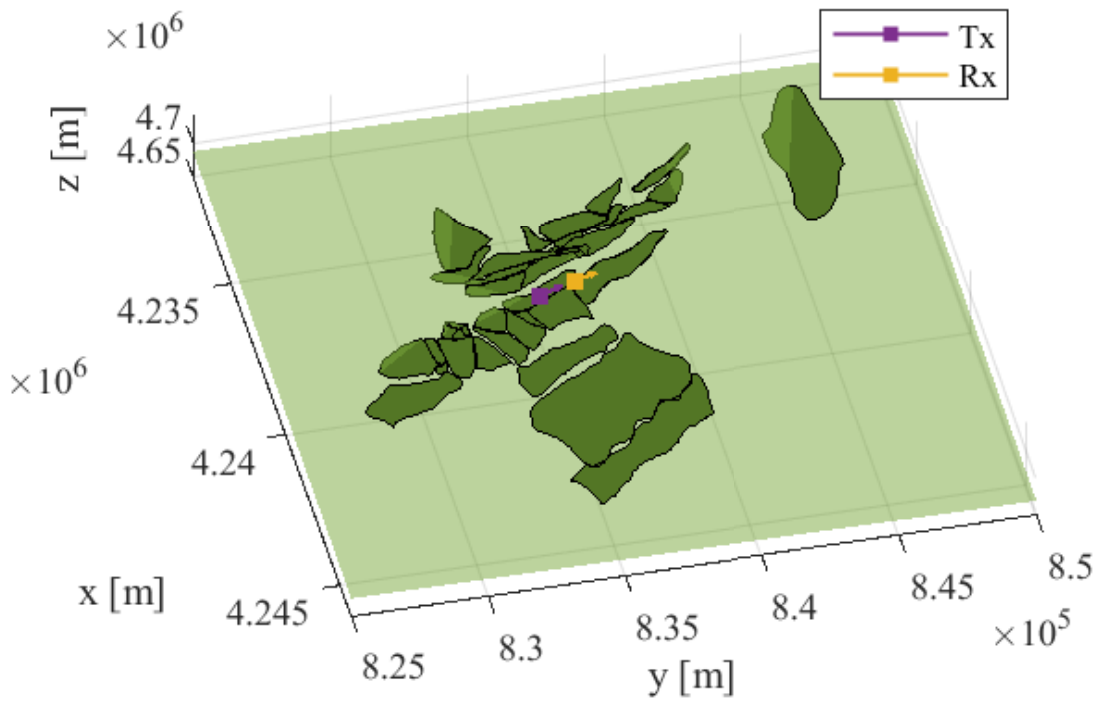

(b) Synthetic A2A valley scenario approximated by 38 finite scattering planes in ECEF coordinates to recreate the real mountain environment. From [114].

Figure 4.2: Flight route and theoretical abstraction of the geometry of the valley.

aircraft and their surroundings. Thus, the presented channel model is not only capable of predicting the channel accurately as shown here, but it can also identify which parts of the channel response correspond to specific reflectors, e.g., a specific mountainside, which can be very helpful for many applications.

The channel behavior at different time instants is described in the following. Similarities and differences in the channel experienced by the aircraft when entering the valley in Fig. 4.3c and in the middle of the valley approximately 30 seconds later in Fig. 4.3d can be observed.

First, the LoS and SR components did not change radically, as the distance between both aircraft and to the ground was kept constant. In addition, the overall shape of the scattering and the limiting frequencies did not change significantly between both positions, since the geometry directly in front and towards the back of the aircraft is mainly given by the ground plane, which did not change. The distribution of the scattering components within its limiting frequencies however changed noticeably between both points (Fig. 4.3c and Fig. 4.3d).

In fact, the movement of the scattering components led to the noticeable change in the channel between Fig. 4.3c and Fig. 4.3d, given that the geometry, i.e. the position of the mountainsides with respect to the aircraft changed slowly but consistently between both time instants. Thus, it is verified that the proposed theoretical channel model is capable of recreating the measured channel at any time instant and confirmed the non-stationary behavior of the channel. If the environment in the form of finite and infinite planes is known or extracted from map data, a simulation of the A2A channel a low-level flight from take off to landing can be conducted.

## 4.4 Communication challenges

Reliable A2A communication is a cornerstone of modern aviation safety and mission effectiveness, yet it remains vulnerable to a variety of flight-dependent challenges. Unlike conventional terrestrial communication systems, aeronautical communication must contend with constantly changing geometries, rapidly changing LoS conditions, and highly dynamic environments.

The communication challenges for aircraft includes the following scenarios, where the channel becomes more challenging for communication compared to an en-route scenario according to [117] and [122].

- **LoS obstruction during banking:** One significant issue arises during banking maneuvers. As an aircraft rolls, the fuselage and wings can obstruct the LoS path between antennas, creating transient shadowing and multipath interference. At high bank angles, antennas mounted on the underside of the fuselage may lose visibility altogether, producing short but potentially critical communication dropouts. As observed in [117], the banking has a significant effect on both the LoS and SR components. Due to the banking, the antenna pattern is also rotated and therefore, the received power drops considerably. The banking affects the LoS and SR components differently, since both components depend very strongly on the overall geometry. For this reason, the relative

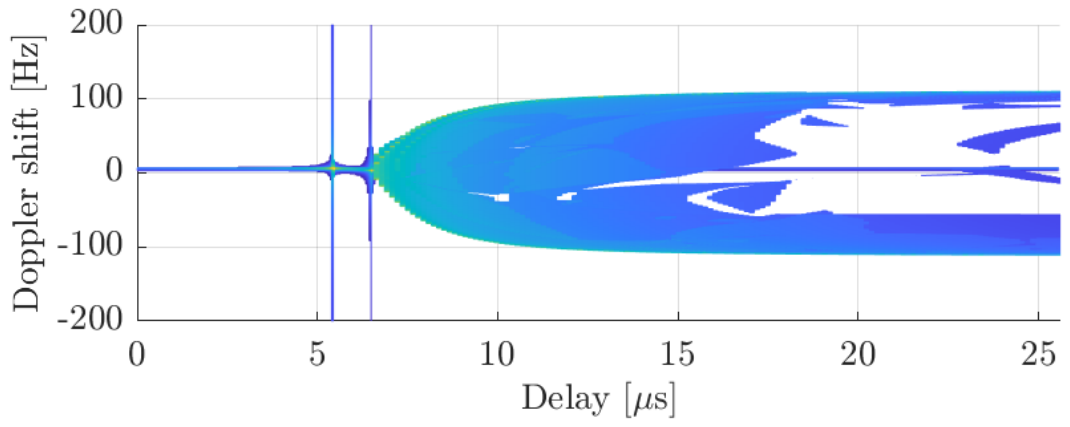

(a) Simulation model results (first location).

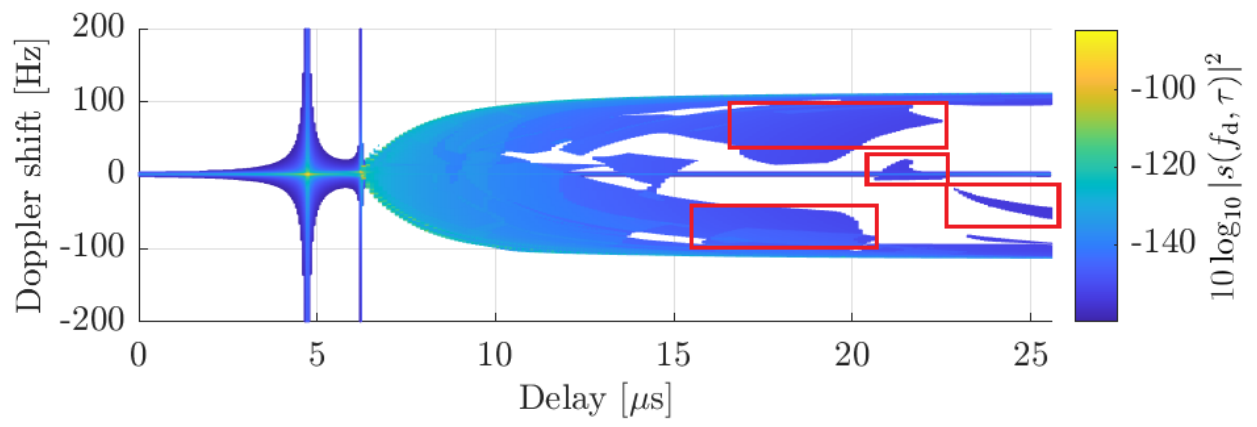

(b) Simulation model results (second location, 30 seconds afterwards).

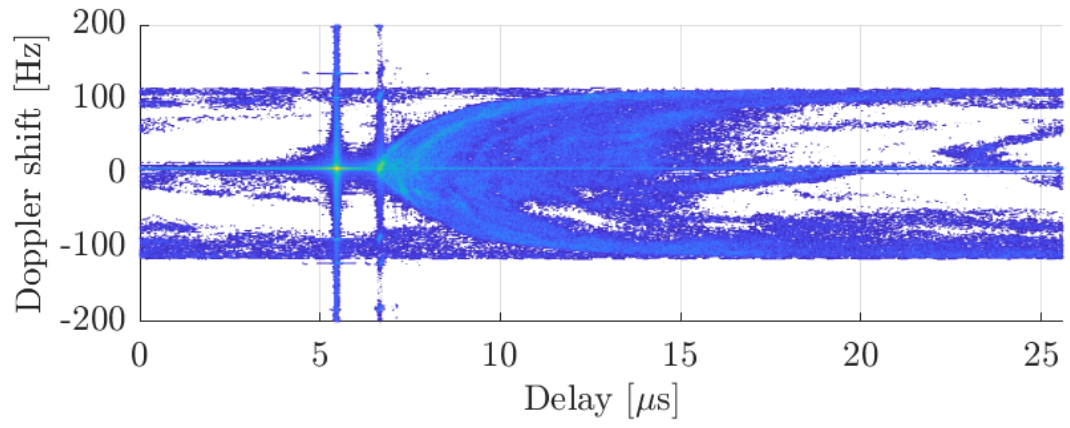

(c) Measurements (first location).

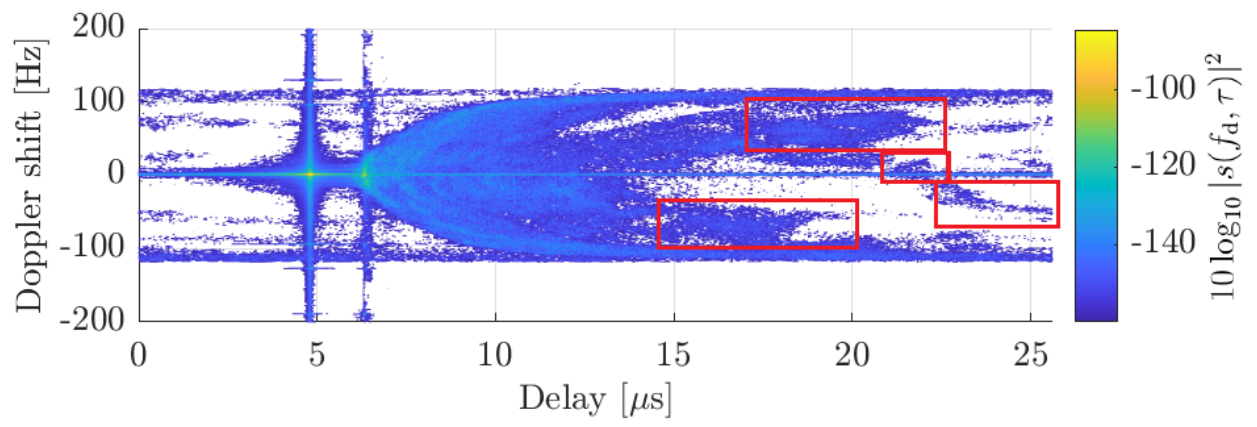

(d) Measurements (second location, 30 seconds afterwards).

Figure 4.3: Channel obtained using the proposed channel modeling approach (higher row: figures a) and b)) and through measurements (lower row: figures c) and d)) in the valley scenario [115]. The results for two locations are shown, with a separation of roughly 30 seconds between them. The red rectangles highlight some clusters of scattering components. The channel changes between both measurement locations and our technique can track these changes accurately.

power difference between LoS and SR component is decreased or in rare cases the SR component could even be stronger than the LoS during banking maneuvers.

- **Impact of Terrain and Take-Off/Landing Scenarios:** The other challenging scenarios for an A2A channel are shown in [122]. During take-off and approach, the challenges are of a different nature. The proximity to the ground introduces scattering, reflection, and attenuation from terrain, buildings, and atmospheric conditions similar to terrestrial communication channels. For civil aircraft, the multipath coming from airports and ground infrastructure only becomes a problem when they are flying at low level like during take-off and landing at the airport [122]. The problem intensifies in military low-altitude flight regimes, where the influence of the terrain is a constant threat. Valleys or urban structures can obstruct signals entirely, while high maneuver rates cause rapid shifts in antenna orientation and impose high Doppler shifts and spreads. Furthermore, the SR component and the scattering components can become very strong.

Mitigation strategies were developed to address these challenges. For example antenna diversity, i.e., placing multiple antennas across an airframe, reduces the risk of complete shadowing during maneuvers. Error correction coding and adaptive power control strengthen link reliability under conditions of attenuation and multipath fading.

Taken together, these approaches represent an ongoing effort to reconcile the physics of radio propagation with the operational demands of flight. The challenge is not merely to preserve connectivity, but to ensure that communication remains dependable in the exact moments when it is most critical—during maneuvering, in congested airspace, and at low altitudes where safety margins are at their thinnest.

## 4.5 Conclusions

The A2A channel models are an important part of the ITS framework providing the foundation for reliable, secure, real-time communication in challenging conditions. Recent advancements in non-stationary 3D geometric-stochastic modeling by using finite and infinite planes have significantly improved model fidelity and allowed usage in high mobility, non-stationary conditions for all flight phases in a low level flight. The generality of the presented model allows the application to any A2A channel.

# Chapter 5: V2V and V2X Communication

## 5.1 Overview

By **Ruisi He**

With the rapid development of autonomous driving and ITS, V2V and V2X communications have emerged as key technologies for enhancing traffic safety, efficiency, and intelligence [123] [124]. V2V and V2X communications enable wireless connections among vehicles, infrastructure, pedestrians, and cloud networks, facilitating information sharing and cooperative perception among traffic participants. This lays a solid foundation for the construction of future-oriented transportation systems.

V2V communication primarily supports low-latency information exchange between vehicles, enabling applications such as emergency braking warnings, cooperative lane changes, and platooning. In contrast, V2X is a broader concept that encompasses V2V, Vehicle-to-Infrastructure (V2I), Vehicle-to-Pedestrian (V2P), and Vehicle-to-Network (V2N) communications. By establishing a highly connected perception–decision–control loop, V2X plays a crucial role in enabling Advanced Driving Assistance Systems (ADAS) and Autonomous Driving Systems (ADS).

Although V2X technology has made rapid progress in recent years, large-scale deployment still faces numerous challenges, including the compatibility and standardization across multiple communication technologies, as well as the lack of comprehensive V2V and V2X channel measurements and models [125] [126] [127] [128]. In the future, with the continued advancement of 5G, edge computing, AI-based perception, and cooperative control technologies, V2X is expected to evolve in terms of system intelligence, autonomy, and environmental adaptability, and play a critical role in emerging domains such as digital twins for cities, vehicle-road cooperative control, and green mobility.

Characterization and modeling of V2X channels are crucial for evaluating performance of V2X communication systems. Currently, there have been several researches on V2X channel measurement and modeling. These include both scenario-specific and frequency-specific case, such as measurements for urban [129], rural [130], highway [131], and intersection [132] environments, as well as for various frequency bands like Sub-6 GHz [133], millimeter-wave [134], and THz [135].

COST INTERACT has actively contributed to the advancement of V2X channel models. For example, a realistic V2X real-time simulation framework named SIVERT is introduced in [129], specifically addressing the challenges of urban environments. This framework integrates the VENERIS project implemented in Unity3D and Network Simulator 3 (NS-3), adopting the Geometry-based Stochastic Channel Model (GSCM) approach to ensure spatially consistent channel modeling in complex urban multi-path propagation environments. The framework supports accurate modeling of urban scenarios where buildings, intersections, and dynamic vehicle interactions contribute to highly variable channel conditions.

Moreover, it allows for detailed antenna modeling through the Effective Aperture Distribution Function (EADF), essential for capturing the intricate propagation effects typical of urban environments. The study in [133] discusses the wireless channel measurements for V2V, V2I, and V2P communication, which were conducted at two frequency bands: 3.2 GHz and 5.81 GHz. The dataset includes synchronized sensor data such as Radar, LiDAR, and GPS. The measurement system employs an OFDM-based multi-band channel sounder, with an effective bandwidth of 150.25 MHz and 601 subcarriers spaced by 250 kHz. The dataset also provides Frame Error Rate (FER) measurements based on IEEE 802.11p communication systems.

The result allows for the detailed study of V2X channel characteristics and their interaction with environmental factors, providing valuable insights for the development and testing of future V2X communication systems. In [134], a joint Sub-6 GHz and mmWave V2I MIMO channel measurement campaign was conducted with a particular focus on characterizing the mmWave band at 26.5 GHz. This frequency band is of particular interest due to its potential to provide ultra-high data rates and low-latency communication required for advanced V2X systems. The measurements were performed in an urban environment using two real-time channel sounders operating simultaneously at 5.93 GHz (Sub-6 GHz) and 26.5 GHz (mmWave).

The experimental setup involved distributed antennas for the Sub-6 GHz band and a single omnidirectional antenna for the mmWave band, specifically designed to evaluate Power Delay Profile, Doppler Spectral Density, and path loss.

However, existing research is insufficient for addressing vehicular sensing channel. As a key application in 6G networks, ISAC-aided V2X communication plays an essential role in enhancing vehicle safety and performance. For example, lots of communication transceivers and sensors are equipped on autonomous vehicles, and it is possible to reuse the current dense deployment Road-Side Units (RSUs) for sensing environments with only minor modifications in hardware, signaling strategy, and standards. To fully realize the potential of ISAC-aided V2X communications, a deep understanding of the radio propagation mechanisms in vehicular sensing channel is essential.

Different from conventional communication channels, sensing channel typically describes the propagation of echoes from vehicles to scatterers and then back to vehicles within the same environment, due to the monostatic ISAC setup. The sensing channels differ significantly from the traditional channel in many aspects. For example, it is only related to the presence of scattering objects in the surrounding environment and is independent of LoS and NLoS communication scenarios. Therefore, vehicular sensing channels are more sensitive to the driving environment, such as surrounding vehicles, pedestrians, and other scatterers.

This results in abundant sensing multipaths, with these paths undergoing a more rapid birth and death process. Moreover, owing to the dynamics of these scatterers and the mobility of vehicles, time-varying behavior is a crucial characteristic of vehicular sensing channels, which should be sufficiently incorporated into channel modeling. Although the aforementioned V2X channel measurements cover most dynamic scenarios, these measurements and models cannot be generalized to vehicular sensing channels.

Due to the sensitivity to the environment, dynamic vehicular sensing channel modeling is a challenging task. The novel features of vehicular sensing channels, such as sensing multipath, clutter multipath, semantic information, etc., are expected to be captured in the measurement [136]. Besides, the typical features such as multipath birth-death, sparsity, non-stationarity, etc., differ significantly from conventional vehicular channels, which is necessary to study.

The dynamic vehicular sensing channel measurement is conducted in the Songshanhu district of Dongguan, located in Guangdong, China, at a frequency band of 28 GHz as in [137]. The mean vehicle speed during the measurements is 30 km/h. The vehicular sensing channel measurement system based on Vector Signal Transceiver (VST) is designed, including a signal generator-based TX, a signal digitizer-based RX, and a power supply based on the vehicular battery. The TX and RX use a directional horn antenna and a 4x8 array antenna, respectively. The sounding signals are multicarrier signals with 1 GHz bandwidth, and are transmitted with the maximum power of 28 dBm. During the measurement, RGB video data from the camera and point-cloud data from the radar are collected and stored simultaneously to aid in mapping between channel multipaths and environmental scatterers.

Based on actual measurements, a dynamic sensing channel TDL model is presented in [137], which is composed of Sensing MPCs (S-MPCs) and Clutter MPCs (C-MPCs). For S-MPCs, they typically exhibit higher power and more pronounced clustering, attributed to the strong reflection from sensing targets, and occur continuously over time. In contrast, C-MPCs have lower power and are generally distributed across the entire delay domain due to widespread scatterers and noise, and occur randomly over time. A tracking algorithm is used to identify S-MPCs and C-MPCs, and a series of characterization, including the number of new S-MPCs, lifetimes, initial power and delay positions, dynamic variations within their lifetimes, clustering, power decay and fading of C-MPCs, is statistically modeled to enable simulated channel with dynamic evolution process [137].

## 5.2 Vehicle-to-Vulnerable Road Users Channel Model

By **Ibrahim Rashdan**

Direct Vehicle-to-Vulnerable Road Users (V2VRUs) communication can prevent accidents by providing 360° awareness and improving detection, localization, and tracking of both vehicles and Vulnerable Road Users (VRUs). Having a realistic propagation channel is a prerequisite for developing a reliable V2VRU communication system. The following presents the first complete parameterization and validation for WINNER-type V2VRU GSCM at 5.2 GHz based on channel measurements. In order to parameterize the GSCM channel model, the Large Scale Parameters (LSPs) and their correlations are estimated in the log domain.

The results show that the log-normal distribution provides a good fit to the distributions of the LSPs. The proposed model is then validated by comparing the distributions of the model parameters, extracted from the simulated channels, with their counterparts extracted from the measured channels. The validation shows that the proposed model provides a very good representation for the V2VRU propagation channel in the considered scenarios. This channel model can be used in simulations to develop and evaluate V2VRU communication and collision avoidance algorithms in critical accident scenarios.

### 5.2.1 Measurements

Channel measurements were conducted in October 2018 in Göthestraße in the city of Germering near Munich (see Fig. 5.1) where 3 – 6 story buildings lined up along the street on one side, while separated by green area on the other side. The street consists of one lane for each direction with parked cars on both sides. It is 12 m wide, with 3 m wide sidewalks. More details on the measurement campaign can be found in [138].

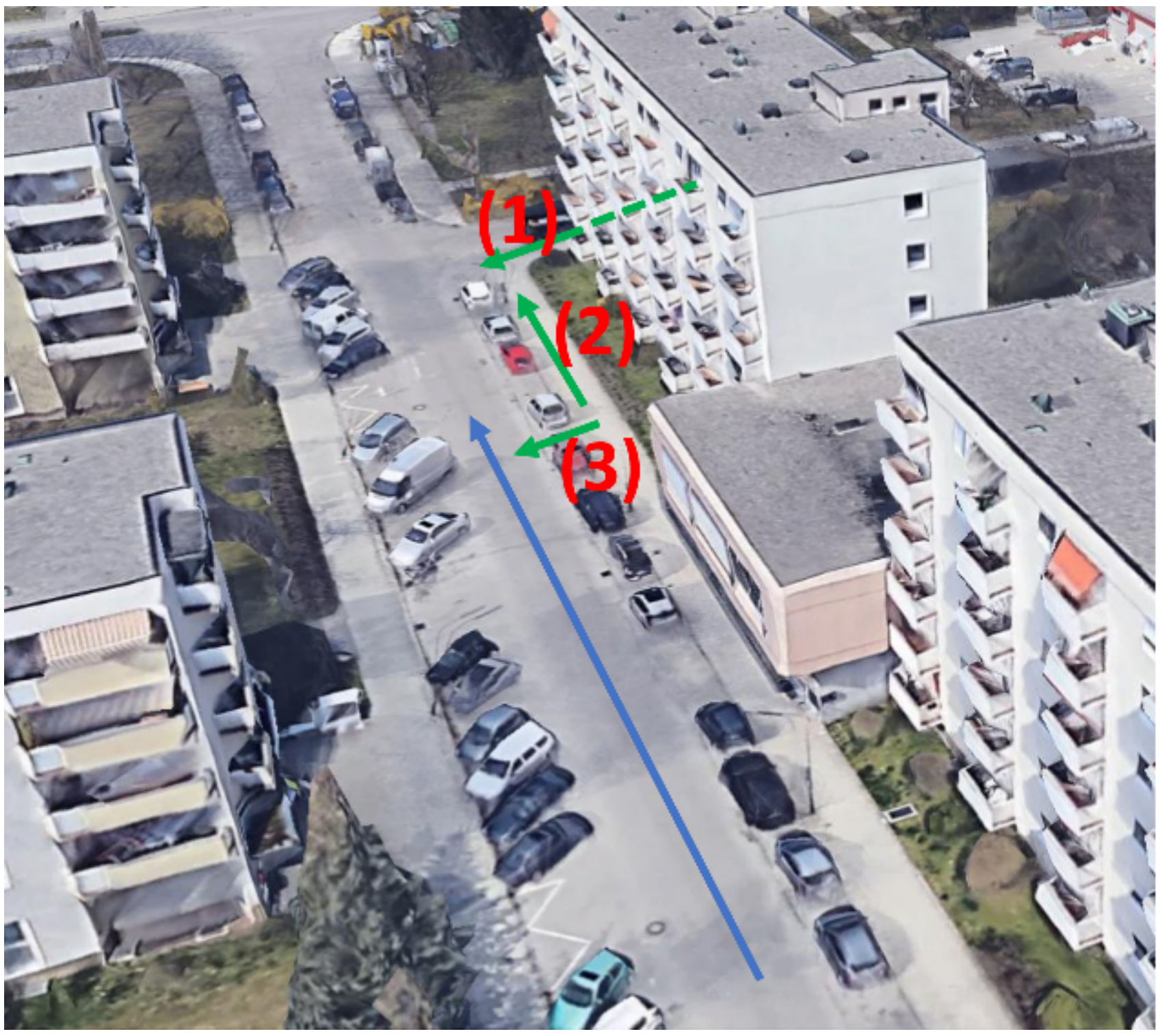

Figure 5.1: Aerial view of the measurement scenarios in the urban environment showing the trajectory of the vehicle (in blue), and the trajectory of the cyclist/pedestrian (in green with the number of the scenario) towards the imaginary collision point. (Google Earth 2018 Geobasis-DE/BKG).

The V2VRU measurement campaign was carried out using the DLR RUSK channel sounder. The DLR RUSK channel sounder is organized into separate TX and RX units, as illustrated in the block diagram in Fig. 5.2. On the TX side, a power unit and battery supply the modules. The Reference Frequency Network (RFN) distributes the timing and frequency signals. The Local Oscillator (LO) provides a 5.2 GHz carrier, which is fed into the Portable Transmitting Station (PTS). The generated OFDM-like signal is then amplified by the Power Amplifier (PA) and transmitted through the antenna.

On the RX side, the signal first passes through a Low-Noise Amplifier (LNA), followed by the Radio Frequency Tuner (RFT) which down-converts the signal. The Digital Receiving Unit (DRU) is responsible for the analog to digital conversion, Automatic Gain Control (AGC), and recording snapshots of the Channel Impulse Response (CIR) every 1.024 ms. Both TX and RX share a 10 MHz reference clock and a calibration link to ensure synchronization and system accuracy. Together, these blocks allow precise characterization of the propagation channel independent of the antenna effects [139].

The measurement signals were transmitted at a carrier frequency of $f_c = 5.2$ GHz, which is close to the 5.9 GHz ITS band. The bandwidth was $B = 120$ MHz, and thus providing a delay resolution of $\Delta\tau = 8.33\,\text{ns}$. During the measurements, the time-variant channel transfer function was recorded every $T_g = 1.024\,\text{ms}$, allowing to resolve a maximum Doppler shift of $f_{d\,\max} = 488\,\text{Hz}$. In order to record the position of the TX and the RX antennas, Global Navigation Satellite System (GNSS) receivers were used. The antennas' positions on the TX and RX sides can be seen in Fig. 5.3.

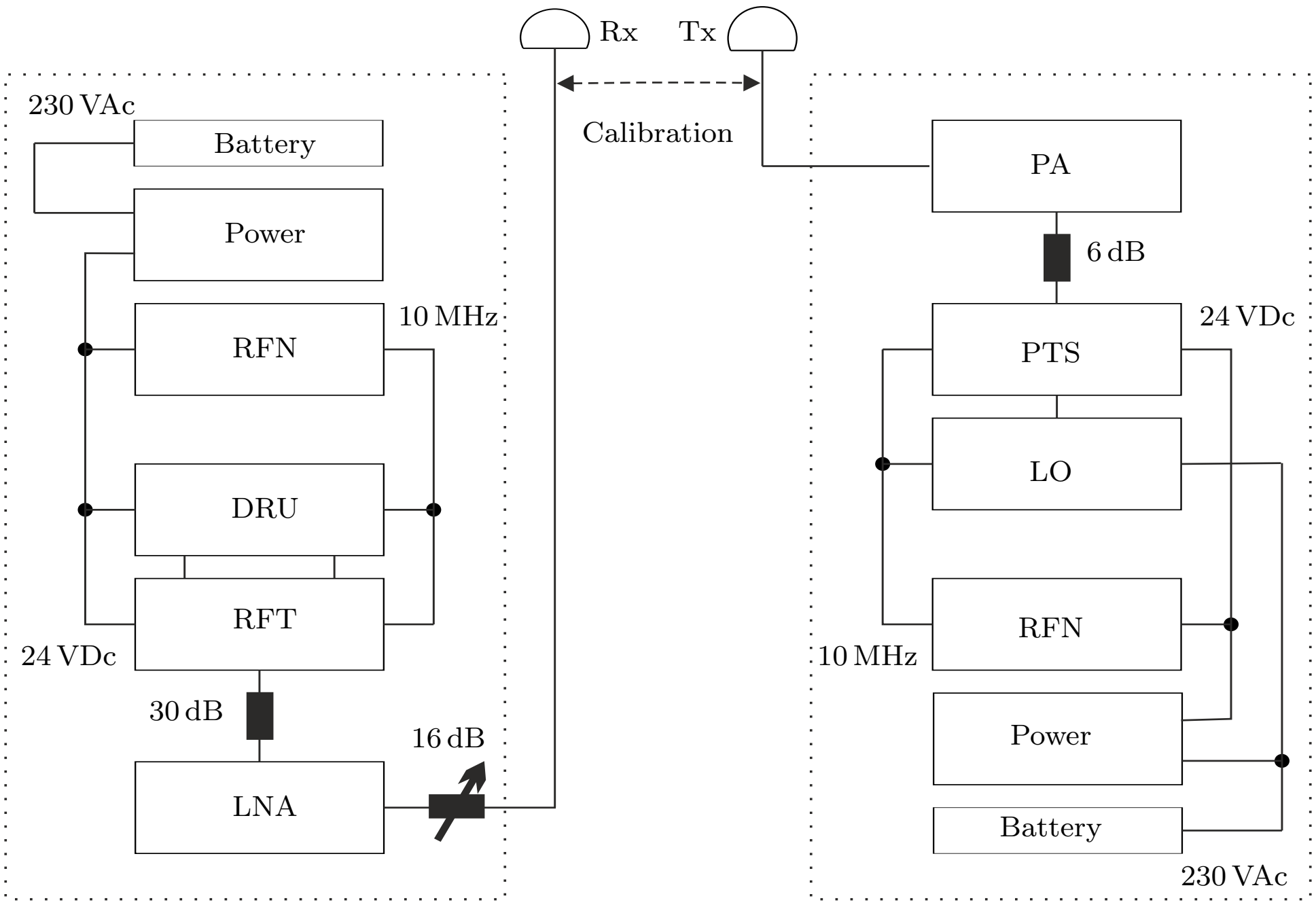

Figure 5.2: Block diagram of the channel sounder, left the RX and right the TX [139].

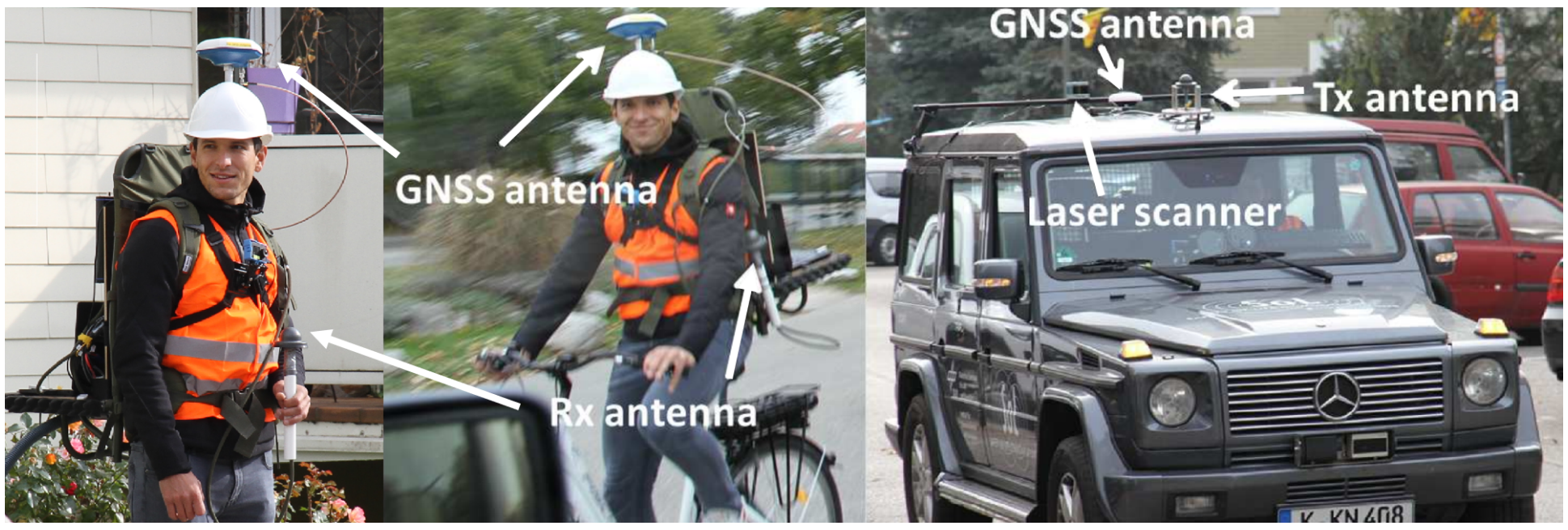

Figure 5.3: Antenna positions on the pedestrian, the cyclist, and the test vehicle.

Fig. 5.1 displays the trajectories of the test vehicle, the cyclist, and the pedestrian on an aerial view for three accident scenarios. Here, the channel model for Scenario 2 will be presented. In this scenario, the cyclist is moving parallel to the vehicle toward the collision point at the intersection. The LoS between the vehicle and the cyclist repeatedly transits between LoS and Obstructed LoS (OLoS) situations due to obstruction by parked vehicles.

## 5.2.2 Model

For developing a GSCM for V2VRU communication by adapting the WINNER II channel model, the LSPs need to be estimated. The LSPs describe the distribution of the power over the delay and angular dimensions, and control the evolution of the channel model. The LSPs are estimated in the power and delay domain, i.e., the Shadow Fading (SF), Delay Spread (DS), K-Factor (KF), and in the angular domain, i.e., Azimuth Spread of Departure (ASD), and Azimuth Spread of Arrival (ASA).

In order to maintain the spatial correlation of the LSPs observed in the measured channel, the autocorrelation of these LSPs is analyzed and the correlation distances are calculated. Furthermore, to ensure the spatial consistency, the cross-correlation coefficients among the LSPs are calculated. The model parameters can then be used as an input to the channel simulator. The first- and second-order statistics of the LSPs are estimated from the parameters of the specular MPCs, i.e., the power, delay, and angles.

Due to the non-stationarity of the channel, the channel is divided into regions, within which the WSS assumption holds. The LSPs and their correlations are evaluated within these regions using a sliding average window of length of 15 $\lambda$ (0.86 m). The length of the window, i.e., the stationarity distance, is found to be valid for all scenarios based on the estimated stationarity distance in [138].

The modeling approach in this work follows the QuaDRiGa modeling approach [140]. The QuaDRiGa model extends the well-known WINNER II [141] and WINNER+ [142] models by adding time evolution and 3D propagation. The log-normal distribution is widely used to model the LSPs in models like WINNER II, 3GPP, and Quadriga. Following the WINNER II approach, the LSPs are fitted to log-normal distributions, ensuring compatibility with standards and comparability with existing literature. Root Mean Square Error (RMSE) is used to quantify the root average squared difference between the empirical and theoretical Cumulative Distribution Functions (CDFs) as in [143], [144], and [145], providing an evaluation of the fit. The RMSE also enables straightforward evaluation of the similarity between the CDFs of LSPs derived from measured and simulated channels, reinforcing the validity of the model.

The model parameters, extracted from the measured channels, are used as an input to the QuaDRiGa channel simulator. The GSCM for V2VRU communications is validated by comparing the distributions of the model parameters, extracted from the simulated channels, with their counterparts extracted from the measured channels, which are referred to as the model input. The validation is performed using a group of 100 simulated channels produced by 100 simulation runs for each scenario. These simulation runs are carried out using TX and RX routes similar to the routes during the channel measurements.

Fig. 5.4a shows the measured path loss variations with the TX-RX distance and the proposed path loss models for Scenario 2. The path loss is modeled by two single-slopes log-distance models. The first slope is for distances up to 9 m and covers the LoS area prior to the collision with an estimated path loss exponent of 1.7. For distances between 9 m and 100 m, the link between the vehicle and the cyclist becomes partially obstructed by parked vehicles, which causes a diffraction loss of maximum 5 dB.

The estimated path loss exponent equals 2.4, which is slightly above the one of the free space model. The local mean path loss, calculated from all 100 simulated channels and depicted by the dashed black curve, is compared with the proposed path loss model. It can be seen that the channel model clearly provides an almost perfect match in terms of path loss.

Figs. 5.4b – 5.4f show the results of parameterization and validation of the channel model in terms of the SF, DS, KF, ASD, and ASA, respectively. Tables 5.1 summarize the comparison results. Detailed parameterization and validation results for Scenario 1 and Scenario 3 can be found in [146].

In terms of SF, the log-normal distribution provides a good approximation of the measured SF distribution with an RMSE value of 0.006. The dotted black curve represents the CDF of the SF derived from 100 simulated channels. A slight mismatch is observed, where the estimated standard deviation from the simulated channels is approximately 1 dB larger than that of the model input resulting in an RMSE of 0.021.

Fig. 5.4c illustrates the RMS DS results divided into LoS and OLoS.The log-normal fit (model input) provides a good approximation of the measured CDF in both LoS and OLoS with RMSE values of 0.032 and 0.020, respectively indicating a good fit. However, the tails of the log-normal fit and the simulated CDFs exhibit slightly larger DS values compared to the measured data. The measured mean DS values are 6.45 ns for LoS and 25.12 ns for NLoS, closely matched by the simulated mean DS of 7.94 ns and 23.99 ns, respectively as detailed in Tables 5.1.

Fig. 5.4d illustrates the CDFs of the measured KF, the log-normal fit (model input), and the simulated channels. The larger KF observed in the LoS condition compared to OLoS is attributed to the presence of a strong, unobstructed LoS path. In the OLoS condition, the LoS path is obstructed by parked vehicles, resulting in attenuation due to diffraction loss. The parametrization results show that the log-normal distribution provides a good fit to the measured KF with RMSE values of 0.014 and 0.019 for LoS and OLoS, respectively. The simulated channels closely match the KF in the measured data with relatively low RMSE values indicating that the proposed model successfully captures the KF characteristics across various propagation conditions.

Figs. 5.4e and 5.4f illustrate the results of ASD and ASA. The parametrization results demonstrate that the measured ASD and ASA closely follow the log-normal fit with minor mismatches and a maximum RMSE of 0.014. The validation results reveal that the simulated channels align well with the measured ASD and ASA, showing good agreement in terms of both the mean and the standard deviation. Detailed parametrization and validation results can be found in [146].

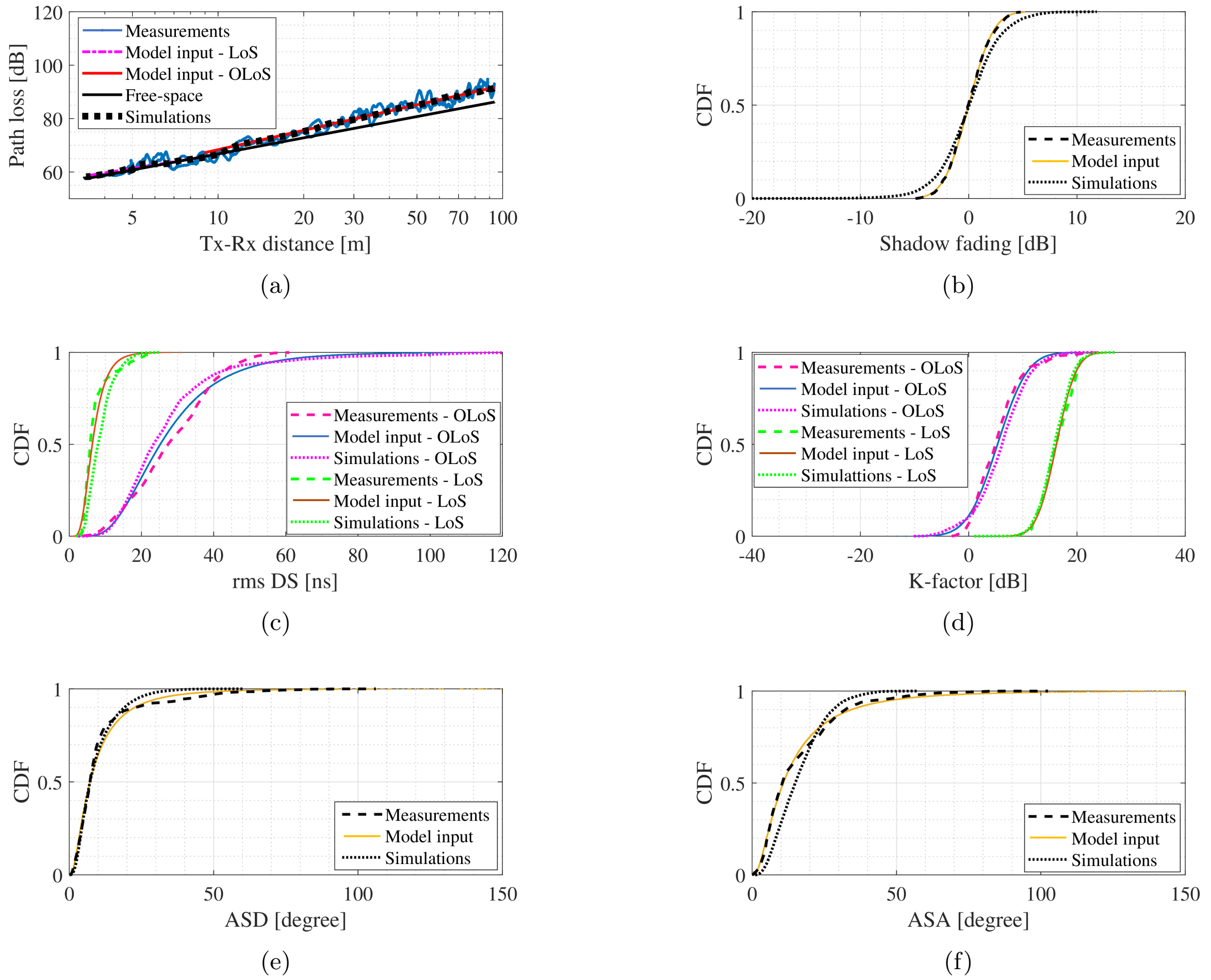

Figure 5.4: Parameterization and validation of the channel model.

### 5.2.3 Link-level Performance Evaluation

In order to develop a reliable communication system and evaluate its performance, realistic channel models are required. The following will investigate this statement through link-level simulations.

The aim of the simulations is:

1. To show the effect of the channel model on the link-level performance.
2. To further validate the proposed V2VRU channel model.

The link-level simulator developed by u-blox, known as the Ublox-V2X simulator [147], is utilized. This simulator is an open-source MATLAB model designed to assess the performance of IEEE 802.11p in terms of PER. The evaluation is conducted across various Signal-to-Noise Ratio (SNR) levels, employing two Modulation and Coding Schemes (MCSs): BPSK 3/4 and 16-QAM 3/4. For the simulations, a packet size of 350 bytes is chosen following the ETSI in [148], with further details available in [149].

PER is defined as the ratio of the number of incorrectly received data packets to the total number of transmitted packets. Lower PER values indicate better performance. In our simulations, 100,000 frames were transmitted for each SNR value. For instance, a PER curve for 10 SNR values is based on 1 million transmitted frames, with a new channel realization with unity gain generated for each frame.

To demonstrate the impact of the channel model on link-level performance, communication performance is analyzed using our WINNER-type GSCMs developed for V2VRU communication, along with two TDL channels and an Additive White Gaussian Noise (AWGN) channel. The two TDL models were developed for urban V2V scenarios and have been adopted by ETSI for V2V communications [150]. The first scenario involves an urban LoS, where two vehicles approach each other in an urban environment with maintained LoS. The second scenario is an urban NLoS, where two vehicles approach an intersection from different directions with LoS blocked by buildings. More details about these TDL models can be found in [151].

| Parameters | | Model input | | Simulated channels | |
|---|---|---|---|---|---|
| | | **LoS** | **OLoS** | **LoS** | **OLoS** |
| SF [dB] | $\mu$ | 0 | 0 | 0 | 0 |
| | $\sigma$ | 1.80 | 1.80 | 2.82 | 2.82 |
| Corr. distance [m] | $d_c$ | 2.27 | 2.27 | 5.54 | 5.54 |
| RMSE | | 0.006 | 0.006 | 0.021 | 0.021 |
| K-factor [dB] | $\mu$ | 16.22 | 5.34 | 15.90 | 5.93 |
| | $\sigma$ | 2.88 | 4.38 | 2.87 | 4.80 |
| Corr. distance [m] | $d_c$ | 4.42 | 30.58 | 4.57 | 46.05 |
| RMSE | | 0.014 | 0.019 | 0.015 | 0.038 |
| DS [$\log_{10}(s)$] | $\mu$ | –8.19 | –7.60 | –8.10 | –7.62 |
| | $\sigma$ | 0.19 | 0.21 | 0.18 | 0.21 |
| Corr. distance [m] | $d_c$ | 4.42 | 70 | 4.57 | 59.45 |
| RMSE | | 0.032 | 0.020 | 0.090 | 0.034 |
| ASD [$\log_{10}$ (°)] | $\mu$ | 0.86 | 0.86 | 0.86 | 0.86 |
| | $\sigma$ | 0.39 | 0.39 | 0.32 | 0.32 |
| Corr. distance [m] | $d_c$ | 63.23 | 63.23 | 66.03 | 66.03 |
| RMSE | | 0.014 | 0.014 | 0.021 | 0.021 |
| ASA [$\log_{10}$ (°)] | $\mu$ | 1.03 | 1.03 | 1.14 | 1.14 |
| | $\sigma$ | 0.39 | 0.39 | 0.26 | 0.26 |
| Corr. distance [m] | $d_c$ | 66.17 | 66.17 | 71.48 | 71.48 |
| RMSE | | 0.010 | 0.010 | 0.057 | 0.057 |

Table 5.1: Comparison of LSPs for Scenario 2.

To further validate our proposed channel model, the PER was also evaluated using the measured V2VRU channel and compared with the PER achieved using the proposed model.

In Fig. 5.5, we plot the PER versus SNR for all channel models and with two MCSs for Scenario 2. In this scenario, the cyclist, as illustrated in Fig. 5.1, moves alongside the vehicle towards the intersection's collision point. The PER is inversely proportional to the SNR. It is observed that PER performance worsens, when a higher MCS is used while keeping the SNR constant. Consequently, a higher SNR is needed to achieve the same performance when transitioning from a lower to a higher MCS order. Compared to BPSK 3/4, PER decreases more slowly with increasing SNR for 16-QAM 3/4.

As expected, the AWGN channel (dotted line) outperforms the fading channels. When comparing performance using the Urban LoS and Urban NLoS V2V TDL channels (circled and crossed lines), it is clear that the Urban LoS channel is less challenging, providing better performance due to the strong LoS and lower delay spread. The LoS channel's performance is similar to that of the GSCM at low SNR levels, specifically below 5 dB and 17 dB with BPSK 3/4 and 16-QAM 3/4, respectively.

However, the GSCMs reveal that PER does not decrease with increasing SNR, resulting in an error floor. Unlike AWGN and TDL channels, an error-free link is unattainable for the V2VRU GSCMs regardless of SNR. This error floor may be mitigated by enhancing receiver design and implementing more efficient channel estimation techniques.

When comparing the PER of the measured channel with that of other channel models, our GSCM provides the closest match. This emphasizes the importance of using a dedicated and realistic GSCM to accurately evaluate communication system performance.

### 5.2.4 Conclusion

A full parameterization for the WINNER-type GSCM was proposed. The LSPs were estimated in the power and delay domain, i.e., SF, DS, narrowband KF, and in the angular domain, i.e., ASD, ASA. The log-normal distribution was found to provide a good fit to the distributions of the LSPs.

The proposed model parameters are used as an input to the QuaDRiGa channel simulator. The proposed channel model is then validated by comparing the simulated channels with the measured channels. It can be concluded that the GSCM with the proposed model parameters is able to produce similar statistics as the measured channels. The proposed model provides a good representation for the V2VRU propagation channel in the considered scenarios. Therefore, it can be used by researchers to develop a reliable V2VRUs communication system and to evaluate their collision avoidance algorithms in critical accident scenarios.

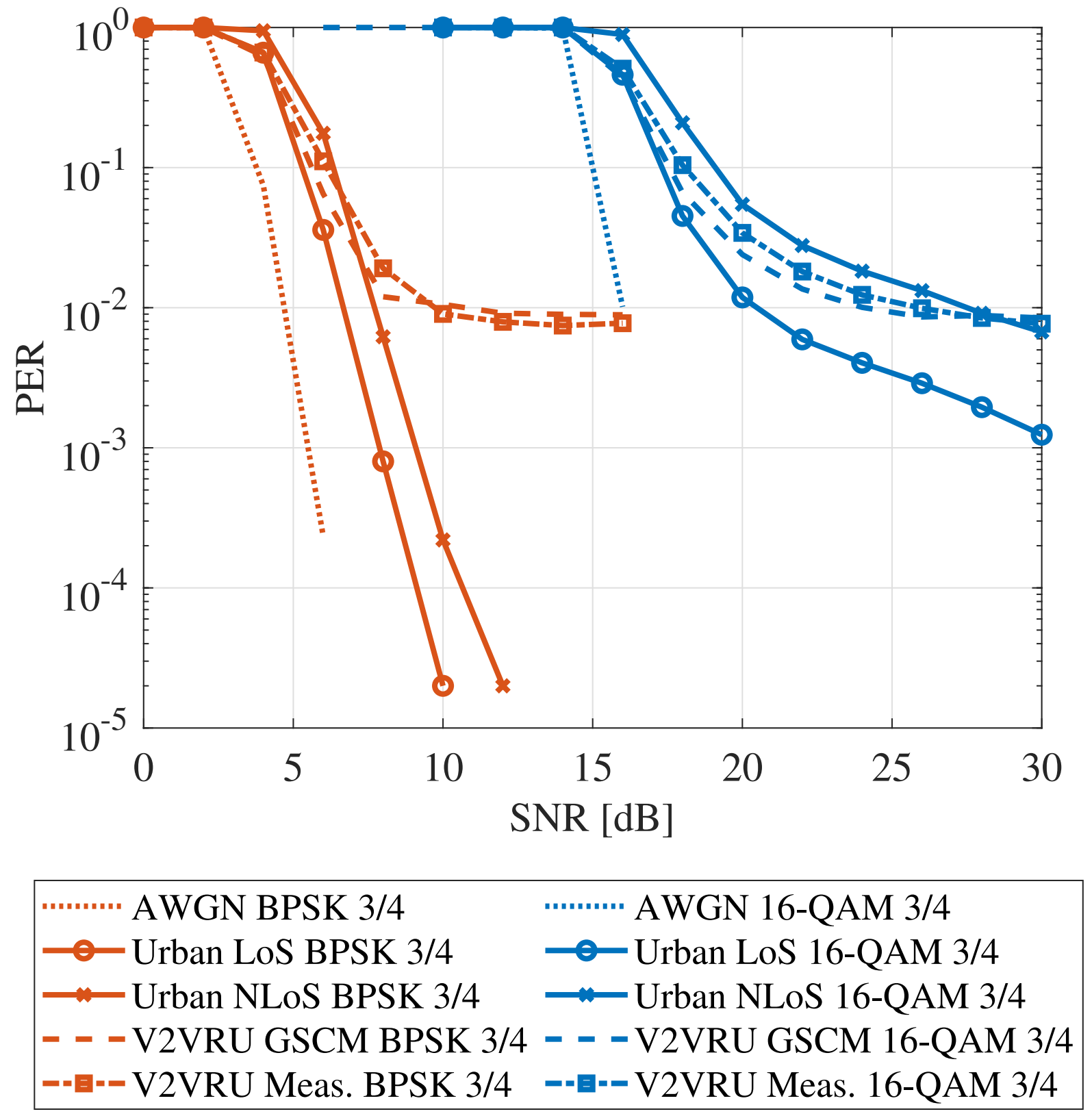

Figure 5.5: Packet error rate versus signal to noise ratio for Scenario 2 for two MCSs and different fading channel models.

Furthermore, the impact of the channel model on the performance of the communication system through link-level simulations is investigated. The results reveal the challenges that V2X communication systems have to face in urban environments: namely an error floor despite the high SNR. Using the proposed GSCM in link-level evaluations provides a more realistic performance compared to the optimistic performance obtained by the TDL channels.

## 5.3 Learning Assisted Content Delivery in Fragmented Vehicular Networks

By **Gianluca Rizzo**

The amount of data generated, exchanged, and consumed in vehicular networks is expected to increase exponentially in the near future [152], driven by the progressive adoption of ITS services and applications and by the diffusion of coordinated autonomous driving. In such a context, direct V2V communications play a pivotal role, enabling low-latency data exchanges while offloading the cellular infrastructure from information exchanges with mainly local relevance.

This includes, for instance, sensor and telemetry data from vehicles exchanged in the context of smart city applications or disaster response [153]. Additionally, bulky 3D models and point clouds are exchanged among vehicles to support autonomous driving, enhance situation awareness for road safety, or facilitate context-based AR vehicular applications [154], [155].

In this section, we focus on the problem of delivering a content object (e.g., a warning message, a map update, or an element of a 3D model, to name a few) via direct V2V communications to those vehicles who request it within a given region of space, and some maximum delay. V2V communications can be inadequate for supporting such a service when relying on host-based protocols such as IP and point-to-point information exchanges [156], [157], [158].

This drawback is particularly evident in highly dynamic scenarios or those settings characterized by (locally) sparse spatial distribution of vehicles, resulting in fragmented and rapidly varying network topologies. Indeed, these features often bring unstable inter-vehicular connectivity, transmission errors, and high and unpredictable latencies, making it difficult to implement reliable and timely exchanges [156], [158], [159].

The Named Data Networking (NDN) [160] communication paradigm, a prominent implementation of Information Centric Networking (ICN), has been proposed to overcome these shortcomings. It is based on in-network caching and information retrieval according to the content name instead of the host location. NDN takes advantage of content redundancy in the network to minimize the number of transmissions required to deliver a given content, potentially improving the effectiveness of content retrieval on the occurrence of network topology changes [161]. Several key issues, however, hamper NDN's efficient and successful use in V2V communications [157], [162].

Specifically, as current NDN approaches typically require the existence of paths between content producers and requesters, their applicability is restricted to connected components in network scenarios that are *fragmented*, i.e., in which at any time instant on average, every node is connected with only a small subset of the other nodes in the region. At the same time, in highly dynamic scenarios, they are severely limited by the instability of those paths [163], [164], [165].

In sparse environments, where store-carry-and-forward is the main mode of communication, several Delay Tolerant Networking (DTN) schemes for location-based probabilistic content caching and delivery have been proposed, such as Floating Content (FC) [166], [167], hovering information [168], and others [169]. They allow tuning opportunistic content replication to minimize the probability of content disappearance from a given region (due to churn) and achieve a target delivery ratio. However, they do not support content retrieval with strict latency requirements, and they are very inefficient (implying many unnecessary content replications) as intra-cluster communication strategies, as well as whenever the content object is of interest to only a fraction of the nodes in the region [170].

A promising approach to address these limitations is to combine NDN for intra-cluster content exchanges (for resource efficiency and low latency) with DTN schemes (for robustness against network fragmentation and sparsity) [162], [171], [172], [173]. Although the majority of existing results are designed for single point-to-point exchanges, in networks with slowly changing, relatively stable topologies and mobility patterns, making them unsuitable for dynamic environments, and very inefficient whenever the same content is of interest to several users in the region.

In what follows, we consider a region of space in which users regularly turn into requesters for content objects, and in which infrastructure support (in the form of orchestration of content replication) is pervasive. All the aforementioned works do not address the following key open issues:

- How to implement a resource-efficient NDN content delivery service with tight delay constraints in a fragmented vehicular network to maximize direct V2V content exchanges?

- How to orchestrate the content delivery process to achieve a target performance (in terms of minimum delivery ratio and maximum latency) in a resource-efficient manner, while ensuring content persistence?

A potential approach to address these issues is the Deep NDN (DeepNDN) scheme [174], which efficiently achieves a target delivery ratio while satisfying latency and persistence constraints. It is based on the joint application of NDN and probabilistic spatial content caching, making it capable of adapting to a wide range of network topologies and dynamic settings. It relies on a learning-based approach for the dynamic management of DeepNDN, which allows for achieving a target minimum delivery ratio in a resource-efficient manner by proactively adapting the content replication and availability to local conditions. This approach employs a Convolutional Neural Network (CNN) architecture to effectively capture the complex relations between spatiotemporal patterns of mobility and the performance of the content delivery service.

### 5.3.1 Assumptions and Notation

We consider a set of mobile nodes (vehicles, pedestrians, bikes) on a road grid, each aware of its position and equipped with wireless interfaces (e.g., IEEE 802.11p, Bluetooth, cellular Device-to-Device (D2D)) for direct communication. Nodes are in *contact* when within transmission range. A cellular interface supports network management, with a SDN controller collecting mobility data to optimize operations.

The focus is the *optimal content delivery problem*: efficiently delivering a content object to all requesting nodes within a *region of interest* and an *observation interval* $T$, minimizing resource costs (cellular, storage, D2D) while meeting latency constraints. The management function monitors requests and offloads delivery tasks from cellular to V2V where possible. Nodes are partitioned into $C$ classes ($c$), road grid into $I$ segments ($i$), and $T$ into $Z$ sub-intervals ($z$). Partitioning balances computational complexity and effectiveness.

Each node is in one of three states: *neutral*, in which it does not have nor request the content; *requester*, in which it has requested but not received the content; and *producer*, in which it possesses the content. Nodes enter the region as neutral. Some may be producers at $T$'s start. A class $c$ node in segment $i$ during $z$ becomes a requester with probability $\mu_{i,c,z}$. If a requester receives the content within a delay $d$, it becomes a producer. Two infrastructure modes are considered: *full D2D* (no cellular fallback; missed requests revert to neutral) and *cloud fallback* (missed requests retrieve content via cellular).

### 5.3.2 DeepNDN Operation

DeepNDN synergizes NDN mechanisms with location-based probabilistic caching to solve the content delivery problem. It uses vehicular NDN with Interest Message (IM) flooding and neighbor-based forwarding. Content objects (or their segments) are uniquely named; nodes have unique IDs for routing.

When a node becomes a requester, it broadcasts an IM every $t_f$ seconds until it receives the content. Recipients check for the content, update their Pending Interest Table (PIT) if not found, and rebroadcast. Producers reply with a Data Message (DM), routed back via PIT entries. Nodes cache received content with a configurable probability, controlling the producer population.

To limit overhead, IMs are forwarded up to $h$ hops and expire after $d$ seconds. When nodes leave the region, they discard cached content and PIT entries.

A DeepNDN scheme is defined by arrays $\mathbf{b} = \{b_{i,c,z}\}$ and $\mathbf{k} = \{k_{i,c,z}\}$, controlling replication and caching per segment, class, and sub-interval. Producers replicate content to neutral nodes on IM receipt, or probabilistically otherwise, and cache with probability $k_{i,c,z}$. Exchanges are unicast, but the scheme can be extended to multicast.

Content disappearance probability (when no producers remain) and delivery ratio (fraction of requesters served within $d$) are key metrics. The management function monitors mobility and requests, forecasts demand, and tunes $\mathbf{b}$ and $\mathbf{k}$ to meet application-specific targets for delivery ratio and delay, optimizing resource use over the observation interval. The DeepNDN scheme's performance depends on replication ($\mathbf{b}$) and caching ($\mathbf{k}$) parameters. We formulate an optimization problem to find parameters that achieve the target delivery ratio $r_0$ while minimizing resource costs and keeping the content disappearance probability below $\epsilon$. The total cost combines storage and communication components:

**Storage Cost:**

$$S(\boldsymbol{b}, \boldsymbol{k}) = L \sum_{i,c,z} W_{i,c,z} \sum_{n \in \mathcal{N}_{i,c,z}} t_{n,z,i} \tag{5.1}$$

where $W_{i,c,z}$=storage cost, $\mathcal{N}_{i,c,z}$=nodes in segment $i$, class $c$ during $z$, $t_{n,z,i}$=storage time.

**Communication Cost:**

$$\Gamma(\boldsymbol{b}, \boldsymbol{k}) = L \sum_{i,c,z} \left[ \left( \gamma^{O}_{i,c,z} + \tfrac{l}{L} \gamma^{IM}_{i,c,z} \right) X^{D2D}_{i,c,z} + \gamma^{G}_{i,c,z} X^{G}_{i,c,z} \right] \tag{5.2}$$

where $\gamma^O, \gamma^{IM}$=D2D content/IM transfers, $\gamma^G$=cellular downloads, $X^{D2D}, X^G$=per-byte costs. Let $P_z(\boldsymbol{b}, \boldsymbol{k})$=content disappearance probability and $r_z(\boldsymbol{b}, \boldsymbol{k})$=delivery ratio in sub-interval $z$. The optimization problem is:

**Problem 1** ***(Optimal DeepNDN)***

$$\begin{aligned} \underset{\boldsymbol{b},\boldsymbol{k}}{\textit{minimize}} \quad & \Gamma + \beta S \\ \textit{s.t.} \quad \forall z : \quad & r_z \geq r_0, \ P_z \leq \epsilon \end{aligned} \tag{5.3}$$

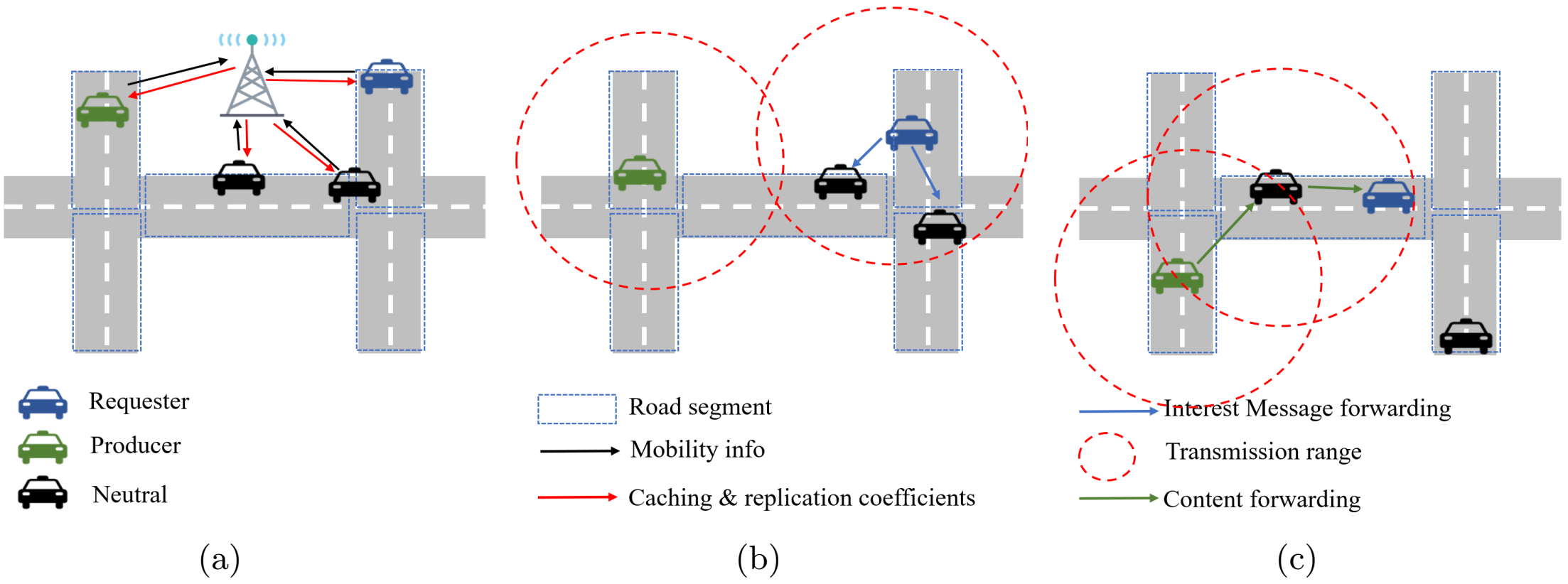

Figure 5.6: Overview. a) Nodes exchange control information with the controller; b) The requester forwards the IM to nodes within its range; c) A producer for the requested content object comes in contact with a neutral node that received the IM, and it forwards the content object to the requester via the neutral node.

where $\beta$ balances cost components, $\epsilon$=max disappearance probability. To solve Problem 1, we adopt a CNN-based approach leveraging spatial correlations in mobility and demand patterns. The architecture efficiently maps system parameters (mobility, requests) to optimal caching/replication strategies $(\mathbf{b}, \mathbf{k})$, minimizing resource costs while meeting delivery ratio $(r_0)$ and content disappearance $(\epsilon)$ constraints. The algorithm goes through three phases of operation (Fig. 5.6):

- **Data Collection**: Aggregates mobility features (node density, speed, contacts) and communication metrics (requesters, producers, delivery ratio) per road segment and time interval. Training data is generated via simulations and runtime measurements, normalized, and enriched via covariate adaptive randomization to ensure balance.
- **Strategy Computation**: Forecasts mobility $(\mathbf{m})$ and requests $(\boldsymbol{\mu})$ are input to the trained CNN, which outputs $(\mathbf{b}, \mathbf{k})$ parameters that satisfy QoS targets. The management function dynamically adjusts strategies if patterns deviate.
- **Deployment**: Parameters $(\mathbf{b}, \mathbf{k})$ are distributed to all nodes. The system handles interval splits or updates if new forecasts arise mid-operation.

The CNNs grid-processing capability captures spatial dependencies between road segments, outperforming tree-based models in accuracy. While architecture optimization is possible, our focus is on the end-to-end workflow. The approach ensures scalability by partitioning the observation interval $T$ into uniform sub-intervals when mobility/request patterns shift significantly.

### 5.3.3 Numerical Evaluation

Network simulation is implemented using Veins/OMNeT++ and mobility modeling using SUMO, with IEEE 802.11p parameters (20 mW TX power, 100 m range). Training utilized $2 \times 10^5$ data points with 10-fold cross-validation, achieving 95% confidence intervals within 3 %. Default parameters included 5 s max delay and 1800 s observation window. Comparative strategies included constrained/unconstrained variants, traditional NDN implementations, and baseline approaches (CEDO [175], SPG-ICN [176]).

**Baseline Manhattan Scenario:** A $10 \times 10$ grid (460 segments) with Manhattan mobility ($30 - 50$ km/h speed, 0.2 turn probability) hosted 100 vehicles on average. It achieved 97% delivery ratio versus 63% for unconstrained NDN (Fig. 5.7), despite 53% lower storage utilization than single-hop configurations (Tab. 5.2). Content disappearance occurred in 11% of NDN cases versus 0% for the single-hop configuration maintained equivalent performance to unconstrained forwarding while reducing IM overhead by 48%.

**Highway Scenario** A 3 km highway (90 km/h speed, 1800 vehicles/hour) tested fragmentation resilience. reduced RSU transmissions by 81% compared to SPG-ICN while doubling D2D content transfers (Tab. 5.3). IM forwarding efficiency improved by 48%, with 37% lower storage duration. Spatial correlation in CNN predictions enabled proactive caching at merging zones, preventing content disappearance during platoon separations.

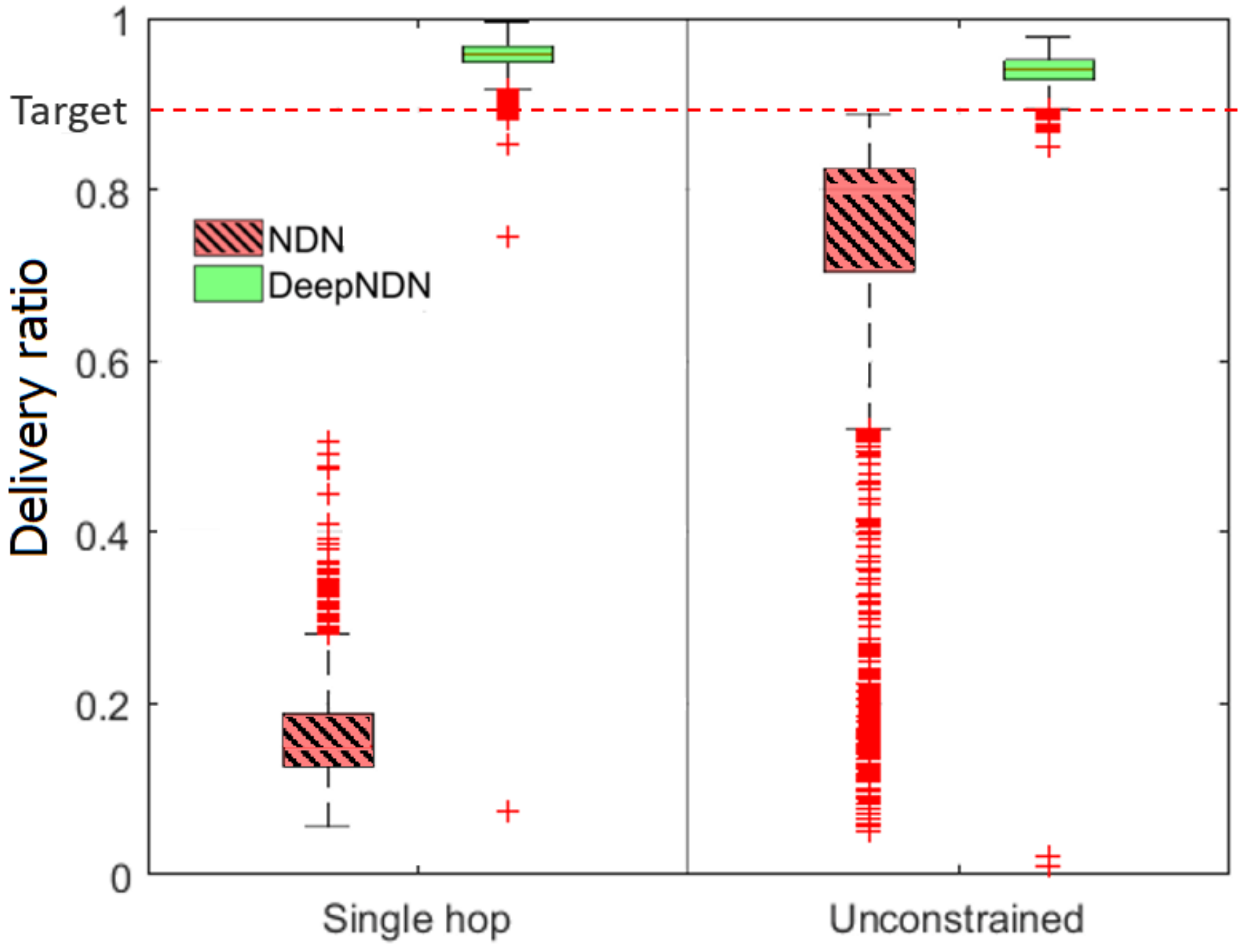

Figure 5.7: Delivery ratio comparison in Manhattan scenario (unconstrained vs. single-hop configurations).

| Metric | Single-hop | Unconstrained |
|---|---|---|
| Content availability | 0.49 | 0.32 |
| Concurrent exchanges | 8.2 | 8.3 |

Table 5.2: Resource utilization in Manhattan scenario.

## 5.4 Testing of Vehicular Communication Systems

By **Thomas Zemen, Anja Dakić**

Connected Cooperative and Automated Mobility (CCAM) can benefit from reliable wireless V2X communication links in safety and time critical situations. The ego vehicle's perception with LIDAR, RADAR and cameras is limited by the LoS. It can be augmented by V2X communication beyond the visible LoS with sensor information from other cooperative vehicles or infrastructure elements in the near vicinity.

Exhaustive testing and validation are crucial to ensure that all elements within such a CCAM system function cohesively, resulting in successful performance across various scenarios. Regarding the wireless communication link two testing methods are possible: i) drive tests with a fully equipped vehicle and ii) (scenario-based) closed Vehicle-in-the-Loop (ViL) tests. Testing and validating ADAS features require numerous kilometers to be driven and tackling various challenging traffic scenarios [177].

| Cost function components | SPG-ICN | DeepNDN |
|---|---|---|
| D2D content transfers | 431 | 778 |
| RSU content transfers | 42 | 8 |
| D2D IM transfers | 3458 | 1796 |
| Content storage [s] | 14664 | 9253 |

Table 5.3: Comparative analysis of DeepNDN and SPG-ICN in the highway scenario.

Therefore, conducting drive tests is not feasible during the design and prototyping phase of ADAS features. However, ViL tests can be applied in different development phases [178] and enable cost-efficient tests in controlled environments that can be easily repeated. The local V2X communication channel quality/reliability is a key information required for such ViL tests.

An example of such a testbed is shown in Fig. 5.8. Here, the ego vehicle, i.e., the vehicle under test, represents the central element of a traffic simulator, whose purpose is to generate a traffic scenario as well as simulate the movement of other vehicles. Additionally, it simulates all available sensors of the ego vehicle. The network simulator models the upper layers of the protocol stack, including the transport, network, and application layers. Finally, to test the performance of a physical communication link, it is necessary to use a realistic physical layer model that includes both a geometry map and a FER model.

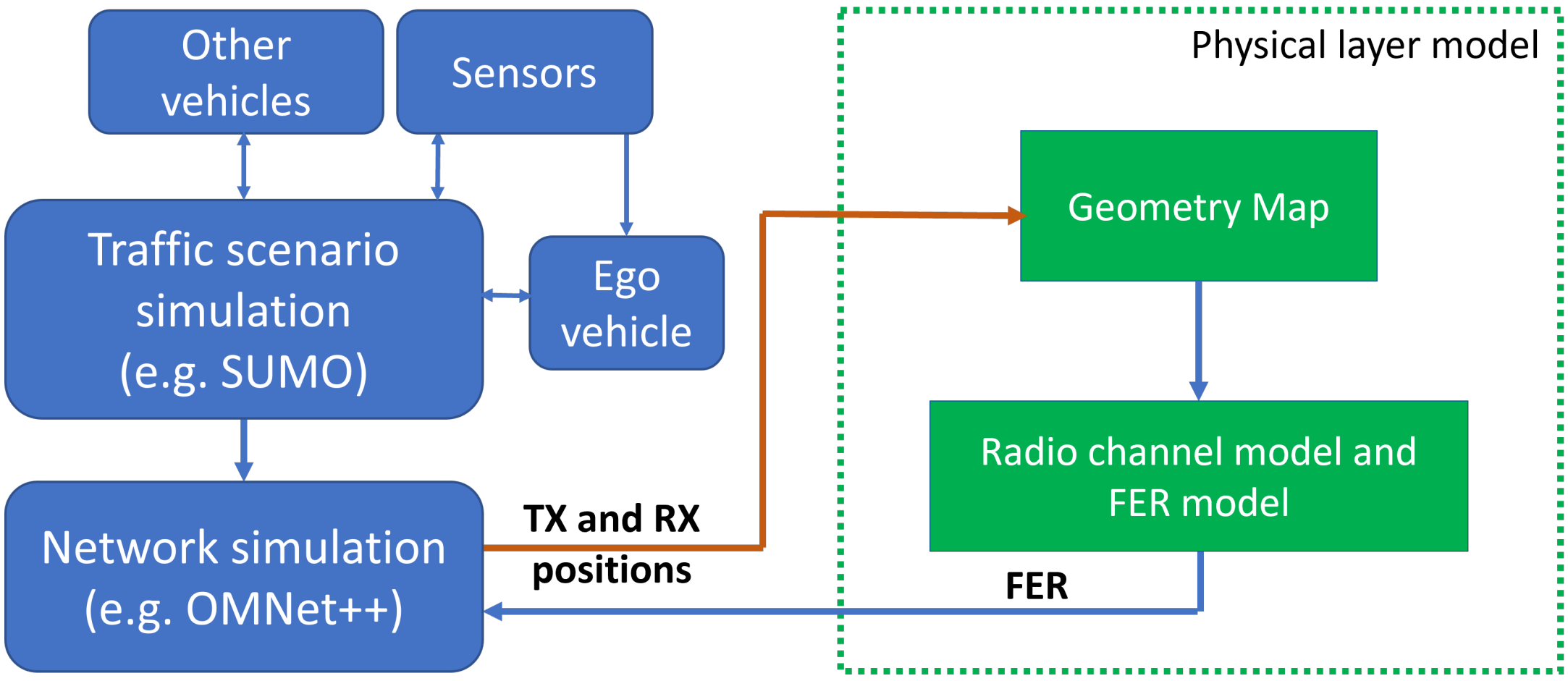

Figure 5.8: ViL testbed architecture.

In order to address information about communication reliability, a methodology for predicting the FER of V2X communication channels is presented in [179]. A Deep Neural Network (DNN) is employed for predicting the expected FER using a *single* received data frame. The DNN is trained in a supervised fashion, where a time-limited sequence of channel frequency responses has been labeled with its corresponding FER values assuming an underlying wireless communication system, e.g., IEEE 802.11p, LTE vehicular or 5G. For generating the training dataset, a GSCM is utilized as well as a hardware-in-the-loop setup. The site-specific channel emulation setup for the hardware-in-the-loop approach is compactly explained in [180].

The ground truth FER is obtained by emulating the time-varying frequency responses using a hardware-in-the-loop setup. The GSCM provides propagation path parameters for each stationarity region. These parameters are used by the channel emulator and remain constant over a chosen time period. This allows transmission of any number of frames and enables FER estimation with an arbitrary level of precision. With this unique dataset [179] achieves an accuracy of 85% for the DNN.

The improved Machine Learning (ML) model for predicting the FER is presented in [181]. In this model, additional convolutional layers are included, and it is trained for a time-varying impulse response. The achieved accuracy on the test dataset is 90.14 %. Both ML models, DNN [179] and [181] are trained with V2I data. To increase the scope of scenarios of the ML model, especially when a small amount of new data is available, we investigate different model adaptation methods in [181] with the idea to efficiently add V2V scenarios to an already pre-trained ML model. The authors of [181] present an adaptive Learning without Forgetting (LwF) method that achieves 82% combined accuracy for the FER prediction within V2V and V2I scenarios.

In [182], additional CCAM use-cases are identified, that benefit from the knowledge about the spatial reliability region of the communication link. FER classes for these regions from the ego cars perspective are provided for the decision-making of the ADAS. The authors of [182] propose a complete testbed architecture for system validation, verification and test scenario generation that integrates FER prediction through the high performance open-source computing reference framework (HOPE), and prove that the measured FER within a city scenario matches very well with the predicted FER obtained with a GSCM that uses OpenStreetMap data enriched with event-specific static objects.

# Chapter 6: Interoperability and Standardization

By **Aniruddha Chandra**

## 6.1 Why do we need an ITS standard?

ITS reduce accident risk for drivers, passengers and VRUs; lower traffic congestion and commute time; improve travel experience with value-added services like infotainment; and contribute towards a sustainable, environment-friendly future [183], [184]. Connected and Automated Vehicles (CAVs) constitute the major component of the ITS, which require V2V, V2I, V2P, V2N, in short, V2X communication capabilities.

V2X standardization is a natural step preceding commercial ITS deployment to ensure interoperability, compatibility and interchangeability of technologies, components, devices, processes and services associated with the ITS framework. For example, adherence to a common V2X standard enables a city municipality to order RSUs from a vendor with confidence, as the RSUs are guaranteed to be compatible with communication On-Board Units (OBUs) fitted in vehicles, irrespective of their Original Equipment Manufacturers (OEMs).

## 6.2 What is an ITS standard?

An engineering standard is a formal document drafted, voted, and recommended by a Standard Developing Organization (SDO), often a not-for-profit organization. The standard is prepared by achieving consensus among the participants of SDO's Working Group (WG), which is comprised of members from the industrial, government, and academic sectors.

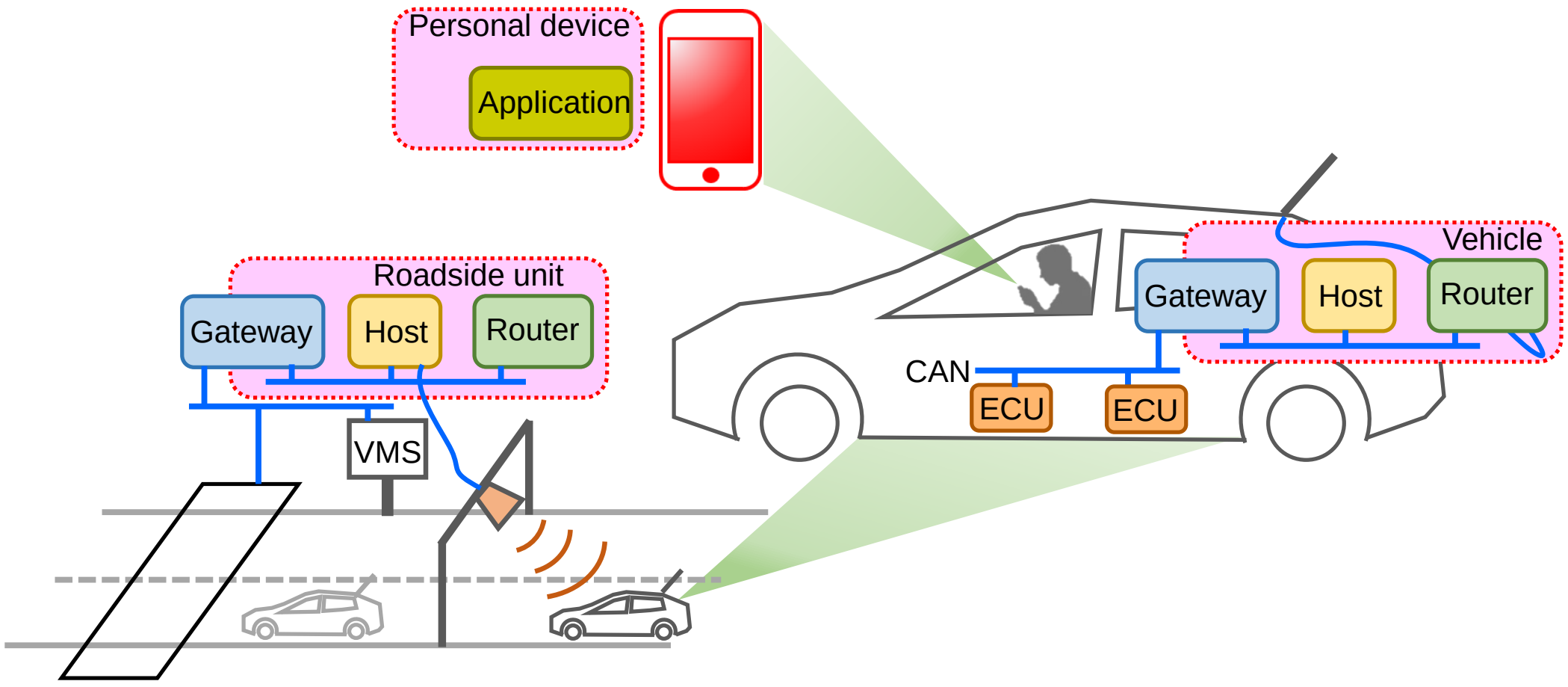

Figure 6.1: Implementation of a V2X architecture conforming to the International Organization for Standardization (ISO) 21217 standard. The ITS Station Unit (ITS-SU) reference architecture is adapted for three levels of hierarchy – ITS-SU at the roadside, ITS-SU in a vehicle, and ITS-SU in a personal device. CAN: Controller Area Network, ECU: Electronic Control Unit, VMS: Variable Message Sign.

The scope of ITS is quite wide as it encompasses organizational, administrative, and engineering aspects from other core branches (say, standards containing pavement height or lane width for optimal safety and congestion). For the current document, our focus is on V2X standards that provide – (a) comprehensive guidelines and unambiguous specifications for developing hardware/software for a full

V2X communication protocol stack, (b) seamless integration with the latest cellular, Wi-Fi, and other wireless technologies, (c) unified communication framework for road, rail and non-terrestrial transport, and (d) help for rolling out ISAC, Simultaneous Localization/Mapping (SLAM), augmented/extended reality (AR/XR) based new use cases. Fig. 6.1 demonstrates how a standard provides vision for a V2X architecture, where we considered ISO 21217 standard [185] as a typical example.

In summary, an ITS standard lists minimum required features and maximum personal/environmental hazards to protect end users. It ensures interoperability between different stakeholders, including the automotive companies (e.g., Audi), telecom operators or telcos (e.g., Orange), original equipment manufacturers or OEMs (e.g., Nokia), cloud and network service providers (e.g., Amazon), software application developers (e.g., Radisys), government agencies (e.g., US Department of Transport), and regulatory bodies, e.g., Federal Communications Commission (FCC).

## 6.3 Evolution of ITS standards

Initial V2X standards were Wi-Fi-based and developed on IEEE 802.11p PHY layer. DSRC, later named Wireless Access in Vehicular Environment (WAVE), emerged in the US, and Cooperative ITS (C-ITS), as well as ITS-G5, were introduced in the EU.

The next generation V2X standards were cellular-based, i.e. Cellular V2X (C-V2X), with the initial versions named as LTE-V2X, and later ones as New Radio (NR)-V2X, aligning with the underlying 3GPP releases [186].

### 6.3.1 Wi-Fi-based standards – DSRC, WAVE, C-ITS

The work on Wi-Fi-based V2X standards started after the FCC in the US allocated 75 MHz of spectrum in the 5.9 GHz band in 1999. In 2004, the IEEE 802.11p task group was formed, and in 2010, the DSRC standard was released. While IEEE 802.11p was defining the PHY and MAC layers of the DSRC protocol stack, the IEEE 1609 series formed upper MAC layers, and the new stack was termed as WAVE. The complete stack is shown in Fig. 6.2. C-ITS, used in the EU, was a similar access layer, but used Basic Transport Protocol (BTP) and Geo-Networking (GN) instead of WAVE Short Message Protocol (WSMP) [187]. In 2019, work on a newer version of Wi-Fi-based V2X standard started, and the IEEE 802.11bd task group was formed [188].

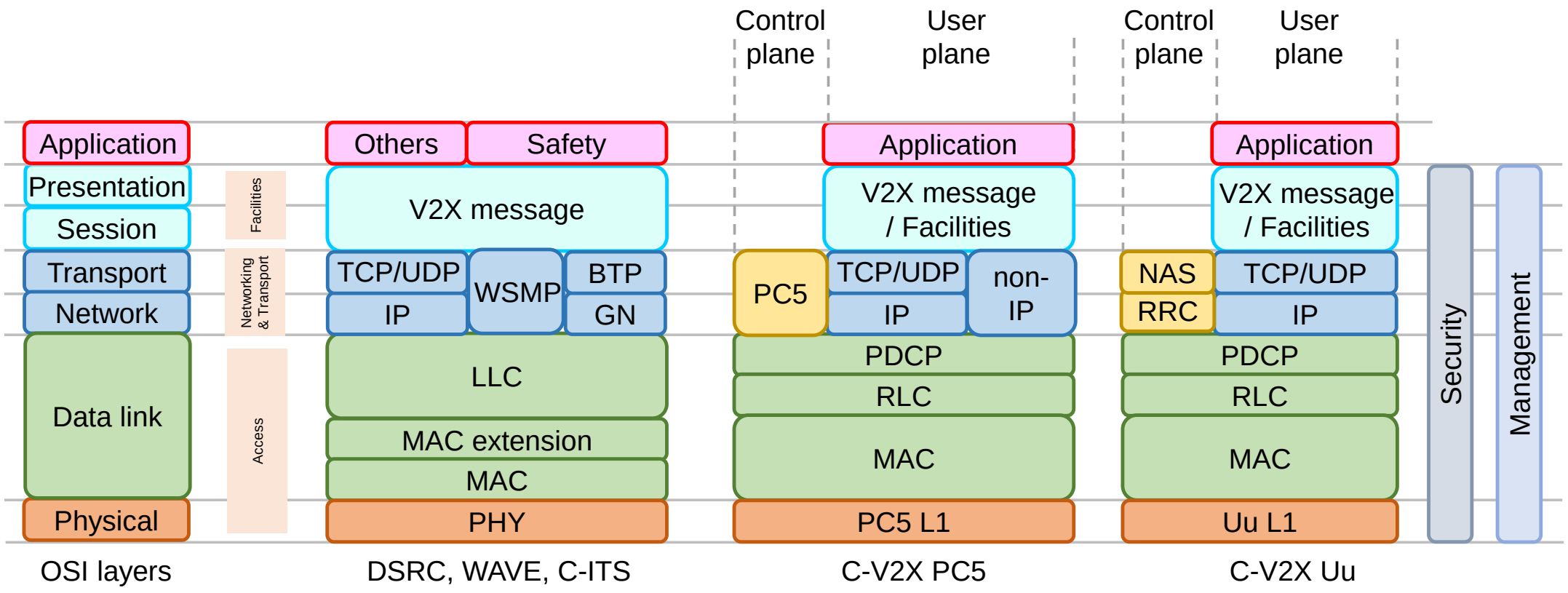

Figure 6.2: Protocol stack for Wi-Fi-based and cellular-based V2X standards. OSI: Open Systems Interconnection, TCP: Transmission Control Protocol, UDP: User Datagram Protocol, IP: Internet Protocol, WSMP: WAVE Short Message Protocol, BTP: Basic Transport Protocol, GN: Geo-Netwoking, LLC: Logical Link Control, MAC: Media Access Control, PHY: Physical, PDCP: Packet Data Convergence Protocol, RLC: Radio Link Control, NAS: Non-Access Stratum, RRC: Radio Resource Control.

### 6.3.2 Cellular-based standards – C-V2X, LTE-V2X , NR-V2X

The term C-V2X was coined by the 5G Automotive Association (5GAA), which was formed in 2016. C-V2X has two provisions – short-range direct communication over the PC5 interface, also known as sidelink, and long-range communication over the Uu interface using cellular infrastructure. While the first one is useful for V2V/V2I applications, the second one can be used for V2P/V2N use cases. Fig. 6.2 shows the control plane (C-plane) and user plane (U-plane) protocol stacks for both PC5 and Uu interfaces. The LTE-V2X stack includes Radio Link Control (RLC) and Packet Data Convergence Protocol (PDCP) on the top of the Media Access Control (MAC) layer, and the NR-V2X stack is identical to LTE-V2X, except an additional Service Data Adaptation Protocol (SDAP) above the PDCP layer (not shown in the figure).

The standardization of C-V2X is aligned with the 3GPP releases; R14 and R15 form long-term-evolution V2X (LTE-V2X), while R16 onwards form the basis for NR-V2X. Although 4G LTE cellular specifications came with 3GPP R8 in 2009, V2X capabilities came with R14 in 2017. In 2018, with R15, unicast and broadcast features were added to LTE-V2X. The general 5G NR specifications also came with R15, but the NR-V2X specifications, such as URLLC features, are available from R16, which was released in 2020. In 2022, R17 was released, and it included multicast and efficient sidelink features.

In the EU, the C-V2X is spearheaded by the ETSI. The ETSI Release 1 EN 303 613 [189] in 2020 for the LTE-V2X sidelink is based on 3GPP R14, and Release 2 EN 303 798 [190] for NR-V2X sidelink is based on 3GPP R16. The two releases include ITS-G5, based on IEEE 802.11p and IEEE 802.11bd, respectively.

## 6.4 Spectrum allocation for ITS

Uniform spectrum allocation is a crucial step for the interoperability of different ITS standards. However, the allocations were staggered in the timeline across different countries, making the unification difficult.

In the US, the FCC allocated the 902 – 928 MHz band in 1985, which is used for V2I services such as electronic toll collection [191]. In 1999, the FCC allocated a whooping 75 MHz band in the 5850 – 5925 MHz range. Two decades later, in 2020, the FCC revoked the use of the lower 45 MHz band (5850 – 5895 MHz) due to lack of use and reallocated the spectrum to Wi-Fi, keeping only the upper 30 MHz band (5895 – 5925 MHz) for possible use. In subsequent proposals, the FCC ruled that the lower 10 MHz band (5895 – 5905 MHz) is reserved for DSRC, and the remaining 20 MHz band (5905 – 5925 MHz) is allocated for C-V2X use-cases [192]. In Canada, earlier, Industry Canada and now Innovation, Science and Economic Development (ISED) followed a path similar to that of the FCC. The current spectrum allocations for North America, along with other regions, can be found in Fig. 6.3.

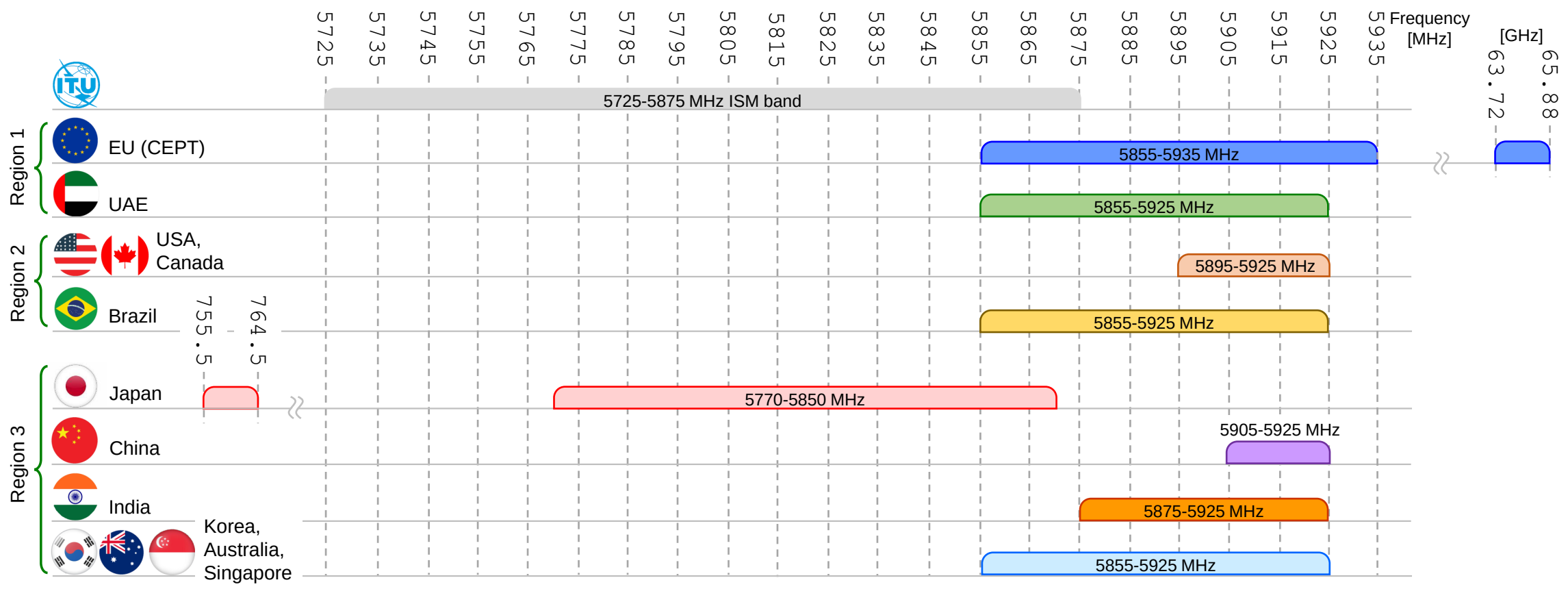

Figure 6.3: Current spectrum allocation for ITS services in the major countries of the three ITU regions. Bands for legacy toll collection services in the US (902 – 928 MHz) and ITS G-5C band (5470 – 5725 MHz) for EU are not shown.

In Europe, the allocations are governed by the European Conference of Postal and Telecommunications Administrations (CEPT). CEPT created ETSI in 1988 as an SDO to produce and maintain various telecommunication-related standards. In response to 802.11p-based DSRC, ETSI came up in 2013 with the ITS in 5 GHz (ITS-G5) standard, having a compatible protocol stack [193]. In total, four bands were specified: the primary two bands are ITS G5 B (5855 – 5875 MHz) for non-safety related applications and ITS G5 A (5875 – 5905 MHz) for road safety applications, ITS G5 D (5905 – 5925 MHz) was for future use, while ITS G5 C (5470 – 5725 MHz) was for radio LAN-based applications which were of limited use.

In 2020, the allocations were revised [194] for C-V2X use cases and the entire spectrum between 5855 – 5935 MHz is kept technology neutral. The 5855 – 5875 MHz is still for non-safety purposes but the band for safety applications is expanded to 5875 – 5935 MHz, out of which 5875 – 5915 MHz is prioritized for road safety, and 5915 – 5935 MHz is prioritized for rail safety, with 5925 – 5935 MHz is exclusively reserved for rail safety. The legacy standards, such as ITS-G5, will continue to operate in the 5895 – 5905 MHz band, and road tolling in the 5795 – 5815 MHz band can continue within a defined protection zone [195]. It is interesting to note that the EU is the only region where one of the mmWave FR2 bands (63.72 – 65.88 GHz) is also allocated for ITS since 2008.

The Association of Radio Industries and Businesses (ARIB), through its ARIB STD-T109 standard in Japan, made the 755.5 – 764.5 MHz band available for ITS in a licence-free and exclusive manner. The 5770 – 5850 MHz is allocated for the Electronic Toll Collection (ETC) systems and follows ARIB STD-T75 standard [191]. The Ministry of Industry and Information Technology (MIIT) in China assigned 5905 – 5925 MHz band for C-V2X pilot in 2016 [196]. The wireless planning and coordination wing of the Department of Telecommunications (DoT) in India, in its National Frequency Allocation Plan for 2022, reserved 5875 – 5925 MHz for V2X technologies [197].

Finally, the Ministry of Science and ICT (MSIT) in Korea, the Australian Communications and Media Authority (ACMA), and the Infocomm Media Development Authority (IMDA) in Singapore standardized the 5855 – 5925 GHz band for ITS in during 2016 – 2017 [198].

## 6.5 Current standardization activities

The current standardization activities involve several SDOs, with the PHY and MAC being governed by 3GPP releases, the network and transport layer standards given by IEEE, while the application layer, such as V2X message set dictionaries, is specified by Society of Automotive Engineers (SAE).

### 6.5.1 3GPP

The 3rd Generation Mobile Group (3GPP) is responsible for developing and maintaining protocols for mobile telecommunications, normally in the form of parallel releases. LTE-V2X was included in the 68th radio access network meeting (RAN#68) as a study item and evolved as a work item in R14 in RAN#70 and RAN#72. In RAN#75, a work item on LTE-eV2X for R15 was included. RAN#80 and RAN#83 in 2018 and 2019, respectively, included NR-V2X as a study item and a work item for R16.

From 2020 to 2022, NR sidelink enhancement work item for R17 was carried out, which was started with RAN#88e, and in 2022 RAN#94e, work item on NR sidelink evolution for R18 started. VRU application scenarios, power saving, and inter-UE coordination mechanisms among terminals for sidelink were considered in R17 NR-V2X. R18 NR-V2X supports increased sidelink data rate, positioning, relay etc. [196] and is frozen in 2024. R19 specifications are being drafted and the work will finish in 2025. It includes items on unmanned aerial systems, FRMCS extensions, and non-public networks.

### 6.5.2 IEEE

The Institute of Electrical and Electronics Engineers (IEEE) is the organization that developed 802 series of standards for Local Area Networks (LANs), and later wireless LAN or Wi-Fi with 802.11. IEEE 802.11p formed the basis for DSRC, which specifies the PHY layer of the Wi-Fi-based V2X protocol stack. In WAVE, the IEEE 1609 family of standards are placed on top of it. With reference to Fig. 6.2, IEEE 1609.0 gives the general architecture, IEEE 1609.2 defines the security vertical with Octet Encoding Rule (OER), IEEE 1609.3 defines the non-IP based WSMP for the network and transport layer, and IEEE 1609.4 defines multi-channel operation in the MAC extension layer [199].

Since 2019, standardization research on IEEE 802.11bd, an evolved version of IEEE 802.11p, has begun [196]. The standard was published in 2024. IEEE 802.11bd defines operation in the 60 GHz band with a 20 MHz bandwidth and is backward compatible to 5.9 GHz band. It also includes midambles for better channel estimation, beamforming capabilities, and Low-Density Parity Check (LDPC) coding at the PHY layer. In addition, IEEE is constantly coming up with intersectoral standards. A typical example is the IEEE 3472 standard group formed in 2024, which aims to provide guidelines for CAVs to be used by persons with disabilities.

### 6.5.3 SAE

The Society of Automotive Engineers (SAE) established its C-V2X Technical Committee in 2017. The TC reports to the V2X Communications Steering Committee of the Motor Vehicle Council. C-V2X TC published two major V2V-oriented standards in 2022, J3161 and J3161/1, which were further revised in 2024. The first one, J3161, is the base standard that provides a reference system architecture for LTE-V2X implementation conforming to 3GPP release 14. It includes C-V2X deployment profiles, radio parameters, MCS choices, message priority, and a description of mode 4 or the LTE-V2X PC5 sideline. In J3161, classification of V2X traffic (based on the priority and direction of the message) and classification of V2X messages (whether for safety or mobility) are also provided [200]. The second one, J3161/1, defines the transmission of V2V Basic Safety Message (BSM), initially described in SAE J2735 data dictionary, and provides guidelines for efficient congestion management [201]. J3161 series specifies operation over the 20 MHz channel in the 5905 – 5925 MHz band (channel 183).

In addition, SAE came up with J3161/2 for LTE-V2X 10 MHz channel in the 5895 – 5905 MHz band, to be used in the United States specifically for public sector applications such as toll gate and platooning [202]. Another related document is J3161/1A, a recommended practice, that describes the testing procedures to determine whether conformity to J3161 and J3161/1 standards is maintained and whether devices from different manufacturers maintain interoperability.

Apart from the V2V-oriented J3161 series, SAE is also responsible for the V2I/Infrastructure-to-Vehicle (I2V)-oriented J2945 family of standards maintaining, standard J3016 defining six levels of driving automation (from level 0/ no automation to level 5/ full automation), and use-case specific standards like road user charging through V2X (J3217), cooperative use of infrastructure sensors (J3224), cooperative use of vehicle manoeuvre information (J3186), among others. New initiatives are in the pipeline, like J3224/1 (profile for Infrastructure sensor sharing), formed in 2025.

### 6.5.4 ISO and other SDOs

The World Trade Organization (WTO) acknowledges international standards developed by the United Nations (UN) agency ITU and the ISO [203]. ITU is responsible for organizing the quadrennial event, the World Radiocommunication Conference (WRC), where existing telecom standards are revised and vision documents are introduced. The next WRC will be in 2027, and there is a likelihood that the 4400 – 4800 MHz band sharing and compatibility (agenda item 1.7) discussions will include V2X or sidelink technologies.

There are two technical committees of ISO that look after V2X standards – Road Vehicles (TC 22) is for CAVs, and ITS (TC 204) is for infrastructure and traffic management [196]. ISO maintains and promotes the development of various safety-related standards as well [204]. One such example is UL 4600, a safety standard for autonomous vehicles that is built upon ISO 26262 and ISO 21448. The latest version 3, released in 2023, includes safety guidelines for both road and non-terrestrial vehicles.

# Chapter 7: Future Trends and Research Directions

## 7.1 Cell-free Massive MIMO Networks

By **Davy Gaillot, Fredrik Tufvesson**

### 7.1.1 Introduction and Definition

MOBILE networks are designed to provide wireless access to data services in wide geographic areas. Initially, these networks were used primarily for voice calls, but modern networks focus on data transmission, with service quality largely determined by data rates. The performance of these networks depends on the propagation of the signal, which weakens with distance. To address this, mobile networks use a system of geographically distributed transceivers, called Access Points (APs), to maintain coverage [205].

Traditional mobile networks operate using a cellular structure, where each UE connects to the AP with the strongest signal, defining a "cell" as typically shown in Fig. 7.1a. These APs are typically placed in elevated locations, such as towers or rooftops, to ensure better signal reach. However, due to the nature of signal propagation and interference, data rates vary across the coverage area, even with advanced hardware such as Massive MIMO. Massive MIMO technology introduced in 5G allows APs to use multiple antennas to serve multiple UEs simultaneously, improving SNR and reducing inter-cell interference [206], [207]. In uplink communication, multiple UEs send data to an AP at the same time, and the AP processes these signals using spatial separation techniques. In the downlink, APs transmit highly directional signals to UEs, further enhancing efficiency.

An alternative to cellular networks is the recently introduced cell-free network architecture, which removes cell boundaries as shown in Fig. 7.1b [6], [208]. In addition, when the number of antennas at the infrastructure side is much larger than the UEs to be served, as an analogy to the conventional Massive MIMO regime in cellular networks, the cell-free massive MIMO networks terminology is prevalent [209], [210].

In this system, multiple APs jointly serve all UEs within a given area, rather than assigning a UE to a single AP. These APs are connected to a Central Processing Unit (CPU) via fronthaul links, which can be wired or wireless. The CPU coordinates AP cooperation, improving network efficiency. The network is divided into an edge and core, similar to cellular networks: APs and CPUs form the edge, while backhaul links connect CPUs to the core network, facilitating internet access and other data services. The fronthaul links serve three key purposes: sharing physical-layer signals for downlink transmission, forwarding uplink data signals to be decoded, and exchanging Channel State Information (CSI) to optimize signal processing.

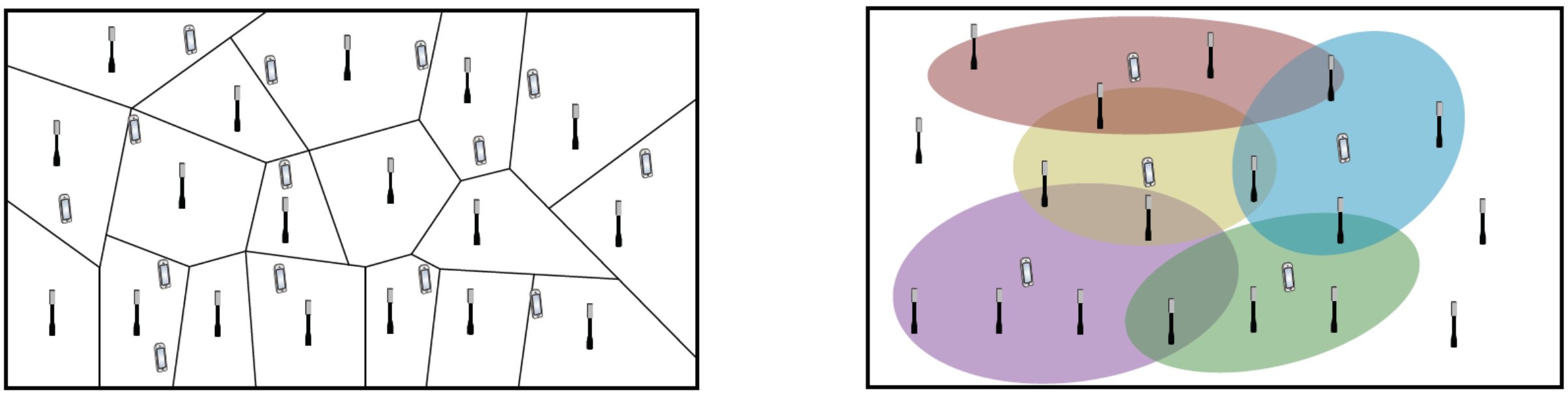

(a) (b)

Figure 7.1: (a) Conventional cellular network, where each UE is only served by one AP. (b) A user-centric implementation of Coordinated Multipoint in a cellular network, where each UE selects a set of preferred APs that will serve it [210].

The cell-free network ensures that all APs with sufficient signal quality contribute to the transmission and reception process. This approach reduces interference and improves network coverage, as multiple APs collaborate rather than competing for signal dominance. A variation of this model is user-centric clustering, where each UE selects a preferred set of APs rather than being restricted to a specific cell. This method guarantees interference control and improves connectivity.

A more advanced implementation of cell-free networks is *Cell-Free Massive MIMO*, first proposed in 2015. Unlike traditional cellular networks, which add multi-cell cooperation to existing infrastructure, Cell-Free Massive MIMO builds a distributed network from the outset. The term "massive" refers to a scenario where there are more APs than UEs, similar to the conventional Massive MIMO approach in cellular networks, where multiple antennas at APs serve fewer UEs. This setup resembles ultra-dense networks but differs in that APs actively cooperate rather than functioning independently.

The main advantage of Cell-Free Massive MIMO is its ability to deliver uniformly high service quality across a coverage area. By using distributed antennas, it eliminates dead zones and provides consistent performance. Early research focused on decentralized operation, where APs handle most signal processing tasks independently, with minimal need for central coordination. The system operates in Time Division Duplex (TDD) mode, meaning uplink and downlink transmissions share the same frequency band but occur at different times. This allows APs to estimate channel conditions by receiving pilot signals from UEs, enabling efficient and adaptive communication.

### 7.1.2 Benefits and Limitations

#### 7.1.2.1 Benefits

The shift from traditional cellular networks to cell-free architectures, especially cell-free massive MIMO, represents a significant advancement in wireless communication. By leveraging distributed APs and cooperative processing, these networks offer manyfold significant advantages over traditional cellular networks. Cell-free systems promise a reduction in transmit power by two orders of magnitude compared to conventional cellular networks while offering more consistent throughput over the entire coverage area. Hence, it provides an improved SNR for enhanced QoS. When packet loss ratio is kept below 2%, these networks show decreased average delay and increased average throughput compared to 4G systems.

Another key benefit is the ability to manage interference by joint processing at multiple APs, which is not done in cellular networks with an equally dense AP deployment. Reliability can also be enhanced thanks to spatial focusing at the UE position. Cell-free networks employ non-uniform power allocation based on collective channel gains of the serving cluster and an equity criterion among all UEs for a better efficiency of network resources. They can support a higher number of UEs compared to 4G systems.

#### 7.1.2.2 Limitations

However, there are some limitations and challenges that need to be considered and addressed. The distributed nature of the system introduces complexity in terms of coordination and synchronization among the various radio units. In addition, implementing a cell-free architecture requires a significant number of distributed antennas and supporting infrastructure, which may be challenging to deploy in some environments. As with any wireless system, channel aging effects need to be considered and managed. While theoretical studies have shown promising results, empirical evidence from real-world deployments is still limited. More extensive testing and measurement campaigns are needed to fully validate the performance of cell-free systems in various scenarios.

The implementation introduces new complexities in network management, such as handling potential overlaps between cooperative clusters and managing the spatial distribution of UE. While cell-free networks show promise in 5G networks, they may not be fully realizable in 4G systems. 4G networks have been shown to lack the capacity to provide certain ultra-low latency services, even with advanced deployment strategies. Finally, the performance analysis of cell-free networks requires sophisticated mathematical tools such as stochastic geometry. This complexity in analysis may pose challenges in network planning and optimization.

### 7.1.3 Cell-Free Massive MIMO Measurements for Vehicular Applications

While not explicitly mentioned for cell-free architectures, 5G networks with appropriate latency reduction strategies can support URLLC services. These services are a requisite for V2I and V2V communication within ITS in contrast to 4G networks. Although the cell-free massive MIMO has recently received a lot of attention at the theoretical level, there are extremely sparse studies in the literature that investigates the attractive features from the point of view of vehicular cell-free massive MIMO [211], [212]. In addition, to consider deploying such networks in the future, it is critical to conduct intensive measurement campaigns to assess its performance on realistic propagation conditions and validate for example AP selection and clustering algorithms as discussed in [213].

#### 7.1.3.1 Cell-Free Radio Channel Sounding Architecture

To this end, novel radio channel sounding architectures must be designed at ITS frequencies such that APs can be fully deployed over an area of interest at the street level, as imagined for RSU with V2I applications. These APs shall be simultaneously connected to one or many UEs to perform radio channel sounding within the same time-frequency resources under receiving mobility. Furthermore, different possible TX antenna system topologies must be investigated, i.e. the number of APs and antennas per AP altogether with their locations.

To this end, the real-time channel sounder MaMIMOSA has been upgraded to measure channels in such distributed configurations. Initially, MaMIMOSA was jointly developed by the University of Lille (France) and Ghent (Belgium) to measure co-located massive MIMO channels for vehicular scenarios [214]. Nominally, it can be set up to measure a $64 \times 16$ radio channel at 5.93 GHz central frequency with an 80 MHz bandwidth. The sounder consists in deporting the transmit antennas using an RF-over-Fiber (RoF) link with 0 dBm output power as shown in Fig. 7.2a.

The output power at the back of each antenna $P_{out}$ depends on the PA characteristics in-between. Each PA is powered by a Li-ion battery. In this way, the real-time channel between a vehicular and a network of distributed antennas can be measured. Based on the MaMIMOSA emitting architecture, up to 64 single-antenna APs or X APs equipped each with a Y-antenna array ($X \times Y = 64$) can be distributed in the environment to emulate a realistic Cell-free massive MIMO network. Both APs and UEs units are powered with Li-ion batteries that provide up to 8 hours of continuous measurements.

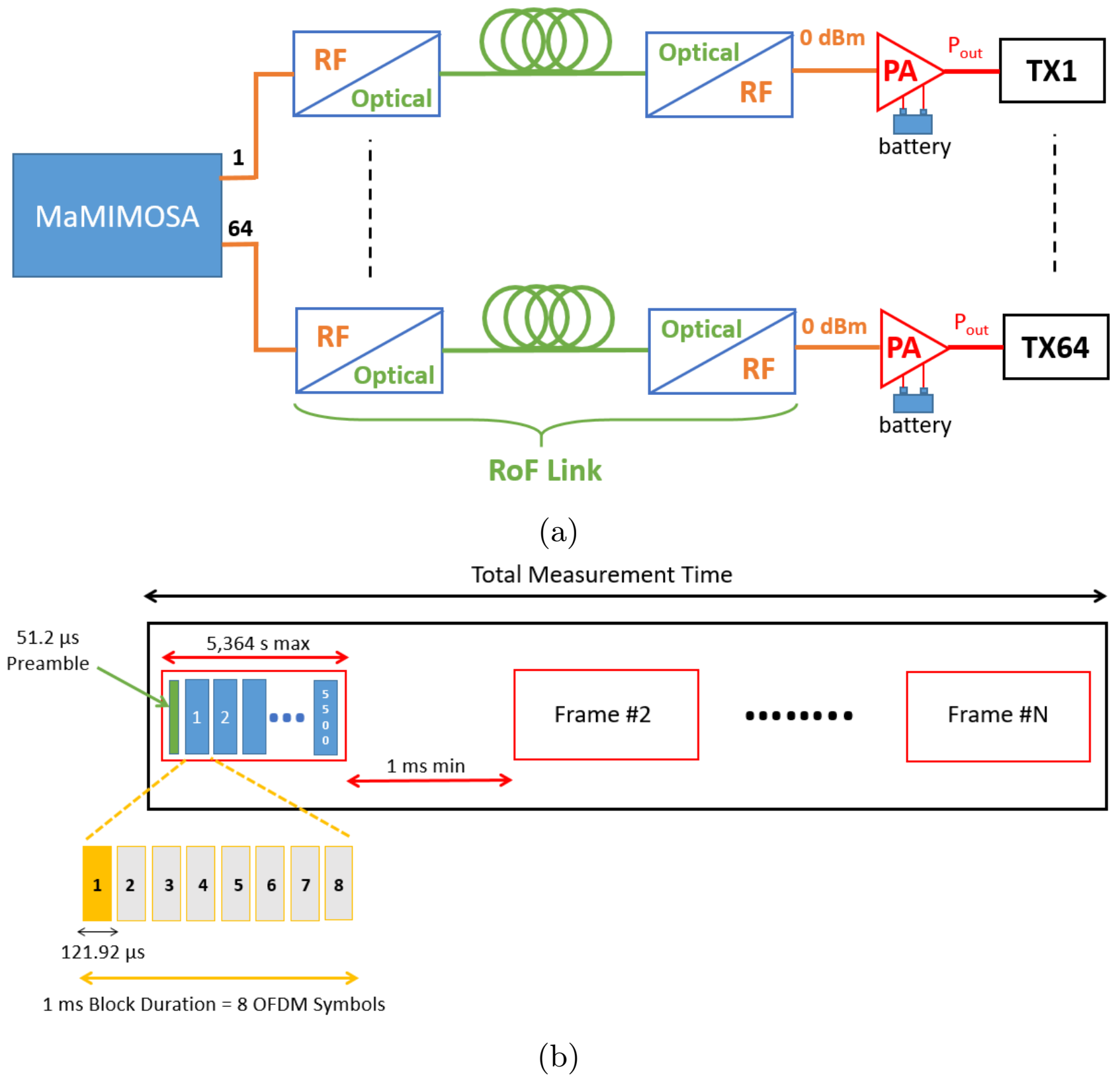

Figure 7.2: MaMIMOSA (a) TX architecture with 64 RoF links and (b) streamshot frame structure.

The streaming mode of MaMIMOSA has been specifically designed for the vehicular context and is fully flexible to account for the Doppler characteristics of the radio channel. It corresponds to the frame structure illustrated in Fig. 7.2b. Data are modulated using interleaved OFDM with 8192 subcarriers and a subcarrier spacing of 12.21 kHz. Each TX antenna transmits a pilot every eight sub-carriers, with one sub-carrier per antenna shift to interleave the TX antennas signals. This defines a single 121.92 µs OFDM symbol with 8 TX. By switching 8 times to another set of 8 antennas, a full block of 64 outputs can be emitted during ∼1 ms. Each frame consists in a maximum of 5500 blocks with a maximum time of 5.364 s and starts with a 51.2 µs preamble for time synchronization purposes. The inter-frame time can be set to the minimum value of 1 ms.

The number of emitting antennas, number of blocks and inter-block time, inter-frame time, and number of frames can be freely selected based upon the investigated scenario. The only limitation resides in the capacity of the system to store the data due to the PCI-Express bandwidth and hard-drive throughput. In any cases, the receiving unit can simultaneously measure up to 16 UEs. This enables the real-time channel between a vehicle and a network of distributed antennas to be measured. The current MaMIMOSA configuration supports up to 8 single-antenna APs or $L$ APs equipped each with an $N$-antenna array, with $L \times N = 8$.

#### 7.1.3.2 Measurements and results

The MaMIMOSA sounder was set up to emulate a small, yet realistic cell-free network and measure the time-varying radio channel at 5.89 GHz with an 80 MHz bandwidth in a suburban environment under various AP and antenna configurations to study the SNR, delay and Doppler spread values, or path loss.

All channel measurements took place on the Cité Scientifique Campus of the University of Lille (France). 6.5 dBi directive patch antennas were used at the transmitting APs, whereas the receiver was a van equipped with four 2 dBi EM6116 omnidirectional antennas on its rooftop spaced apart by more than $10\lambda$.

In the main demonstration campaign, the scenario consisted in driving the van on a section of the campus boulevard from point A to point B over a 275 m distance with a roundabout in-between at an average speed of 25 km/h (Fig.7.3). Four AP configurations were investigated in this campaign from fully co-located to fully distributed antennas. The first one consists in one AP equipped with 8 antennas, the second one 2 APs equipped with 4 antennas each, the third one 4 APs equipped with 2 antennas each, and the fourth one 8 single-antenna APs. These configurations are called Config. 1 to 4, respectively. The radio channel was measured with the streaming mode using ∼70 frames (∼35 s).

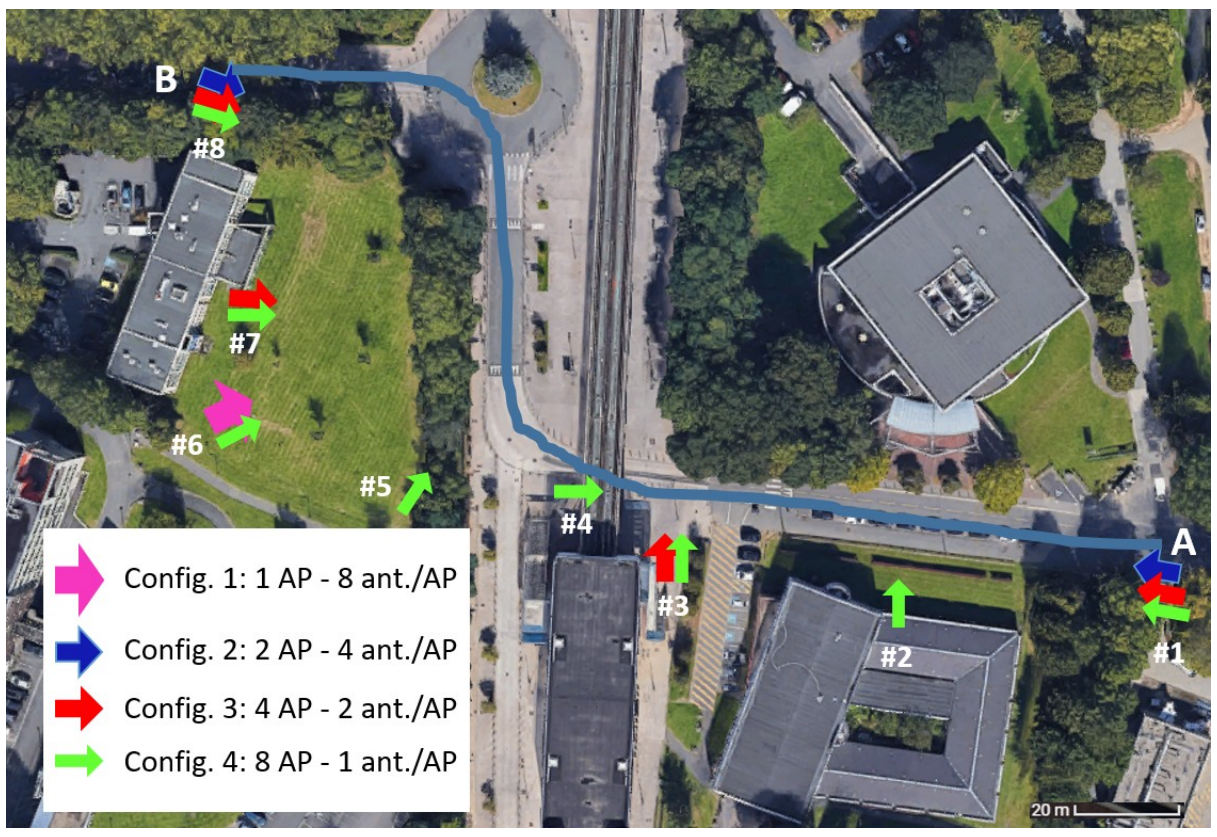

Figure 7.3: Top-view of the measurement campaign on the Lille University Cité Scientifique Campus presenting the four AP configurations.

The advantages of Config. 3 and 4 are highlighted compared to the other two in many aspects in Fig. 7.4. First, nearly 49% (resp. 80%) of the van positions present an SNR higher than 20 dB in Config. 1 (resp. Config. 2) compared to 98.5% and nearly 100% in Config. 3 and Config. 4.

The delay spread values computed from the AP - UE link with largest SNR and for SNR values greater than 10 dB were found to be varying between 12.5 (minimum measurable RMS) to 150 ns and are in line with values reported in the literature for a similar scenario and shadowing conditions. A first observation is that distributing the antennas (ex: Config. 4) naturally produced higher values of the maximum RMS delay spread, compared to less distributed configurations. In addition, the minimum RMS delay spread was the smallest for Config. 4. These low values generally correspond to LoS links between transmit and receive antennas, which are more frequent in distributed scenarios. Finally, the antenna selection criterion based on SNR yields very close RMS delay spread values to those obtained with a decision based on the observed RMS delay spread.

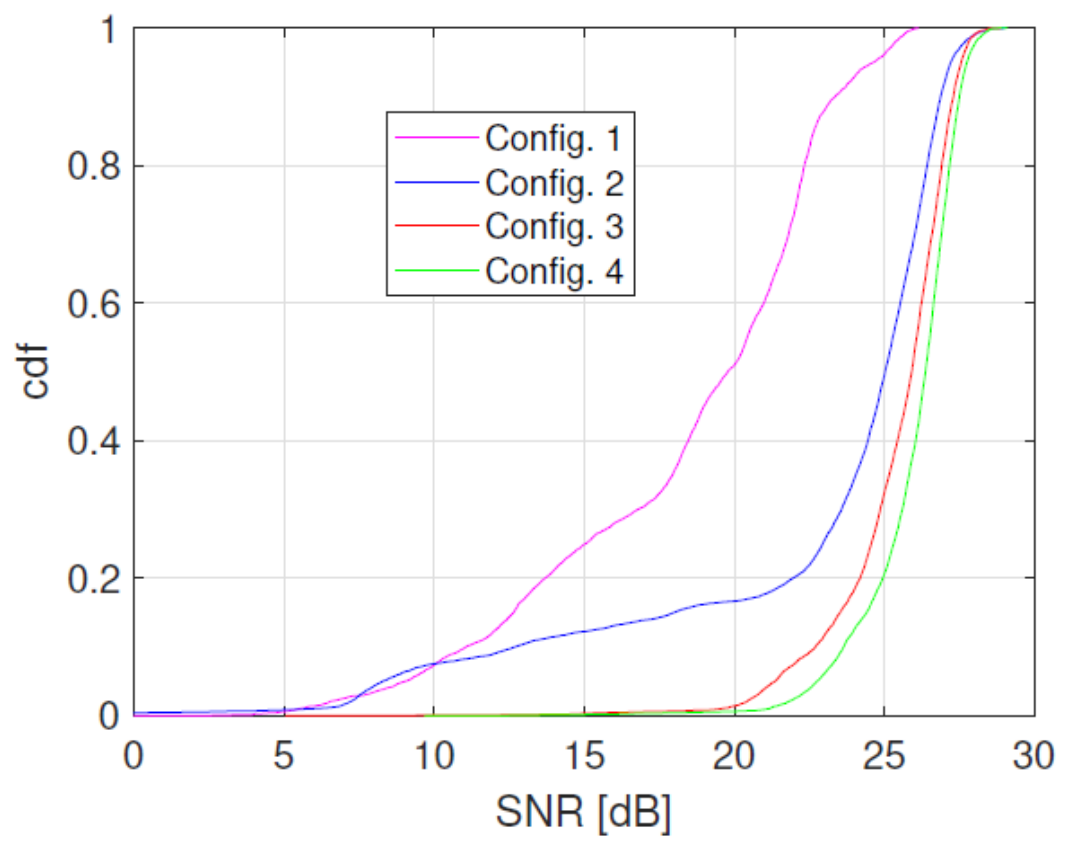

(a)

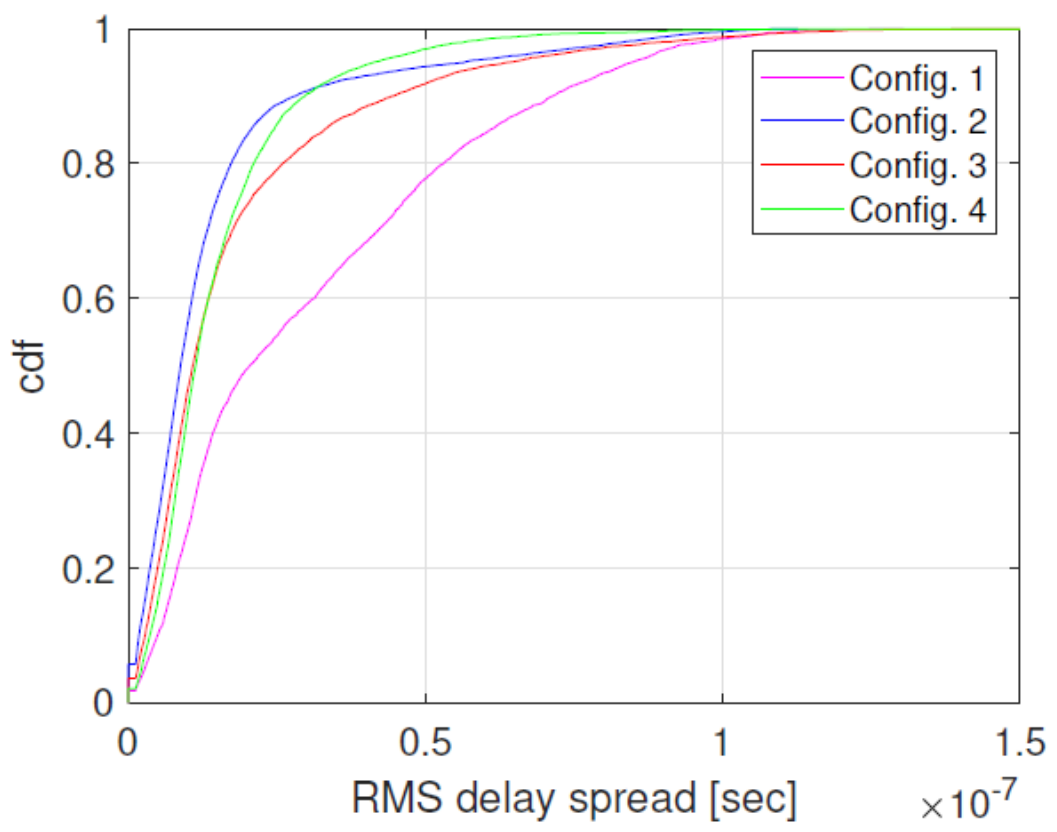

(b)

Figure 7.4: For the four AP configurations, CDF of (a) the SNR and (b) the RMS delay spread.

Optimized deployment strategies of the distributed cell-free network are currently missing in the literature for a realistic complex propagation scenario. Here, the dependence of the AP location in the scenario on the propagation fading properties was investigated for the fully distributed Config. 4. To this end, the path loss exponent $n$ and lognormal shadow fading $\chi_g$ were derived for each AP location from the received gain averaged across the receivers, when driving from B to A. $n$ values varying between 1.85 and 4.89 were derived, with $\chi_g$ values being below 3 dB. It is shown that the links with a strong LoS (like AP1 and AP8) exhibit $n$ values close to 2 (free space propagation). In contrast, for the other OLoS shadowing conditions, larger $n$ values are obtained.

#### 7.1.3.3 Main Conclusions

The results showed a clear advantage of the three distributed configurations with respect to the centralized Config. Although a limited number of transmit antennas (eight) was considered, which is below the expected number in an operational distributed massive MIMO network, the study experimentally validates the promises of cell-free networks compared to co-localized antennas, with a higher and spatially more uniform SNR. The analysis of the path loss characteristics for the cell-free network also provides valuable guidelines for deploying APs in the scenario.

## 7.2 Spectrum Management and Frequency Bands

By **Michał Sybis, Fernando J. Velez, Nila Bagheri, Jon M. Peha**

The advancement of ITS fundamentally depends on robust wireless connectivity. As modern vehicles become interconnected elements of a smart mobility ecosystem, the need for effective spectrum management becomes critical to ensure reliable V2X communication. The proliferation of connected vehicles, the growing deployment of RSUs, and the rapid expansion of mobility services all contribute to the increased demand for limited radio frequency resources. This intense competition introduces significant challenges related to interference mitigation and spectrum allocation, directly affecting transportation safety and the overall efficiency of ITS infrastructures [215].

To address these challenges, this study builds on the concept of Vehicular Dynamic Spectrum Access (VDSA) [216], proposing a method to offload a portion of V2V traffic – particularly in autonomous platooning scenarios – from the congested 5.9 GHz band to underutilized frequency bands, such as TV White Spaces (TVWS). Within the VDSA framework, autonomous platooning functions as a secondary service, which must coexist without disrupting incumbent primary users. To improve the effectiveness of spectrum access in this context, we employ an Edge Intelligence System (EIS) [217], which facilitates real-time decision-making and dynamic optimization of communication in TVWS.

The television spectrum is especially well-suited for this application because of the consistent and stable nature of its primary signals. The positions and configurations of the Digital Terrestrial Television (DTT) transmitters, along with the associated frequency allocations, remain relatively unchanged over time. This predictability allows accurate measurements of signal strength along transportation corridors, which can be stored in a Context Database (CDB) with high temporal reliability. Leveraging these stable data through the EIS enables autonomous platooning services to make informed use of the available TVWS spectrum, ensuring efficient communication and protection of DTT receivers from harmful interference.

### 7.2.1 Frequency Allocation

The effective deployment of ITS depends on the availability of dedicated and/or shared spectrum bands that support low-latency, high-reliability communication for V2V, V2I, and broader V2X applications. To meet the growing demands for connected vehicle spectrum, a novel approach is proposed that enables V2X technologies (such as C-V2X and NR-V2X) to share spectrum with Wi-Fi and other unlicensed devices [218], [219]. This approach does not require changes to existing Wi-Fi technology, thus preserving current deployments and demanding only modest adaptations to V2X, which reduces overall complexity and cost. It uses a backward compatible form of beaconing, with dynamically adjustable resource allocation to improve efficiency. Importantly, it operates without the need for a cellular operator or centralized controller.

This method is especially promising for the spectrum bands adjacent to the ITS band, where it can support both connected vehicles and Wi-Fi 6. The simulation results show that it can protect the quality of service for both V2X and Wi-Fi, while significantly improving spectrum efficiency. Furthermore, standards bodies (such as IEEE 802.11 and 3GPP) and spectrum regulators could play a pivotal role in enabling the adoption of this approach [218], [219].

**5.9 GHz Band (5.895 – 5.925 GHz)**: FCC has adopted a set of rules aimed at maximizing the efficient use of the 30 MHz of spectrum allocated to ITS within the 5.895 – 5.925 GHz band by formally adopting C-V2X. These rules codify key technical specifications such as power limitations, emission constraints, and message prioritization and offer operational flexibility by allowing the spectrum to be utilized as three separate 10 MHz channels, a combination of one 20 MHz and one 10 MHz channel, or a single contiguous 30 MHz channel. The proposal emphasizes the prioritization of safety-critical communications and does not impose retroactive modifications on existing C-V2X deployments operating under waiver. Additionally, FCC outlines a two-year transition period to phase out legacy systems DSRC [220]. In Europe, the 5.9 GHz band is harmonized for ITS according to ECC Decision (08)01, covering the frequency range of 5855 – 5935 MHz [221].

This band is subdivided into three primary segments to support different ITS functions. The range 5855 – 5875 MHz is designated for non-safety-related road ITS applications and is shared with non-specific Short-Range Devices (SRDs). The core segment, 5875 – 5915 MHz, is prioritized for safety-related roads ITS, enabling reliable communications between vehicles and infrastructure. Finally, the 5915 – 5935 MHz portion is reserved for critical safety rails ITS, which support communications within railway transport systems. Although the band currently relies on channelization 10 MHz, ongoing discussions are considering the adoption of channels 20 MHz to improve spectral efficiency and accommodate the evolving requirements of C-ITS. This allocation reflects Europe's commitment to a technology-neutral regulatory approach, which allows the coexistence of both ITS-G5 and C-V2X technologies within the harmonized spectrum framework ITS [222], as illustrated in Fig. 7.5, which outlines frequency segmentation and regulatory provisions.

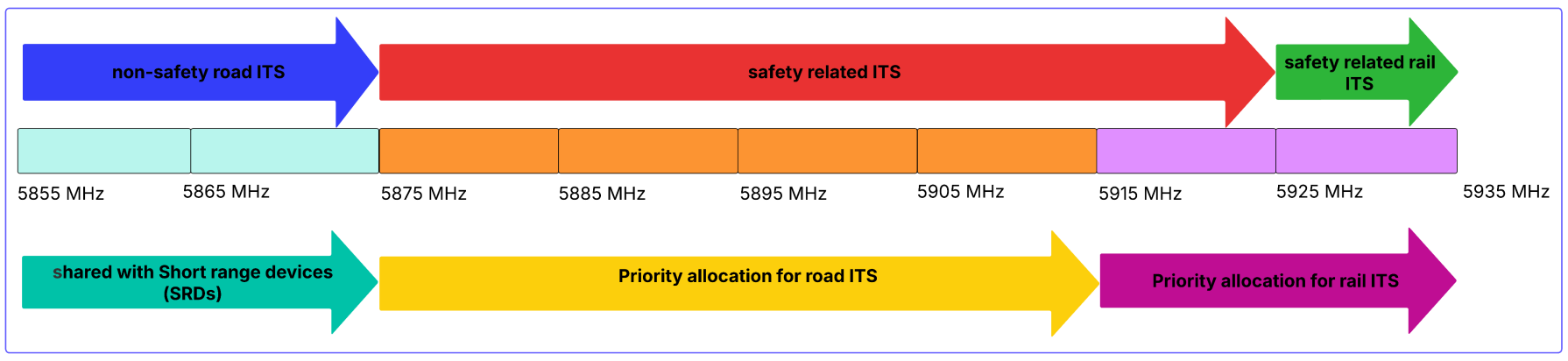

Figure 7.5: European 5.9 GHz band allocation for ITS applications.

In October 2018, China's MIIT designated the 5905 – 5925 MHz frequency band specifically for LTE-V2X (PC5) communication. This allocation is divided into two 10 MHz sub-bands: the lower portion is intended for V2V interactions, while the upper portion supports V2I and I2V communications. Operational deployment is managed through a mixed regulatory model, where OBUs operate under a license-exempt regime, while RSU require formal licensing [194], [223].

**mmWave Bands (FR2 frequency range)**: mmWave frequencies can support exceptionally high bandwidth, enabling ultra-fast data transmission rates when high-bandwidth allocations are made. The deployment of ITS requires a robust and high-speed V2X communication infrastructure. To support this demand, mmWave communication will play a vital role by enabling fast and reliable links for V2V and V2I communications. Due to the limited spectrum resources available below 6 GHz, meeting the high data rate requirements of advanced V2X systems is increasingly difficult. This has led to growing interest in mmWave frequencies, which offer significantly broader bandwidths capable of supporting next-generation vehicular applications. 5G wireless technology has emerged as a key enabler, offering enhanced data throughput, ultra-low latency, and improved QoS to meet the stringent demands of vehicular networks [224].

The communication capacity of 5G networks can be substantially improved by using mmWave technology, which is well-suited to meet the growing demands for high data traffic in vehicular environments.

However, due to their limited propagation range and susceptibility to signal attenuation, the mmWave frequencies are suitable for short-range, high-capacity applications, such as certain V2I scenarios. As there is a denser deployment of users in dense urban areas, a higher system capacity is required, and the highest bandwidth offered by mmWaves may be beneficial. Nevertheless, mmWaves can also be effectively considered in other types of deployment settings within ITS infrastructure. Although the mmWave spectrum is subject to challenges such as increased atmospheric attenuation and higher path losses, these drawbacks can be addressed through directional transmission techniques such as beam forming.

By steering the signal toward the intended receivers, beamforming improves communication reliability and compensates for propagation losses [225].

### 7.2.2 Possible Solutions

To effectively support high-mobility and interference-prone environments typical of ITS, it is essential to adopt adaptive and intelligent spectrum management strategies. Traditional Dynamic Spectrum Management (DSM) techniques – originating from the cellular and cognitive radio domains – often depend on precise CSI, centralized control, or spectrum sensing, which introduce significant computational demands and are best suited for static or slowly changing networks. These constraints render conventional DSM less effective in vehicular contexts, where network topologies evolve rapidly and real-time responsiveness is critical.

This challenge becomes even more pronounced in dense Internet of Things (IoT)-style ITS scenarios, where numerous devices must access spectrum resources in a decentralized, low-latency, and scalable manner. To address these limitations, this section explores two complementary approaches: one based on EIS, and the other leveraging machine learning techniques to access dynamic spectrum in autonomous vehicle networks.

#### 7.2.2.1 EIS-Based Approach

To enhance VDSA in autonomous platooning scenarios, we propose leveraging an EIS supported by data stored in a CDB [226]. The CDB is assumed to contain essential information such as:

- Measured DTT signal strength in protected reception areas (location-dependent),
- DTT signal levels along road segments that may interfere with V2V communication (location-dependent),
- Real-time data on nearby platoons, including frequency bands and transmission power used (time-dependent),
- Regulatory guidelines at national and regional levels concerning transmission policies.

The construction of CDBs is based on empirical measurements as described in [226].Depending on where data processing occurs and how information is exchanged, three architectural models can be distinguished for EIS:

- **Centralized:** A single entity processes the information for all platoons, enabling full coordination.
- **Distributed:** Each platoon handles its own channel selection based on locally available data, with minimal and delayed inter-platoon communication.
- **Hybrid:** Combines elements of both approaches, with detailed and up-to-date information shared among nearby platoons, and generalized data for more distant ones.

Each architecture presents different trade-offs, particularly with respect to latency and coordination. In our analysis, we treat the centralized setup as an ideal benchmark with perfect information availability, while for distributed and hybrid models, communication delays are factored into the evaluation.

The simulations were conducted in a motorway scenario with four platoons traveling in opposing directions, implemented using a custom C++ tool. The evaluation metrics include:

- **Reception Success Rate:** Defined as the ratio of successfully received messages from the platoon leader to the total number of transmissions. The analysis focuses on the leader-to-member link, often the weakest due to longer distances.
- **DTT Protection (Signal to interference ratio (SIR) Distribution):** The SIR is examined in the DTT receivers, with a threshold of 39.5 dB considered the minimum for acceptable operation.
- **Frequency Switching Rate:** The number of band changes per simulation run, which is associated with operational overhead due to the need for message dissemination and equipment reconfiguration.

The reliability of intra-platoon communications in terms of successful reception rate and the average number of frequency band changes is shown in Fig. 7.6 in Fig. 7.7, respectively. The results indicate that the centralized architecture, despite its coordination, leads to the weakest communication for tail-end vehicles. This is because it prioritizes primary system protection, often leading to frequency selections that require minimal transmission power. In contrast, the hybrid model achieves the best balance-reducing interference through localized coordination without imposing strict constraints on distant platoons.

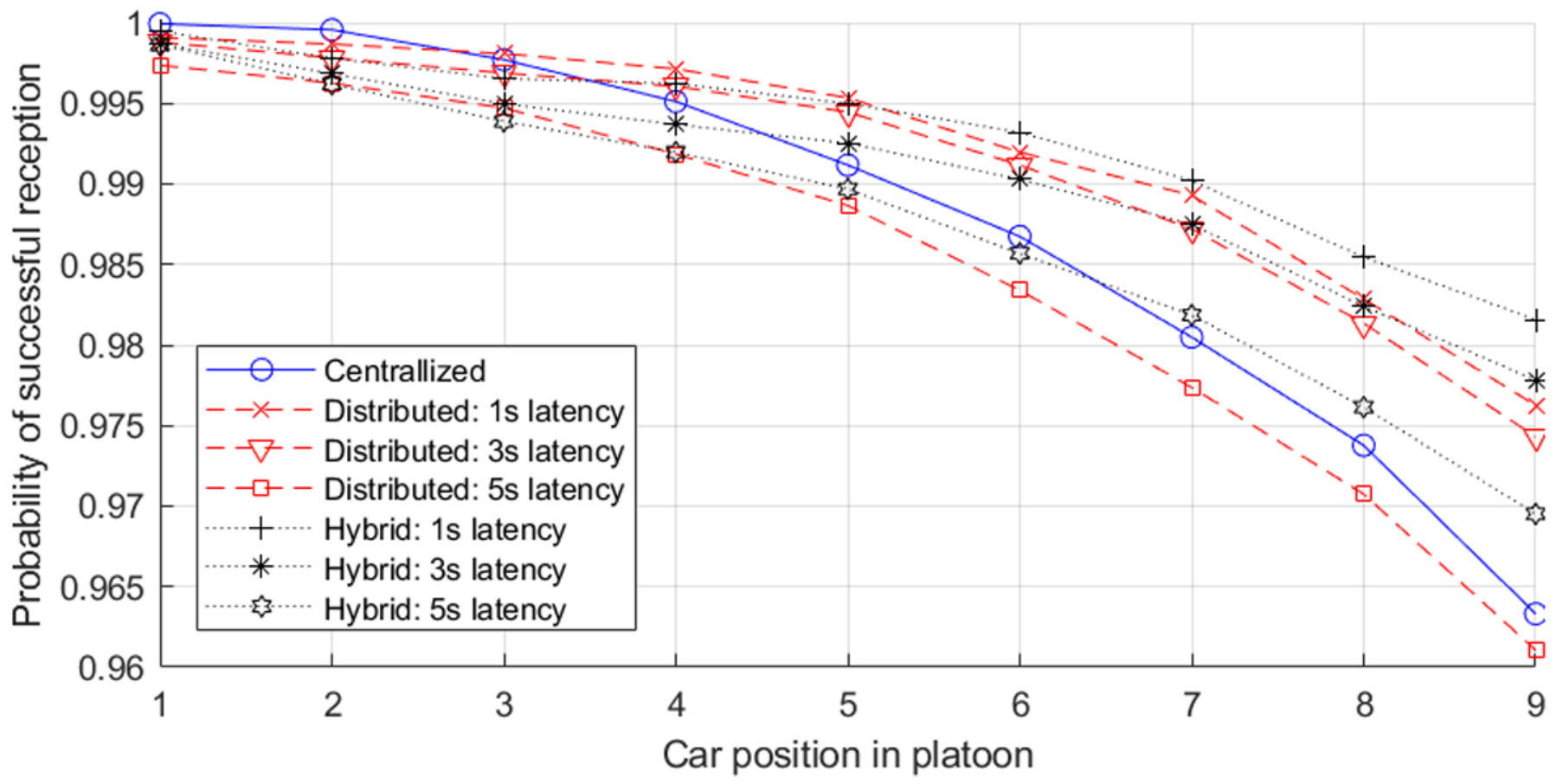

Figure 7.6: Estimated probability of successful reception of leader's packets vs. car position in platoon.

**In summary**, the hybrid architecture emerges as the most effective solution. Maintain high communication reliability and protects DTT systems while minimizing spectrum reallocation overhead. The distributed approach, in contrast, struggles with coordination delays and insufficient protection for the primary system.

#### 7.2.2.2 Machine Learning-Based Approach

To address the challenge of accurately estimating SINR, we introduce a reinforcement learning strategy-specifically the Q-learning algorithm – as an alternative to traditional methods [227]. This approach is used to determine the most suitable frequency bands for platoons. For each possible number of platoons within a coordination zone, separate Q-tables are generated to reflect varying levels of inter-platoon interference depending on their proximity.

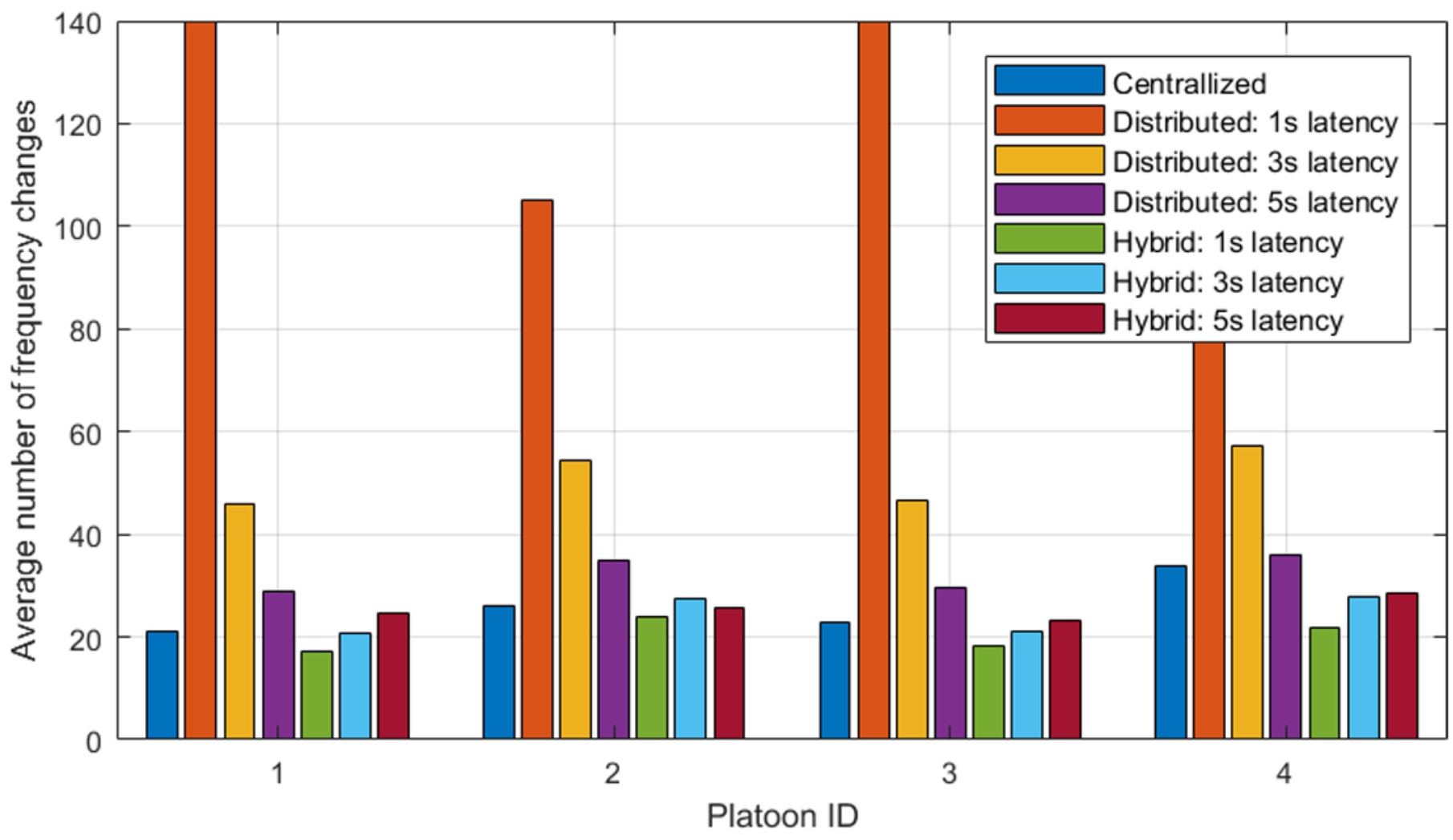

Figure 7.7: Average number of frequency changes per platoon in a single simulation run (140s of motorway movement).

The state of the system is defined by the spatial positions of the platoons, represented as their geometric centers. These positions are discretized into $N$ intervals by dividing the length of the road segment $D$ into equal sections. Hence, the total number of states is given by:

$$S = N^M$$

where $M$ denotes the number of platoons.

Each action in the Q-learning process corresponds to an allocation of $K_{\text{VDSA}}$ available frequency channels among the $M$ platoons, resulting in a total number of actions equal to:

$$A = (K_{\text{VDSA}})^M$$

The learning objective is to maximize a reward $r$, defined as the ratio of successfully delivered intra-platoon messages to the total transmitted. The Q-table is updated using the standard rule:

$$Q(s_t, a_t) \leftarrow Q(s_t, a_t) + \alpha \left[ r + \gamma \max_a Q(s_{t+1}, a) - Q(s_t, a_t) \right]$$

where:

- $s_t$ and $a_t$ represent the current state and action at time $t$,
- $\alpha$ is the learning rate,
- $\gamma$ is the discount factor for future rewards.

At each iteration, the algorithm selects an action randomly with probability $\varepsilon$ or chooses the action with the highest Q-value (greedy selection) with probability $1 - \varepsilon$.

The Q-learning-based VDSA method is evaluated against a standard distributed algorithm. For Q-learning, we set the exploration rate to $\varepsilon = 0.01$ and train the model over 2000 simulation runs. In the distributed case, communication delays of 1 s and 5 s are considered when exchanging information between platoons. The evaluation criteria are consistent with those used in the previous section.

The simulation scenario includes three platoons: two traveling in the same direction (with slightly different speeds), and one in the opposite direction. This setup creates prolonged interference between the first two platoons and shorter, more intense interference for the third.

Fig. 7.8 presents the success rates of receiving the leader message. The greedy Q-learning method performs comparable to the low-latency distributed solution and is significantly better than its high-latency counterpart.

Fig. 7.9 highlights the average number of channel changes. The distributed system with 1 s latency shows the highest switching frequency due to the “ping-pong” effect, where the platoons repeatedly choose the same channels and then switch to resolve conflicts. This effect is less pronounced at 5 s latency. The greedy Q-learning model exhibits the lowest switching rate (below 4), thanks to centralized processing that allows stable channel assignments.

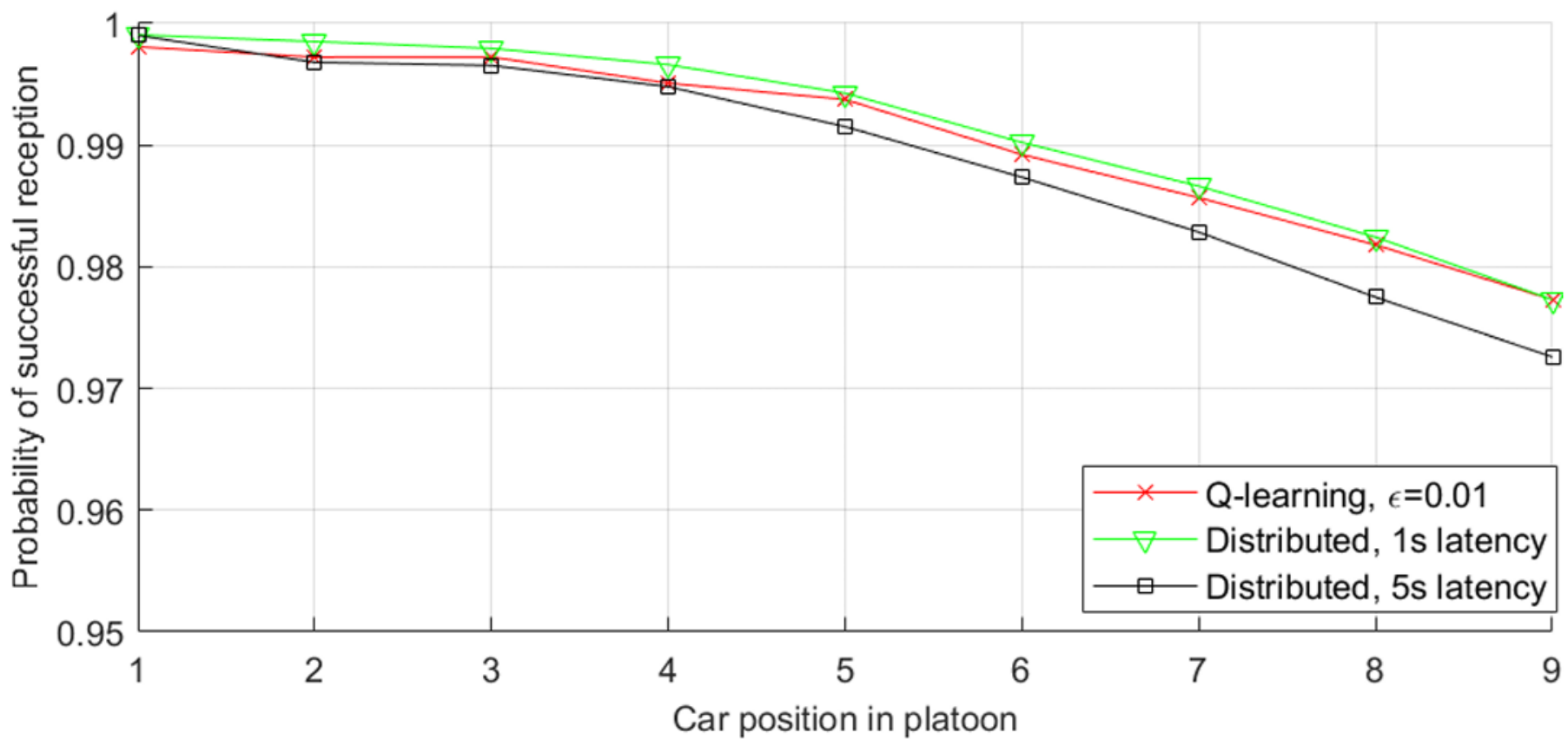

Figure 7.8: Estimated probability of successful reception of leader's packets vs. car position in platoon.

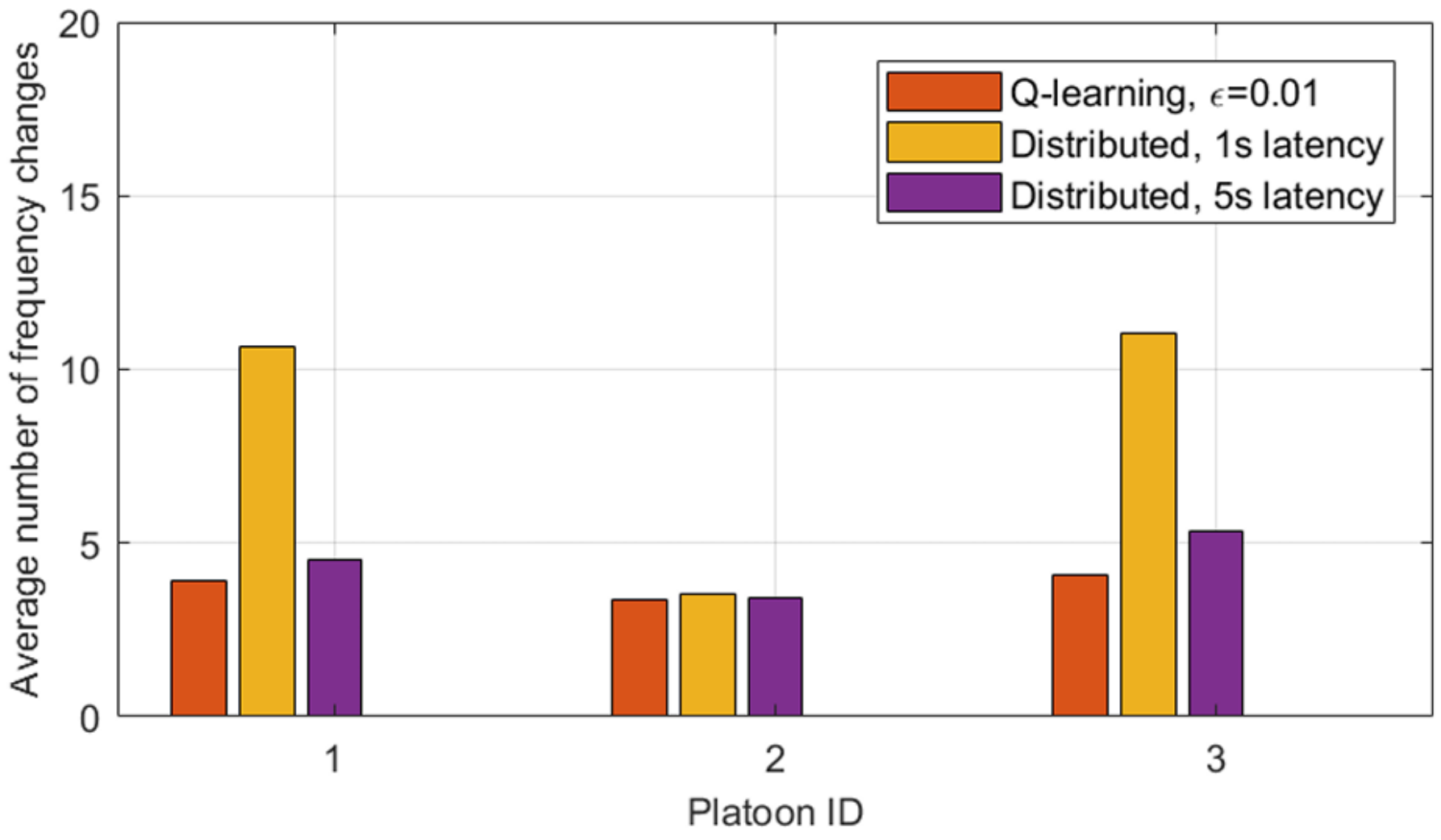

Figure 7.9: Average number of frequency changes per platoon in a single simulation run (140s of motorway movement).

In summary, the Q-learning approach matches or exceeds the performance of the distributed strategy in terms of communication reliability and interference management. It also minimizes frequency reassignment overhead. However, the scalability of this method is limited by the exponential growth of states and actions. Moreover, real-time training may degrade performance at high exploration rates, highlighting the need for offline pre-training in practical applications.

### 7.2.3 Main Conclusions

The dynamic spectrum sharing is a foundational requirement for the reliable operation of ITS. It enables diverse communication systems, such as V2V, V2I, and roadside units, to coexist and coordinate over shared wireless spectrum, ensuring efficient use of limited frequency resources. Rather than assigning fixed channels, these systems must operate concurrently in shared or adjacent frequency bands, making coordinated access essential to avoid interference.

However, the spatiotemporal variability inherent in ITS – due to high vehicular mobility, changing traffic densities, and heterogeneous physical environments – makes spectrum access unpredictable and increasingly prone to interference. Traditional static or model-based allocation techniques struggle to adapt to such dynamic conditions.

To address these challenges, artificial intelligence, particularly deep learning and reinforcement learning, has emerged as a transformative enabler. These techniques enable spectrum management systems to predict spectrum demand, monitor interference patterns, and allocate channels in real time, even with incomplete or delayed channel state information. By leveraging traffic data, mobility patterns, and edge computing capabilities, AI-driven frameworks, e.g., based on Q-learning, enable low-latency, scalable, and interference-aware spectrum access.

As ITS networks continue to expand in complexity and density, the integration of AI into spectrum sharing frameworks will be critical – not only for optimizing communication efficiency but also for ensuring the resilience, safety, and responsiveness of next-generation transportation ecosystems[228].

## 7.3 Intelligent Surfaces and Antennas

By **Fernando J. Velez, Nila Bagheri, Jon M. Peha**

### 7.3.1 Introduction

The future of ITS depends on advanced wireless technologies that address challenges such as signal blockages, high mobility, and energy efficiency. Intelligent surfaces and antennas including Reconfigurable Intelligent Surfaces (RIS), active smart surfaces, and AI-optimized antenna arrays are emerging as promising solutions that dynamically control electromagnetic waves to extend coverage and optimize connectivity without energy-intensive infrastructure.

### 7.3.2 Reconfigurable Intelligent Surfaces for ITS

V2X communication is central to ITS, especially for CAVs, but link reliability suffers from SNR degradation and fading in dynamic environments. RIS is currently being actively studied as a potential solution, particularly in the shift toward 5G-Advanced V2X. The evolving 3GPP standards (Releases 14 – 18) have strengthened device-to-device communication and created opportunities for RIS integration [8], [229].

In addition, For the V2V and V2I modes, RIS directly influences the channel conditions and improves link performance. In contrast, in V2N scenarios where communication relies on base stations, the evolution of the Radio Access Network (RAN), such as Cloud RAN (C-RAN) and AI-enhanced RAN can complement RIS by enhancing coordination between cellular infrastructure and vehicular links. 3GPP Release 18 places a particular emphasis on incorporating artificial intelligence and machine learning (AI/ML) into the C-RAN, aiming to optimize and improve network efficiency. Crucially, RIS remains backward-compatible with current NR-V2X infrastructure while being adaptable for 5G-Advanced.

RIS can enable dynamic beam steering and reflection optimization, while most AI-driven beam management research has focused on base stations and roadside units, RIS-assisted approaches may also extend to vehicular links in future deployments. Its passive operation boosts energy efficiency while improving V2V and V2I sidelink performance, extending coverage, and supporting both relaying and Reduced Capability (RedCap) devices. RIS technology provides several mechanisms to improve the energy efficiency of vehicular networks. By steering radio waves toward intended directions, RIS lowers the transmission power required for communication, which is especially valuable in V2X scenarios, where vehicles frequently exchange short and bursty messages. RIS can also strengthen weak signals by enhancing reception in areas with poor coverage, thereby reducing the power demand on vehicles when receiving V2X messages.

Furthermore, by creating optimized signal paths and effectively reducing the travel distance of radio waves, RIS can decrease communication latency, an essential requirement for real-time applications such as collision avoidance. Recent studies have further confirmed that RIS integration supports energy-efficient beamforming and contributes to reducing the overall energy consumption of vehicular networks [230], [231], [232].

As standardization progresses, RIS may play an enabling role in more reliable, efficient, and scalable next-generation V2X communications [233].

Recent studies highlight that RIS is being investigated as a promising solution for vehicular wireless networks. RIS consists of reconfigurable elements, mostly passive, but in some cases active that adjust the phase, amplitude, and sometimes the polarization of incoming signals, enabling flexible and controlled channel reconfiguration. While architecturally different from massive MIMO, RIS is expected to play for 6G a role similar to that of massive MIMO in 5G in improving link reliability and spectral efficiency [234]. Applied to vehicular communications, RIS has been shown to enhance V2I capacity, extend coverage, and mitigate blind spots, while consuming far less power than relay-based schemes, making promising options to support the sustainability objectives of future ITS.

Moreover, RIS deployment can enhance connectivity under NLoS conditions, thereby extending the effective communication range [235]. RIS enhances LoS communications by producing an artificial scattering environment, which improves spatial multiplexing and enables the concurrent transmission of multiple independent data streams [236]. In dense urban areas, deploying RIS panels on roadside poles or building facades can boost vehicular connectivity by improving signal reception, mitigating interference, and extending communication range.

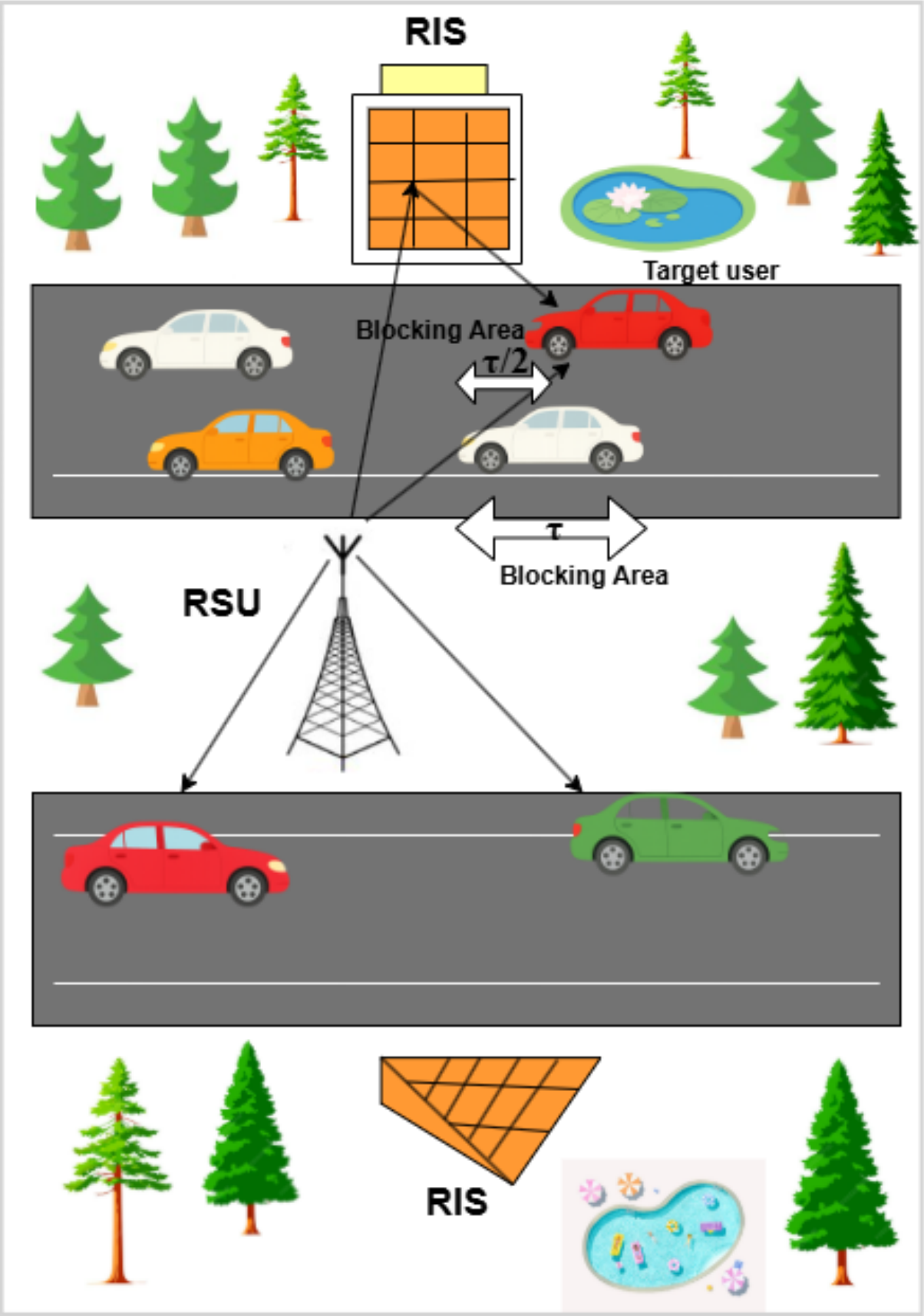

Figure 7.10: RIS-assisted vehicular communication scenario with RSUs, target user, blocking area, and RIS providing an alternative reflected path from the other side of the road ($\tau/2$ denotes half of the length of a vehicle).

To further enhance reliability in autonomous transportation networks, researchers have proposed architectures that integrate real-time, software-defined RIS control [237]. Beyond reliability, RIS also contributes to security by shaping reflection paths that allow only authorized vehicles to access the channel, thereby reducing the risk of spoofing and eavesdropping.

According to studies in [238], [239], RIS has been introduced as an emerging approach to improve the performance of wireless networks. The focus of this work is on evaluating the Secrecy Outage Probability (SOP) in RIS-assisted vehicular communications, where SOP is defined as the probability that the instantaneous secrecy capacity of a communication link falls below the required secrecy rate. Two communication models are examined: in V2V communications, the RIS operates as a relay between vehicles, while in V2I communications, the RIS functions as the receiver. In both cases, the presence of a passive eavesdropper is assumed, attempting to intercept the transmitted information. Closed-form analytical expressions for SOP are derived and validated, and the results reveal that employing RIS can considerably enhance the secrecy performance of both V2V andV2I links [238], [239].

Recent research has also investigated resource-allocation strategies in RIS-assisted vehicular systems. For example, one study maximized V2I throughput, while safeguarding V2V SINR constraints by formulating joint optimization over RIS phase shifts, bandwidth distribution, and transmit power, solved via a two-stage framework: (i) RIS reflection and power allocation, followed by (ii) bandwidth assignment [240]. In a related work, a RIS-enabled vehicular network with RSUs was explored, where RIS provided indirect connectivity to blocked areas. The authors formulated a joint optimization problem aiming to maximize the minimum user rate, incorporating both RSUs resource scheduling and RIS phase-shift design [241].

As demonstrated by the authors of [242], RIS-assisted vehicular communication can effectively reduce outage probability, particularly for vehicles near the RIS. The study also shows that outage performance is strongly influenced by the vehicle density moving on the road and number of reflecting elements of the RIS.

As shown in Fig. 7.10, the considered scenario involves a two-lane road in each direction, with a single-antenna RSUs deployed on the separating strip at a distance $x = x_0$, while the RIS is located at a distance $x = 2.2x_0$. The target user (red vehicle) is positioned in the outer lane, while vehicles in the inner lane can act as dynamic obstacles that block the LoS link between the RSUs and the user. Vehicles are modeled as rectangles with length $\tau$, and the average inter-vehicle distance $d$ is expressed as multiples of $\tau$ (e.g., $d = \tau, 5\tau, 10\tau$). To overcome blockage, a RIS with $M$ passive reflecting elements is deployed on the other side of the road, providing an alternative reliable reflected path to improve coverage and link reliability, as shown in Fig. 7.10.

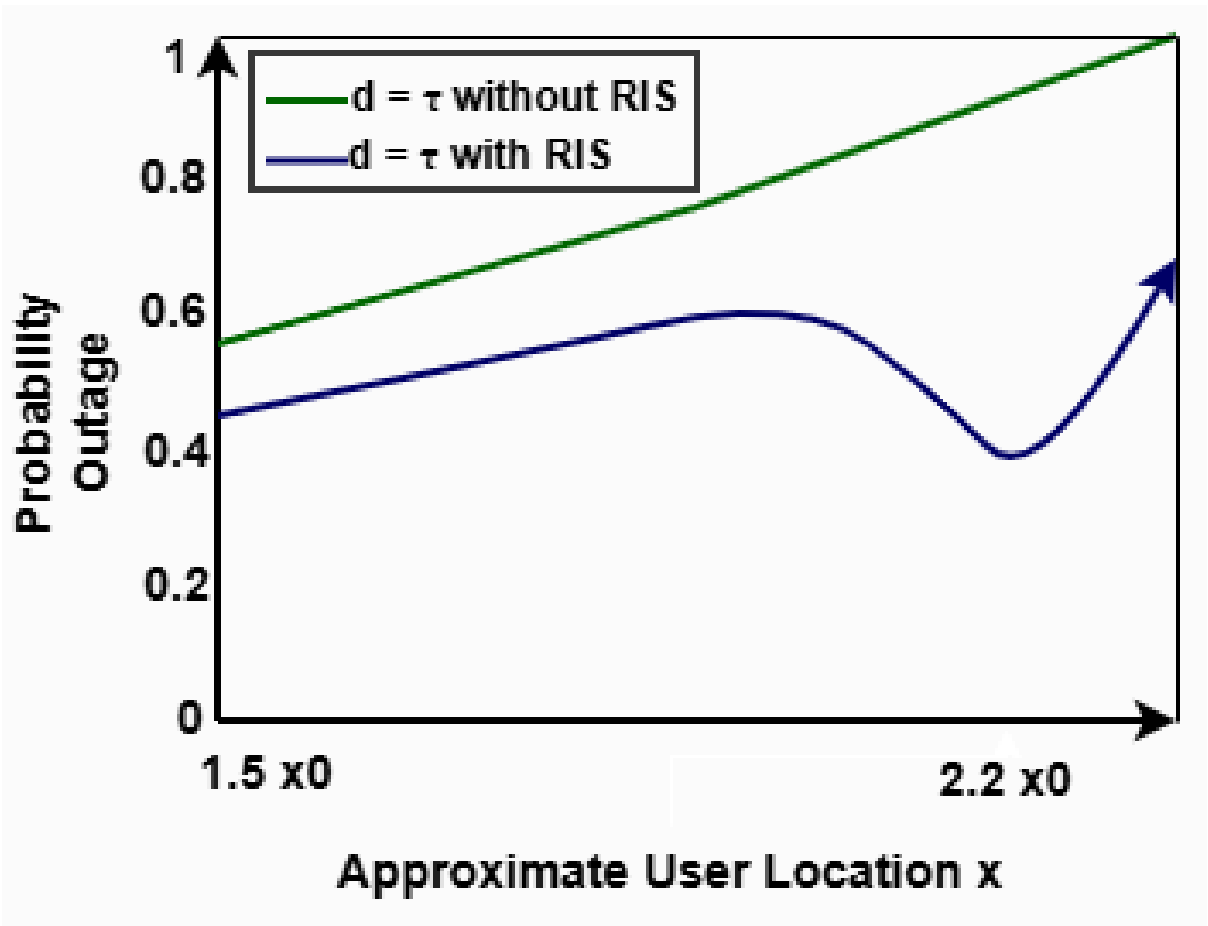

Figure 7.11: Approximate outage probability versus user location for inter-vehicle distances $d \approx \tau$ (without RIS) and $d \approx \tau$ (with RIS). The RSU located at $x = x_0$, and minimum outage is observed near the RIS position at $x \approx 2.2x_0$.

To further evaluate the role of RIS, Fig. 7.11 illustrates the approximate outage probability with and without RIS. The green curve corresponds to the baseline case without RIS, while the RIS-assisted case demonstrates a clear reduction in outage probability, especially near the RIS position at $x \approx 2.2x_0$. Moreover, according to [242], when the inter-vehicle distance increases to $d \approx 11\tau$, the probability of blockage decreases, leading to further performance improvement. These results indicate that RIS deployment, together with larger vehicle spacing, can substantially enhance link reliability.

For performance evaluation, authors from [242] assumed the lane width is set to 4 m (with two lanes), and each vehicle is modeled as a rectangle with a length of $\tau$ =5 m. The carrier frequency is 28 GHz, and the RSU transmit power is 27 dBm. According to the results reported in [242], the outage probability decreases as the number of RIS elements increases, with the most significant improvements observed near the RIS location.

For example, with $M = 500$, the outage probability around the RIS becomes even lower than that near the RSU, since a large RIS can effectively reflect signals and extend coverage. Compared to the baseline without RIS ($M = 0$), the region where the outage probability $OP < 0.3$ expands by nearly three times when $M = 500$. Although the effective range of a single RIS is relatively limited, these results highlight the potential of deploying multiple cooperative RIS panels along the roadside to ensure robust coverage across the entire area. Furthermore, as the number of RIS elements grows, the advantage of outage reduction extends over a broader range of distances, resulting in wider and more reliable coverage.

Fig. 7.12 shows that the outage probability decreases as the user approaches the RIS, reaching, according to [242], its minimum at the RIS position ($x \approx 2.2x_0$).

On the analytical side, closed-form expressions for outage probability and bit error rate have been derived for both single- and multi- RIS deployments under Rayleigh fading (including aggregated in-terference). Monte Carlo simulations validate these results, demonstrating that increasing either the number of RIS panels or the reflecting elements significantly enhances coverage reliability and communication safety [243]. Beyond conventional RIS, Star-RIS architectures which simultaneously reflect and transmit have been explored to build stronger vehicular links and to integrate with federated learning at the network edge; coupled with deep reinforcement learning (e.g., DDQN) for phase-shift, power, and beamforming optimization, these hybrids improve adaptability, reduce latency, and enhance QoS in highly dynamic vehicular environments [244].

Overall, RIS shows potential to enhance efficiency, coverage, and reliability in ITS, and may provide a foundation for more sustainable, 6G-enabled vehicular networks. Ongoing research highlights both the substantial performance benefits and the remaining challenges, particularly deployment under high mobility, scalability in dense urban environments, and robustness against eavesdropping and malicious interference.

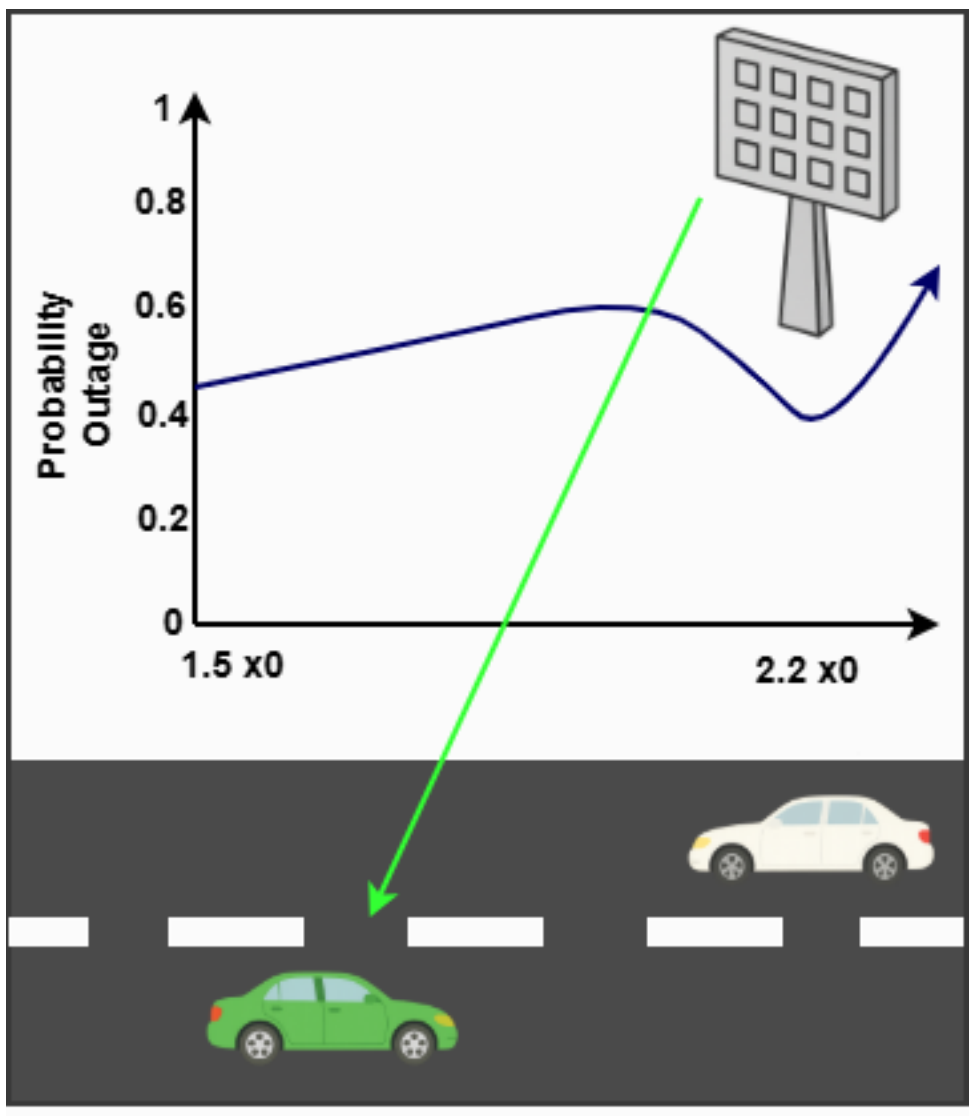

Figure 7.12: Approximate outage probability versus user location, with the lowest value observed near the RIS position at $x \approx 2.2x_0$ due to the reflected signal path.

### 7.3.3 Antennas

Antennas constitute a fundamental component in enabling reliable communication within ITS, which are designed to enhance transportation safety, operational efficiency, and overall traffic management through the integration of advanced communication technologies. Functioning as the critical interface between electronic subsystems and the wireless communication environment, antennas facilitate real-time data exchange among vehicles, roadside infrastructure, satellites, and vulnerable road users, thereby supporting the seamless operation of connected and autonomous transportation ecosystems.

In V2X communication, antennas enable seamless interactions between vehicles (V2V), infrastructure (V2I), pedestrians (V2P), and the network (V2N). These communications occur over multiple frequency bands such as the 5.9 GHz band for DSRC, sub-6 GHz and mmWave bands for C-V2X and 5G NR), requiring tailored antenna solutions that offer high gain, directivity, and interference mitigation. A variety of antenna types are employed in ITS deployments:

- **Omnidirectional antennas:** radiate signals equally in all horizontal directions, making them ideal for urban and dense traffic environments, where vehicles, infrastructure, and pedestrians are constantly moving in different directions. Their ability to cover all directions removes the need for precise alignment or beam steering, which simplifies installation on vehicles, RSUs, and other ITS components. In V2V and V2I communication, omnidirectional antennas ensure that nearby vehicles and infrastructure nodes can consistently communicate, regardless of their relative position or orientation. They are also generally cheaper, lighter, and more compact than directional or phased array antennas, making them suitable for large-scale deployment in vehicles and urban infrastructure [245], [246], [247], [248].

- **Directional antennas:** offer several advantages in ITS applications by concentrating signal energy in a specific direction. This focused transmission results in higher gain, which translates to extended communication range and improved signal quality. Such characteristics are particularly beneficial in V2I communication on highways, where vehicles move at high speeds and need to establish reliable links with distant roadside units or base stations. By narrowing the radiation pattern, directional antennas also help to reduce interference from unwanted sources, enhancing the SNR and ensuring more stable data transmission. This is especially important in densely populated environments where multiple wireless systems operate simultaneously.

  Moreover, directional antennas contribute to energy efficiency, as the transmitted power is not wasted in unnecessary directions. This can be critical in infrastructure units that are solar-powered or energy-constrained. In combination with beam steering technologies, such as those used in phased array systems, directional antennas can dynamically track moving vehicles, maintaining high-quality links without mechanical movement [249], [250], [251]. Phased array antennas represent an advanced class of directional antenna systems that are revolutionizing V2X communications. Unlike conventional directional antennas with fixed radiation patterns, phased arrays dynamically control their beam direction and shape through electronic phase shifting across multiple antenna elements. This capability makes them uniquely suited to the demanding requirements of intelligent transportation systems.

  As directional antennas phased arrays offer several distinct advantages. They can electronically steer their beams without any mechanical movement, allowing for rapid beam switching in microseconds. This enables continuous tracking of fast-moving vehicles, while maintaining optimal signal strength. The system can generate multiple simultaneous beams, permitting concurrent communication with various endpoints like vehicles, infrastructure, and vulnerable road users.

  Advanced beamforming techniques provide precise control over radiation patterns, enabling focused energy transmission and sophisticated interference mitigation through null steering. The directional characteristics of phased arrays address critical V2X communication challenges. Their focused beams maintain reliable links with moving vehicles even in complex urban environments with multiple obstacles. The technology's adaptive directionality allows for real-time optimization based on vehicle movements and changing surroundings. Compared to omnidirectional solutions, phased arrays offer significantly extended communication ranges, while reducing interference in congested spectral environments [252], [253].

- **MIMO antennas:** significantly enhance data throughput and communication reliability in 5G-based ITS by using multiple transmitting and receiving antennas simultaneously. This allows the system to transmit parallel data streams over the same frequency band, increasing spectral efficiency, reducing latency, and improving resilience against signal fading and interference, which are critical factors for real-time V2X communication in dynamic traffic environments [254], [255].

- **Fractal Patch Antenna (FPA) based on Photonic Crystal (PhC):** will improve communication within ITS, and realize the growing interest in leveraging alternative frequency ranges (beyond 5.9 GHz) for advanced V2X applications [256]. The use of mmWave bands is identified as having an enormous potential for performance enhancement of these systems. According to [257], integrating a microstrip patch antenna on a PhC substrate improves antenna bandwidth and gain, providing miniaturization. In this research, as an example, the designed antenna resonates at 31.42 GHz, 37.76 GHz and 38.92 GHz. In [257], by using PhC characteristics, the proposed nonagonal PhC fractal antenna reaches a maximum gain of 10.88 dBi at 38.92 GHz. Photonic Band Gap structures use periodic arranged materials to regulate wave propagation, reducing surface waves and expanding bandwidth.

  Through simulation and analysis, including radiation pattern, gain, and reflection coefficient plot assessment, the antenna performance was thoroughly evaluated. The study highlighted the potential of the proposed FPA-PhC antenna configuration to enhance communication networks within the V2X, significantly advancing the V2X with support from the mmWave bands.

The structure of the proposed FPA-PhC is illustrated in Fig. 7.13, where (a) shows the geometry and key design parameters of the antenna, and (b) presents the square-lattice photonic crystal used for the substrate and ground layer. The radiation characteristics of the antenna are analyzed in 7.14, which shows the gain plots at 31.42 GHz, 37.76 GHz, and 38.92 GHz, corresponding to the resonant frequencies of the design. The proposed FPA-PhC achieves a maximum gain of 10.88 dBi at 38.92 GHz with stable radiation performance across all bands, demonstrating the effectiveness of the photonic crystal substrate in reducing surface waves and improving bandwidth and efficiency.

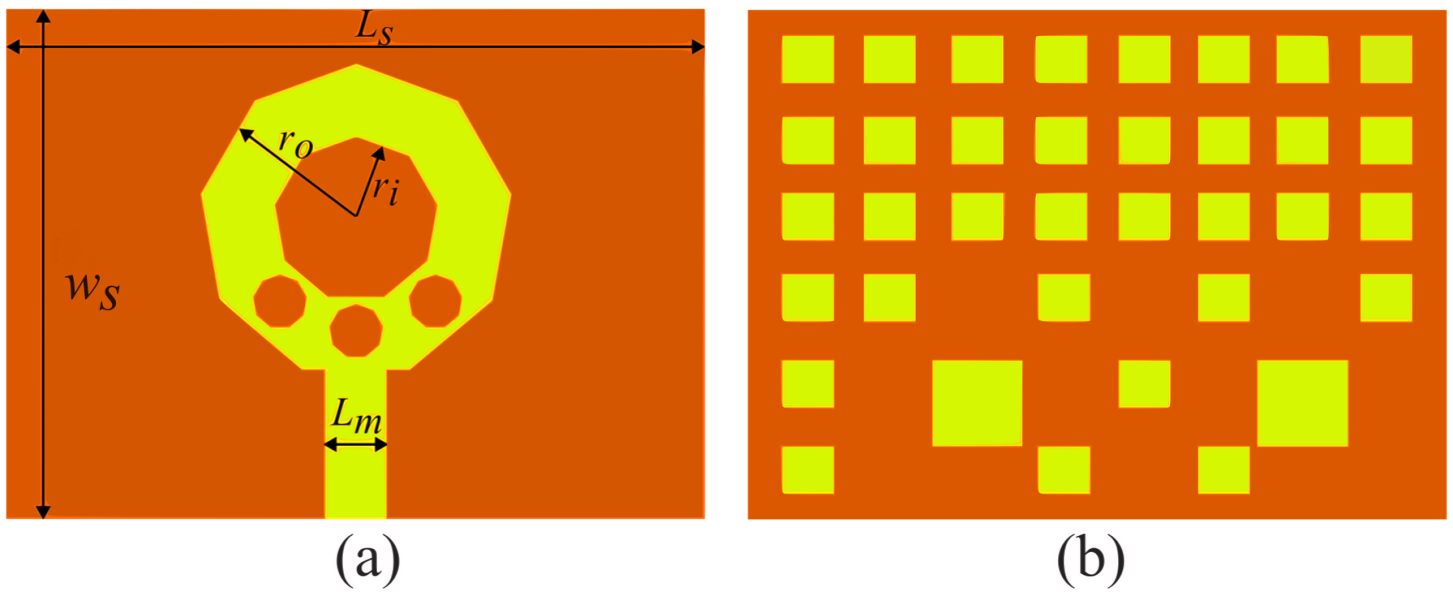

Figure 7.13: (a) Configuration of the proposed microstrip patch antenna, illustrating its dimensions and layout. (b) Structure of the square lattice photonic crystal used for the ground and substrate of the antenna [257].

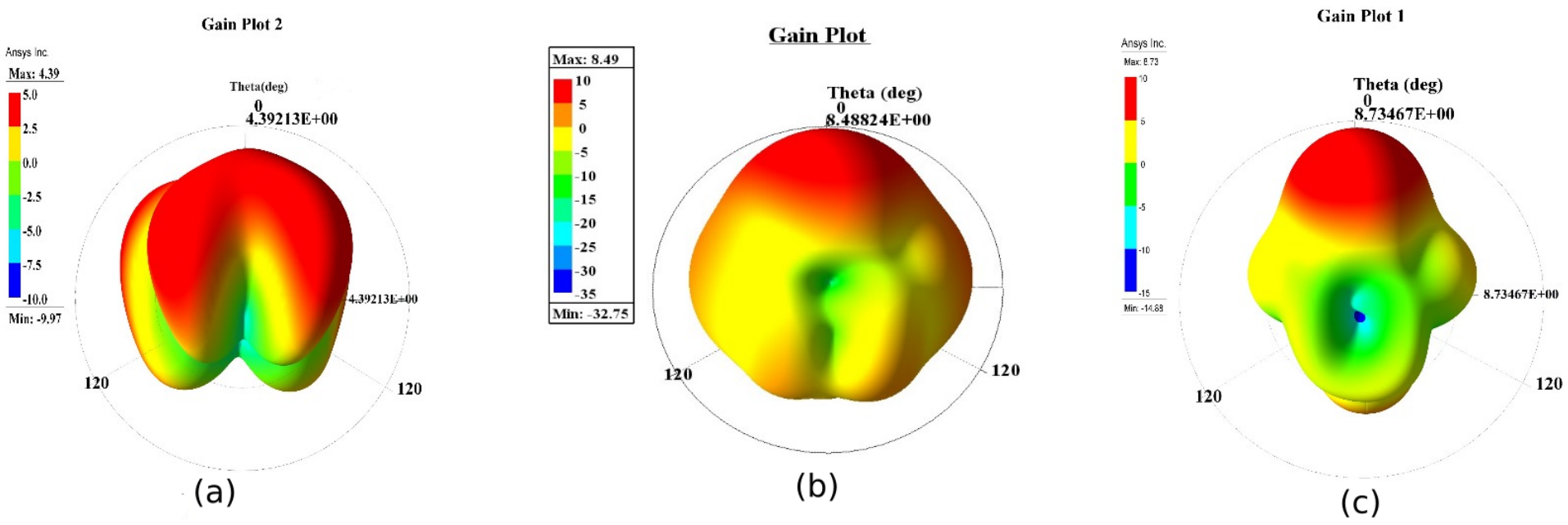

Figure 7.14: Gain plot for the proposed photonic crystals microstrip patch antenna with square lattice, at (a) 31.42 GHz, (b) 37.76 GHz, (c) 38.92 GHz.[257].

## 7.4 Digital Twin for Wireless V2X Communications

By **Francesco Linsalata**

The evolution of vehicular networks towards 6G aims to revolutionize mobility by enhancing wireless connectivity in automotive applications. V2X communications, operating at mmWave and sub-THz frequencies, enable high data rates and low-latency information exchange, surpassing traditional sensor-based detection methods. However, the high-frequency nature of these bands makes wireless links highly susceptible to LoS blockages, particularly in dense urban environments where buildings and moving vehicles disrupt connectivity. Ensuring continuous and reliable connectivity in such dynamic scenarios is a fundamental challenge for next-generation vehicular networks.

To address these issues, the concept of Digital Twin (DT) has emerged as a key enabler for 6G networks, facilitating the seamless integration of the physical and digital worlds. A DT is a high-fidelity virtual representation of the physical environment, continuously updated with real-time data to enhance situational awareness and optimize network performance. In the context of V2X communications, DTs leverage high-resolution 3D maps, multi-modal sensor data, and advanced wireless propagation models to provide accurate channel estimations and predictive network optimization.

The implementation of DTs in vehicular networks involves multi-modal sensing at both infrastructure nodes (e.g., road-side units) and connected vehicles. This includes LiDAR, radar, cameras, and inertial measurement units, which collectively capture environmental dynamics. By integrating sensor data with ray-tracing-based wireless channel simulations, a DT can dynamically adapt to changing communication conditions, mitigating the impact of blockages and enabling proactive network reconfiguration.

A fundamental advantage of DT-enabled V2X systems lies in their ability to reduce the decision time required for critical tasks such as beam selection, channel estimation, and handover management. Unlike traditional exhaustive search methods, which require testing multiple beam pairs, DTs can leverage data-driven models to predict optimal beamforming configurations with reduced overhead. This is particularly beneficial for high-mobility scenarios where rapid topology changes necessitate real-time network adjustments.

The integration of DTs with automotive simulation platforms further enhances their capability to model realistic vehicular scenarios. By incorporating tools such as high-fidelity 3D urban reconstructions and real-time ray tracing, DTs can provide detailed insights into wireless propagation characteristics under various environmental conditions. This facilitates the development of robust V2X communication strategies that account for dynamic obstacles, multipath effects, and time-varying channel conditions.

As vehicular networks transition towards 6G, DT technology is expected to play a crucial role in achieving URLLC for intelligent transportation systems. Future advancements in DT-driven V2X communications will focus on improving model accuracy, reducing computational complexity, and integrating AI-driven predictive analytics to further enhance network resilience and efficiency.

Overall, digital twins are becoming an important tool for developing and testing V2X communication systems. In [258], the authors present a setup that combines a geometry-based stochastic channel model with hardware-in-the-loop to estimate packet error rates under changing conditions. Using this, they show that relay-assisted communication can significantly reduce latency and error rates in a safety-critical overtaking scenario. For simulation-based modeling, ray tracing remains essential. A measurement-calibrated RT simulator is used in [259] to study V2I handovers at mmWave frequencies, focusing on how factors like route shape and base station spacing affect signal quality. To speed up these simulations, Brennan et al. propose an interpolation method that reuses ray path data across a vehicle's trajectory, reducing computational load without a big drop in accuracy [260]. [261] takes a different approach, estimating reflection points directly from power variations in the received signal, without needing detailed maps or prior knowledge of the environment. Together, these efforts show how digital twins and improved simulation techniques can support more reliable and efficient V2X system design.

## 7.5 Integrated Sensing and Communication

By **Martin Schmidhammer, Haibin Zhang**

Integrated Sensing and Communications (ISAC) combines communication and sensing functionalities within a unified system, addressing the requirements of both data exchange and environmental awareness [10], [262]. With its combination of communication and sensing functions, ISAC offers considerable potential as a spectrum, energy, hardware, and cost-efficient solution for ITS, which rely heavily on robust communication, as well as timely and reliable information about the environment [9]. However, the integration of ISAC into ITS faces several challenges, including coping with highly dynamic propagation environments of the radio channel, ensuring reliable performance in congested networks, meeting stringent latency and reliability requirements of ITS applications, etc. [263], [264], [265].

Overcoming these challenges is crucial for realizing the benefits of ISAC in ITS applications. In the following, we provide an overview of the opportunities and challenges of ISAC for ITS, give an overview of the current technological trends in the field of ISAC for ITS, and outline future developments and prospects.

### 7.5.1 Potentials and Challenges of ISAC in ITS

With regard to the main objectives of ITS, namely increasing overall safety for all road users, improving efficiency and traffic flow, optimizing energy consumption and providing a more comfortable and convenient experience for travelers [266], ISAC can contribute to use-cases for safety-critical applications, traffic management and efficiency, environmental monitoring and infrastructure health, and user services and convenience. Examples of use-cases include [267], [268]

- Pedestrian/animal intrusion detection on a highway
- Sensing-assisted automotive maneuvering and navigation
- Sensing at crossroads with/without obstacle (see e.g. Fig. 7.15)
- Sensing for parking space determination
- Vehicles sensing for ADAS
- Blind spot detection

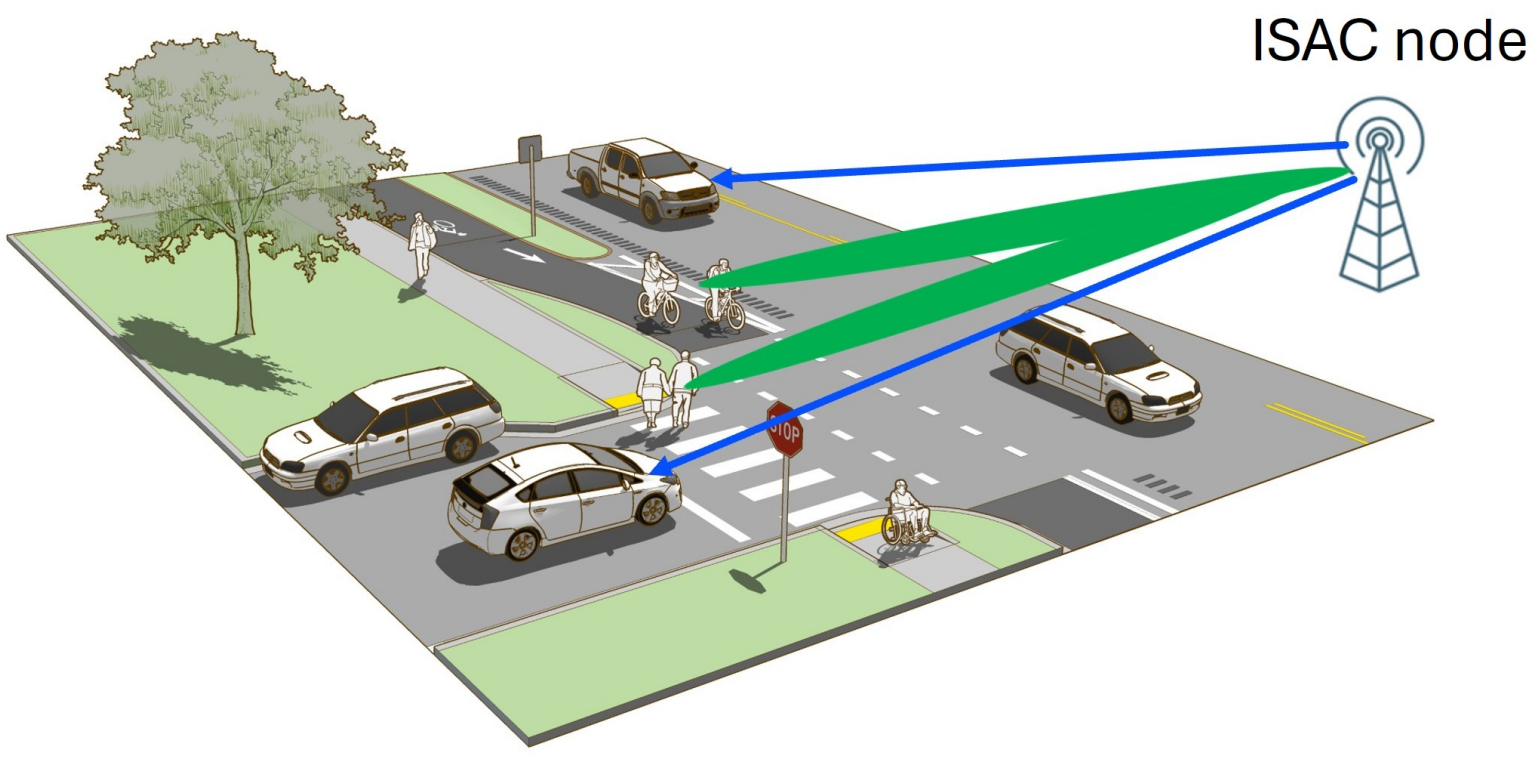

Figure 7.15: Illustration of ISAC application at road intersections (edited on an intersection figure of www.ruraldesignguide.com).

- Vision aided smart traffic management
- Automated guided vehicles travelling in airports
- Emergency vehicle route planning

The challenges of ISAC depend largely on the level of communication-sensing integration, whether through shared sites, RF units, spectrum, waveforms, or signals. It may occur at different layers of the network:

- From the perspectives of the physical layer, What's the likelihood and importance of LoS? In case of monostatic sensing, self-interference cancellation remains a challenge. While in the case of bistatic and distributed deployment, inter-node synchronization should be guaranteed.
- From the perspectives of radio network management, how to jointly schedule communication traffic and sensing tasks? How to avoid or mitigate communication–sensing interference in case different signals are used for communication and sensing. Sensing targets may be mobile (e.g. in ITS), and in this case how to dynamically assign network nodes for the purpose of seamless provision of sensing services?
- From the perspectives of network architectures, What is the potential architecture of future network of sensing? How to enable integration of non-3GPP sensory data?

Vehicular environments are characterized by high dynamics as well as a large number of moving users and objects. Accordingly, both sensing and communication functionalities of ISAC systems are further challenged by:

- high mobility and dynamics
- multi-entity interference and spectrum congestion
- environmental challenges
- latency and synchronization.

### 7.5.2 Channel and Propagation Models

The reliable operation of ISAC in ITS fundamentally relies on an accurate understanding of the radio propagation environment. Channel models in this context must account for dynamic mobility, diverse frequency ranges, and complex urban topologies. These models are essential not only for low-level signal processing – such as beamforming and localization – but also for higher-level system functions like resource allocation, link adaptation, and cooperative perception.

Vehicular environments, especially in dense urban settings, pose unique challenges to propagation modeling. Higher-frequency signals, including mmWave and sub-THz, exhibit strong directionality and are highly susceptible to blockage. For example, road bridges can cause additional propagation losses of up to 23 dB. To better predict such effects, empirical models like the single road bridge model have been proposed to reduce errors in link performance estimation under obstruction.

To navigate these propagation challenges, waveform design is essential in adapting to environmental dynamics. While the design space is broad, current approaches generally fall into three categories: communication-centric, sensing-centric, and joint waveform design. Communication-centric waveforms – especially OFDM – remain dominant due to their compatibility with wireless communication standards (e.g. 5G NR or IEEE 802.11 WLAN), spectral efficiency, and support for high data rates. Recent enhancements, such as constellation shaping, aim to improve radar performance by minimizing delay and Doppler estimation errors [269], [270]. Alternatives like Orthogonal Time Frequency Space (OTFS) and Orthogonal Chirp Division Multiplexing (OCDM) offer stronger resilience to Doppler spread and multipath interference, making them particularly useful in high-speed vehicular environments [271], [272], [273], [274].

On the sensing side, radar-centric waveforms such as chirp sequences and pseudorandom-coded signals (e.g., PMCW) are being adapted to carry embedded data. These offer low sidelobes and strong resolution but are often limited in communication capacity [275]. In applications where sensing dominates – such as V2V coordination and collision avoidance – modulation schemes like CAESAR or frequency-hopping MIMO facilitate data transfer without compromising detection fidelity.

Joint waveform designs attempt to balance communication and sensing performance by optimizing metrics like SINR, Cramér-Rao Lower Bound (CRLB), and mutual information. OFDM and OCDM can be co-optimized across domains such as beamforming, subcarrier selection, and time-frequency framing, allowing simultaneous target detection and data exchange – a capability that is becoming increasingly important for cooperative driving systems [269].

Understanding how these waveforms propagate in the real world is crucial. Measurement campaigns have shown that propagation characteristics at different frequency bands can exhibit high correlation. Time-synchronized channel measurements at 3.2, 34.3, and 62.35 GHz suggest that path information at the lower bands can be used to guide link establishment at higher frequencies – particularly valuable for beam alignment in fast-changing vehicular scenarios [276].

At lower frequencies, especially in sub-6 GHz bands, multipath is abundant and traditionally considered a distortion. However, modern approaches leverage it for localization and sensing. In multipath-enhanced device-free localization, user-induced fluctuations in the power of reflected components reveal positional information [277]. The underlying empirical fading model is derived based on an extensive set of wideband and ultra-wideband measurements in both indoor and urban environments [278].

Machine learning further enhances propagation modeling and signal interpretation. For instance, automotive radar systems can estimate angle-of-arrival using neural networks trained on large synthetic datasets—yielding accuracy comparable to classical methods like MUSIC, while being better suited to real-time deployment under noise and clutter.

The complexity of ISAC systems also requires models that move beyond basic parameters like path loss or SNR. Reflectivity patterns and micro-Doppler effects provide critical information for classification tasks in distributed sensing networks. Measurements of these signatures from road users such as cyclists and pedestrians are increasingly being used to inform the design of integrated communication and sensing platforms.

In sum, modern channel and waveform models are converging toward a holistic view that integrates empirical measurement, advanced signal processing, and learning-based inference. This evolution is shaping the foundation of scalable and resilient ISAC technologies needed for next-generation ITS.

### 7.5.3 Example

The authors in [279] investigated a resource allocation framework that enables a RSU to perform sensing and provide vehicles with a comprehensive view of the intersection through sidelink transmission. In the addressed scenarios, communication and sensing tasks share the same sidelink radio resource in a time-frequency-multiplexed manner. Two different communication-sensing joint resource allocation strategies have been proposed and their performance difference has been analyzed via system-level simulation: so-called "Shared Resource Pool (Shared)" where communication and sensing share the same time-frequency resource pool, and "Sensing-based Resource Division (SBRD)", where the RSU dynamically configures and partitions the resource pool between sensing and communication tasks. Fig. 7.16 compares the performance (in average probability of detection) between the two resource allocation strategies and their dependence on the number of vehicles (NV).

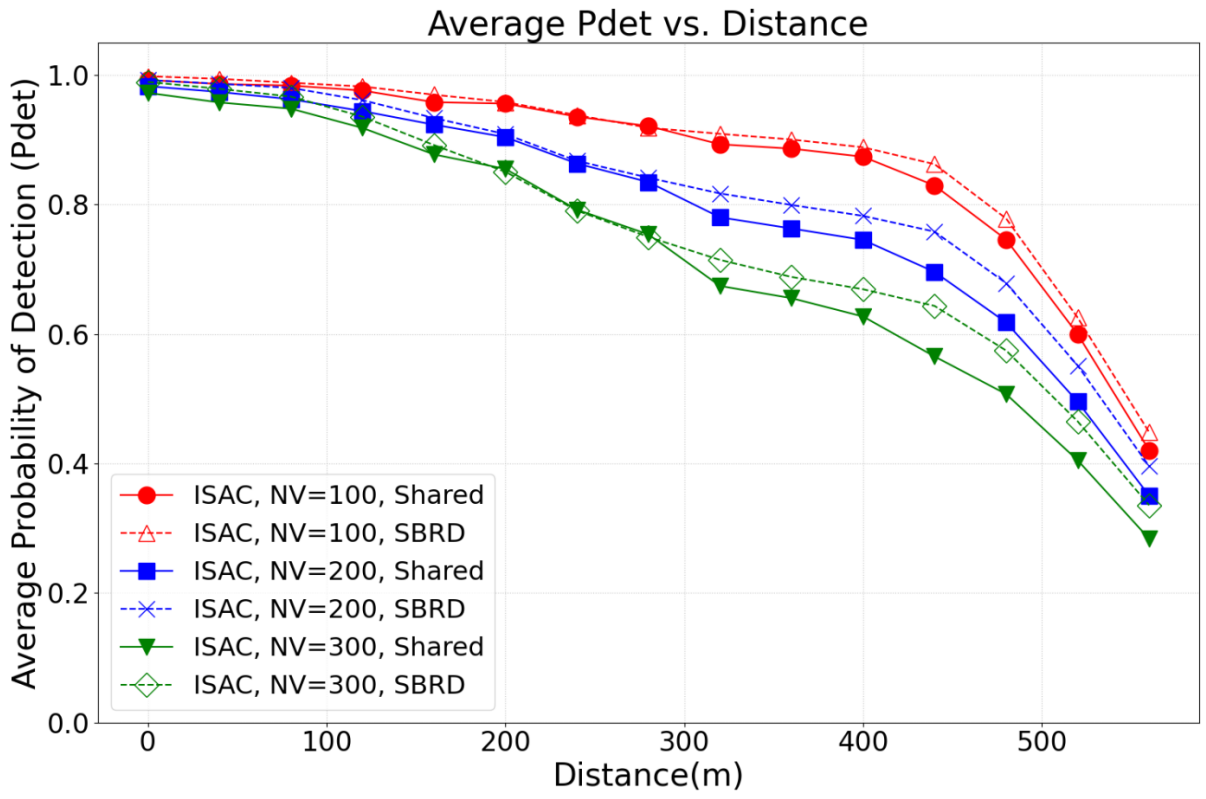

Figure 7.16: performance comparison between two communication-sensing joint resource allocation strategies [279].

### 7.5.4 Future Developments and Prospects

Ongoing developments suggest that ISAC could evolve in several important directions in ITS, with a focus on improved integration with emerging technologies, closer alignment with existing sensor systems, as well as addressing critical aspects of security, reliability, and standardization.

The advancement of 5G-Advanced and the development of 6G networks are expected to create new opportunities for ISAC within ITS. These communication systems aim to provide highly reliable, low-latency connectivity alongside high data rates and dense coverage, which may support more precise and continuous joint sensing and communication functions. Combined with such infrastructure, ISAC could enable more responsive interactions between vehicles and road infrastructure, particularly in complex urban and highway scenarios. At the same time, the use of artificial intelligence in ISAC systems could allow for more adaptive and predictive functionalities, for example, by identifying unusual traffic patterns or anticipating hazardous situations in real time.

ISAC is unlikely to replace established sensing systems such as LiDAR, radar, or cameras, but rather to enhance them as part of a layered sensing approach. This will be essential to maintain robustness under varying environmental conditions and across different use-cases. Moreover, the combination of ISAC with broader ITS components, including the IoT, cloud services, and edge computing, offers the potential to create a more distributed and efficient transport management system. ISAC technologies are also beginning to exploit more complex aspects of radio propagation, such as using multipath reflections for device-free localization [262], [277], which may help detect pedestrians or obstacles without requiring them to carry active devices. Further developments include the use of bistatic micro-Doppler and reflectivity measurements for distributed ISAC, as well as methods to estimate bistatic channels through monostatic observations, indicating the breadth of techniques currently being explored.

As these technologies move closer to deployment in safety-critical ITS applications, concerns around security and reliability become more pressing. Since ISAC systems combine sensing and communication functions over the same radio interface, they are exposed to threats such as spoofing and jamming [280], which could compromise both the perception and connectivity layers simultaneously.

Addressing these risks will require not only technical safeguards at the signal processing level but also system-level strategies and monitoring mechanisms capable of detecting and responding to anomalies promptly.

Alongside technical innovation, standardization plays a key role in enabling ISAC to scale within ITS infrastructures. Organizations like IEEE are currently developing standards such as IEEE 802.11bf, which aims to support WLAN-based sensing in both bi-static and multi-static modes.

Similarly, 3GPP has initiated efforts to define ISAC use cases and requirements as part of the 5G-Advanced evolution, with early features expected in Release 20. These activities will likely lay the groundwork for more integrated ISAC capabilities in 6G networks, where sensing and communication are expected to be designed from the outset as part of the same system.

Finally, ISAC offers the potential to make ITS systems more resource-efficient by reducing the duplication of infrastructure and making more effective use of existing network components and spectrum. That integration aligns well with broader goals of sustainable and environmentally conscious transport systems, while also offering practical benefits in terms of deployment and operational efficiency.

## 7.6 Integration with Open RAN

By **Francesco Linsalata**

The proliferation of CAVs is driving the evolution of vehicular communications, with V2X technologies poised to become a cornerstone of 6G networks. These systems promise to enable cooperative sensing, low-latency safety services, and autonomous driving capabilities. However, the highly dynamic nature of vehicular environments, combined with stringent performance requirements, presents significant challenges in terms of network adaptability, latency, and reliability.

Traditional RAN architectures, characterized by rigid and vertically integrated components, often fall short in addressing the flexibility and responsiveness required by V2X systems. To overcome these limitations, the Open Radio Access Network (O-RAN) paradigm has been introduced as a next-generation RAN architecture. O-RAN disaggregates hardware and software, promotes open interfaces, and integrates softwarized control loops through intelligent, modular components known as RAN Intelligent Controllers (RICs). This architectural transformation enables centralized and data-driven orchestration across heterogeneous network deployments [281], [282].

In this section, we explore the convergence of O-RAN and V2X communications, focusing on how the programmability and intelligence offered by O-RAN can enhance vehicular connectivity. In particular, we investigate the role of RICs in enabling fine-grained control of vehicular links, resource slicing, and direct V2V communication. We also highlight the architectural innovation of *mobi-O RAN*, a framework that extends O-RAN's capabilities to directly manage CAV communication stacks and facilitate hybrid network topologies [281], [282].

### O-RAN Architecture Overview

At its core, the O-RAN architecture separates control and data planes and introduces a layered control framework comprising:

- **Non-Real-Time RIC (Non-Real-Time (RT) RIC)**: operating on time scales of seconds to minutes, responsible for long-term policy optimization and training of machine learning models (via *rApps*).
- **Near-Real-Time RIC (Near-RT RIC)**: operating on 10 ms to 1 s scales, executing real-time optimization and orchestration tasks through modular control applications (*xApps*).
- **dApps**: distributed applications running on edge nodes with real-time capabilities, still under standardization.

These components interact with network infrastructure using standardized interfaces:

- E2 (for data/control exchange),
- O1/O2 (for management and orchestration),
- Open APIs for xApp and rApp integration.

Given the distinctive characteristics of vehicular networks, extensions to the baseline O-RAN architecture are required for practical deployment. The proposed mobi-ORAN framework addresses these needs [281], [282]. Figure 7.17 illustrates an implementation using 5G as the radio access technology, featuring a typical O-RAN setup with both the near-RT RIC and non-RT RIC deployed as software modules within the edge cloud controller. A DT may also be hosted in the same edge cloud controller, ensuring seamless data exchange with the RICs. The O-RAN applications running on both RICs communicate with the network infrastructure via standardized interfaces – E2 for the near-RT RIC and O1/O2 for the non-RT RIC.

Designing an O-RAN-enabled V2X system, namely mobi-ORAN, requires all V2X entities to be reachable by the RICs through these interfaces. Since BSs are typically equipped with these terminations (as shown in Figure 7.17), no architectural modifications are necessary. However, mobi-ORAN introduces extensions to these interfaces to accommodate the specific data collection and control needs of vehicular scenarios.

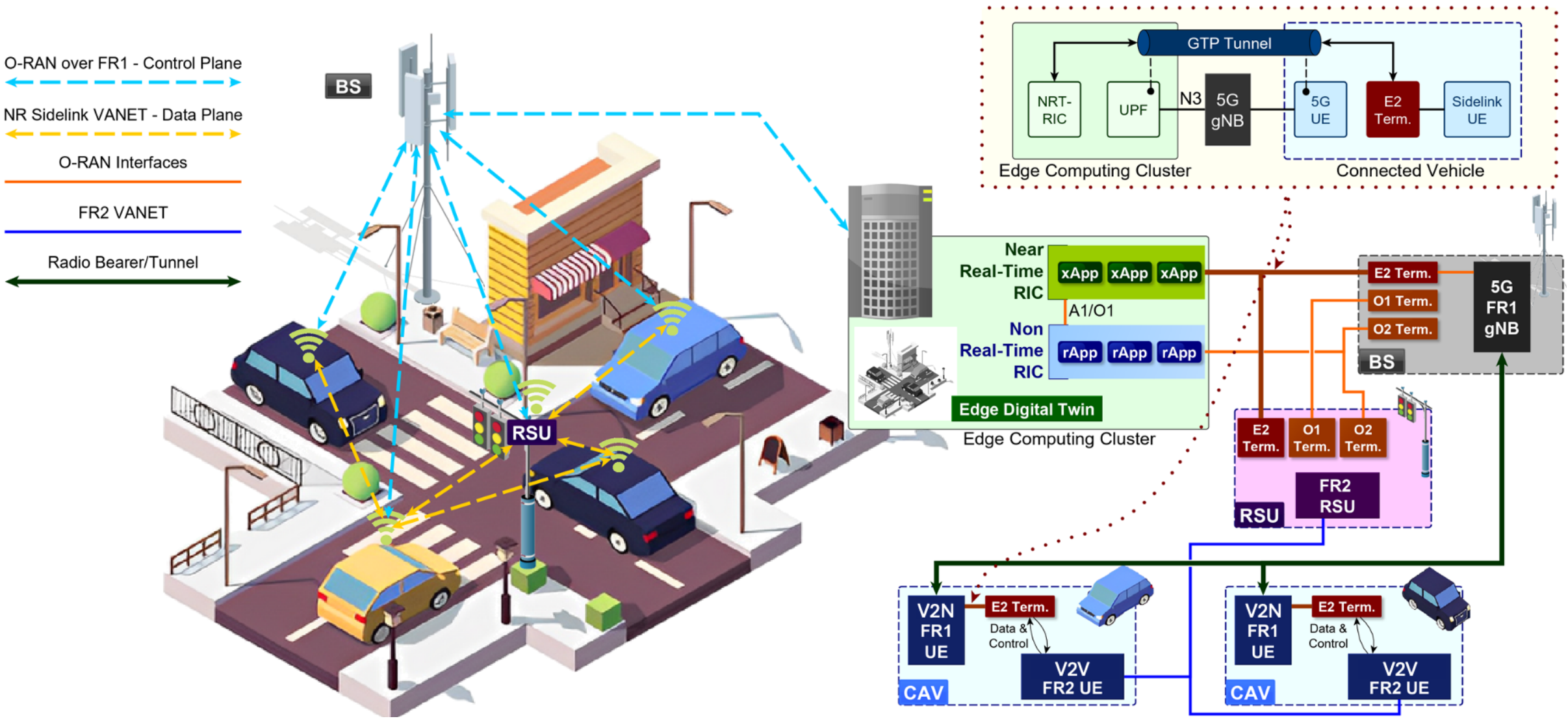

Figure 7.17: Details of the next-generation mobi-ORAN architecture [98].

RSUs play a key role in V2X systems by providing V2I connectivity and enhancing V2V communications. Although O-RAN currently does not specify RSUs, integrating O1, O2, and E2 terminations into them would be straightforward. Communication between RSUs and RICs in the edge cloud controller could reuse existing control-plane links, allowing RSUs to participate in data collection and control through mobi-ORAN interface extensions.

Direct communication between RICs and CAVs, however, poses greater challenges. Still, enabling centralized control of the NR Sidelink stack in CAVs would unlock the ability to orchestrate V2V communications – one of the key challenges in V2X networking [281], [282]. By leveraging this architecture, O-RAN microservices can dynamically adapt to the vehicular context, optimize beam management, predict congestion, and allocate resources across network slices, ensuring low-latency, high-reliability communication [281], [282].

### 7.6.1 Beam Selection and Management

mmWave communication enables high-capacity vehicular links but requires precise beam alignment. Traditional beam management strategies struggle to cope with fast-moving vehicles due to the high overhead of beam training. O-RAN introduces a Non-RT RIC for network-wide beam policy optimization and a Near-RT RIC for real-time beam adaptation. By integrating data from vehicle sensors, urban maps, and historical beam alignment information, O-RAN facilitates predictive beamforming, reducing alignment latency and improving link reliability.

### 7.6.2 Resource Allocation for V2X Slicing

ITS applications encompass a diverse range of services, from URLLC for safety applications to enhanced mobile broadband for infotainment. Efficient resource allocation is critical to support these heterogeneous requirements. O-RAN-based slicing allows dynamic allocation of spectrum and computing resources based on real-time vehicular traffic patterns. xApps running on Near-RT RIC can monitor network congestion and proactively adjust resource allocation, while rApps in Non-RT RIC optimize long-term resource distribution strategies using machine learning techniques.

### 7.6.3 V2V Connectivity and Network Management

Direct V2V communication mitigates infrastructure dependency and enhances network efficiency. However, link stability in high-mobility scenarios remains a challenge. O-RAN RICs can optimize V2V connectivity by dynamically selecting link candidates based on predicted channel conditions and mobility patterns. Additionally, intelligent relay assignment mechanisms utilizing edge computing can further enhance V2V link reliability.

## 7.7 Long-term vision for ITS in the 6G era

By **Laura Finarelli**

### 7.7.1 Introduction

The evolution from 5G to 6G represents a transformative leap for ITS, enabling a hyper-connected, intelligent, and autonomous mobility ecosystem. 6G technologies such as THz communication and visible light communication, ultra-massive MIMO, ISAC, and AI-driven edge intelligence are poised to redefine how vehicles, infrastructure, and users interact in real time [283], [284]. With these advancements, ITS will support ultra-reliable, low-latency communication, centimeter-level positioning, and pervasive intelligence, laying the groundwork for seamless autonomous driving, smart traffic management, and sustainable transportation solutions.

This section explores how 6G's core capabilities will shape the future of ITS, building on the foundations laid by 5G, while expanding its reach through dynamic spectrum access and resource management strategies. This approach transcends traditional cell-based architectures, facilitating a more virtualized and adaptive communication environment.

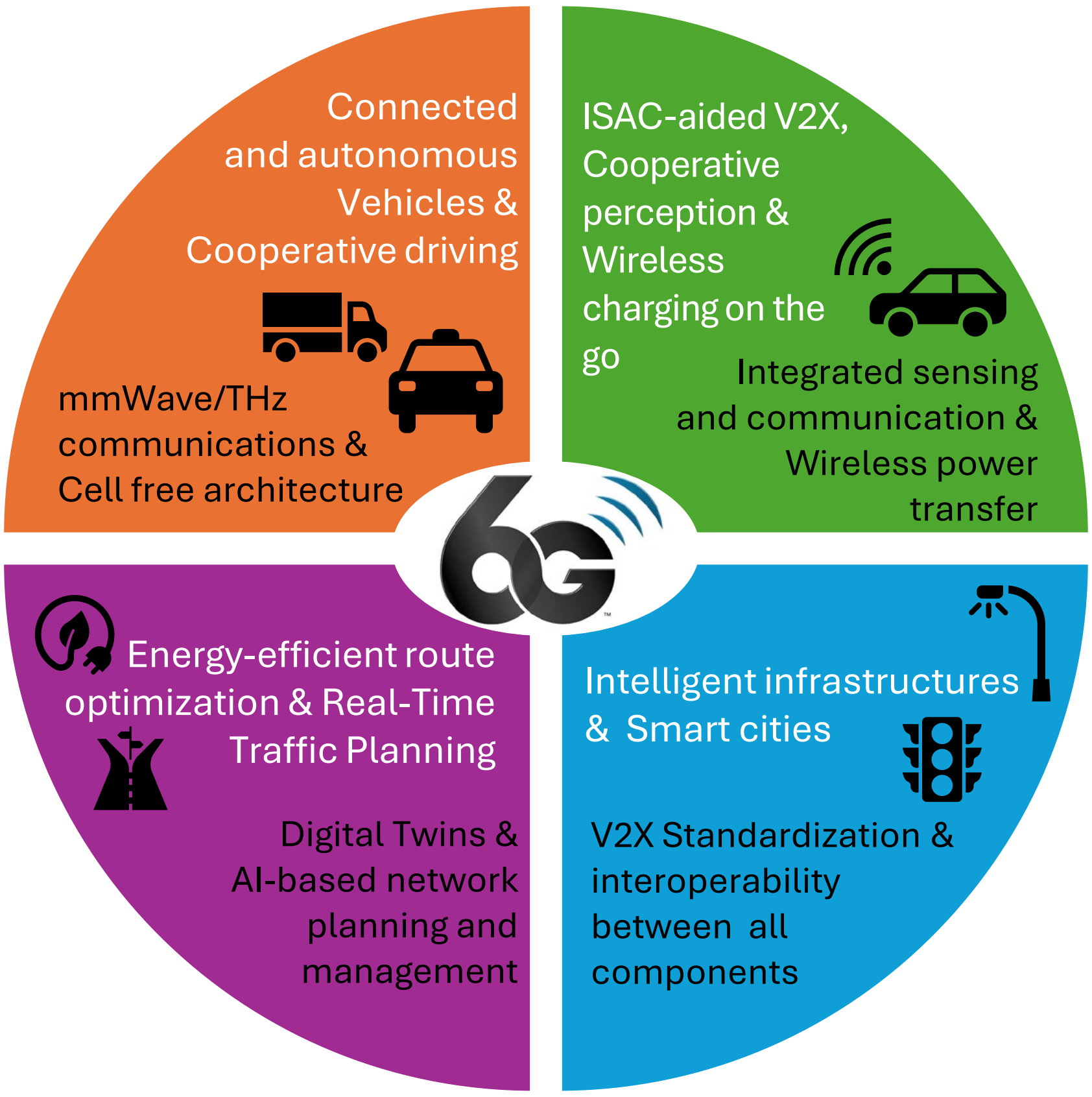

Figure 7.18: Use cases of intelligent transportation systems in the 6G era and their key enablers.

### 7.7.2 CAV and Cooperative Driving

The automotive industry is undergoing significant transformation, particularly in the domain of smart vehicles. The global smart car market is expected to reach USD 265.7 billion by 2032 [285]. Additionally, more than half of all vehicles are expected to fall into the Level 1 – 5 range of autonomy as described by the SAE.

Efficient communication is essential for cooperative driving, where decisions are data-driven, thus with high throughput requirements, and with nearly zero latency. At every instant, each vehicle should be able to make its own decision using information from neighbors and the environment, and then broadcast the output to others.

The key enablers to enhance the communication between and within vehicles are expected to be mmWaves and THz Radio, both lying in the FR2 frequency range. Those will ensure a wider and more efficient spectrum allocation to support connected vehicles, roadside units, reconfigurable intelligent surfaces, and all the intelligent infrastructure contributing to the ITS [286].

However, these advantages do not come for free, and some re-engineering at both the physical layer and system level is required [287]. For instance, given the smaller wavelength, highly localized beamforming is required to achieve an acceptable SINR at the user level. Given the mobility of users and vehicles, beam management protocols for the terahertz MAC layer must carefully balance the improvement of the link budget with the need for awareness of both spectrum usage and infrastructure availability.

Another crucial enabler for facilitating URLLC in ITS will be the cell-free paradigm, where all resources are virtualized and APs collaborate instead of competing for those resources [207], [288]. A CPU coordinates APs cooperation by managing the sharing of physical-layer signals for downlink transmission, forwarding uplink data signals for decoding, and exchanging CSI to optimize signal processing. More in general, V2X Communication using FR2 frequency range, together with cell-free architectures, should allow URLLC and optimized resource scheduling and sharing.

### 7.7.3 ISAC-aided V2X, Cooperative Perception and Wireless Charging on the Go

CAVs depend on data gathered through multiple sensors, which creates a need for extremely high data transmission rates often reaching gigabit-per-second levels. For instance, it is estimated that level 5 AV (as described by SAE) would generate from $1-10\,\text{Gbit/s}$ of data collected via cameras, radars, Lidars that also need to be aggregated and elaborated. To support this demand, mmWave communications are poised to play a key role, by enabling fast (latency $\leq 1\,\text{ms}$) and reliable links for V2V, V2I, and internal vehicle communications. To allow vehicles to not only communicate, but also efficiently exchange data, a dedicated integration between communication and sensing links should be aimed at.

ISAC synergistically combines communication and sensing functionalities into a cohesive system, effectively addressing the demands of both data transmission and sharing, and environmental awareness, which are crucial for cooperative perception. Moreover, the potential of exploiting the same channel also for wireless power transfer should be investigated. This is already the case of Simultaneous Wireless Information and Power Transfer (SWIPT) networks [289] where the resources are shared to satisfy both requests. Developing an efficient model for this channel is imperative and can significantly enhance the battery life of electric vehicles. Peer-to-peer or infrastructure-to-vehicle energy sharing has the potential to become a significant and viable component of future green ITS.

### 7.7.4 Energy-efficient Route Optimization and Real-time Traffic Planning

A better handling of city traffic and congestion requires coordination between multiple agents and fusion of different data sources. All the information should be processed and predictions are necessary to decrease collisions and hazards towards a safer environment for all road users. The improvements in the ITS field are also expected to boost users' quality of life. Broadly speaking, ITS aims to enhance road safety for both VRUs and car users. Safety being ensured, it is important to consider the sustainability and greenness of ITS. Energy efficiency is expected to be one of the main Key Performance Indicator (KPI) of incoming 6G networks [290]. Traffic optimization and CAV should aim to enhance sustainability by reducing carbon emissions, improving vehicle punctuality, and creating more efficient travel routes [291].

To achieve these objectives, the implementation of digital twins and, more broadly, the AI paradigm proposed by 6G is recommended. Digital twins will facilitate the seamless integration of physical and digital layers, enabling hyper-realistic simulations and rapid AI-assisted decision making.

Creating a digital version of traffic scenarios and paths will enable offline simulations that can enhance real-life environments. This approach will also facilitate the collection of synthetic data for training AI models to predict hazards and determine optimal routes based on current or forecasted traffic conditions.

### 7.7.5 Intelligent Infrastructures and Smart Cities

ITSs are already utilizing roadside units and sensors installed along the roads to gather and communicate information. In a smart city, where streetlights and traffic signals are also interconnected, every component can contribute valuable data and serve as a communication relay. Not only will smart infrastructures be capable of making autonomous, data-driven decisions using AI, but they will also participate in the communication and sensing ecosystem, enhancing the use cases and applications previously proposed [292].

To support this diverse array of infrastructures, standardization is essential to ensure cooperation, interoperability, and reliability among all components. Future 6G network architectures should aim to integrate a unified Core Network (CN) that can connect multiple Radio Access Technologies (RATs) simultaneously, allowing for seamless switching [286]. Moreover, the short-range limitation of the FR2 frequency range will no longer be an issue in an environment where everything is connected. While each use-case is examined individually, they are fundamentally interconnected. Progress in one area can facilitate advancements in others, leading to a synergistic enhancement of overall outcomes.

Despite significant advancements in V2X communications, critical gaps remain. It is increasingly evident that no single communication technology can fully ensure the safety, dependability, and reliability required for ITS. The path to a truly **smart transportation system** requires a multi-network approach that integrates the strengths of various technologies, including 5G-V2X, mmWave, IEEE 802.11p, and its successor, IEEE 802.11bd, which is still undergoing standardization. While multi-technology communication is crucial for dependable cooperative driving, economic stakeholders continue to compete for dominance, each striving to establish their technology as the industry standard. This competition risks delaying progress and missing a critical opportunity to advance ITS as a whole. Adherence to a common communication standard would enable compatibility, confidence, and trustability of the overall ITS.

### 7.7.6 Emerging Business Opportunities in 6G-Enabled ITS

As is often the case with the advent of a disruptive technological paradigm, 6G is expected to attract new stakeholders while opening up fresh business avenues for both established industries and emerging players [290].

From a purely vehicular point of view, 6G will enable large-scale Autonomous Mobility-as-a-Service (MaaS), supporting robo-taxis (as it is already the case in China), self-driving deliveries, and ride-sharing models reminiscent of Uber. This emerging market calls for in-depth studies on the economic trade-offs involved and the development of appropriate regulatory policies to ensure fair pricing, grid stability, and secure transactions.

On the user side, centimeter-level localization and THz communication will enhance augmented reality services for navigation, tourism, and entertainment [293]. Furthermore, dynamic insurance models could adapt to real-time user behavior and context to ensure fair and personalized policies.

# Chapter 8: Conclusion

By **Adrian Kliks, Shrief Rizkalla**

The white paper consolidates current knowledge and future perspectives on communication technologies enabling next-generation mobility systems. The findings presented in the document highlight the need for multidisciplinary innovation, cross-domain integration, and standardization to achieve safe, efficient, and intelligent transport infrastructures. Based on the analyses and discussions contained in the report, the following key conclusions can be drawn:

- First, reliable, low-latency, and high-capacity communication links are the cornerstone of future ITS, enabling automation, cooperative control, and safety across all transport modes.
- Second, accurate channel modeling and large-scale measurement campaigns are fundamental to further investigating communication systems, especially in dynamic propagation conditions, such as high-mobility and non-stationary scenarios like high-speed trains and UAV networks.
- T2G communications are evolving from GSM-R-based systems to intelligent, multi-bearer all-IP networks, with emerging technologies such as FRMCS, 5G/6G, satellite links, and AI-driven networking paving the way for safer, faster, and more efficient rail transport.
- UAV-assisted communication and aerial base stations offer a promising solution to extend coverage, ensure service continuity, and support emergency or temporary network deployments in ITS.
- A2A channel models form a key component of the ITS framework, enabling reliable real-time communication in dynamic flight conditions through advanced non-stationary 3D geometric-stochastic modeling applicable to all flight phases in a low level flight.
- V2X communication plays a crucial role in enable ADAS, where it can also prevent accidents and improve detection, localization and tracking of both vehicles and VRUs.
- In that context, the practical validation of proposed models and technologies in near-real operational environments significantly enhances their implementation readiness and accelerates technology adoption.
- Next, emerging technologies, such as cell-free massive MIMO, reconfigurable intelligent surfaces, and integrated sensing and communication, represent transformative enablers for reliable, adaptive, and intelligent transportation communication networks.
- Experimental studies confirm the advantages of cell-free networks compared to co-localized antennas for V2I communication, in terms of higher and spatially more uniform SNR.
- Effective spectrum management and standardization across IEEE, 3GPP, ISO, and ETSI are prerequisites for harmonized ITS operation and global interoperability of communication technologies.
- The integration of AI-driven control, edge computing, and digital twin models can facilitate predictive, self-optimizing, and resilient network management for real-time ITS applications.
- Finally, the long-term vision for ITS within the 6G era envisions seamless, cooperative, and sustainable mobility ecosystems, where communication, sensing, and computation are tightly coupled to support autonomous, efficient, and safe transportation systems.

# List of Acronyms

**2D** Two-dimensional

**3D** Three-dimensional

**3GPP** 3rd Generation Mobile Group

**4G** Fourth Generation

**5G** Fifth Generation

**5G NR** 5G New Radio

**5G-R** 5G-Railway

**5GAA** 5G Automotive Association

**6G** Sixth Generation

**6DoF** Six Degrees of Freedom

**A2G** Air-to-Ground

**A2A** Aircraft-to-Aircraft

**AAM** Advanced Air Mobility

**AB** Alpha-Beta

**ABS** Aerial Base Station

**ACMA** Australian Communications and Media Authority

**ADAS** Advanced Driving Assistance Systems

**ADS** Autonomous Driving Systems

**AGC** Automatic Gain Control

**AI** Artificial Intelligence

**AP** Access Point

**AR** Augmented Reality

**ARIB** Association of Radio Industries and Businesses

**ASA** Azimuth Spread of Arrival

**ASD** Azimuth Spread of Departure

**ATO** Automatic Train Operation

**AWGN** Additive White Gaussian Noise

**BB** BroadBand

**BS** Base Station

**BLE** Bluetooth Low Energy

**BH** BackHauling

**BSA** Basic Set of Applications

**BSM** Basic Safety Message

**BTP** Basic Transport Protocol

**CAPEX** Capital Expenditure

**C-RAN** Cloud RAN

**C-MPC** Clutter MPC

**C-ITS** Cooperative ITS

**C-V2X** Cellular V2X

**CA** Cooperative Awareness

**CAM** Cooperative Awareness Message

**CAV** Connected and Automated Vehicle

**CBTC** Communication based Train Control

**CCDF** Complementary Cumulative Distribution Functions

**CCAM** Connected Cooperative and Automated Mobility

**CCTV** Closed-Circuit Television

**CDF** Cumulative Distribution Function

**CDB** Context Database

**CEPT** Conference of Postal and Telecommunications Administrations

**CN** Core Network

**CNN** Convolutional Neural Network

**CPU** Central Processing Unit

**CSI** Channel State Information

**CIR** Channel Impulse Response

**CRLB** Cramér-Rao Lower Bound

**DNN** Deep Neural Network

**DeepNDN** Deep NDN

**D2D** Device-to-Device

**DLR** German Aerospace Center

**DM** Data Message

**DRU** Digital Receiving Unit

**DSRC** Dedicated Short Range Communication

**DoT** Department of Telecommunications

**DTN** Delay Tolerant Networking

**DTT** Digital Terrestrial Television

**DT** Digital Twin

**DS** Delay Spread

**DSM** Dynamic Spectrum Management

**eVTOL** Electric Vertical Take-Off and Landing

**ECC** Electronic Communications Committee

**EADF** Effective Aperture Distribution Function

**ERTMS** European Rail Traffic Management System

**ETSI** European Telecommunications Standards Institute

**ETC** Electronic Toll Collection

**EIS** Edge Intelligence System

**FC** Floating Content

**FCC** Federal Communications Commission

**FER** Frame Error Rate

**FRMCS** Future Railway Mobile Communication System

**FSPL** Free Space Path Loss

**FMCW** Frequency Modulated Continuous Wave

**FPA** Fractal Patch Antenna

**G2A** Ground-to-Air

**GSCM** Geometry-based Stochastic Channel Model

**GN** Geo-Networking

**GoA4** Grade of Automation 4

**GSM-R** Global System for Mobile Communications – Railway

**GPS** Global Positioning System

**GNSS** Global Navigation Satellite System

**IAB** Integrated Access and Backhaul

**ICN** Information Centric Networking

**ICT** Information and Communication Technology

**IEEE** Institute of Electrical and Electronics Engineers

**IMDA** Infocomm Media Development Authority

**ISAC** Integrated Sensing and Communications

**IoT** Internet of Things

**ITS** Intelligent Transportation System

**ITS-G5** ITS – 5 GHz

**ITS-SU** ITS Station Unit

**ITU** International Telecommunications Unit

**ISED** Innovation, Science and Economic Development

**ISO** International Organization for Standardization

**IM** Interest Message

**I2V** Infrastructure-to-Vehicle

**KF** K-Factor

**KPI** Key Performance Indicator

**LAN** Local Area Network

**LDPC** Low-Density Parity Check

**LSP** Large Scale Parameter

**LwF** Learning without Forgetting

**LoS** Line-of-Sight

**LNA** Low-Noise Amplifier

**LO** Local Oscillator

**LTE** Long Term Evolution

**MaaS** Mobility-as-a-Service

**M2M** Mobile-to-Mobile

**MAC** Media Access Control

**MCS** Modulation and Coding Scheme

**M2M** Mobile-to-Mobile

**MIMO** Multiple-Input Multiple-Output

**MIIT** Ministry of Industry and Information Technology

**ML** Machine Learning

**MF** Matched Filtering

**MBS** Mobile Base Station

**MN** Moving Network

**mmWave** millimeter wave

**MSIT** Ministry of Science and ICT

**MPC** MultiPath Component

**NASA** National Aeronautics and Space Administration

**NDN** Named Data Networking

**NLoS** Non Line-of-Sight

**NS-3** Network Simulator 3

**NR** New Radio

**OBU** On-Board Unit

**OEM** Original Equipment Manufacturer

**OER** Octet Encoding Rule

**OCDM** Orthogonal Chirp Division Multiplexing

**OFDM** Orthogonal Frequency Division Multiplexing

**OPEX** Operational Expenditure

**OOS** Operator-Oriented Services

**OTFS** Orthogonal Time Frequency Space

**OLoS** Obstructed LoS

**O-RAN** Open Radio Access Network

**PA** Power Amplifier

**PhC** Photonic Crystal

**PER** Packet Error Rate

**PL** Path Loss

**PIT** Pending Interest Table

**PDCP** Packet Data Convergence Protocol

**PDP** Power Delay Profile

**PHY** Physical layer

**POS** Passenger Oriented Services

**PTS** Portable Transmitting Station

**QoS** Quality of Service

**RATs** Radio Access Technologies

**RAN** Radio Access Network

**RIC** RAN Intelligent Controller

**RoF** RF-over-Fiber

**RedCap** Reduced Capability

**RFN** Reference Frequency Network

**RFT** Radio Frequency Tuner

**RIS** Reconfigurable Intelligent Surfaces

**RLC** Radio Link Control

**RSS** Receiver Signal Strength

**RSSI** Receiver Signal Strength Indicator

**RSU** Road-Side Unit

**RX** Receiver

**RMS** Root Mean Square

**RMSE** Root Mean Square Error

**RMa** Rural Macro

**RT** Real-Time

**S2S** Ship-to-Ship

**SAR** Synthetic Aperture Radar

**SAE** Society of Automotive Engineers

**SBS** Static Base Station

**SBRD** Sensing-based Resource Division

**SDAP** Service Data Adaptation Protocol

**SDO** Standard Developing Organization

**SDFW** Self Driving Freight Wagons

**S-MPC** Sensing MPC

**SF** Shadow Fading

**SINR** Signal to Interference plus Noise Ratio

**SIR** Signal to interference ratio

**SL** Side Link

**SLAM** Simultaneous Localization/Mapping

**SNR** Signal-to-Noise Ratio

**SOS** Safety-Oriented Services

**SOP** Secrecy Outage Probability

**SR** Specular Reflection

**SWaP** Size, Weight, and Power

**SWIPT** Simultaneous Wireless Information and Power Transfer

**SRD** Short-Range Device

**TDD** Time Division Duplex

**TDL** Tapped-Delay Line

**T2G** Train-to-Ground

**T2T** Train-to-Train

**TBS** Transmission-Based Signaling

**TCMS** Train Control and Management System

**TDBP** Time-Domain Back Projection

**TX** Transmitter

**TVWS** TV White Spaces

**UAS** Unmanned Aerial Systems

**UAV** Unmanned Aerial Vehicle

**U2U** UAV-to-UAV

**UE** User Equipment

**URLLC** Ultra-Reliable Low Latency Communication

**UIC** Union Internationale des Chemins

**UMa** Urban Macro

**UMi** Urban Micro

**UN** United Nations

**UWB** Ultra WideBand

**V2I** Vehicle-to-Infrastructure

**V2N** Vehicle-to-Network

**V2P** Vehicle-to-Pedestrian

**V2V** Vehicle-to-Vehicle

**V2VRU** Vehicle-to-Vulnerable Road User

**V2X** Vehicle-to-Everything

**ViL** Vehicle-in-the-Loop

**VCTS** Virtually Coupled Train Sets

**VST** Vector Signal Transceiver

**VRU** Vulnerable Road User

**VDSA** Vehicular Dynamic Spectrum Access

**WAVE** Wireless Access in Vehicular Environment

**WG** Working Group

**WSS** Wide-Sense Stationary

**WSSUS** Wide-Sense Stationary Uncorrelated Scattering

**WPS** Weighted Processor Sharing

**WRC** World Radiocommunication Conference

**WSMP** WAVE Short Message Protocol

**WTO** World Trade Organization

**XR** Extended Reality

**ZF** Zero Forcing